\documentstyle[12pt,epsf]{report}
\begin{document}
\baselineskip 0.6cm
\abovedisplayskip 0.5cm
\belowdisplayskip 0.5cm

%
%

\newcommand{\ie}{{\it i.e. }}
\newcommand{\beq}{ \begin{eqnarray} }
\newcommand{\eeq}{ \end{eqnarray} }
\newcommand{\beqstar}{ \begin{eqnarray*} }
\newcommand{\eeqstar}{ \end{eqnarray*} }
\newcommand{\gsim}{ \mathop{}_{\textstyle \sim}^{\textstyle >} }
\newcommand{\lsim}{ \mathop{}_{\textstyle \sim}^{\textstyle <} }
\newcommand{\vev}[1]{ \left\langle {#1} \right\rangle }
\newcommand{\lsp}{ \left ( }
\newcommand{\rsp}{ \right ) }
\newcommand{\lmp}{ \left \{ }
\newcommand{\rmp}{ \right \} }
\newcommand{\llp}{ \left [ }
\newcommand{\rlp}{ \right ] }
\newcommand{\labs}{ \left | }
\newcommand{\rabs}{ \right | }
\newcommand{\pslash}{{\not{p}}}
\newcommand{\qslash}{{\not{q}}}
\newcommand{\dslash}{{\not{\partial}}}
\newcommand{\K}{ {\rm K} }
\newcommand{\EV}{ {\rm eV} }
\newcommand{\KEV}{ {\rm keV} }
\newcommand{\MEV}{ {\rm MeV} }
\newcommand{\GEV}{ {\rm GeV} }
\newcommand{\TEV}{ {\rm TeV} }
\newcommand{\mgra}{ m_{3/2} }
\newcommand{\epho}{ \epsilon_{\gamma} }
\newcommand{\ebg}{ \bar{\epsilon}_{\gamma} }
\newcommand{\eele}{ E_{e} }
\newcommand{\enu}{ E_{\nu} }
\newcommand{\enubg}{ \bar{E}_{\nu} }
\newcommand{\bQ}{ \bar{Q} }
\newcommand{\btheta}{ \bar{\theta} }
\newcommand{\dalpha}{ \dot{\alpha} }
\newcommand{\dbeta}{ \dot{\beta} }
\newcommand{\dgamma}{ \dot{\gamma} }
\newcommand{\HII}{${\rm H_{II}}$}
\newcommand{\ff}[1]{ \labs f^{abc} \rabs^{2} }
\renewcommand{\tt}[1]{ \labs T^{a}_{ji} \rabs^{2} }
\newcommand{\eps}{\delta }
\newcommand{\gb}[1]{ A^{#1} }
\newcommand{\gf}[1]{ \lambda^{(#1)} }
\newcommand{\cb}[1]{ \phi_{#1} }
\newcommand{\cf}[1]{ \chi_{#1} }

%
%

\begin{titlepage}
\begin{flushright}
TU-479\\
March, 1995
\end{flushright}
\vskip 4cm
\begin{center}
{\Large Ph.D thesis}
\vskip 0.5cm
{\Large \bf Effects of the Gravitino on the Inflationary Universe}
\vskip 4cm
{\large Takeo~Moroi}
\vskip 0.5cm
{\it Department of Physics, Tohoku University, Sendai 980-77, Japan}
\end{center}
\end{titlepage}

%
%

\pagenumbering{roman}
\setcounter{page}{1}
\tableofcontents
\newpage
\pagenumbering{arabic}
\setcounter{page}{1}

%
%

\chapter{Introduction}
\label{chap:intro}

\section{Overview}

\hspace*{\parindent}
When we think of new physics beyond the standard model, supersymmetric
(SUSY) extension~\cite{NPB70-39} of the standard model is one of the
most attractive candidates.  Cancellation of quadratic divergences in
SUSY models naturally explains the stability of the electroweak scale
against radiative corrections~\cite{Maiani,Veltman}.  Furthermore, if we
assume the particle contents of the minimal SUSY standard model (MSSM),
the three gauge coupling constants in the standard model meet at $\sim
10^{16}$GeV~\cite{PRD44-817,PLB260-447}, which strongly supports
grand unified theory (GUT) based on SUSY~\cite{NPB193-150,ZPC11-153}.

In spite of these strong motivations, no direct evidence of SUSY
(especially superpartners) has been discovered yet. Therefore, the SUSY
is broken in nature, if it exists. Although many efforts have been made
to understand the origin of the SUSY breaking, no convincing scenario of
SUSY breaking has found yet. Nowadays, many people expect the existence
of {\it local} SUSY (\ie supergravity)~\cite{NPB212-413} and try to find
a mechanism to break SUSY spontaneously in the framework. In the broken
phase of the supergravity, super-Higgs effect occurs and gravitino,
which is the superpartner of graviton, acquires mass by absorbing the
Nambu-Goldstone fermion associated with SUSY breaking. In this case,
the gravitino mass $\mgra$ is expected to give us some
informations about the SUSY breaking mechanism.  For example, in models
with the minimal kinetic term, the following (tree level) super-trace
formula among the mass matrices ${\bf M}_J^2$'s holds;
\begin{eqnarray}
{\rm Str}{\bf M}^{2} \equiv \sum_{{\rm spin}~J} (-1)^{2J} (2J+1)
{\rm tr}{\bf M}^{2}_{J} \simeq
2(n_\phi-1) \mgra^{2},
\end{eqnarray}
where $n_\phi$ is the number of the chiral multiplets in the
spontaneously broken local SUSY model. In this case, all the SUSY
breaking masses of squarks and sleptons are equal to the gravitino mass
at the Planck scale. Meanwhile in models with ``no-scale like'' K\"ahler
potential~\cite{PLB143-410,PLB147-99}, SUSY breaking masses for
sfermions vanish at the gravitational scale and are induced by radiative
corrections, and hence the gravitino mass is not directly related to the
scale of the SUSY breaking in the observable sector (which contains
ordinary particles in the standard model and their superpartners).  In
order to understand the physics of the SUSY breaking, it is significant
to clarify the property of the gravitino.  But contrary to our
theoretical interests, we have no hope to see the gravitino in collider
experiments since its interaction is extremely weak.

On the contrary, cosmological arguments provide us some informations
about the gravitino. In general, cosmology severely constrains
properties of exotic particles. Let us review the constraints derived
from cosmology.
\begin{itemize}
\item The first is on the mass density of the exotic particle during
the big-bang nucleosynthesis. If it is too large, it speeds up the
expansion rate of the universe during that epoch and results in too many
$^4$He.
\item  The second is on the entropy production by the decay of the
exotic particle. If the decay of the exotic particle releases a large
amount of entropy, the baryon-to-photon ratio may become much well below
what is observed today.
\item The third arises from the effects of the decay products on the
big-bang nucleosynthesis. If the photon or some charged particle is
produced by the decay of the exotic particle after the big-bang
nucleosynthesis has started, energetic photons induced by the decay
products may destruct light nuclei (D, $^3$He, $^4$He) and destroy the
great success of the big-bang nucleosynthesis.
\item Furthermore, one can obtain the fourth constraints by considering
the cosmic microwave background distortion by the exotic particle with
lifetime larger than $\sim 10^{10}{\rm sec}$.
\item If exotic particle is stable, its present mass density provides
us a fifth constraint.
\end{itemize}
In fact, the most severe constraints on models based on supergravity are
derived from the light element photo-dissociation and the present mass
density of the universe.

Following the above arguments, we can obtain stringent constraints on the
gravitino mass in the standard big-bang cosmology. If the gravitino is
unstable, it may decay after the big-bang nucleosynthesis and releases
tremendous amount of entropy, which may conflict with the big-bang
nucleosynthesis scenario.  As Weinberg first pointed
out~\cite{PRL48-1303}, the gravitino mass should be larger than $\sim
10\TEV$ so that the gravitino can decay before the big-bang
nucleosynthesis starts.  Furthermore in SUSY models with $R$-parity
invariance, unstable gravitino produces heavy stable particle (\ie the
lightest superparticle) in its decay processes, which results in
unacceptably high mass density of the present
universe~\cite{NPB277-556}. In order to reduce the number density of the
lightest superparticle through pair annihilation processes, gravitinos
should decay when the temperature of the universe is higher than (1 --
10)GeV. This requires that the gravitino mass should be larger than
($10^6$ -- $10^7$)GeV, which seems to be disfavored from the naturalness
point of view. In the case of stable gravitino, the gravitino mass
larger than $\sim 1\KEV$ is excluded since the present mass density of
the gravitino exceeds the critical density of the present
universe~\cite{PRL48-223}.  The above constraints on the gravitino mass
seem to be very stringent especially for models with the minimal
K\"ahler potential, since in such models the gravitino mass is expected
to give the scale of the SUSY breaking in observable sector.

In the inflationary universe~\cite{PRD23-347}, however, situation
changes~\cite{PLB118-59}.  In this case, the initial abundance of the
gravitino is diluted during the inflation, and hence the number density
of the gravitino becomes much less than that in the case of the standard
big-bang cosmology. But even in the inflationary universe, the gravitino
may cause the cosmological problems mentioned above since secondary
gravitinos are produced through scattering processes off the background
radiations or decay processes of superparticles.  As we will see later,
number density of the secondary gravitinos is approximately proportional
to the reheating temperature after the inflation and hence the
upperbound on the reheating temperature is derived.

In this thesis, we study details on the gravitino production in early
universe and on its effects in the inflationary universe. Compared with
the previous works, we have made an essential improvement on the
following points.
\begin{itemize}
\item Gravitino production cross sections are calculated by using
full relevant terms in the local SUSY lagrangian.
\item High energy photon spectrum induced by the gravitino decay is
obtained by solving the Boltzmann equations numerically.
\item Time evolutions of the light nuclei (D, $^3$He, $^4$He) with
non-standard energetic photons are calculated by modifying Kawano's
computer code.
\end{itemize}
In our analysis, we assume that the light elements are synthesized
through the (almost) standard scenario of the big-bang nucleosynthesis
(with baryon-to-photon ratio $10^{-9}$ -- $10^{-10}$), and take the
reheating temperature as a free parameter.

\section{Organization of this thesis}

\hspace*{\parindent}
The outline of this thesis is as follows.  The former half of this
thesis is devoted to the review of related topics, especially that of
the gravitino properties. In Chapter~\ref{chap:mssm}, we review the
motivation of SUSY. In Chapter~\ref{chap:sugra}, the gravitino field
which is the gauge field associated with local SUSY transformation is
introduced. Furthermore, lagrangian based on local SUSY is also shown in
Chapter~\ref{chap:sugra} and the super-trace formula in that framework
is derived. Conventions used in Chapter~\ref{chap:sugra} (and in other
chapters) are shown in Appendix~\ref{ap:notation}. In
Chapter~\ref{chap:feynman}, we quantize a massive gravitino field and
derive Feynman rules for gravitino.

In the latter half of this thesis, we study the cosmology with the
gravitino in detail. Overview of phenomenology with the gravitino is
given in Chapter~\ref{chap:overview}. In Chapter~\ref{chap:heavy},
effects of unstable gravitino in the inflationary cosmology are analyzed
in detail.  In deriving constraints, we first derive photon spectrum
induced by the decay of the gravitino. The procedure to obtain the
photon spectrum is given in Appendix~\ref{ap:spectrum}. Then, we
calculate the time evolution of light nuclei with the obtained high
energy photon spectrum, and we derive constraints on the reheating
temperature and on the gravitino mass. In our analysis, we assume the
standard big-bang nucleosynthesis scenario which is reviewed in
Appendix~\ref{ap:bbn}.  The case of stable gravitino is discussed in
Chapter~\ref{chap:light}.  Chapter~\ref{chap:summary} is devoted to
discussions.

%
%

\chapter{Motivations of supersymmetry}
\label{chap:mssm}

\section{Hierarchy problem in the standard model}

\hspace*{\parindent}
For particle physicists, symmetries in nature are significant guiding
principles.  Especially, interactions of elementary particles (like
quarks and leptons) can be understood by using the concept of the local
gauge symmetry. Strong interaction is expected to originate to ${\rm
SU(3)}_{C}$ gauge group, and its theoretical predictions (like three
gluon vertex and asymptotic free nature of its gauge coupling constant)
have been confirmed experimentally. Meanwhile, results of recent
electroweak precision measurements are in good agreements with the
predictions of the spontaneously broken ${\rm SU(2)}_{L}$ $\times $ ${\rm
U(1)}_{Y}$ gauge theory. Accompanied by theoretical and experimental
successes, the standard model, based on the ${\rm SU(3)}_{C}$ $\times$
${\rm SU(2)}_{L}$ $\times $ ${\rm U(1)}_{Y}$ gauge group, is regarded as
the established one which describes particle interactions below the
energy scale $\sim$ 100GeV.

But once we look up high energy scale, one unpleasant problem, which is
called hierarchy problem, appears in the standard model. In the standard
model, existence of the elementary scalar boson, \ie Higgs boson, is
assumed in order to cause a spontaneous breaking of the gauge symmetry
${\rm SU(2)}_{L}$ $\times $ ${\rm U(1)}_{Y}$ $\rightarrow$ ${\rm
U(1)}_{em}$.  This is the origin of the hierarchy problem. As one can
easily see, radiative corrections to the Higgs boson mass squared
$\delta m_{H}^{2}$ are quadratically divergent. Therefore, if one
assumes the existence of the cut-off scale of the standard model
$\Lambda_{CUT}$ at which the parameters in the standard model are set by
a more fundamental theory, $\delta m_{H}^{2}$ $\sim$ $O(\alpha
\Lambda_{CUT}^{2})$ where $\alpha$ represents the coupling factor. The
relation between the bare mass squared $m_{H,B}^{2}$ and the renormalized
one $m_{H,R}^{2}$ is written in the following way;
\beq
m_{H,R}^{2} = m_{H,B}^{2} + \delta m_{H}^{2}.
\label{fine-tune}
\eeq
In order to give the electroweak scale correctly, the renormalized mass
squared $m_{H,R}^{2}$ should be $O((100\GEV)^{2})$. On the other hand,
if we assume a larger value of $\Lambda_{CUT}$, $\delta m_{H}^{2}$
increases quadratically and a fine tuning of $m_{H,B}^{2}$ is needed so
that the renormalized mass squared $m_{H,R}^{2}$ remains
$O((100\GEV)^{2})$.  For example, if we assume the cut-off scale of the
standard model to be at the Planck scale $\sim$ $10^{19}\GEV$, both
$m_{H,B}^{2}$ and $\delta m_{H}^{2}$ are $O((10^{19}\GEV)^{2})$ for
$\alpha$ $\sim$ $O(1)$, and they should be chosen as
\beq
m_{H,B}^{2} + \delta m_{H}^{2} \sim
O((10^{19}\GEV)^{2}) - O((10^{19}\GEV)^{2}) \sim
O((100\GEV)^{2}).
\eeq
This is a terrific fine tuning. Therefore, if one assume that the
cut-off scale of the standard model is much larger than the electroweak
scale, we have to accept an unbelievable fine tuning of Higgs boson
mass. This is the hierarchy problem. In fact, this problem stems from
the fact that there is no symmetry which stabilizes the electroweak
scale~\cite{yanagida@yamagata,hep-ph/9410285}.  In order to solve this
problem, we hope that some new physics (in other words, some new
symmetry) in which quadratic divergences do not exist at all, appears at
a energy scale $O(100\GEV-1\TEV)$ and solve this difficulty.

\section{Supersymmetric extension of the standard model}

\hspace*{\parindent}
One of the most attractive solution to the hierarchy problem is
SUSY~\cite{NPB70-39}. SUSY is a symmetry which transforms bosons into
fermions and vice versa. Therefore, in SUSY models the number of bosonic
degrees of freedom is equal to that of fermionic ones. As we will
see later, quadratic divergence of the Higgs (and other) boson masses
are canceled out between the contributions from boson and fermion loops.

Experimentally, however, we have not found any superpartners of the
observed particles. This fact indicates that SUSY is broken in nature,
if it exists. In order to solve the hierarchy problem, the SUSY must be
broken softly~\cite{NPB159-429} so that quadratic divergences do not
exist at all. Usually, such a softly broken global SUSY model is
regarded as a low energy effective theory of the spontaneously broken
local SUSY model. We will comment on this point in the next chapter and
here, we consider a phenomenologically acceptable (softly broken) SUSY
model.

When we extend the standard model to the supersymmetric one, we usually
add ``superpartners'' for the ordinary particles existed in the standard
model. In Table~\ref{table:MSSM-particles}, we show the particles
in the minimal SUSY standard model (MSSM) and their gauge quantum
numbers.  Along with the existence of the superpartners, one big
difference between the standard model and the SUSY one is the number of
Higgs doublets, \ie the MSSM requires two Higgs doublets (see
Table~\ref{table:MSSM-particles}). In the SUSY standard model, Higgs
bosons are accompanied by their fermionic superpartners which have the
same gauge quantum numbers as the Higgs bosons. In this case, anomaly
cancellation is not guaranteed if both of $H_{1}$ and $H_{2}$ are not
included.  Furthermore, in order to give fermion masses to up-type
quarks as well as down-type quarks and leptons from Yukawa couplings of
Higgs bosons, at least two chiral superfields $H_{1}$ and $H_{2}$ are
needed. Mainly from the above two reasons, two Higgs doublets with
representation $({\bf 1}, {\bf 2}, -1/2)$ and $({\bf 1}, {\bf 2}, 1/2)$
are introduced into the MSSM.
%
%
\begin{table}[t]
\begin{center}
\begin{tabular}{l|cc}
\multicolumn{3}{l}
{\bf Gauge sector} \\ \hline \hline
{Representation} & {Boson~$(R=+1)$} & {Fermion~$(R=-1)$}
\\ \hline
{$({\bf 8},{\bf 1},0)$} &
{$G_{\mu}$} & {$\tilde{g}$}
\\
{$({\bf 1},{\bf 3},0)$} &
{$W_{\mu}$} & {$\tilde{w}$}
\\
{$({\bf 1},{\bf 1},0)$} &
{$B_{\mu}$} & {$\tilde{b}$}
\\ \hline \hline
\end{tabular}

\vspace*{3mm}

\begin{tabular}{l|cc}
\multicolumn{3}{l}
{\bf Higgs sector} \\ \hline \hline
{Representation} & {Boson~$(R=+1)$} & {Fermion~$(R=-1)$}
\\ \hline
{$({\bf 1},{\bf 2},-1/2)$} &
{$H_{1}$} & {$\chi_{H_1}$}
\\
{$({\bf 1},{\bf 2},1/2)$} &
{$H_{2}$} & {$\chi_{H_2}$}
\\ \hline \hline
\end{tabular}

\vspace*{3mm}

\begin{tabular}{l|cc}
\multicolumn{3}{l}
{\bf Quark / lepton sector} \\ \hline\hline
{Representation} & {Boson~$(R=-1)$} & {Fermion~$(R=+1)$}
\\ \hline
{$({\bf 3},{\bf 2},1/6)$} &
{$\tilde{q}_{i}$} & {$q_{i}$}
\\
{$({\bf 3^{*}},{\bf 1},-2/3)$}  &
{$\tilde{u}^{c}_{i}$} & {$u^{c}_{i}$}
\\
{$({\bf 3^{*}},{\bf 1},1/3)$} &
{$\tilde{d}^{c}_{i}$} & {$d^{c}_{i}$}
\\
{$({\bf 1},{\bf 2},-1/2)$}  &
{$\tilde{l}_{i}$} & {$l_{i}$}
\\
{$({\bf 1},{\bf 2},1)$} &
{$\tilde{e}^{c}_{i}$} & {$e^{c}_{i}$}
\\ \hline \hline
\end{tabular}
\caption
{Particle content of the minimal SUSY standard model. Index $i$ is the
generation index which runs from 1 to 3. For each particles,
representation of the ${\rm SU(3)}_{C}$ $\times$ ${\rm SU(2)}_{L}$
$\times $ ${\rm U(1)}_{Y}$ gauge group is also shown.}
\label{table:MSSM-particles}
\end{center}
\end{table}
%
%

Next, we will see the lagrangian of the MSSM. As a first step, we
comment on $R$-parity. If we assume a particle content of the MSSM shown
in Table~\ref{table:MSSM-particles}, we can write down interactions
which violate baryon- or lepton-number conservations. For example,
interactions such as $u^{c}d^{c}\tilde{d}^{c}$ or $d^{c}q\tilde{l}$
cannot be forbidden by gauge invariance or renormalizability. But
phenomenologically, strength of these interactions is severely
constrained since they may induce unwantedly high rate of nucleon decay
and neutron-anti-neutron oscillation, and they wash out baryon number in
the early universe~\cite{PLB256-457}.  Rather than assuming extremely
small coupling constants for them, we usually forbid these dangerous
terms by introducing a discrete symmetry, that is called $R$-parity.
$R$-parity assigns $+1$ for ordinary particles in the standard model and
$-1$ for their superpartners. One can see that if we require the
invariance under the $R$-parity, baryon- and lepton-numbers are conserved
under the restriction of renormalizability. In this thesis, we adopt
the $R$-invariance below.  Notice that $R$-invariance also guarantees
the stability of the lightest $R$-odd particle, \ie the lightest
superparticle (LSP).

Assuming the $R$-invariance, the superpotential of the MSSM is given by
\begin{eqnarray}
W_{\rm MSSM} &=& y^{(u)}_{ij} u^c_i q_j H_2  +
y^{(d)}_{ij} d^c_i q_j H_1 +
y^{(e)}_{ij} e^c_i l_j H_1
\nonumber \\ &&
+ \mu_H H_1 H_2,
\end{eqnarray}
where $i$ and $j$ are generation indices, and we have omitted the group
indices for simplicity. Here, $y^{(u)}$, $y^{(d)}$ and $y^{(e)}$ are the
Yukawa coupling constants of the up-, down- and lepton-sector,
respectively.

Since the SUSY should be  broken in nature, SUSY breaking terms
are also necessary in lagrangian. In order not to induce quadratic
divergences, SUSY should be broken softly. In general, soft SUSY
breaking terms are gaugino mass terms, scalar mass terms and trilinear
coupling terms for scalar bosons of same chirality~\cite{NPB159-429}.
In the MSSM, SUSY breaking terms are given by
\begin{eqnarray}
{\cal L}_{\rm soft} &=&
- m^2_{\tilde{q}ij} \tilde{q}^{*}_{i} \tilde{q}_{j}
- m^2_{\tilde{u}ij} \tilde{u}^{c*}_{i} \tilde{u}^c_{j}
- m^2_{\tilde{d}ij} \tilde{d}^{c*}_{i} \tilde{d}^c_{j}
- m^2_{\tilde{l}ij} \tilde{l}^{*}_{i} \tilde{l}_{j}
- m^2_{\tilde{e}ij} \tilde{e}^{c*}_{i} \tilde{e}^c_{j}
\nonumber \\ &&
- \lsp A^{(u)}_{ij} \tilde{u}^{c}_{i} \tilde{q}_{j} H_2
+ A^{(d)}_{ij} \tilde{d}^{c}_{i} \tilde{q}_{j} H_1
+ A^{(e)}_{ij} \tilde{e}^{c}_{i} \tilde{l}_{j} H_1 + h.c. \rsp
\nonumber \\ &&
- m_{H_1}^2 \labs H_1 \rabs^2 - m_{H_2}^2 \labs H_2 \rabs^2
- \lsp m_3^2 H_1 H_2 + h.c. \rsp
\nonumber \\ &&
- \lsp m_{G3} \tilde{g}\tilde{g}
+ m_{G2} \tilde{w}\tilde{w}
+ m_{G1} \tilde{b}\tilde{b} + h.c. \rsp,
\end{eqnarray}
where $m^2_{\tilde{q}}$ -- $m^2_{\tilde{e}}$ are the squark
and slepton masses, and $m_{G3}$ -- $m_{G1}$ the gauge fermion
masses. In the next chapter, we will see that these SUSY breaking
parameters can be obtained in a low energy effective theory of local
SUSY models.

%
%
\begin{figure}[t]
\epsfxsize=13cm

\centerline{\epsfbox{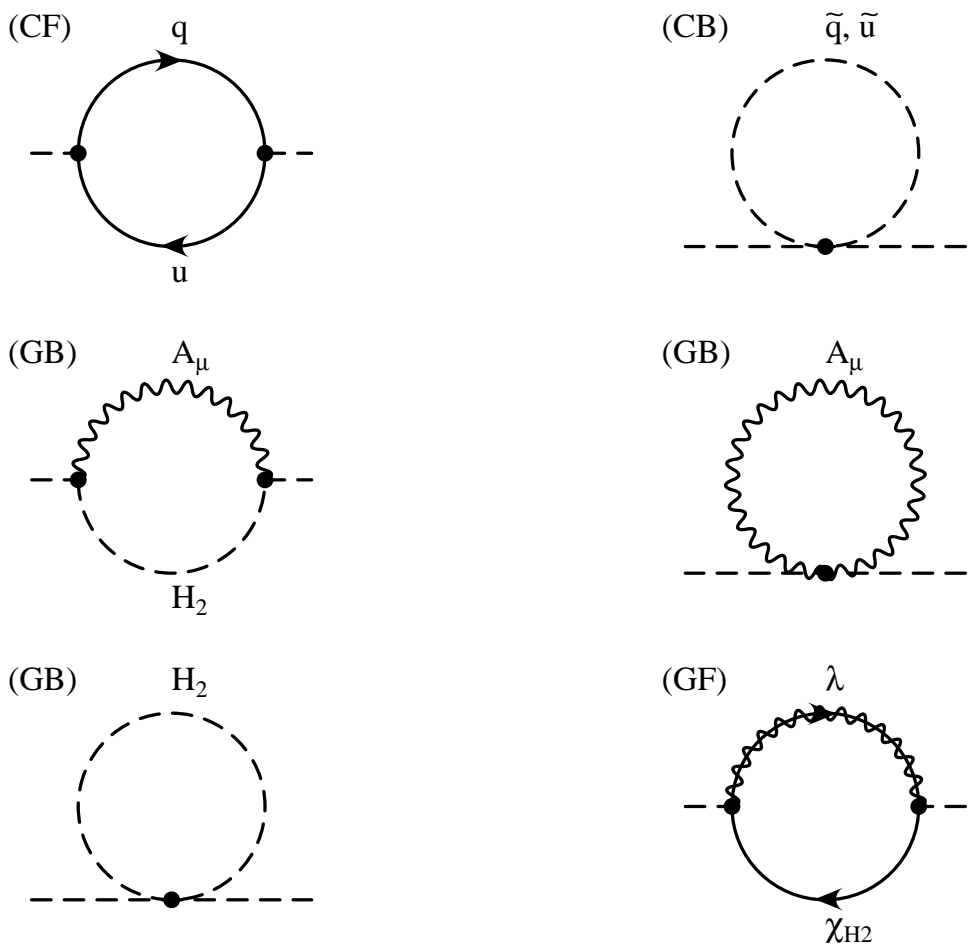}}

\caption{Quadratically divergent Feynman diagrams for the $H_{2}$
mass. Diagram with CF (CB, GB, GF) is the contribution from chiral
fermion (chiral boson, gauge boson, gauge fermion) loop. Dashed lines in
external lines represent $H_{2}$. Notice that the diagram with $H_2$
loop (lower-left) originates to the gauge $D$-term, and hence we
classify it as the contribution of gauge boson.}
\label{fig:quadra}

\end{figure}
%
%
As mentioned before, quadratic divergence of the two point functions of
scalar bosons disappears in SUSY models. At the one loop level, this can
be easily seen. For example, we will see the cancellation in the mass of
$H_2$. Feynman diagrams which give quadratically divergent radiative
corrections to the $H_{2}$ mass are shown in Fig.~\ref{fig:quadra}. Each
of them are quadratically divergent;
\begin{eqnarray}
\left. \delta m_{H_{2}}^{2} \rabs_{\rm CB} &\simeq&
\frac{1}{8\pi^2} y^{(u)}_{ij} y^{(u)*}_{ij} \Lambda_{CUT}^2
+\cdot\cdot\cdot,
\label{cancel-cb} \\
\left. \delta m_{H_{2}}^{2} \rabs_{\rm CF} &\simeq&
- \frac{1}{8\pi^2} y^{(u)}_{ij} y^{(u)*}_{ij} \Lambda_{CUT}^2
+\cdot\cdot\cdot,
\label{cancel-cf} \\
\left. \delta m_{H_{2}}^{2} \rabs_{\rm GB} &\simeq&
\frac{1}{4\pi^2} \lsp \frac{1}{2} g_2^2 + \frac{1}{4} g_1^2 \rsp
\Lambda_{CUT}^2 +\cdot\cdot\cdot,
\label{cancel-gb} \\
\left. \delta m_{H_{2}}^{2} \rabs_{\rm GF} &\simeq&
- \frac{1}{4\pi^2} \lsp \frac{1}{2} g_2^2 + \frac{1}{4} g_1^2 \rsp
\Lambda_{CUT}^2 +\cdot\cdot\cdot,
\label{cancel-gf}
\end{eqnarray}
where CF (CB, GB, GF) represents the contribution from chiral fermion
(chiral boson, gauge boson, gauge fermion), $\cdot\cdot\cdot$
the terms which do not contain quadratic divergences and $\Lambda_{CUT}$
the cut-off. These quadratic divergences, however, cancel out between
boson and fermion loops.  Quadratic divergences in other scalar masses
also disappear in the same way, and hence the hierarchy problem can be
solved by extending the standard model to the supersymmetric one.

\begin{figure}[p]
\epsfxsize=14cm

\centerline{\epsfbox{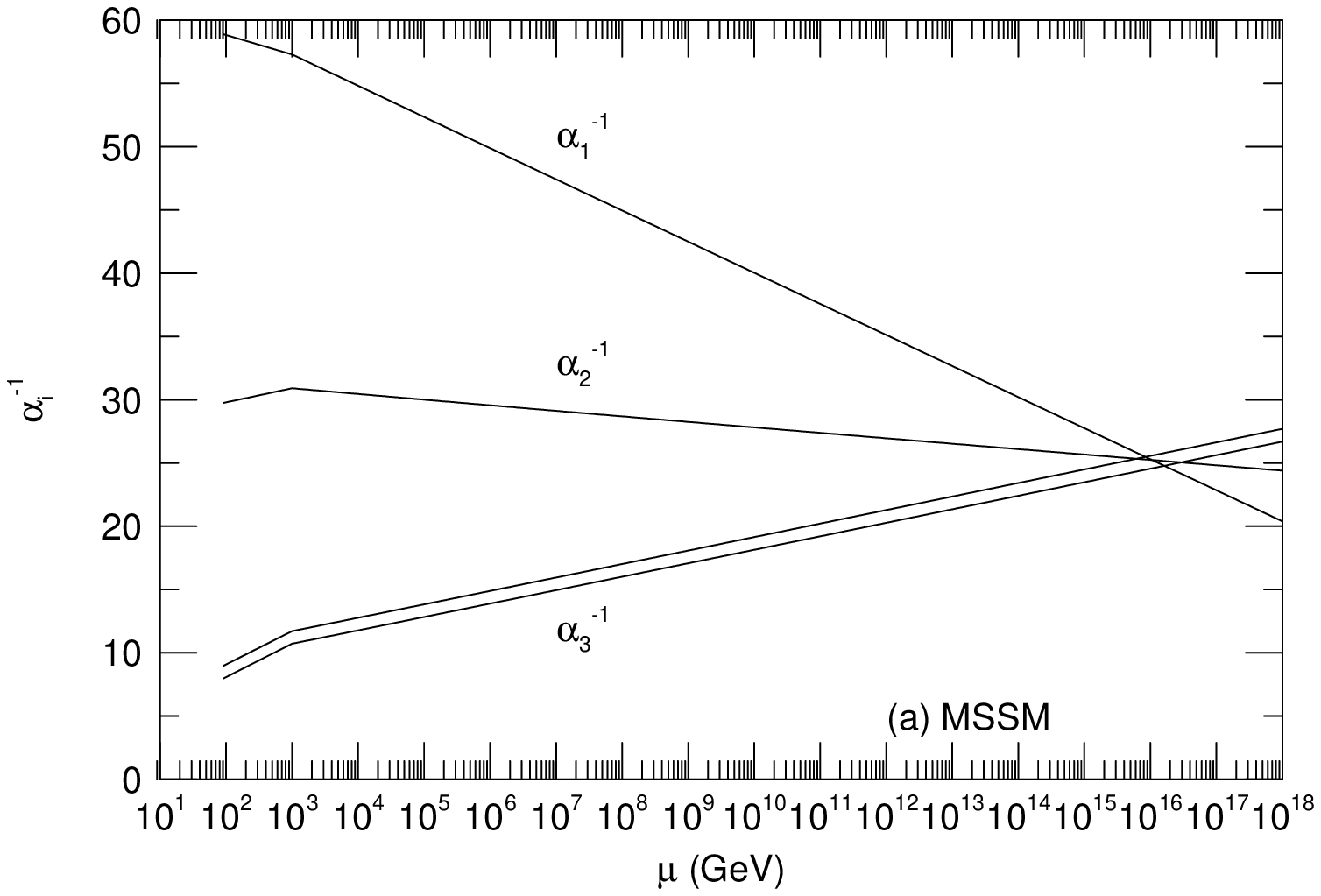}}

\vspace{1cm}

\centerline{\epsfbox{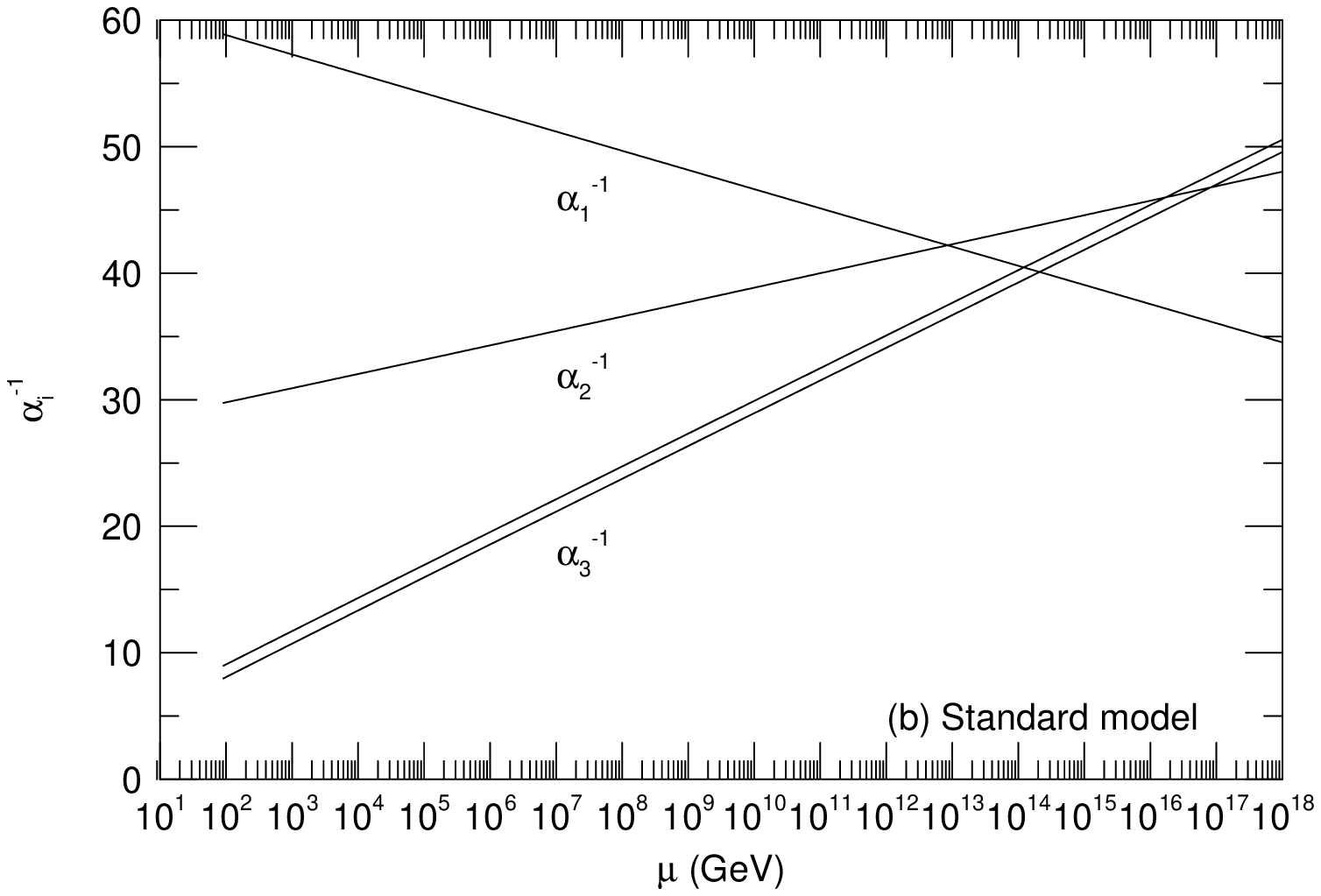}}

\caption
{Renormalization group flow of the coupling constants of ${\rm
SU(3)}_{C}$, ${\rm SU(2)}_{L}$ and ${\rm U(1)}_{Y}$ gauge group for the
case of (a) the MSSM, and (b) the standard model. Here, we use two loop
renormalization group equations, and take the SUSY scale at 1TeV for
the MSSM case.}
\label{fig:gcc-flow}
\end{figure}

In the MSSM, other interesting new physics, \ie the grand unified theory
(GUT)~\cite{PRL25-438}, is suggested from the renormalization group
analysis~\cite{PRD44-817,PLB260-447}. As mentioned before, SUSY
extension of the standard model increases the number of particles, which
changes the renormalization group equation of the gauge coupling
constants. In Fig.~\ref{fig:gcc-flow}, we show the renormalization
group flow of ${\rm SU(3)}_{C}$, ${\rm SU(2)}_{L}$ and ${\rm U(1)}_{Y}$
gauge coupling constants in the MSSM case and in the standard model
case. In the MSSM case, three gauge coupling constants meet at the
energy scale $\sim 10^{16}\GEV$ which may be identified with the
GUT scale, while in the standard model case, the renormalization group
flow of the gauge coupling constants conflicts with the gauge coupling
unification.
%
%
%
%

\begin{figure}[t]
\epsfxsize=14cm

\centerline{\epsfbox{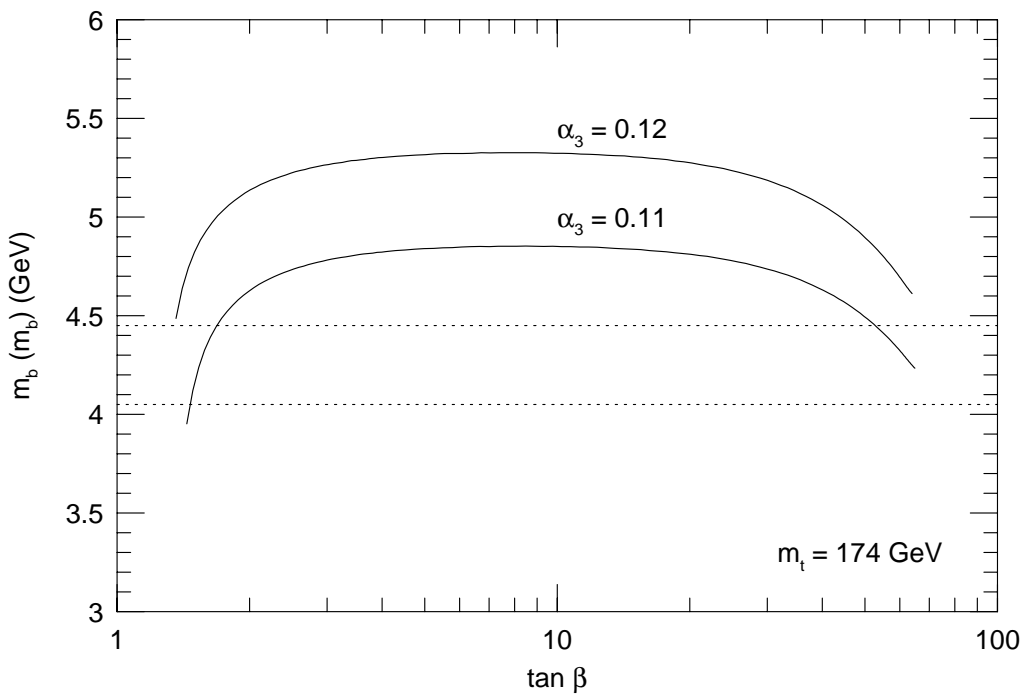}}

\caption
{The predicted value of the running bottom-quark mass $m_b(m_b)$ is
shown as a function of $\tan\beta$ for $\alpha_3(m_Z)=0.11$ and 0.12.
Here, we take the (on-shell) top-quark mass at 174GeV.}
\label{fig:mb}
\end{figure}

Another indirect evidence of SUSY GUT is the bottom-tau mass
ratio~\cite{PLB247-387,PRD47-1093,PRD49-1454}. In SU(5) or SO(10) GUT,
Yukawa coupling constants of down- and lepton-sectors to the Higgs boson
are also expected to be unified at the GUT scale, and hence we can get a
relation between the bottom- and tau-Yukawa coupling constants at the
electroweak scale. By using this relation, the mass of the bottom
quark is obtained once the tau-lepton mass ($\simeq
1.777\GEV$~\cite{PDG}) is fixed. In Fig.~\ref{fig:mb}, we show the
predicted bottom-quark mass as a function of $\tan\beta
=\vev{H_1}/\vev{H_2}$. Notice that the maximal and minimal values for
$\tan\beta$ are determined so that all the Yukawa coupling constants do
not blow up below the GUT scale. As one can see, SUSY GUT predicts the
bottom-quark mass to be (4 -- 6)GeV which is close to that determined
from experiments; $m_b(m_b)=(4.25\pm 0.2)\GEV$~\cite{PRepC87-77} (where
we have doubled the uncertainty), especially when $\tan\beta$ approaches
its maximal or minimal value.
%
%
%
%

Contrary to those attractive features, no direct evidence of SUSY has
been found, which certainly indicates that the SUSY is a (softly) broken
symmetry. The physics of SUSY breaking is, however, still an open
question and we have not understood it yet. Especially in the framework
of global SUSY, it seems to be very much difficult to construct a
phenomenologically favorable model. One of the reason is that there
exists a mass formula in the global SUSY model;
\beq
{\rm Str}{\bf M}^2 \equiv
\sum_{J=0}^{1} (-1)^J {\rm tr}{\bf M}_J^2 = 0,
\label{str_global}
\eeq
which prevents all the squarks and sleptons from having masses larger
than those of quarks and leptons. To avoid this constraint, many people
extend the global SUSY to the local one and consider the physics of SUSY
breaking in the framework of supergravity. In the next chapter, we will
investigate local SUSY model and see how the mass formula in global SUSY
models (\ref{str_global}) is modified.

%
%

\chapter{Review of supergravity}
\label{chap:sugra}

\hspace*{\parindent}
In this chapter, we will introduce a lagrangian which is invariant under
the local SUSY transformation, and derive super-trace formula in that
framework. Conventions used in this chapter are essentially equal to
those used in ref.\cite{Wess&Bagger} except that we use the metric (in
flat space-time) as $g_{\mu\nu}\simeq{\rm diag}(1,-1,-1,-1)$.  For our
convention, see also Appendix~\ref{ap:notation}.

\section{Heuristic approach to supergravity lagrangian}

\hspace*{\parindent}
Compared with the global SUSY, one of the characteristics of the local
one is the existence of a gauge field associated with the local SUSY,
which is called gravitino. As in the case of ordinary gauge theories,
the gravitino couples to a Noether current of SUSY and maintains the
invariance under the local SUSY transformation.  In this section, we
will briefly review the role of the gravitino in the local SUSY theory
by using the simplest model, which is the Wess-Zumino
model~\cite{NPB70-39} without interactions.

Let us begin with the global case. In this case, total lagrangian
contains only two terms, one is the kinetic term of a massless complex
scalar boson $\phi$ and the other is that of a massless chiral fermion
$\chi$;
\beq
{\cal L}_{\rm WZ} = \partial^\mu \phi \partial_\mu \phi^*
+ i \overline{\chi} \bar{\sigma}^{\mu} \partial_{\mu} \chi.
\label{L_WZ}
\eeq
Up to total derivative, this lagrangian is invariant under the following
global SUSY transformation,
\beq
\delta \phi &=& \sqrt{2} \xi \chi,
\label{dphi} \\
\delta \chi &=& - i \sqrt{2} \sigma^{\mu} \overline{\xi}
(\partial_{\mu}\phi),
\label{dchi}
\eeq
where $\xi$ is the infinitesimal Grassmann-odd parameter.

If $\xi$ has space-time dependence, lagrangian (\ref{L_WZ}) is not
invariant but extra terms which are proportional to $\partial\xi$ or
$\partial\overline{\xi}$ appear with the supertransformation
(\ref{dphi}) and (\ref{dchi});
\beq
\delta {\cal L}_{\rm WZ}
&=& \sqrt{2} \lmp
( \partial_{\mu} \xi ) \sigma^{\nu} \bar{\sigma}^{\mu} \chi
( \partial_{\nu} \phi^* ) +
\overline{\chi} \bar{\sigma}^{\mu} \sigma^{\nu}
( \partial_{\mu} \overline{\xi} )
( \partial_{\nu} \phi ) \rmp + ({\rm total~derivative})
\nonumber \\
&\equiv& i ( \partial_{\mu} \xi ) J^{\mu} + h.c.
+ ({\rm total~derivative}).
\label{dL_WZ}
\eeq
where
\beq
J^\mu \equiv
-i \sqrt{2} \sigma^{\nu} \bar{\sigma}^{\mu} \chi ( \partial_{\nu} \phi^* ).
\label{J_chiral}
\eeq
Notice that $J^\mu$ is the Noether current of SUSY, which is called
supercurrent. In order to keep invariance, we introduce a gauge field
$\psi_{\mu}$. As in the cases of ordinary gauge theories, the gauge
field $\psi_\mu$ couples to the supercurrent in the following way;
\beq
{\cal L}_{\psi J} = - \frac{i}{2} G_S \psi_{\mu} J^\mu + h.c.~,
\label{L_GJ}
\eeq
where $G_S$ is the ``coupling constant'' which we will determine later.
Since the charge of SUSY has a Grassmann-odd nature with spin index, the
gauge field $\psi_{\mu}$ associated with SUSY is a spin $\frac{3}{2}$
fermion. Varying eq.(\ref{L_GJ}), one obtains
\beq
\delta {\cal L}_{\psi J} =
- \frac{i}{2} G_S \lmp
(\delta\psi_{\mu}) J^\mu + \psi_{\mu} (\delta J^\mu) \rmp +  h.c.
\label{dL_GJ}
\eeq
Therefore, if $\psi_{\mu}$ transforms as
\beq
\delta\psi_{\mu} \sim \frac{2}{G_S} \partial_{\mu} \xi,
\label{dpsi_naive}
\eeq
the first term in eq.(\ref{dL_GJ}) cancels out the contribution from
eq.(\ref{dL_WZ}).

Next, we will consider the second term in eq.(\ref{dL_GJ}).
Supertransformation of the supercurrent (\ref{J_chiral}) gives
energy-momentum tensor ${T_\mu}^\nu$ of the chiral multiplet $(\phi,
\chi)$;
\beq
\lmp \overline{Q}_{\dot{\alpha}} , J^{\mu}_{\alpha} \rmp &=&
-2 \sigma^{\nu}_{\alpha\dot{\alpha}} {T_\nu}^\mu
+ ({\rm total~derivative}),
\nonumber \\
\lmp Q^{\alpha} , \overline{J}^{\mu\dot{\alpha}} \rmp &=&
-2 \bar{\sigma}^{\nu\dot{\alpha}\alpha} {T_\nu}^\mu
+ ({\rm total~derivative}),
\eeq
where $Q$ and $\overline{Q}$ are the generators of the SUSY
transformation, and hence the second term in eq.(\ref{dL_GJ}) becomes
\beq
- \frac{i}{2} G_S \lmp \psi_{\mu} (\delta J^\mu) \rmp +  h.c.
=
\frac{i}{2} G_S \lmp
\psi_\mu \sigma_\nu \overline{\xi} + \psi_\nu \sigma_\mu \overline{\xi}
+ \overline{\psi}_\mu \bar{\sigma}_\nu \xi
+ \overline{\psi}_\nu \bar{\sigma}_\mu \xi
\rmp T^{\mu\nu}.
\label{psi*dj}
\eeq

In order to cancel out these terms, we rewrite the lagrangian
(\ref{L_WZ}) by explicitly expressing the metric tensor $g_{\mu\nu}$;
\beq
{\cal L}_{\rm WZ} \rightarrow
\sqrt{-g} g_{\mu\nu} \lsp
\partial^\mu \phi \partial^\nu \phi^*
+ i \overline{\chi} \bar{\sigma}^{\mu} \partial^{\nu} \chi \rsp,
\label{L_WZ2}
\eeq
where $g\equiv\det{g_{\mu\nu}}$, and use the fact that the metric is the
gauge field associated with the energy-momentum tensor (which is the
Noether current of the space-time translation), that is, the
energy-momentum tensor ${T_\mu}^\nu$ is obtained if one varies
lagrangian by $g_{\mu\nu}$;
\beq
\frac{\partial {\cal L}}{\partial g_{\mu\nu}} \sim 2 T^{\mu\nu}.
\label{t_mn}
\eeq
Then, the metric tensor $g_{\mu\nu}$ (\ie graviton) can be regarded as a
superpartner of the gravitino field $\psi_\mu$, and its transformation
law is determined so that the local SUSY invariance is maintained;
\beq
\delta g_{\mu\nu} \frac{\delta {\cal L}_{\rm WZ}}{\delta g_{\mu\nu}}
\sim - \frac{i}{2} G_S \lmp
\psi^\mu \sigma^\nu \overline{\xi} + \psi^\nu \sigma^\mu \overline{\xi}
+ \overline{\psi}^\mu \bar{\sigma}^\nu \xi
+ \overline{\psi}^\nu \bar{\sigma}^\mu \xi
\rmp T_{\mu\nu}.
\label{dL_WZ2}
\eeq
Combining eq.(\ref{t_mn}) with eq.(\ref{dL_WZ2}), one can obtain the
transformation law of the metric tensor;
\beq
\delta g_{\mu\nu} \sim - i G_S \lsp
\psi_\mu \sigma_\nu \overline{\xi} + \psi_\nu \sigma_\mu \overline{\xi}
+ \overline{\psi}_\mu \bar{\sigma}_\nu \xi
+ \overline{\psi}_\nu \bar{\sigma}_\mu \xi
\rsp.
\label{dg}
\eeq

In the following arguments, in fact, it is more convenient to use the
vierbein ${e_\mu}^a$ rather than the metric tensor
$g_{\mu\nu}=\eta_{ab}{e_\mu}^a{e_\nu}^b$, where $\eta_{ab} = {\rm
diag}(1,-1,-1,-1)$ is the metric tensor in flat space-time. In
supergravity models, transformation law of the vierbein ${e_\mu}^a$ is
defined as
\beq
\delta {e_\mu}^a = - i G_S \lsp \xi \sigma^a \overline{\psi}_\mu +
\overline{\xi} \bar{\sigma}^a \psi_\mu \rsp,
\label{de_naive}
\eeq
with $G_S=M^{-1}$. This transformation law gives eq.(\ref{dg}). Under
the local SUSY transformation, the vierbein ${e_\mu}^a$ and the
gravitino $\psi_\mu$ (and some other auxiliary fields) make up a
multiplet, which we call a supergravity multiplet.

As we have seen, if we extend the global SUSY to the local one, the
metric tensor $g_{\mu\nu}$ automatically comes into the theory, and
hence we must consider gravity. This is the reason why the local SUSY is
sometimes called supergravity.

\section{Minimal supergravity model}

\hspace*{\parindent}
In the previous section, we have introduced the gravitino field
$\psi_\mu$ in order to keep the local SUSY invariance. As we have seen,
the gravitino field couples to the supercurrent, but the strength $G_S$
of the coupling between the gravitino and the supercurrent has not been
determined yet. In order to determine the coupling strength $G_S$, we
must see the invariance of the kinetic terms of ${e_\mu}^a$ and
$\psi_\mu$ under the local SUSY transformation.

In this section, we will explicitly investigate the local SUSY
invariance of the minimal supergravity model which contains only the
graviton ${e_\mu}^a$ and the gravitino field $\psi_\mu$. As a result, we
will see the coupling strength $G_S$ should be equal to the inverse of the
gravitational scale.

The lagrangian of the minimal supergravity model is given by
\beq
{\cal L}_{\rm MSG} ={\cal L}_{\rm EH} + {\cal L}_{\rm RS},
\label{L_MSG}
\eeq
where ${\cal L}_{\rm EH}$ is the usual Einstein-Hilbert lagrangian, and
${\cal L}_{\rm RS}$ is the Rarita-Schwinger lagrangian which is
essentially the kinetic term of the spin $\frac{3}{2}$ gravitino field.
The Einstein-Hilbert lagrangian is given by
\beq
{\cal L}_{\rm EH} = -\frac{M^2}{2} e R,
\eeq
with
\beq
R \equiv {e_a}^\mu {e_b}^\nu \lsp
\partial_\mu {\omega_\nu}^{ab} - \partial_\nu {\omega_\mu}^{ab} -
{\omega_\mu}^{ac} {\omega_{\nu c}}^{b}
+ {\omega_\nu}^{ac} {\omega_{\mu c}}^{b} \rsp,
\eeq
where ${\omega_{\mu}}^{ab}$ denotes the spin connection and $e\equiv\det
{e_a}^\mu$. On the other hand, the Rarita-Schwinger lagrangian can be
written as
\beq
{\cal L}_{\rm RS} = e \epsilon^{\mu\nu\rho\sigma}
\overline{\psi}_\mu \bar{\sigma}_\nu \tilde{{\cal D}}_\rho \psi_\sigma,
\eeq
where $\epsilon^{\mu\nu\rho\sigma}$ is the totally anti-symmetric tensor
($\epsilon_{0123}=-1$ in flat space-time) and the covariant derivative
of the gravitino field is given by
\beq
\tilde{\cal D}_\mu \psi_\nu \equiv \partial_\mu \psi_\nu
+ \frac{1}{2} {\omega_\mu}^{ab} \sigma_{ab} \psi_\nu.
\eeq
In the following, we will see the invariance of the minimal supergravity
lagrangian (\ref{L_MSG}) under the local SUSY transformation;
\beq
&& \delta {e_\mu}^a = - i G_S \lsp \xi \sigma^a \overline{\psi}_\mu
+ \overline{\xi} \bar{\sigma}^a \psi_\mu \rsp,
\label{de} \\
&& \delta \psi_\mu = \frac{2}{G_S} \tilde{\cal D}_\mu \xi \equiv
\frac{2}{G_S} \lsp \partial_\mu \xi
+ \frac{1}{2} {\omega_\mu}^{ab} \sigma_{ab} \xi \rsp,
\label{dpsi}
\eeq
where parameter $G_S$ will be determined so that the local SUSY
invariance is maintained (see eq.(\ref{dpsi_naive}) and
eq.(\ref{de_naive})).

Before checking the invariance of the minimal supergravity lagrangian
(\ref{L_MSG}), we will comment on the spin connection
${\omega_\mu}^{ab}$. As we will see below, ${\omega_\mu}^{ab}$ is
represented as a function of the vierbein ${e_\mu}^a$ and the gravitino
$\psi_\mu$ by solving its field equation;
\beq
\frac{\delta{\cal L}_{\rm MSG}}{\delta {\omega_\mu}^{ab}} =
\frac{\partial {\cal L}_{\rm MSG}}{\partial {\omega_\mu}^{ab}} -
\partial_\nu \frac{\partial{\cal L}_{\rm MSG}}
{\partial \lsp \partial_\nu {\omega_\mu}^{ab} \rsp} = 0,
\label{eq4w}
\eeq
and hence ${\omega_\mu}^{ab}$ is regarded as an auxiliary field. From
eq.(\ref{eq4w}), one can obtain
\beq
M^2 \lmp \lsp \partial_\mu {e_\nu}^a + {\omega_\mu}^{ab} e_{\nu b} \rsp -
\lsp \partial_\nu {e_\mu}^a + {\omega_\nu}^{ab} e_{\mu b} \rsp \rmp =
- \frac{i}{2} \lsp \overline{\psi}_{\mu} \bar{\sigma}^a \psi_\nu -
\overline{\psi}_{\nu} \bar{\sigma}^a \psi_\mu \rsp .
\label{eq4w2}
\eeq
By solving this equation, explicit form of the spin connection is given
by
\beq
\omega_{\mu\rho\sigma}
&\equiv& e_{\rho a} e_{\sigma b} {\omega_\mu}^{ab}
\nonumber \\
&=&
\frac{1}{2} \lmp
e_{\sigma a} \lsp \partial_\mu {e_\rho}^a
- \partial_\rho {e_\mu}^a \rsp+
e_{\rho a} \lsp \partial_\sigma {e_\mu}^a
- \partial_\mu {e_\sigma}^a \rsp -
e_{\mu a} \lsp \partial_\rho {e_\sigma}^a
- \partial_\sigma {e_\rho}^a \rsp
\rmp
\nonumber \\
&&
- \frac{i}{4M^2}
e_{\sigma a} \lsp \psi_\rho \sigma^a \overline{\psi}_\mu
- \psi_\mu \sigma^a \overline{\psi}_\rho \rsp
- \frac{i}{4M^2}
e_{\rho a} \lsp \psi_\mu \sigma^a \overline{\psi}_\sigma
- \psi_\sigma \sigma^a \overline{\psi}_\mu \rsp
\nonumber \\
&&
+ \frac{i}{4M^2}
e_{\mu a} \lsp \psi_\sigma \sigma^a \overline{\psi}_\rho
- \psi_\rho \sigma^a \overline{\psi}_\sigma \rsp.
\label{omega}
\eeq

Now let us see the invariance of the lagrangian (\ref{L_MSG}).  With the
help of chain rule, variation of the total lagrangian is given by
\beq
\delta {\cal L}_{\rm MSG} &=&
\delta {e_\mu}^a
\left.\frac{\delta {\cal L}_{\rm MSG}}{\delta {e_\mu}^a}
\rabs_{\psi,\omega} +
\delta \psi_\mu \left.\frac{\delta {\cal L}_{\rm MSG}}{\delta \psi_\mu}
\rabs_{e,\omega}
\nonumber \\
&&
+\lsp \delta {e_\mu}^a \frac{\delta {\omega_\mu}^{ab}}{\delta{e_\mu}^a}
+\delta \psi_\mu \frac{\delta {\omega_\mu}^{ab}}{\delta\psi_\mu} \rsp
\left.\frac{\delta {\cal L}_{\rm MSG}}
{\delta {\omega_\mu}^{ab}}
\rabs_{e,\psi}~,
\label{dL_MSG}
\eeq
where we have used the fact that the spin connection ${\omega_\mu}^{ab}$
is a function of the vierbein ${e_\mu}^a$ and the gravitino $\psi_\mu$.
The important point is that the last two terms in eq.(\ref{dL_MSG})
which are proportional to $(\delta{\cal L}_{\rm MSG}/\delta\omega)$
vanish since the spin connection obeys its field equation $(\delta{\cal
L}_{\rm MSG} / \delta\omega)=0$. Therefore, we only have to vary
${e_\mu}^a$ and $\psi_\mu$ (with ${\omega_\mu}^{ab}$ fixed) in order to
obtain $\delta{\cal L}_{\rm MSG}$. (We denote this operation $\Delta$.)

Variation of the Einstein-Hilbert lagrangian ${\cal L}_{\rm EH}$ gives
the Einstein tensor multiplied by $\delta{e_\mu}^a$;
\beq
\Delta {\cal L}_{\rm EH} &\equiv&
\delta {e_\mu}^a
\left.\frac{\delta {\cal L}_{\rm EH}}{\delta {e_\mu}^a}
\rabs_{\psi,\omega} +
\delta \psi_\mu \left.\frac{\delta {\cal L}_{\rm EH}}{\delta \psi_\mu}
\rabs_{e,\omega}
\nonumber \\
&=&
i G_S M^2 e
\lsp {R_a}^\mu
- \frac{1}{2} {e_a}^\mu R \rsp \overline{\psi}_\mu \sigma^a \xi
+ h.c.~,
\label{dL_E}
\eeq
with
\beq
{R_\mu}^a \equiv {e_b}^\nu \lsp
\partial_\mu {\omega_\nu}^{ab} - \partial_\nu {\omega_\mu}^{ab} -
{\omega_\mu}^{ac} {\omega_{\nu c}}^{b}
+ {\omega_\nu}^{ac} {\omega_{\mu c}}^{b} \rsp.
\eeq
On the other hand, after some straightforward calculations, $\Delta{\cal
L}_{\rm RS}$ becomes the following form;
\beq
\Delta {\cal L}_{\rm RS} &\equiv&
\delta {e_\mu}^a
\left.\frac{\delta {\cal L}_{\rm RS}}{\delta {e_\mu}^a}
\rabs_{\psi,\omega} +
\delta \psi_\mu \left.\frac{\delta {\cal L}_{\rm RS}}{\delta \psi_\mu}
\rabs_{e,\omega}
\nonumber \\
&=&
\lmp - \frac{i}{G_S} M^2 e
\lsp {R_a}^\mu
- \frac{1}{2} {e_a}^\mu R \rsp \overline{\psi}_\mu \bar{\sigma}^a \xi
+ h.c. \rmp
\nonumber \\
&&
- e \epsilon^{\mu\nu\rho\sigma}
\lmp \frac{2}{G_S}
\lsp \partial_\mu {e_{\nu}}^a + {\omega_\mu}^{ab} e_{\nu b} \rsp
+iG_S \lsp \overline{\psi}_\mu \bar{\sigma}^a \psi_\nu \rsp \rmp
\overline{\xi} \bar{\sigma}_a \tilde{\cal D}_\rho \psi_\sigma.
\label{dL_RS}
\eeq
By setting $G_S=M^{-1}$, the first line in eq.(\ref{dL_RS}) is equal
to $-\Delta {\cal L}_{\rm EH}$, the second line vanishes due to
eq.(\ref{eq4w2}), and hence $\delta{\cal L}_{\rm MSG}$ vanishes;
\beq
\delta{\cal L}_{\rm MSG} =
\Delta{\cal L}_{\rm EH} + \Delta{\cal L}_{\rm RS} = 0.
\eeq
This is the end of the proof of the invariance. We have seen the
invariance of the minimal supergravity lagrangian (\ref{L_MSG}) under
the local SUSY transformation (\ref{de}) and (\ref{dpsi}) with
$G_S=M^{-1}$.

This fact suggests that the coupling strength $G_S$ between the
gravitino field $\psi_\mu$ and the supercurrent $J_\mu$ is not a free
parameter but a model independent constant which is determined by the
requirement that the kinetic term of the supergravity multiplet
($\psi_\mu$, ${e_\mu}^a$) is invariant under the local SUSY
transformation. In the next section, we can see that general
supergravity lagrangian contains interaction terms between the gravitino
field $\psi_\mu$ and the supercurrent $J_\mu$ with definite coupling
strength (\ie $G_S=M^{-1}$);
\beq
{\cal L}_{\psi J} = - \frac{i}{2M} \psi_{\mu} J^\mu + h.c.
\label{J-psi_int}
\eeq
As we will see in the following chapters, such interaction terms become
very important in investigating phenomenology with the gravitino.

\section{General supergravity lagrangian}

\hspace*{\parindent}
In this section, we will extend the minimal supergravity model to
general one. Derivation of the general supergravity lagrangian is given
in elsewhere~\cite{NPB212-413,Wess&Bagger}, but it is very much
complicated task. Therefore, we only give a final form here by following
ref.\cite{Wess&Bagger}. In this section, we use the $M=1$ unit for
simplicity.

The general supergravity lagrangian is essentially characterized by
three functions; K\"ahler potential $K(\phi,\phi^*)$, superpotential
$W(\phi)$, and kinetic function $f(\phi)$ for vector multiplets. Notice
that the K\"ahler potential $K(\phi,\phi^*)$ is a function of scalar
fields $\phi$ and $\phi^*$, while the superpotential $W(\phi)$ and the
kinetic function $f(\phi)$ depend scalar fields with definite chirality.

By using these functions, the general form of the supergravity
lagrangian, which contains scalar field $\phi$, chiral fermion $\chi$,
gauge boson $A_\mu$, and gauge fermion $\lambda$ as well as the vierbein
${e_\mu}^a$ and the gravitino $\psi_\mu$ can be written as
\beq
{\cal L}_{\rm SUGRA} &=&
- \frac{1}{2} eR
+ e g_{ij^*} \tilde{\cal D}_\mu \phi^i  \tilde{\cal D}^\mu \phi^{*j}
- \frac{1}{2} e g^2 D_{(a)} D^{(a)}
\nonumber \\ &&
+ i e g_{ij^*}
\overline{\chi}^j \bar{\sigma}^\mu \tilde{\cal D}_\mu \chi^i
+ e \epsilon^{\mu\nu\rho\sigma}
\overline{\psi}_\mu \bar{\sigma}_\nu \tilde{{\cal D}}_\rho \psi_\sigma
\nonumber \\ &&
-\frac{1}{4} e f^R_{(ab)} F_{\mu\nu}^{(a)} F^{\mu\nu(b)}
+\frac{1}{8} e \epsilon^{\mu\nu\rho\sigma} f^I_{(ab)}
F_{\mu\nu}^{(a)} F_{\rho\sigma}^{(b)}
\nonumber \\ &&
+\frac{i}{2} e \llp
\lambda_{(a)} \sigma^\mu \tilde{\cal D}_\mu \overline{\lambda}^{(a)}
+ \overline{\lambda}_{(a)} \bar{\sigma}^\mu
\tilde{\cal D}_\mu \lambda^{(a)}
\rlp
-\frac{1}{2} f^I_{(ab)} \tilde{\cal D}_\mu
\llp e \lambda^{(a)} \sigma^\mu \overline{\lambda}^{(b)}
\rlp
\nonumber \\ &&
+\sqrt{2} e g g_{ij^*} X^{*j}_{(a)} \chi^i \lambda^{(a)}
+\sqrt{2} e g g_{ij^*} X^{i}_{(a)}
\overline{\chi}^j \overline{\lambda}^{(a)}
\nonumber \\ &&
-\frac{i}{4} \sqrt{2} e g \partial_{i} f_{(ab)}
D^{(a)} \chi^i \lambda^{(b)}
+\frac{i}{4} \sqrt{2} e g \partial_{i^*} f^*_{(ab)}
D^{(a)} \overline{\chi}^i \overline{\lambda}^{(b)}
\nonumber \\ &&
-\frac{1}{4} \sqrt{2} e \partial_{i} f_{(ab)}
\chi^i \sigma^{\mu\nu} \lambda^{(a)} F_{\mu\nu}^{(b)}
-\frac{1}{4} \sqrt{2} e \partial_{i^*} f^*_{(ab)}
\overline{\chi}^i \bar{\sigma}^{\mu\nu} \overline{\lambda}^{(a)}
F_{\mu\nu}^{(b)}
\nonumber \\ &&
+\frac{1}{2} e g D_{(a)} \psi_\mu \sigma^\mu \overline{\lambda}^{(a)}
-\frac{1}{2} e g D_{(a)}
\overline{\psi}_\mu \bar{\sigma}^\mu \lambda^{(a)}
\nonumber \\ &&
-\frac{1}{2} \sqrt{2} e g_{ij^*} \tilde{\cal D}_{\nu} \phi^{*j}
\chi^i \sigma^\mu \bar{\sigma}^\nu \psi_\mu
-\frac{1}{2} \sqrt{2} e g_{ij^*} \tilde{\cal D}_{\nu} \phi^{i}
\overline{\chi}^j \bar{\sigma}^\mu \sigma^\nu \overline{\psi}_\mu
\nonumber \\ &&
-\frac{i}{4} e \llp
\psi_\mu \sigma^{\nu\rho} \sigma^\mu \overline{\lambda}_{(a)}
+ \overline{\psi}_\mu \bar{\sigma}^{\nu\rho}
\bar{\sigma}^\mu \lambda_{(a)} \rlp \llp
F_{\nu\rho}^{(a)} + \hat{F}_{\nu\rho}^{(a)} \rlp
\nonumber \\ &&
+ \frac{1}{4} e g_{ij^*} \llp
i \epsilon^{\mu\nu\rho\sigma} \psi_\mu \sigma_\nu \overline{\psi}_\rho
+ \psi_\mu \sigma^\sigma \overline{\psi}^\mu \rlp
\chi^i \sigma_\sigma \overline{\chi}^i
\nonumber \\ &&
-\frac{1}{8} e \llp g_{ij^*} g_{kl^*} - 2 R_{ij^*kl^*} \rlp
\chi^i \chi^k \overline{\chi}^j \overline{\chi}^l
\nonumber \\ &&
+ \frac{1}{16} e \llp 2 g_{ij^*} f^R_{(ab)}
+ f^{R(cd)-1} \partial_i f_{(bc)} \partial_{j^*} f^*_{(ad)} \rlp
\overline{\chi}^j \bar{\sigma}^\mu \chi^i
\overline{\lambda}^{(a)} \bar{\sigma}_\mu \lambda^{(b)}
\nonumber \\ &&
+\frac{1}{8} e \nabla_i \partial_j f_{(ab)}
\chi^i \chi^j \lambda^{(a)} \lambda^{(b)}
+\frac{1}{8} e \nabla_{i^*} \partial_{j^*} f^*_{(ab)}
\overline{\chi}^i \overline{\chi}^j \overline{\lambda}^{(a)}
\overline{\lambda}^{(b)}
\nonumber \\ &&
+\frac{1}{16} e f^{R(cd)-1} \partial_{i} f_{(ac)} \partial_{j} f_{(bd)}
\chi^i \lambda^{(a)} \chi^j \lambda^{(b)}
\nonumber \\ &&
+\frac{1}{16} e f^{R(cd)-1} \partial_{i^*}
f^*_{(ac)} \partial_{j^*} f^*_{(bd)}
\overline{\chi}^i \overline{\lambda}^{(a)} \overline{\chi}^j
\overline{\lambda}^{(b)}
\nonumber \\ &&
-\frac{1}{16} e g^{ij^*} \partial_i f_{(ab)} \partial_{j^*} f^*_{(cd)}
\lambda^{(a)} \lambda^{(b)} \overline{\lambda}^{(c)} \overline{\lambda}^{(d)}
\nonumber \\ &&
+\frac{3}{16} e \lambda_{(a)} \sigma^\mu \overline{\lambda}^{(a)}
\lambda_{(b)} \sigma_\mu \overline{\lambda}^{(b)}
\nonumber \\ &&
+\frac{i}{4} \sqrt{2} e \partial_i f_{(ab)}
\llp \chi^i \sigma^{\mu\nu} \lambda^{(a)}
\psi_\mu \sigma_\nu \overline{\lambda}^{(b)}
-\frac{1}{4} \overline{\psi}_\mu \bar{\sigma}^\mu \chi^i
\lambda^{(a)} \lambda^{(b)} \rlp
\nonumber \\ &&
+\frac{i}{4} \sqrt{2} e \partial_{i^*} f^*_{(ab)}
\llp \overline{\chi}^i \bar{\sigma}^{\mu\nu} \overline{\lambda}^{(a)}
\overline{\psi}_\mu \bar{\sigma}_\nu \lambda^{(b)}
-\frac{1}{4} \psi_\mu \sigma^\mu \overline{\chi}^i
\overline{\lambda}^{(a)} \overline{\lambda}^{(b)} \rlp
\nonumber \\ &&
-e e^{K/2} \lmp
W^* \psi_\mu \sigma^{\mu\nu} \psi_\nu
+ W \overline{\psi}_\mu \bar{\sigma}^{\mu\nu} \overline{\psi}_\nu \rmp
\nonumber \\ &&
+ \frac{i}{2} \sqrt{2} e e^{K/2} \lmp
D_i W \chi^i \sigma^{\mu} \overline{\psi}_\mu
+ D_{i^*} W^* \overline{\chi}^i \bar{\sigma}^{\mu} \psi_\mu \rmp
\nonumber \\ &&
-\frac{1}{2} e e^{K/2} \lmp
{\cal D}_i D_j W \chi^i \chi^j
+ {\cal D}_{i^*} D_{j^*} W^* \overline{\chi}^i \overline{\chi}^j \rmp
\nonumber \\ &&
+ \frac{1}{4} e e^{K/2} g^{ij^*} \lmp
D_{j^*} W^* \partial_i f_{(ab)}
\lambda^{(a)} \lambda^{(b)}
+ D_i W \partial_{j^*} f^*_{(ab)}
\overline{\lambda}^{(a)} \overline{\lambda}^{(b)} \rmp
\nonumber \\ &&
-e e^K \llp
g^{ij^*} \lsp D_i W \rsp \lsp D_{j^*} W^* \rsp - 3 W^* W \rlp,
\label{L_SUGRA}
\eeq
where $f^R\equiv{\rm Re}f$ and $f^I\equiv{\rm Im}f$. Indices
$i,j,\cdot\cdot\cdot$ represent species of chiral multiplets, and $(a),
(b), \cdot\cdot\cdot$ are indices for adjoint representation of gauge
group (with gauge coupling constant $g$ and structure constant
$f^{abc}$) which are raised and lowered with $f^R_{(ab)}$ and its
inverse. Notice that the K\"ahler potential $K$ and the superpotential
$W$ in the total lagrangian (\ref{L_SUGRA}) can be arranged into the
following form;
\beq
G \equiv K + \ln \lsp W^* W \rsp.
\label{G}
\eeq

Covariant derivatives are defined as
\beq
\tilde{\cal D}_\mu \phi^i &\equiv&
\partial_\mu \phi^i
- g A_\mu^{(a)} X^i_{(a)},
\\
\tilde{\cal D}_\mu \chi^i &\equiv&
\partial_\mu \chi^i + \frac{1}{2} {\omega_{\mu}}^{ab}\sigma_{ab} \chi^i
+\Gamma^i_{jk} \tilde{\cal D}_\mu \phi^j \chi^k
- g A_\mu^{(a)} \frac{\partial X^i_{(a)}}{\partial \phi^j}\chi^j
\nonumber \\ &&
- \frac{1}{4} \lsp K_j \tilde{\cal D}_\mu \phi^j -
K_{j^*} \tilde{\cal D}_\mu \phi^{*j} \rsp \chi^i
-\frac{i}{2} g A_\mu^{(a)} {\rm Im} F_{(a)} \chi^i,
\\
\tilde{\cal D}_\mu \lambda^{(a)} &\equiv&
\partial_\mu \lambda^{(a)}
+ \frac{1}{2} {\omega_{\mu}}^{ab}\sigma_{ab} \lambda^{(a)}
- g f^{abc} A_\mu^{(b)} \lambda^{(c)}
\nonumber \\ &&
+ \frac{1}{4} \lsp K_j \tilde{\cal D}_\mu \phi^j -
K_{j^*} \tilde{\cal D}_\mu \phi^{*j} \rsp \lambda^{(a)}
+\frac{i}{2} g A_\mu^{(b)} {\rm Im} F_{(b)} \lambda^{(a)},
\\
\tilde{\cal D}_\mu \psi_\nu &\equiv&
\partial_\mu \psi_\nu
+ \frac{1}{2} {\omega_{\mu}}^{ab}\sigma_{ab} \psi_\nu
\nonumber \\ &&
+ \frac{1}{4} \lsp K_j \tilde{\cal D}_\mu \phi^j -
K_{j^*} \tilde{\cal D}_\mu \phi^{*j} \rsp \psi_\nu
+\frac{i}{2} g A_\mu^{(b)} {\rm Im} F_{(b)} \psi_\nu.
\eeq
$F_{\mu\nu}^{(a)}$ is the field strength tensor for the gauge boson
$A_\mu^{(a)}$, and $\hat{F}_{\mu\nu}^{(a)}$ is defined as
\beq
\hat{F}_{\mu\nu}^{(a)} \equiv
F_{\mu\nu}^{(a)} - \frac{i}{2} \lsp
\psi_\mu \sigma_\nu \overline{\lambda}^{(a)}
+ \overline{\psi}_\mu \bar{\sigma}_\nu \lambda^{(a)}
+ \psi_\nu \sigma_\mu \overline{\lambda}^{(a)}
+ \overline{\psi}_\nu \bar{\sigma}_\mu \lambda^{(a)} \rsp.
\eeq

Differentiation by the scalar field $\phi^i$ is symbolically represented
by the index $i$;
\beq
(\cdot\cdot\cdot)_i \equiv
\partial_i (\cdot\cdot\cdot) \equiv
\frac{\partial (\cdot\cdot\cdot)}{\partial \phi^i},~~~
(\cdot\cdot\cdot)_{i^*} \equiv
\partial_{i^*} (\cdot\cdot\cdot) \equiv
\frac{\partial (\cdot\cdot\cdot)}{\partial \phi^{*i}}.
\label{deriv_i}
\eeq
With the help of eq.(\ref{deriv_i}), ``derivatives'' of the
superpotential are defined as
\beq
D_i W &\equiv& W_i + K_i W,
\\
{\cal D}_i D_j W &\equiv& W_{ij} + K_{ij} W
+ K_i D_j W + K_j D_i W - K_i K_j W - \Gamma^k_{ij}D_k W.
\eeq
``Metric of the K\"ahler manifold'' $g_{ij^*}$ is defined by varying the
K\"ahler potential $K$ by the scalar fields $\phi^i$ and $\phi^{*j}$;
\beq
g_{ij^*} \equiv \frac{\partial^2 K}{\partial \phi^i \partial \phi^{*j}},
\eeq
and $g^{ij^*}$ is its inverse,
\beq
g_{ij^*} g^{ik^*} = \delta_{j^*}^{k^*},
{}~~~g_{ij^*} g^{kj^*} = \delta_i^k.
\eeq
{}From this metric, ``connection'' $\Gamma^k_{ij}$ and ``curvature''
$R_{ij^*kl^*}$ is given by
\beq
\Gamma^k_{ij} &\equiv&
g^{kl^*} \frac{\partial}{\partial \phi^i} g_{jl^*},
\label{gamma_K} \\
R_{ij^*kl^*} &\equiv&
\frac{\partial}{\partial \phi^i} \frac{\partial}{\partial \phi^{j^*}}
g_{kl^*}
-
g^{mn^*}
\lsp \frac{\partial}{\partial \phi^{j^*}} g_{ml^*} \rsp
\lsp \frac{\partial}{\partial \phi^i} g_{kn^*} \rsp.
\eeq
By using the connection given in eq.(\ref{gamma_K}), ``covariant
derivative'' $\nabla_i$ is defined as
\beq
\nabla_i V_j \equiv \frac{\partial}{\partial \phi^i} V_j
- \Gamma^k_{ij} V_k.
\eeq
(Here, $V_i$ is a function of scalar fields with index $i$.)

Next, we will comment on $X^{(a)i}$ and $D^{(a)}$. $X^{(a)i}(\phi)$ is
the Killing vector associated with K\"ahler metric $g_{ij^*}$. That is,
with the field transformation
\beq
\phi^i &\rightarrow& \phi^{i\prime} = \phi^i + X^{(a)i}(\phi) \epsilon,
\\
\phi^{i*} &\rightarrow&
\phi^{i*\prime} = \phi^{i*} + X^{*(a)i}(\phi^*) \epsilon,
\eeq
(where $\epsilon$ is a infinitesimal parameter), the Lie derivative
$\delta^{(L)}_X$ of $g_{ij^*}$ vanishes;
\beq
\delta^{(L)}_X g_{\tilde{i}\tilde{j}} &\equiv&
X^{(a)\tilde{k}} \frac{\partial}{\partial \phi^{\tilde{k}}}
g_{\tilde{i}\tilde{j}} +
g_{\tilde{i}\tilde{k}} \frac{\partial}{\partial \phi^{\tilde{j}}}
X^{(a)\tilde{k}} +
g_{\tilde{j}\tilde{k}} \frac{\partial}{\partial \phi^{\tilde{i}}}
X^{(a)\tilde{k}}
\nonumber \\
&=&
\nabla_{\tilde{i}} X^{(a)}_{\tilde{j}} +
\nabla_{\tilde{j}} X^{(a)}_{\tilde{i}}
\nonumber \\
&=& 0,
\label{killing_cond}
\eeq
where $X^{(a)}_{\tilde{i}}\equiv g_{\tilde{i}\tilde{j}}
X^{(a)\tilde{j}}$, and the index~$\tilde{i}$ represents both $i$ and
$i^*$.  From the above equation, we obtain two equations for the Killing
vectors $X^{(a)i}(\phi)$ and $X^{*(a)i}(\phi^*)$;
\beq
\nabla_i X^{(a)}_j + \nabla_j X^{(a)}_i &=& 0,
\\
\nabla_i X^{(a)}_{j^*} + \nabla_{j^*} X^{(a)}_i &=& 0.
\label{kill_eq}
\eeq
The former equation is automatically satisfied, while the latter
allows one to write down the Killing vectors $X^{(a)i}(\phi)$ and
$X^{*(a)i}(\phi^*)$ as derivatives of some function $D^{(a)}$ (which is
called Killing potential);
\beq
X^{(a)i}(\phi) &=& -i g^{ij^*} \frac{\partial}{\partial \phi^{*j}}
D^{(a)},
\label{kill_pot} \\
X^{*(a)i}(\phi^*) &=& i g^{ij^*} \frac{\partial}{\partial \phi^i}
D^{(a)}.
\label{kill_pot*}
\eeq
By solving eq.(\ref{kill_eq}), eq.(\ref{kill_pot}) and
eq.(\ref{kill_pot*}), the Killing vectors $X^{(a)i}(\phi)$,
$X^{*(a)i}(\phi^*)$, and the Killing potential $D^{(a)}$ can be
obtained.  $F^{(a)}$, which is a analytic function of $\phi^i$, is
defined as
\beq
F^{(a)} \equiv -i g^{ij^*} \frac{\partial D^{(a)}}{\partial \phi^{*j}}
\frac{\partial K}{\partial \phi^{i}} + i D^{(a)}.
\label{f^(a)}
\eeq

For example for the minimal K\"ahler potential $K^{min}=\phi^i
\phi^{*i}$, the Killing vectors and the
Killing potential take the following forms;
\beq
X^{(a)i} (\phi) &=& -i T^a_{ij} \phi^j,
\\
X^{*(a)i} (\phi^*) &=& i \phi^{*j} T^a_{ji},
\\
D^{(a)} &=& \phi^{*i} T^a_{ij} \phi^j,
\label{Dminimal}
\eeq
where $T^a_{ij}$ is a generator of gauged Lie group.

For the local SUSY transformation, variations of each component field
are given by
\beq
\delta {e_\mu}^a &=& - i \lsp \xi \sigma^a \overline{\psi}_\mu
+ \overline{\xi} \bar{\sigma}^a \psi_\mu \rsp,
\\
\delta \phi^i &=& \sqrt{2} \xi \chi^i,
\\
\delta \chi^i &=&
i \sqrt{2} \sigma^\mu
\lsp \tilde{\cal D}_\mu \phi^i - \frac{1}{\sqrt{2}} \psi_\mu \chi^i \rsp
- \Gamma^i_{jk} \delta \phi^j \chi^k
\nonumber \\ &&
+ \frac{1}{4} \lsp K_j \delta\phi^j - K_{j^*} \delta\phi^{j^*} \rsp
\chi^i
- \sqrt{2} F^i  \xi
\nonumber \\ &&
+ \frac{1}{2\sqrt{2}} \xi g^{ij^*} \partial_{j^*} f^*_{(ab)}
\overline{\lambda}^{(a)} \overline{\lambda}^{(b)},
\label{dchi_full} \\
\delta A_\mu^{(a)} &=& i \lsp \xi \sigma_\mu \overline{\lambda}^{(a)}
+ \overline{\xi} \bar{\sigma}_\mu \lambda^{(a)} \rsp,
\\
\delta \lambda^{(a)} &=&
\hat{F}_{\mu\nu}^{(a)} \sigma^{\mu\nu} \xi
- \frac{1}{4} \lsp K_j \delta\phi^j - K_{j^*} \delta\phi^{j^*} \rsp
\lambda^{(a)}
- i g D^{(a)} \xi
\nonumber \\ &&
+ \frac{1}{2\sqrt{2}} \xi f^{R(ab)-1} \partial_i f_{(bc)}
\chi^i \lambda^{(c)}
- \frac{1}{2\sqrt{2}} \xi f^{R(ab)-1} \partial_{i^*} f_{(bc)}^*
\overline{\chi}^i \overline{\lambda}^{(c)},
\label{dlambda_full}\\
\delta \psi_\mu &=&
2 \tilde{\cal D}_\mu \xi
- \frac{i}{2} \sigma_{\mu\nu} \xi
g_{ij^*} \chi^i \sigma^\nu \overline{\chi}^j
+ \frac{i}{2} \lsp {e_\mu}^a e_{\nu a} + \sigma_{\mu\nu} \rsp \xi
\lambda_{(a)} \sigma^\nu \overline{\lambda}^{(a)}
\nonumber \\ &&
- \frac{1}{4} \lsp K_j \delta\phi^j - K_{j^*} \delta\phi^{j^*} \rsp
\psi_\mu
+ i e^{K/2} W \sigma_\mu \overline{\xi}.
\eeq
with
\beq
F^i \equiv e^{K/2} g^{ij^*} D_{j^*} W^*.
\label{aux_F}
\eeq
Notice that $F^i$ given in eq.(\ref{aux_F}) corresponds to the auxiliary
field in chiral multiplet in the global SUSY case. (Do not confuse $F^i$
with $F^{(a)}$ defined in eq.(\ref{f^(a)}).)

As mentioned in the previous section, the total lagrangian
(\ref{L_SUGRA}) contains interaction terms between the gravitino field
$\psi_\mu$ and the supercurrent $J_\mu$ in the form of
eq.(\ref{J-psi_int}).  For a later convenience, we note the interaction
terms here;
\beq
{\cal L}_{\psi J} &=&
-\frac{1}{\sqrt{2}M} e g_{ij^*} \tilde{\cal D}_{\nu} \phi^{*j}
\chi^i \sigma^\mu \bar{\sigma}^\nu \psi_\mu
-\frac{1}{\sqrt{2}M} e g_{ij^*} \tilde{\cal D}_{\nu} \phi^{i}
\overline{\chi}^j \bar{\sigma}^\mu \sigma^\nu \overline{\psi}_\mu
\nonumber \\ &&
-\frac{i}{2M} e \lsp
\psi_\mu \sigma^{\nu\rho} \sigma^\mu \overline{\lambda}_{(a)}
+ \overline{\psi}_\mu \bar{\sigma}^{\nu\rho}
\bar{\sigma}^\mu \lambda_{(a)} \rsp
F_{\nu\rho}^{(a)},
\label{L_J-psi}
\eeq
where we have explicitly written down the $M$-dependence.

\section{Super-Higgs mechanism}

\hspace*{\parindent}
With the general supergravity lagrangian (\ref{L_SUGRA}), we now can
investigate the super-Higgs mechanism~\cite{NPB212-413} which is closely
related to the spontaneous breaking of the local SUSY. If the global
SUSY is broken spontaneously, massless Nambu-Goldstone fermion which is
called goldstino appears, while in spontaneously broken local SUSY
models, this goldstino component is absorbed by the gravitino through
the super-Higgs mechanism. In this section, we will discuss this
mechanism in more detail.  (As in the previous section, we take $M=1$
unit in this and the next sections.)

We will begin by considering the SUSY breaking in supergravity. In
global SUSY models, a spontaneous breaking occurs if some auxiliary
field, which is obtained by the SUSY transformation of some chiral
fermion $\chi$ or gauge fermion $\lambda$, receives non-vanishing
vacuum-expectation values. In the local SUSY case, we also use
$\delta\chi$ and $\delta\lambda$ as order parameters of the SUSY
breaking from the reasons mentioned below. As one can see in
eq.(\ref{dchi_full}) and eq.(\ref{dlambda_full}), the local SUSY
transformations of spin $\frac{1}{2}$ fermions contain terms which are
similar to the auxiliary fields obtained in the global case. Assuming
(local) Lorentz invariance of the ground state and no fermion-fermion
condensation (like gaugino-gaugino condensation), $\vev{\delta\chi}\neq
0$ and $\vev{\delta\lambda}\neq 0$ reduce to the following conditions;
\beq
&& \langle \delta\chi^i \rangle
= - \sqrt{2} \langle F^i \rangle \xi\neq 0,
\label{f_neq_0} \\
&& \langle \delta\lambda^{(a)} \rangle
= - i g \langle D^{(a)} \rangle \xi \neq 0.
\label{d_neq_0}
\eeq
As we will see later, if $\delta\chi$ or $\delta\lambda$ has
non-vanishing vacuum-expectation value, the gravitino, which is the
gauge field associated with the local SUSY, absorbs a massless
eigenstate of fermion mass matrix and acquires a non-vanishing mass
(which is called super-Higgs mechanism). Furthermore, the
vacuum-expectation value of $F^i$ or $D^{(a)}$ may provide a mass
splitting of bosonic and fermionic states (\ie ${\rm Str}{\bf M}^2 \neq
0$, see the next section).  These phenomena indicate a spontaneous
breaking of SUSY and hence in this thesis, we regard $\delta\chi$ and
$\delta\lambda$ as order parameters of SUSY even in local SUSY case.

Before investigating detail of the fermion mass matrix, we give some
comments. The first comment is on kinetic terms of chiral and vector
multiplets.  In supergravity lagrangian (\ref{L_SUGRA}), these kinetic
terms are characterized by the K\"ahler potential $K$ and the kinetic
function $f$, respectively. The kinetic terms are properly normalized if
$g_{ij^*}\equiv K_{ij^*}=\delta_{ij^*}+\cdot\cdot\cdot$, and
$f_{(ab)}=\delta_{ab}+\cdot\cdot\cdot$, where $\cdot\cdot\cdot$
represents the higher order terms which produce interactions of higher
dimensions. For simplicity, we ignore these higher order terms and use
the minimal K\"ahler potential $K^{min}$ and the minimal kinetic
function $f^{min}$;
\beq
K^{min}=\sum_i \phi^i\phi^{*i},~~~f_{(ab)}^{min}=\delta_{ab}.
\eeq

Another comment is on the cosmological constant. Since we are interested
in field theories in the Minkowski space-time, we require the
cosmological constant to vanish. In the supergravity models, the
vanishing cosmological constant is obtained if the following condition
is satisfied;
\beq
e^K \lmp
\lsp D_i W \rsp \lsp D_{i^*} W^* \rsp
- 3 W^* W \rmp
+ \frac{1}{2} g^2 D^{(a)} D^{(a)}=0.
\label{cc=0}
\eeq
If $e^{K/2}D_iW$ or $D^{(a)}$ has a non-vanishing vacuum-expectation
value, the superpotential $W$ should have a non-vanishing
vacuum-expectation value in order to satisfy the condition (\ref{cc=0}).
As we will see later, the gravitino mass is given by $e^{K/2}W$, and hence
the condition for the vanishing cosmological constant (\ref{cc=0})
requires non-vanishing gravitino mass.

Now, let us discuss the super-Higgs mechanism. We first write down the
quadratic part of the fermion terms without derivative in the lagrangian
(\ref{L_SUGRA}) by using $G$ defined in eq.(\ref{G});
\beq
{\cal L}_{\rm F}^{(2)} &=&
- e^{G/2}
\lsp \psi_\mu \sigma^{\mu\nu} \psi_\nu
+ \overline{\psi}_\mu \bar{\sigma}^{\mu\nu} \overline{\psi}_\nu \rsp
\nonumber \\ &&
+\frac{1}{2} g D^{(a)} \psi_\mu \sigma^{\mu} \overline{\lambda}^{(a)}
-\frac{1}{2} g D^{(a)} \overline{\psi}_\mu \bar{\sigma}^{\mu} \lambda^{(a)}
\nonumber \\ &&
- e^{G/2}
\lsp \frac{i}{\sqrt{2}} G_i \chi^i \sigma^\mu \overline{\psi}_\mu
+ \frac{i}{\sqrt{2}} G_{i^*} \overline{\chi}^i \bar{\sigma}^\mu \psi_\mu \rsp
\nonumber \\ &&
-e^{G/2}
\lmp \frac{1}{2} \lsp G_{ij} + G_i G_j \rsp \chi^i \chi^j
+ \frac{1}{2} \lsp G_{i^*j^*} + G_{i^*} G_{j^*} \rsp
\overline{\chi}^i \overline{\chi}^j \rmp
\nonumber \\ &&
+ \sqrt{2} g \lsp
-i D^{(a)}_i \chi^i \lambda^{(a)}
+i D^{(a)}_{i^*} \overline{\chi}^i \overline{\lambda}^{(a)} \rsp.
\eeq
Notice that if $e^{K/2}D_iW$ or $D^{(a)}$ has a non-vanishing
vacuum-expectation value, the gravitino field $\psi_\mu$ mixes with spin
$\frac{1}{2}$ fermion $\chi$ or $\lambda$. These mixing terms can be
eliminated by a shift of the gravitino field $\psi_\mu$;
\beq
{\cal L}_{\rm F}^{(2)} &=&
-e^{G/2}
\lsp \psi_\mu + \frac{1}{3} \overline{\eta} \bar{\sigma}_\mu \rsp
\sigma^{\mu\nu}
\lsp \psi_\nu - \frac{1}{3} \sigma_\nu \overline{\eta} \rsp
\nonumber \\ &&
-e^{G/2}
\lsp \frac{1}{2} G_{i^*j^*} + \frac{1}{6} G_{i^*} G_{j^*} \rsp
\overline{\chi}^i \overline{\chi}^j
\nonumber \\ &&
- \frac{1}{6} e^{-G/2}
g^2 D^{(a)} D^{(b)} \overline{\lambda}^{(a)} \overline{\lambda}^{(b)}
\nonumber \\ &&
+ i \lsp
\sqrt{2} g D^{(a)}_{i^*} - \frac{\sqrt{2}}{3} g G_{i^*}  D^{(a)} \rsp
\overline{\chi}^i \overline{\lambda}^{(a)} +h.c.~,
\eeq
with
\beq
\overline{\eta} \equiv  \frac{i}{\sqrt{2}} G_{i^*} \overline{\chi}^i
+ \frac{1}{2} e^{-G/2} g D^{(a)} \overline{\lambda}^{(a)}.
\label{def-eta}
\eeq

{}From the above lagrangian, we can read off the gravitino mass $\mgra$
as\footnote
{For simplicity, in the following arguments in this and the next
sections, we omit the bracket $\langle\hat{\cal O}\rangle$ which
represents the vacuum-expectation value of $\hat{\cal O}$.}
\beq
\mgra = e^{G/2} \equiv |e^{K/2} W|.
\eeq
On the other hand, masses of the spin $\frac{1}{2}$ fermions $(\chi^i,
\lambda^{(a)})$ can be obtained from the following mass matrix
\beq
\lsp
\begin{array} {cc}
m_{ij}^{(1/2)} & m_{ib}^{(1/2)} \\
m_{ja}^{(1/2)} & m_{ab}^{(1/2)}
\end{array}
\rsp ,
\label{fermion_mass}
\eeq
whose each components are given by
\beq
m_{ij}^{(1/2)} &\equiv& e^{G/2}
\lsp G_{ij} + \frac{1}{3} G_i G_j \rsp,
\label{m_ij}
\\
m_{ab}^{(1/2)} &\equiv& \frac{1}{3} e^{-G/2}
g^2 D^{(a)} D^{(b)},
\label{m_ab}
\\
m_{ia}^{(1/2)} &\equiv& i \sqrt{2}
\lsp
- g D^{(a)}_i + \frac{1}{3} g G_i D^{(a)} \rsp.
\label{m_ia}
\eeq
One can easily check that the fermionic field $\eta$ defined in
eq.(\ref{def-eta}) is a massless eigenstate of the mass matrix
(\ref{fermion_mass}) and hence it is a goldstino component. Therefore,
the gravitino field $\psi_\mu$ absorbs the goldstino component $\eta$
and acquires a non-vanishing mass $\mgra=e^{G/2}$.

\section{Super-trace formula in supergravity}

\hspace*{\parindent}
Having seen the structure of the mass matrix of fermionic sector in the
previous section, we will next investigate mass differences between
bosonic and fermionic states. For this purpose, we use the super-trace
of the mass-squared matrices
\beq
{\rm Str}{\bf M}^2 \equiv
\sum_{J=0}^{3/2} (-1)^J {\rm tr}{\bf M}_J^2,
\eeq
since it represents some information on the mass splitting of bosons and
fermions.

First, let us consider trace of the scalar mass matrix. In deriving
masses of scalar bosons (and other fields), we expand the scalar
potential $V$ at a stationary point in scalar field. From the
supergravity lagrangian (\ref{L_SUGRA}), scalar potential $V$ is given
by
\beq
V(\phi,\phi^*)= e^G \lsp G_i G_{i^*} - 3 \rsp
+ \frac{1}{2} g^2 D^{(a)} D^{(a)}.
\eeq
Stationary condition can be written as
\beq
0 &=& \frac{\partial V}{\partial \phi^j}
\nonumber \\
&=& G_j e^G \lsp G_i G_{i^*} - 3 \rsp
+ e^G \lsp G_{ij} G_{i^*} + G_{i} G_{i^*j} \rsp
+ g^2 D^{(a)} D^{(a)}_j.
\eeq
Multiplying this equation by $G_{j^*}$, and using the identity
$D^{(a)}_j G_{j^*}=D^{(a)}$, one can obtain
\beq
e^G G_j G_{j^*} \lsp G_i G_{i^*} - 3 \rsp
+ e^G \lsp G_{ij} G_{i^*} G_{j^*} + G_{i} G_{i^*} \rsp
+ g^2 D^{(a)} D^{(a)} = 0.
\label{stationary}
\eeq
Furthermore, we demand the cosmological constant to vanish at the
stationary point.  The condition for vanishing cosmological constant has
been derived in eq.(\ref{cc=0}), which becomes
\beq
e^G \lsp G_i G_{i^*} - 3 \rsp
+ \frac{1}{2} g^2 D^{(a)} D^{(a)} = 0.
\label{cc=02}
\eeq

The mass-squared matrix for scalar fields can be obtained from the
second derivative of the scalar potential $V$. Especially, its diagonal
part is given by
\beq
\frac{\partial^2 V}{\partial \phi^j \partial \phi^{*k}} &=&
e^G \biggl\{ G_j G_{k^*} \lsp G_i G_{i^*} - 3 \rsp
+ \delta_{jk^*} \lsp G_i G_{i^*} - 3 \rsp + G_{ij} G_{i^*k^*}
\nonumber \\ &&
+ G_j \lsp G_{k^*} + G_{i} G_{i^*k^*} \rsp
+ G_{k^*} \lsp G_{ij} G_{i^*} + G_j \rsp
+ \delta_{jk^*} \biggl\}
\nonumber \\ &&
+ g^2 \lsp D^{(a)}_{j} D^{(a)}_{k^*} + D^{(a)} D^{(a)}_{jk^*} \rsp.
\eeq
With the stationary condition (\ref{stationary}) and the condition for
vanishing cosmological constant (\ref{cc=02}), the trace of the scalar
mass matrix ${\bf M}^2_0$ can be obtained as
\beq
{\rm tr}{\bf M}^2_0 &=&
2 \frac{\partial^2 V}{\partial \phi^j \partial \phi^{*j}}
\nonumber \\
&=& \lsp - n_\phi -1 \rsp g^2 D^{(a)} D^{(a)}
- \frac {1}{2} e^{-G} \lsp g^2 D^{(a)} D^{(a)} \rsp^2
\nonumber \\ &&
+ 2 e^G \lsp G_{ij} G_{i^*j^*} + n_\phi \rsp
+ 2 g^2 \lsp D^{(a)}_{i} D^{(a)}_{i^*} + D^{(a)} D^{(a)}_{ii^*} \rsp,
\label{trM0}
\eeq
where $n_\phi \equiv \sum_i g_{ii^*}$ is the number of chiral
multiplets.

Next, we will consider the mass matrix for vector bosons. Mass terms
of vector bosons come from the covariant derivatives of scalar fields;
\beq
\frac{1}{2} m_A^{2(ab)} A_\mu^{(a)} A^{(b)\mu} =
g^2 \phi^i T^a_{ij} T^b_{jk} \phi^k A_\mu^{(a)} A^{(b)\mu}.
\eeq
{}From the above equation, one can easily read off the mass-squared matrix
for vector bosons, and its trace is given by
\beq
{\rm tr}{\bf M}^2_1 &=& 3 \times 2 g^2
\phi^{*i} T^a_{ij} T^a_{jk} \phi^k
\nonumber \\
&=& 6 g^2 D^{(a)}_{i} D^{(a)}_{i^*},
\label{trM1}
\eeq
where the Killing potential $D^{(a)}$ for the case of the minimal
K\"ahler potential is given in eq.(\ref{Dminimal}). Notice that the
prefactor 3 in the right-hand side of eq.(\ref{trM1}) corresponds to the
number of polarization of a massive vector boson.

The mass matrix of fermionic sector has been derived in the previous
section, and we can easily obtain the trace of it. Masses of spin
$\frac{1}{2}$ fermions are obtained from the matrix
(\ref{fermion_mass}), and the trace of the spin $\frac{1}{2}$ fermion
mass-squared matrix is given by
\beq
{\rm tr}{\bf M}_{1/2}^2 &=& 2 \times
\lsp \sum_{ij} \labs m_{ij}^{(1/2)} \rabs^2
+ \sum_{ab} \labs m_{ab}^{(1/2)} \rabs^2
+ 2 \sum_{ia} \labs m_{ia}^{(1/2)} \rabs^2 \rsp
\nonumber \\
&=& 2 e^G \lsp G_{ij} G_{i^*j^*} - 1 \rsp
- 2 g^2 D^{(a)} D^{(a)}
\nonumber \\ &&
- \frac{1}{2} e^{-G} \lsp g^2 D^{(a)} D^{(a)} \rsp^2
+ 8 g^2 D^{(a)}_{i} D^{(a)}_{i^*}.
\label{trM1/2}
\eeq

In supergravity, spin $\frac{3}{2}$ particle is only the gravitino, and
the trace of its mass matrix is given by
\beq
{\rm tr}{\bf M}_{3/2}^2 = 4 e^G = 4 \mgra^2.
\label{trM3/2}
\eeq

Combining eq.(\ref{trM0}), eq.(\ref{trM1}), eq.(\ref{trM1/2}) and
eq.(\ref{trM3/2}), the super-trace of the mass-squared matrices are
given by
\beq
{\rm Str}{\bf M}^2 = 2 \lsp n_\phi - 1 \rsp \mgra^2
- \lsp n_\phi - 1 \rsp g^2 D^{(a)} D^{(a)}
+ 2 g^2 D^{(a)} D^{(a)}_{ii^*}.
\label{supertrace_result}
\eeq
In SUSY models with $e^{K/2}D_iW\neq 0$ and $D^{(a)}=0$ (\ie if the SUSY
is broken by the $F$-term condensation), the SUSY breaking is
characterized by gravitino mass $\mgra$, and scalar particles are
expected to become heavier than their superparticles. This is
phenomenologically favorable.  In the next section, we will see an
example for such models.

\section{Model with Polonyi's superpotential}

\hspace*{\parindent}
In constructing a phenomenologically acceptable model based on
supergravity, we mostly consider a model which contains two sectors; a
so-called hidden sector responsible for the spontaneous breaking of
SUSY, and an observable sector which contains ordinary particles such as
quarks, leptons, Higgses, gauge fields, and their superpartners. The
strength of interactions between the hidden and the observable sectors
are at the same order of gravitational one, and hence these two sectors
couple very weakly for each other. In this section, we will investigate
a simple model for the hidden sector proposed by Polonyi~\cite{Polonyi}.
As we will see below, this model is simple but very suggestive.

We first discuss the hidden sector. The simplest hidden sector, proposed
by Polonyi, contains only one chiral multiplet $(\phi_P, \chi_P)$, which
takes the following superpotential $W_{\rm P}$;
\beq
W_{\rm P} = \mu^2 \lsp \phi_P + \omega \rsp,
\label{W_P}
\eeq
where $\mu$ and $\omega$ are the free parameters which we will determine
by the phenomenological requirements. For the K\"ahler potential for the
Polonyi field $\phi_P$, we use the minimum one, \ie $K=\phi^*_P\phi_P$.
In this model, the order parameter of SUSY breaking is given by
\beq
F_P =
\mu^2 \lmp \frac{\phi_P^*}{M^2} \lsp \phi_P + \omega \rsp + 1 \rmp
e^{\phi_P\phi_P^*/2M^2}.
\eeq
For the case $|\omega|<2M$, solution to the equation $F_P=0$ does not
exist, and the SUSY is expected to be spontaneously broken. The scalar
potential for the Polonyi field $\phi_P$ is obtained as
\beq
V \lsp \phi_P \rsp =
\mu^4 \lmp \labs \frac{1}{M^2} \phi_P^* \lsp \phi_P + \omega \rsp + 1
\rabs^2
- \frac{3}{M^2} \labs \phi_P + \omega \rabs^2 \rmp
e^{\phi_P\phi_P^*/M^2}.
\label{V-of-phi_P}
\eeq
The minimum of this potential is given by the solution to the following
equation;
\beq
0 &=&
\frac{\partial V}{\partial \phi_P}
\nonumber \\
&=&
\frac{\mu^4}{M^2}
\Biggl[ \lsp \phi_P + \omega \rsp
\lmp \frac{1}{M^2} \phi_P^* \lsp \phi_P + \omega \rsp + 1 \rmp
\nonumber \\ &&
+ \phi_P
\lmp \frac{1}{M^2} \phi_P \lsp \phi_P^* + \omega \rsp + 1 \rmp
- 3 \lsp \phi_P^* + \omega \rsp \Biggr]
e^{\phi_P\phi_P^*/M^2}
\nonumber \\ &&
+
\frac{\mu^4}{M^2}
\lmp \labs \frac{1}{M^2} \phi_P^* \lsp \phi_P + \omega \rsp + 1 \rabs^2
- \frac{3}{M^2} \labs \phi_P + \omega \rabs^2 \rmp
\phi_P^* e^{\phi_P\phi_P^*/M^2}.
\label{stationary_P}
\eeq
Furthermore, since we are interested in field theories in flat
space-time, we demand the cosmological constant to vanish at the
potential minimum;
\beq
\vev{V} =
\vev{
\mu^4 \lmp \labs \frac{1}{M^2} \phi_P^* \lsp \phi_P + \omega \rsp + 1
\rabs^2
- \frac{3}{M^2} \labs \phi_P + \omega \rabs^2 \rmp
e^{\phi_P\phi_P^*/M^2}
} =0.
\label{cc=0_P}
\eeq
The parameter $\omega$ is chosen so that eq.(\ref{stationary_P}) and
eq.(\ref{cc=0_P}) have a solution simultaneously. Then $\omega$ is
determined to be $\omega = \pm (2-\sqrt{3}) M$.\footnote
{In fact, even for $\omega = \pm (2+\sqrt{3}) M$,
eq.(\ref{stationary_P}) and eq.(\ref{cc=0_P}) have a solution
simultaneously. In this case, however, the solution does not correspond
to the absolute minimum of the potential $V$ and the potential at the
true minimum becomes negative.  Therefore, we do not consider this case.}
Hereafter, we use the branch $\omega = (2-\sqrt{3}) M$.

By solving eq.(\ref{stationary_P}), vacuum-expectation value of the
Polonyi field $\phi_P$ is given by
\beq
\vev{\phi_P} = \lsp \sqrt{3} - 1 \rsp M,
\eeq
and one can obtain the vacuum-expectation value of the superpotential
$W_{\rm P}$ and the order parameter $F_P$;
\beq
\vev{W_{\rm P}} &=& \mu^2 M,
\label{vev-W_P} \\
\vev{F_P} &=& \sqrt{3} e^{2-\sqrt{3}} \mu^2 .
\label{vev-F_P}
\eeq
{}From the vacuum-expectation value of $W_{\rm P}$, the gravitino acquires
a non-vanishing mass
\beq
\mgra = e^{2-\sqrt{3}} \frac{\mu^2}{M}.
\eeq
Notice that for the given gravitino mass $\mgra$, the scale of the
condensation of $F_P$ (which is the same order of $\mu^2$) is
determined; $\vev{F_P}\sim O(\mgra M)$.

The mass eigenvalues $m_{\phi_{P1}}$ and $m_{\phi_{P2}}$ for the scalar
fields are obtained by expanding Polonyi field $\phi_P$ around its
vacuum-expectation value;
\beq
\phi_P = (\sqrt{3}-1)M +
\frac{1}{\sqrt2}\lsp \phi_{P1} + i\phi_{P2}\rsp .
\eeq
Substitute this to the potential (\ref{V-of-phi_P}), the masses are
given by
\beq
m_{\phi_{P1}}^2 = 2\sqrt{3} \mgra^2,~~~
m_{\phi_{P2}}^2 = (4 - 2\sqrt{3}) \mgra^2,
\eeq
while the superpartner of $\phi_P$ is a goldstino and absorbed in
the gravitino.

Now, let us consider the coupling between hidden and observable
sectors. The hidden sector couples to the observable sector if we
introduce a superpotential $W_{\rm obs}$ for the observable sector;
\beq
W=W_{\rm P}+W_{\rm obs}(\phi_{\rm obs}),
\eeq
where we denote the particles in the observable sector by $\phi_{\rm
obs}$.  Assuming the minimal K\"ahler potential for $\phi_{\rm obs}$,
the scalar potential is given by
\beq
V \lsp \phi_P , \phi_{\rm obs} \rsp &=&
\exp \lsp \frac{\labs \phi_P \rabs^2
+ \sum_i \labs \phi_{\rm obs}^i \rabs^2}{M^2} \rsp
\nonumber \\ &&
\times
\lsp \labs \frac{\phi_P^*}{M^2} W +
\frac{\partial W_{\rm P}}{\partial\phi_P} \rabs^2
+ \sum_i \labs \frac{\phi_{\rm obs}^{i*}}{M^2} W +
\frac{\partial W_{\rm obs}}{\partial\phi_{\rm obs}^i} \rabs^2 -
\frac{3}{M^2} |W|^2 \rsp .
\label{V_P_tot}
\eeq
In the potential (\ref{V_P_tot}), all interactions between the hidden
and the observable sectors are suppressed by powers of $M^{-1}$.

In order to derive a low energy effective theory for the observable
sector, we take a limit $M\rightarrow \infty$ with gravitino mass $\mgra
=\langle e^{K/2M^2}W/M^2\rangle$ fixed. (This is sometimes called flat
limit.)  Then, one can obtain a potential for the observable sector;
\beq
V(\phi_{\rm obs}) &\simeq&
\sum_i \labs
\frac{\partial \tilde{W}_{\rm obs}}{\partial\phi_{\rm obs}^i} \rabs^2
+ \mgra^2 \sum_i \labs \phi_{\rm obs}^i \rabs^2
\nonumber \\ &&
+
\mgra \llp
\sum_i \phi_{\rm obs}^i
\frac{\partial \tilde{W}_{\rm obs}}{\partial\phi_{\rm obs}^i}
+ \lmp b^* (a+b) -3 \rmp \tilde{W}_{\rm obs} + h.c. \rlp,
\label{flat_lim}
\eeq
where
\beq
&& \tilde{W}_{\rm obs}\equiv\vev{e^{K/2M^2}}W_{\rm obs},
\\
&& a^* \equiv
\vev{\frac{(\partial W_{\rm P}/\partial \phi_P)M}{W_{\rm P}}}=1,
\\
&& b \equiv \vev{\frac{\phi_P}{M}}=\sqrt{3}-1.
\label{V_flat}
\eeq
The above potential is nothing but a potential for the softly broken
(global) SUSY. The first term in eq.(\ref{V_flat}) is a scalar potential
in a global supersymmetric model, while the rest can be regarded as soft
SUSY breaking terms for scalar bosons. Especially, in this model all the
scalars in the observable sector receive an universal SUSY breaking mass
which is fixed by gravitino mass $\mgra$.

In order to keep the masses of squarks and sleptons at
$O(100\GEV-1\TEV)$ so that the hierarchy problem is solved, the
gravitino mass $\mgra$ is taken smaller than $O(1\TEV)$. From this
phenomenological requirement, $\vev{F_P}$ is determined;
\beq
\vev{F_P} \sim ( 10^{11} - 10^{12} ) \GEV,
\eeq
provided $\mgra\sim O(100\GEV-1\TEV)$.

Finally we will comment on gauge fermion mass. In a model with the
minimal kinetic function for gauge multiplet; $f_{(ab)}=\delta_{(ab)}$,
gauge fermion remains massless. To generate a non-vanishing gauge fermion
mass term, we assume a non-minimal kinetic function; $(\partial
f_{(ab)}/ \partial\phi_P )\neq 0$. Then with a SUSY breaking in the
hidden sector, gauge fermion acquires mass from the interaction in the
total supergravity lagrangian (\ref{L_SUGRA});
\beq
{\cal L}_{\lambda\lambda} =
\frac{1}{4} \vev{F_P \frac{\partial f_{(ab)}}{\partial\phi_P}}
\lambda^{(a)} \lambda^{(b)} +h.c.
\eeq
The simplest example for the non-minimal kinetic function is
\beq
f_{(ab)} = \delta_{(ab)}
\lmp 1 + \frac{\kappa}{M} \lsp \phi_P - \vev{\phi_P} \rsp \rmp,
\eeq
where $\kappa$ is a dimensionless coupling parameter. Then, the gauge
fermion mass is given by $\kappa\vev{F_P}/2M$, which is of the order of
the gravitino mass $\mgra$ for $\kappa\sim O(1)$.

\section{Mass of a scalar field in the hidden sector}

\hspace*{\parindent}
As we have seen in the previous section, masses of the Polonyi fields
$\phi_{P1}$ and $\phi_{P2}$ are of the order of the gravitino mass
$\mgra$. In fact, this is not an accidental case, \ie in supergravity
there exists a scalar field with mass of order $\mgra$ in a wide class
of models with the following features.
\begin{itemize}
\item At the stationary point of the scalar potential $V$, cosmological
the constant vanishes.
\item SUSY is broken by a condensation of a $F$-term in the hidden
sector. (In this case, the vacuum-expectation value of $\langle
F\rangle$ is $O(\mgra M)$ so that the cosmological constant vanishes.)
\item Cut-off scale of the model is of the order of the gravitational
scale $M$.
\end{itemize}
As we will see below, a scalar boson with mass of order $\mgra$ exists
in a model with these conditions.

The existence of such a scalar field can be seen by using only simple
order estimations. Assuming the cosmological constant to vanish, the
condensation of $F$-terms are constrained as\footnote
{In order to obtain a correct normalization of kinetic terms, we take
$\langle G_{ij^*}\rangle = \delta_{ij^*}$.}
\beq
\langle F^i \rangle \lsim O(\langle e^{G/2M^2} W/M\rangle )
\sim O(\mgra M)~~~({\rm for~all}~i).
\eeq
Combining this condition with the stability condition $\langle\partial
V/\partial\phi\rangle = 0$, we get~\cite{PRD49-779,TU462}
\beq
\vev{Me^{G/2M^2}G_{ij}F^j} \lsim O(\mgra^2 M),
\label{V'=0&cc=0}
\eeq
where we have used that the cut-off scale of the model is of order $M$,
\ie $\langle G_{ijk^*}\rangle\lsim M^{-1}$.

The mass squared matrix of the scalar fields are given by the second
derivative of the scalar potential $V$. Especially, the diagonal part is
given by
\beq
\vev{\frac{\partial^2 V}{\partial\phi^i\partial\phi^{j^*}}} &=&
\langle Me^{G/2M^2}G_{ik} \rangle \langle g^{kl^*}\rangle
\langle Me^{G/2M^2}G_{l^*j^*}\rangle
\nonumber \\ &&
- \lsp \langle Me^{G/2M^2}G_{ik} \rangle
\langle g^{kl^*}K_{l^*j^*m}\rangle \langle F^m\rangle + h.c. \rsp
\nonumber \\ &&
+ O(\mgra^2).
\eeq
Denoting the chiral multiplet which breaks SUSY $(\phi^X, \chi^X, F^X)$,
the vacuum expectation value of $F^X$ is given by $\langle F^X\rangle \sim
O(\mgra M)$. Then, constraint (\ref{V'=0&cc=0}) leads to
\beq
\vev{Me^{G/2M^2}G_{Xi}} \lsim O(\mgra),
\eeq
and hence
\beq
\vev{\frac{\partial^2 V}{\partial\phi^X\partial\phi^{X^*}}} \lsim
O(\mgra^2).
\eeq
{}From the above, we can conclude that the mass-squared matrix of scalar
fields has a eigenvalue of order $\mgra^2$ or less. We should note here
that such a scalar field may cause serious cosmological
difficulty~\cite{PLB131-59} (so-called Polonyi problem) which will be
discussed in Chapter~\ref{chap:overview}.

%
%

\chapter{Feynman rules for the gravitino}
\label{chap:feynman}

\hspace*{\parindent}
In the previous chapter, we have seen that the gravitino, which is the
gauge field associated with local SUSY invariance, plays a crucial role
in supergravity. In order to see physics of the gravitino, we must
discuss a field theory for it.

In supergravity, however, there exist non-renormalizable interactions
which are suppressed by $M^{-1}$ (or its higher power), and hence we
cannot calculate loop effects by using the full supergravity lagrangian
given in the previous chapter.  Furthermore, because of these
non-renormalizable interactions, the Born-unitarity breaks down at high
energy scales of order $M$.  Therefore, we regard the supergravity as a
low energy effective theory which is appropriate for energy scales much
below the gravitational one, and only consider tree level processes.

In this chapter, we derive Feynman rules for the massive gravitino field
by using the supergravity lagrangian (\ref{L_SUGRA}). For this purpose,
we first solve the field equation for massive Rarita-Schwinger field,
and then we discuss the quantization of the free gravitino field.
Hereafter, we only consider the nature of the gravitino in (nearly) flat
space-time, and hence we restrict the background metric in that of flat
space-time $g_{\mu\nu}={\rm diag}(1,-1,-1,-1)$.

\section{Four-component notation for fermions}

\hspace*{\parindent}
For our following arguments, it is more convenient to use a
four-component spinor for a fermionic field rather than two-component
one used in the previous chapter. Therefore, we briefly introduce it
before quantizing the gravitino field. (For our notations and
conventions, see Appendix~\ref{ap:notation}).

The four-component spinors $\psi$ (in chiral representation) can be
constructed from two component spinors $\xi$ and $\eta$;
\beq
\psi \sim
\lsp \begin{array}{c}
\xi_\alpha \\ \overline{\eta}^{\dot{\alpha}}
\end{array} \rsp,~~~
\overline{\psi} \sim \lsp \eta^\alpha, \overline{\xi}_{\dot{\alpha}} \rsp.
\eeq
Corresponding to this, $\gamma$-matrix consists of $2\times 2$ matrices
$\sigma^\mu$ and $\bar{\sigma}^\mu$;
\beq
\gamma^\mu =
\lsp \begin{array}{cc}
0 & \bar{\sigma}^\mu \\ \sigma^\mu & 0
\end{array} \rsp.
\eeq

In the above representation, we define four-component spinors for the
gravitino, $\psi_\mu^{(M)}$, gaugino, $\lambda^{(M)}$, and chiral
fermion, $\chi_R^{(D)}$;
\beq
\psi_\mu^{(M)} \equiv
\lsp \begin{array}{c}
\psi_\mu \\ \overline{\psi}_\mu
\end{array} \rsp &,~~~&
\overline{\psi}_{\mu}^{(M)} \equiv \lsp \psi_\mu, \overline{\psi}_\mu \rsp,
\\
\lambda^{(M)} \equiv
\lsp \begin{array}{c}
\lambda \\ \overline{\lambda}
\end{array} \rsp &,~~~&
\overline{\lambda}^{(M)} \equiv \lsp \lambda, \overline{\lambda} \rsp,
\\
\chi^{(D)}_R \equiv
\lsp \begin{array}{c}
\chi \\ 0
\end{array} \rsp &,~~~&
\overline{\chi}_L^{(D)} \equiv \lsp 0, \overline{\chi} \rsp.
\eeq
Notice that $\psi_\mu^{(M)}$ and $\lambda^{(M)}$ satisfies the Majorana
condition
\beq
\psi_\mu^{(M)} = \psi_\mu^{(M)C} \equiv C \overline{\psi}_\mu^{(M)T},~~~
\lambda^{(M)} = \lambda^{(M)C} \equiv C \overline{\lambda}^{(M)T},
\eeq
where $C$ is the charge conjugation matrix (see
Appendix~\ref{ap:notation}), while $\chi_R^{(D)}$ is chiral fermion with
a definite chirality;
\beq
\frac{1}{2} \lsp 1+\gamma_5 \rsp \chi_R^{(D)} = \chi_R^{(D)},~~~
\frac{1}{2} \lsp 1-\gamma_5 \rsp \chi_R^{(D)} = 0,
\eeq
where $\gamma_5 \equiv i \gamma^0 \gamma^1 \gamma^2 \gamma^3$.

\section{Wave function for massive gravitino}

\hspace*{\parindent}
We will start by discussing the wave function for massive gravitino. The
lagrangian for the free gravitino field is given in the full
supergravity lagrangian (\ref{L_SUGRA}). In the four-component notation,
it can be written as
\beq
{\cal L}_{\rm MRS} = -\frac{1}{2} \epsilon^{\mu\nu\rho\sigma}
\overline{\psi}_\mu \gamma_5 \gamma_\nu \partial_\rho \psi_\sigma
- \frac{1}{4} \mgra
\overline{\psi}_\mu \llp \gamma^\mu , \gamma^\nu \rlp \psi_\nu,
\label{L_MRS}
\eeq
where we have dropped the index ($M$) for the gravitino field, for
simplicity. (Hereafter, we use only the four-component notation
otherwise mentioned, and drop indices ($M$) and ($D$) in order to avoid
complications due to too many indices.) By using the Majorana condition
for the gravitino field; $\psi_\mu = C\overline{\psi}_\mu^T$, the
lagrangian (\ref{L_MRS}) becomes
\beq
{\cal L}_{\rm MRS} = \frac{1}{2} \epsilon^{\mu\nu\rho\sigma}
\psi^T_\mu C^{\dagger} \gamma_5 \gamma_\nu \partial_\rho \psi_\sigma
+ \frac{1}{4} \mgra
\psi^T_\mu C^{\dagger} \llp \gamma^\mu , \gamma^\nu \rlp \psi_\nu.
\label{L_MRS2}
\eeq
This lagrangian is our starting point.

Varying the above lagrangian for $\psi_\mu$, we can obtain the field
equation for the free gravitino field;
\beq
0 &=&
\lmp \lsp \frac{\partial}{\partial \psi_\mu} \rsp
-\partial_\nu
\lsp \frac{\partial}{\partial \lsp \partial_\nu \psi_\mu \rsp} \rsp \rmp
{\cal L}_{\rm MRS}
\nonumber \\
&=&
\epsilon^{\mu\nu\rho\sigma}
C^{\dagger} \gamma_5 \gamma_\nu \partial_\rho \psi_\sigma
+ \frac{1}{2} \mgra
C^{\dagger} \llp \gamma^\mu , \gamma^\nu \rlp \psi_\nu.
\label{eq_psi}
\eeq
{}From this, the following two equations are derived;
\beq
\mgra \lsp \dslash \gamma^\nu \psi_\nu
- \gamma^\nu \dslash \psi_\nu \rsp &=& 0,
\label{eq_psi1} \\
2 i \lsp \partial_\lambda \gamma^\mu \psi_\mu
- \dslash \psi_\lambda \rsp
+ \mgra \lsp \gamma_\lambda \gamma^\mu \psi_\mu
+ 2 \psi_\lambda \rsp &=& 0.
\label{eq_psi2}
\eeq
Notice that the former (latter) equation can be obtained by operating
$\partial_\mu$ ($\gamma_\lambda \gamma_\mu$) on eq.(\ref{eq_psi}). For
a later convenience, we derive one more equation by multiplying
eq.(\ref{eq_psi2}) by $\gamma^\lambda$;
\beq
i \lsp \dslash \gamma^\mu \psi_\mu
- \gamma^\lambda \dslash \psi_\lambda \rsp
+ 3 \mgra \gamma^\mu \psi_\mu = 0.
\label{eq_psi3}
\eeq
For the {\it massive} gravitino ($\mgra\neq 0$), eq.(\ref{eq_psi1}) --
eq.(\ref{eq_psi3}) yields the following simple equations;
\beq
\gamma^\mu \psi_\mu &=& 0,
\label{eq_g=0}
\\
\partial^\mu \psi_\mu &=& 0,
\label{eq_d=0}
\\
\lsp i \dslash - \mgra \rsp \psi_\mu &=& 0.
\label{eq_D=0}
\eeq

The solutions to the above equations can be constructed by using the
wave function $u$ for spin $\frac{1}{2}$ field and the polarization
vector $\epsilon_\mu$ for spin 1 field. In solving the field equations
for the gravitino field $\psi_\mu$, it is convenient to consider the
wave function $\tilde{\psi}_\mu$ in the momentum space, $\psi_\mu \sim
e^{-ipx}\tilde{\psi}_\mu$. Then, the solution to eq.(\ref{eq_g=0})
-- eq.(\ref{eq_D=0}) is given by~\cite{PR145-1152}
\beq
\tilde{\psi}_\mu ({\bf p},\lambda) =
\sum_{s,m}
\left\langle \lsp \frac{1}{2},\frac{s}{2} \rsp \lsp 1,m \rsp ~\right|
\left.  \lsp \frac{3}{2},\lambda \rsp \right\rangle
u({\bf p},s) \epsilon_\mu({\bf p},m),
\label{psi_p}
\eeq
where $\langle (\frac{1}{2},\frac{s}{2})(1,m)~|~(\frac{3}{2},\lambda)
\rangle$ is the Clebsch-Gordan coefficient, whose value is shown in
Table~\ref{table:CG_coef}.
%
%
\begin{table}[t]
\begin{center}
\begin{tabular}{r|ccc} \hline \hline
{}       & {$m=-1$}       & {$m=0$}        & {$m=+1$} \\ \hline
{$s=-1$} & {1}            & {$\sqrt{2/3}$} & {$\sqrt{1/3}$} \\
{$s=+1$} & {$\sqrt{1/3}$} & {$\sqrt{2/3}$} & {1} \\ \hline \hline
\end{tabular}
\caption
{Clebsch-Gordan coefficients for the case $\lambda =\frac{s}{2}+m$;
$\langle (\frac{1}{2},\frac{s}{2})(1,m) ~|~(\frac{3}{2},\frac{s}{2}+m)
\rangle$. Notice that the coefficient with $\lambda \neq \frac{s}{2}+m$
vanishes.}
\label{table:CG_coef}
\end{center}
\end{table}
%
%

The wave function for spin $\frac{1}{2}$ field $u({\bf p},s)$ is the
solution to the ordinary Dirac equation (in momentum space) with a
definite helicity ($s=\pm 1$);
\beq
\lsp \pslash - \mgra \rsp u({\bf p},s) &=& 0,
\label{dirac_eq} \\
\lsp {\bf n}{\bf \Sigma} \rsp u({\bf p},s) &=& s~u({\bf p},s),
\label{helicity}
\eeq
where ${\bf n}{\bf \Sigma}$ is the helicity operator (see
Appendix~\ref{ap:notation}). The explicit form of $u({\bf p},s)$ in the
Dirac representation is given in Appendix~\ref{ap:notation}, and we take
the following normalization condition on $u({\bf p},s)$;
\beq
\overline{u}({\bf p},s) u({\bf p},s^{'}) = 2 \mgra \delta_{ss^{'}}.
\label{norm_u}
\eeq

For the momentum vector $p^\mu = (E,~|{\bf p}|\sin\theta\cos\phi,~|{\bf
p}|\sin\theta\sin\phi,~|{\bf p}|\cos\theta)$ of a massive particle
($p_\mu p^\mu = \mgra^2 \neq 0$), the polarization vectors take the
following forms;
\beq
\epsilon_\mu({\bf p},1) &=& \frac{1}{\sqrt{2}}
(0, ~\cos\theta\cos\phi - i \sin\phi, ~\cos\theta\sin\phi + i \cos\phi,
{}~-\sin\theta ),
\label{pol_+} \\
\epsilon_\mu({\bf p},0) &=& \frac{1}{\mgra}
(|{\bf p}|, ~-E\sin\theta\cos\phi,
{}~-E\sin\theta\sin\phi, ~-E\cos\theta),
\label{pol_0} \\
\epsilon_\mu({\bf p},-1) &=& \frac{-1}{\sqrt{2}}
(0, ~\cos\theta\cos\phi + i \sin\phi,
{}~\cos\theta\sin\phi - i \cos\phi, ~-\sin\theta ),
\label{pol_-}
\eeq
which are normalized as
\beq
\epsilon_\mu^*({\bf p},m)  \epsilon^\mu({\bf p},m^{'})
= - \delta_{mm^{'}}.
\label{norm_e}
\eeq
Notice that the polarization vectors (\ref{pol_+}) -- (\ref{pol_-})
satisfy the following condition;
\beq
p^\mu \epsilon_\mu({\bf p},m) = p^\mu \epsilon_\mu^*({\bf p},m) = 0.
\label{pe=0}
\eeq

By using eq.(\ref{dirac_eq}), eq.(\ref{helicity}) and eq.(\ref{pe=0}),
one can easily check that $\tilde{\psi}_\mu$ defined in eq.(\ref{psi_p})
obeys the following equations;
\beq
\gamma^\mu \tilde{\psi}_\mu ({\bf p},\lambda) &=& 0,
\label{eq_g=0'}
\\
p^\mu \tilde{\psi}_\mu ({\bf p},\lambda) &=& 0,
\label{eq_d=0'}
\\
\lsp \pslash - \mgra \rsp \tilde{\psi}_\mu ({\bf p},\lambda)
&=& 0,
\label{eq_D=0'}
\eeq
and hence the wave function in the coordinate space $\psi_\mu\sim
e^{-ipx}\tilde{\psi}_\mu$ satisfies eq.(\ref{eq_g=0}) --
eq.(\ref{eq_D=0}).

The explicit form of $\tilde{\psi}_\mu({\bf p},\lambda)$ depends on
representations of the $\gamma$-matrix, and we do not write it down
because it is complicated but almost useless. Instead of that, we give
some useful identities for $\tilde{\psi}_\mu({\bf p},\lambda)$, which
hold irrespective of the representation of $\gamma$-matrix.
Normalization of $\tilde{\psi}_\mu ({\bf p},\lambda)$ is fixed by
eq.(\ref{norm_u}) and eq.(\ref{norm_e});
\beq
\overline{\tilde{\psi}}_\mu ({\bf p},\lambda)
\tilde{\psi}^\mu ({\bf p},\lambda^{'}) =
-2 \mgra \delta_{\lambda\lambda^{'}}.
\label{norm_psi}
\eeq
Furthermore, one can obtain the following identity which will be useful
in deriving the momentum operator given in eq.(\ref{momentum_op}) for
the gravitino field;
\beq
\overline{\tilde{\psi}}_\mu ({\bf p},\lambda) \gamma_\nu
\tilde{\psi}^\mu ({\bf p},\lambda^{'}) =
-2 p_\nu \delta_{\lambda\lambda^{'}}.
\eeq
For the helicity sum of the gravitino field, the following formula
exists;
\beq
P_{\mu\nu} ({\bf p}) &\equiv&
\sum_\lambda \tilde{\psi}_\nu ({\bf p},\lambda)
\overline{\tilde{\psi}}_\mu ({\bf p},\lambda)
\nonumber \\
&=&
-\lsp \pslash - \mgra \rsp
\nonumber \\ &&
\times
\lmp \lsp g_{\mu\nu} - \frac{p_\mu p_\nu}{\mgra^2} \rsp
-\frac{1}{3} \lsp g_{\mu\sigma} - \frac{p_\mu p_\sigma}{\mgra^2} \rsp
\lsp g_{\nu\lambda} - \frac{p_\nu p_\lambda}{\mgra^2} \rsp
\gamma^\sigma \gamma^\lambda \rmp.
\label{P_munu}
\eeq
$P_{\mu\nu}({\bf p})$ obeys the following equations which correspond to
eq.(\ref{eq_g=0'}) -- eq.(\ref{eq_D=0'});
\beq
&& \gamma^\mu P_{\mu\nu}({\bf p})
= P_{\mu\nu}({\bf p}) \gamma^\nu = 0,
\\ &&
p^\mu P_{\mu\nu}({\bf p}) = P_{\mu\nu}({\bf p}) p^\nu = 0,
\\ &&
\lsp \pslash - \mgra \rsp P_{\mu\nu}({\bf p}) =
P_{\mu\nu}({\bf p}) \lsp \pslash - \mgra \rsp = 0.
\eeq

General solution to eq.(\ref{eq_g=0}) -- eq.(\ref{eq_D=0}) can be
expanded by the mode function given in eq.(\ref{psi_p}) and its charge
conjugation;
\beq
\psi_\mu (x) = \int \frac{d^3{\bf p}}{(2\pi )^3 2p_0} \sum_\lambda
\lmp e^{i{\bf px}} \tilde{\psi}_\mu ({\bf p},\lambda)
a_{{\bf p}\lambda} (t)
+ e^{-i{\bf px}} \tilde{\psi}_\mu^{C} ({\bf p},\lambda)
a_{{\bf p}\lambda}^\dagger(t) \rmp,
\label{expansion}
\eeq
where $p_0 \equiv \sqrt{|{\bf p}|^2+\mgra^2}$, and $a_{{\bf
p}\lambda}(t)$ is the expansion coefficient whose time dependence is
determined by the field equation (\ref{eq_D=0}); $a_{{\bf
p}\lambda}(t)\sim e^{-ip_0t}$. Notice that because of the Majorana
nature of the gravitino field $\psi_\mu$, the coefficient of the
$\tilde{\psi}_\mu^{C}({\bf p},\lambda)$-term in eq.(\ref{expansion}) is
not an independent variable, but it is to be $a_{{\bf
p}\lambda}^\dagger(t)$. One can easily check that the gravitino field
$\psi_\mu$ in eq.(\ref{expansion}) satisfies the Majorana condition;
$\psi_\mu = \psi_\mu^C$.

\section{Quantization of free massive gravitino field}

\hspace*{\parindent}
In order to obtain Feynman rules for the gravitino, we have to quantize
first the free gravitino field. In this section, we discuss the
canonical quantization of the free gravitino field.  Since in this
thesis, we only consider spontaneously broken local SUSY models in which
the gravitino acquires a non-vanishing mass term, we only deal with the
case of the massive gravitino.

Constraints on the physical mode of the gravitino field are obtained in
the previous section, and they are given in eq.(\ref{eq_g=0}) and
(\ref{eq_d=0}). The mode expansion of the physical degrees of freedom is
given in eq.(\ref{expansion}), and as in the usual quantization
procedure, we regard the coefficients $a_{{\bf p}\lambda}(t)$'s and
$a_{{\bf p}\lambda}^{\dagger}(t)$'s as dynamical variables and derive
commutation relations among them.

As a first step in the quantization procedure, we derive canonical
momenta.  Time derivatives of $a_{{\bf p}\lambda}(t)$ and $a_{{\bf
p}\lambda}^{\dagger}(t)$, which we denote $\dot{a}_{{\bf p}\lambda}(t)$
and $\dot{a}_{{\bf p}\lambda}^{\dagger}(t)$, are present in the
lagrangian as
\beq
L_{\rm MRS} &\equiv& \int d^3 {\bf x} {\cal L}_{\rm MRS}
\nonumber \\
&=& \frac{1}{2} i
\int \frac{d^3{\bf p}}{(2\pi )^3 2p_0} \sum_\lambda
\lmp
a_{{\bf p}\lambda}^\dagger(t) \dot{a}_{{\bf p}\lambda}(t)
+ a_{{\bf p}\lambda}(t) \dot{a}_{{\bf p}\lambda}^\dagger(t)
\rmp
\nonumber \\ &&
+~({\rm terms~without}~\dot{a}_{{\bf p}\lambda},~
\dot{a}_{{\bf p}\lambda}^\dagger).
\label{l_by_a}
\eeq
Differentiating lagrangian (\ref{l_by_a}) with respect to $\dot{a}_{{\bf
p}\lambda}(t)$ or $\dot{a}_{{\bf p}\lambda}^\dagger(t)$, one can obtain
canonical momenta $\Pi_{{\bf p}\lambda}$ and $\overline{\Pi}_{{\bf
p}\lambda}$;
\beq
&& \Pi_{{\bf p}\lambda} \equiv
\frac{\delta L_{\rm MRS}}{\delta \dot{a}_{{\bf p}\lambda}}
= i \frac{1}{2} \frac{1}{(2\pi )^3 2p_0} a_{{\bf p}\lambda}^\dagger(t),
\label{can_mom} \\ &&
\overline{\Pi}_{{\bf p}\lambda} \equiv
\frac{\delta L_{\rm MRS}}{\delta \dot{a}_{{\bf p}\lambda}^\dagger}
= i \frac{1}{2} \frac{1}{(2\pi )^3 2p_0} a_{{\bf p}\lambda}(t).
\label{can_mom_bar}
\eeq
Since $\dot{a}_{{\bf p}\lambda}(t)$ and $\dot{a}_{{\bf
p}\lambda}^\dagger(t)$ cannot be expressed as functions of canonical
momenta, eq.(\ref{can_mom}) and eq.(\ref{can_mom_bar}) are
regarded as primary constraints on this system;
\beq
&& \Phi_{{\bf p}\lambda} \equiv
\Pi_{{\bf p}\lambda} -
i \frac{1}{2} \frac{1}{(2\pi )^3 2p_0} a_{{\bf p}\lambda}^\dagger(t)
=0,
\label{pri_con1} \\ &&
\overline{\Phi}_{{\bf p}\lambda} \equiv
\overline{\Pi}_{{\bf p}\lambda} -
i \frac{1}{2} \frac{1}{(2\pi )^3 2p_0} a_{{\bf p}\lambda}(t) = 0.
\label{pri_con2}
\eeq

For this system, Poisson bracket $\{\cdot\cdot\cdot\}_{\rm P}$ is
defined as in the usual case;
\beq
\lmp F,G \rmp_{\rm P} &\equiv&
\int d^3 {\bf p} \sum_{\lambda} \lmp
\lsp \frac{\delta F}{\delta a_{{\bf p}\lambda}} \rsp_{\rm R}
\lsp \frac{\delta G}{\delta \Pi_{{\bf p}\lambda}} \rsp_{\rm L}
+ \lsp \frac{\delta F}{\delta a_{{\bf p}\lambda}^\dagger}
\rsp_{\rm R}
\lsp \frac{\delta G}{\delta \overline{\Pi}_{{\bf p}\lambda}}
\rsp_{\rm L}
\rmp
\nonumber \\&&
+ \int d^3 {\bf p} \sum_{\lambda}
\lmp \lsp \frac{\delta G}{\delta a_{{\bf p}\lambda}} \rsp_{\rm R}
\lsp \frac{\delta F}{\delta \Pi_{{\bf p}\lambda}} \rsp_{\rm L}
+ \lsp \frac{\delta G}{\delta a_{{\bf p}\lambda}^\dagger} \rsp_{\rm R}
\lsp \frac{\delta F}{\delta \overline{\Pi}_{{\bf p}\lambda}}
\rsp_{\rm L} \rmp,
\eeq
where the index L (R) represents left (right) derivative. Especially for
the dynamical variables $a_{{\bf p}\lambda}(t)$, $a_{{\bf
p}\lambda}^\dagger(t)$ and the canonical momenta $\Pi_{{\bf p}\lambda}$,
$\overline{\Pi}_{{\bf p}\lambda}$, Poisson brackets are given as
\beq
\lmp a_{{\bf p}\lambda}(t), \Pi_{{\bf p}^{'}\lambda^{'}} \rmp_{\rm P}
&=&
\delta_{\lambda \lambda^{'}} \delta({\bf p} - {\bf p}^{'}),
\\
\lmp a_{{\bf p}\lambda}^\dagger(t) ,
\overline{\Pi}_{{\bf p}^{'}\lambda^{'}}
\rmp_{\rm P}
&=&
\delta_{\lambda \lambda^{'}} \delta({\bf p} - {\bf p}^{'}),
\eeq
and those for constraints $\Phi_{{\bf p}\lambda}$ and
$\overline{\Phi}_{{\bf p}\lambda}$ can be obtained as
\beq
C \lsp {\bf p}, \lambda ; {\bf p}^{'}, \lambda^{'} \rsp &\equiv&
\lsp \begin{array}{cc}
\lmp \Phi_{{\bf p}\lambda},
\Phi_{{\bf p}^{'}\lambda^{'}} \rmp_{\rm P}
&
\lmp \Phi_{{\bf p}\lambda},
\overline{\Phi}_{{\bf p}^{'}\lambda^{'}} \rmp_{\rm P}
\\
\lmp \overline{\Phi}_{{\bf p}\lambda},
\Phi_{{\bf p}^{'}\lambda^{'}} \rmp_{\rm P}
&
\lmp \overline{\Phi}_{{\bf p}\lambda},
\overline{\Phi}_{{\bf p}^{'}\lambda^{'}} \rmp_{\rm P}
\end{array}\rsp
\nonumber \\
&=& \frac{i}{( 2 \pi )^{3} 2p_{0} }
\lsp \begin{array}{cc}
0 & \delta_{\lambda \lambda^{'}}\delta({\bf p} - {\bf p}^{'}) \\
\delta_{\lambda \lambda^{'}}\delta({\bf p} - {\bf p}^{'}) & 0
\end{array} \rsp.
\eeq
Notice that the above matrix $C \lsp {\bf p}, \lambda ; {\bf p}^{'},
\lambda^{'} \rsp$ has its inverse, and hence time evolutions of the
primary constraints (\ref{pri_con1}) and (\ref{pri_con2}) do not induce
secondary ones.

In order to quantize this constrained system, we introduce Dirac bracket
$\{\cdot\cdot\cdot\}_{\rm D}$;
\beq
\lmp F , G \rmp_{\rm D} &\equiv& \lmp F , G \rmp_{\rm P} -
\int d^3{\bf p} d^3{\bf p}^{'} \sum_{\lambda\lambda^{'}}
\lmp F , \tilde{\Phi}_{{\bf p}\lambda} \rmp_{\rm P}
C^{-1} \lsp {\bf p}, \lambda ; {\bf p}^{'}, \lambda^{'} \rsp
\lmp \tilde{\Phi}_{{\bf p}^{'}\lambda^{'}} , G \rmp_{\rm P}
\nonumber \\
&=&
\lmp F , G \rmp_{\rm P}
+ \lsp 2 \pi \rsp^3 i \int d^3 {\bf p} \sum_{\lambda} 2 p_0
\lmp F, \Phi_{{\bf p}\lambda} \rmp_{\rm P}
\lmp \overline{\Phi}_{{\bf p} \lambda} , G \rmp_{\rm P}
\nonumber \\ &&
+ \lsp 2 \pi \rsp^3 i \int d^3 {\bf p} \sum_{\lambda} 2 p_0
\lmp F, \overline{\Phi}_{{\bf p}\lambda} \rmp_{\rm P}
\lmp \Phi_{{\bf p} \lambda} , G \rmp_{\rm P},
\eeq
where $\tilde{\Phi}_{{\bf p}\lambda}$ represents both $\Phi_{{\bf
p}\lambda}$ and $\overline{\Phi}_{{\bf p}\lambda}$, and $F$ and $G$
arbitrary variables. Then, this system is quantized by replacing the
Dirac bracket by (anti-)commutator;
\beq
i \lmp F , G \rmp_{\rm D} \rightarrow
\llp F , G \rmp \equiv FG - (-1)^{|F||G|} GF,
\eeq
where $|F|=1$ if $F$ is Grassmann-odd and $|F|=0$ if $F$ is
Grassmann-even. Especially, commutation-relations among $a_{{\bf
p}\lambda}(t)$ and $a_{{\bf p}\lambda}^\dagger (t)$ are obtained as
\beq
&& \lmp a_{{\bf p}\lambda}(t)
, a_{{\bf p}^{'}\lambda^{'}}^\dagger (t) \rmp =
\lsp 2 \pi \rsp^3 2 p_0
\delta_{\lambda \lambda^{'}}\delta({\bf p} - {\bf p}^{'}),
\\ &&
\lmp a_{{\bf p}\lambda}(t)
, a_{{\bf p}^{'}\lambda^{'}}(t) \rmp =
\lmp a_{{\bf p}\lambda}^\dagger(t)
, a_{{\bf p}^{'}\lambda^{'}}^\dagger(t) \rmp = 0.
\eeq

As in the ordinary procedure, we can get a hamiltonian for this system;
\beq
H &=& \int d^3 {\bf p} \sum_{\lambda} \lsp
\Pi_{{\bf p}\lambda} a_{{\bf p}\lambda} +
\overline{\Pi}_{{\bf p}\lambda} a_{{\bf p}\lambda}^\dagger \rsp
-\int d^3{\bf x} {\cal L}_{\rm MRS}
\nonumber \\
&=& \int d^3 {\bf x}
\lsp \frac{i}{2} \overline{\psi}_\mu \gamma^i \partial_i \psi^\mu
- \frac{1}{2} \mgra \overline{\psi}_\mu \psi^\mu \rsp
\nonumber \\
&=& \int \frac{d^3 {\bf p}}{(2\pi)^3 2p_0} \sum_{\lambda}
p_0 a_{{\bf p}\lambda}^\dagger(t) a_{{\bf p}\lambda}(t).
\eeq
By using this hamiltonian, equations of motion for $a_{{\bf
p}\lambda}(t)$ and $a_{{\bf p}\lambda}^\dagger (t)$ are derived;
\beq
&& i \frac{d}{dt} a_{{\bf p}\lambda}(t)
= \llp a_{{\bf p}\lambda}(t) , H \rlp
= p_0 a_{{\bf p}\lambda}(t),
\\ &&
i \frac{d}{dt} a_{{\bf p}\lambda}^\dagger (t)
= \llp a_{{\bf p}\lambda}^\dagger (t) , H \rlp
= - p_0 a_{{\bf p}\lambda}^\dagger (t),
\eeq
and these equations can be easily solved;
\beq
a_{{\bf p}\lambda}(t) = a_{{\bf p}\lambda} e^{-ip_0t},~~~
a_{{\bf p}\lambda}^\dagger(t) = a_{{\bf p}\lambda}^\dagger
e^{ip_0t},
\eeq
where we denote $a_{{\bf p}\lambda}(t=0)$ and $a_{{\bf
p}\lambda}^\dagger(t=0)$ as $a_{{\bf p}\lambda}$ and $a_{{\bf
p}\lambda}^\dagger$.

Then, by using $a_{{\bf p}\lambda}$ and $a_{{\bf p}\lambda}^\dagger$,
the field operator $\psi_\mu$ can be expanded as%
\beq
\psi_\mu = \int \frac{d^3{\bf p}}{(2\pi )^3 2p_0} \sum_\lambda
\lmp e^{-ipx} \tilde{\psi}_\mu ({\bf p},\lambda)
a_{{\bf p}\lambda}
+ e^{ipx} \tilde{\psi}_\mu^{C} ({\bf p},\lambda)
a_{{\bf p}\lambda}^\dagger \rmp.
\label{expansion2}
\eeq
Furthermore, the momentum operator $P_\mu$ for the gravitino field is
given by
\beq
P_\mu =
\int \frac{d^3 {\bf p}}{(2\pi)^3 2p_0} \sum_{\lambda}
p_\mu a_{{\bf p}\lambda}^\dagger a_{{\bf p}\lambda}.
\label{momentum_op}
\eeq

Fock space for the gravitino field is constructed by operating the
creation operator $a_{{\bf p}\lambda}^\dagger$ on vacuum state
$|0\rangle$ which satisfies the condition $a_{{\bf p}\lambda} |0\rangle
=0$. Especially, one particle state of the gravitino with momentum $p$
and helicity $\lambda$ is defined as
\beq
\left| p,\lambda \right\rangle \equiv
a_{{\bf p}\lambda}^\dagger |0\rangle .
\label{1pt_state}
\eeq
As one can see, this state satisfies the following relations;
\beq
\langle p,\lambda | p^{'} ,\lambda^{'} \rangle &=&
\lsp 2 \pi \rsp^3 2 p_0
\delta_{\lambda\lambda^{'}} \delta ( {\bf p} - {\bf p}^{'} ),
\label{norm_1pt}
\\
P_\mu \left| p,\lambda \right\rangle &=&
p_\mu \left| p,\lambda \right\rangle,
\label{mom_eigen}
\eeq
where $P_\mu$ is the momentum operator defined in
eq.(\ref{momentum_op}). Notice that the normalization condition
(\ref{norm_1pt}) on the one particle state (\ref{1pt_state}) is Lorentz
invariant since $p_0 \delta ({\bf p}-{\bf p}^{'})$ is a Lorentz
invariant variable.

{}From the above arguments, we obtain Feynman rules for the external
gravitinos; for the gravitino with momentum $p$ and helicity $\lambda$
in initial the state, we should assign $\tilde{\psi}_\mu({\bf
p},\lambda)$ or $\overline{\tilde{\psi}^C}_\mu({\bf p},\lambda)$, and
that in the final state, $\overline{\tilde{\psi}}_\mu({\bf p},\lambda)$
or $\tilde{\psi}_\mu^C({\bf p},\lambda)$.

\section{Interactions of the gravitino}

\hspace*{\parindent}
In the previous section, we have quantized the massive free gravitino field,
and obtained Feynman rules for the gravitino in external line. Our next
purpose is to discuss the Feynman rules for the interactions of the
gravitino field. Many interaction terms concerning the gravitino field
$\psi_\mu$ exist in the full supergravity lagrangian (\ref{L_SUGRA}),
but most of them are irrelevant for our present purpose. This fact arises
from the following two reasons.
\begin{itemize}
\item
Since we are considering the processes with the energy scale much below
the gravitational one; $\sqrt{s}\ll M$, contributions of the Feynman
diagrams with higher dimensional operators are suppressed by factor
$\sim \sqrt{s}/M \ll 1$ (or its higher power). As we will see later, the
dimension of relevant operators is always five, and interaction terms
with the dimension higher than six are not important for us.

\item
In our analysis in Chapter~\ref{chap:heavy} and
Chapter~\ref{chap:light}, gravitinos only appear at external lines in
the Feynman diagram. Then, interaction terms which contain
$\gamma^\mu\psi_\mu$ or $\overline{\psi}_\mu\gamma^\mu$ are not
necessary since they vanish due to
eq.(\ref{eq_g=0'}).
\end{itemize}
In the following arguments, we take only the relevant interaction terms
into account and ignore contributions from other terms.

The most important interaction terms come from the coupling between the
gravitino field and the supercurrent, which is given in
eq.(\ref{L_J-psi}).  In the four-component notation (and in the flat
space-time), these terms are written as
\beq
{\cal L}_{\psi J} &=&
-\frac{1}{\sqrt{2}M} \tilde{\cal D}_{\nu} \phi^{*i}
\overline{\psi}_\mu \gamma^\nu \gamma^\mu \chi^i_R
-\frac{1}{\sqrt{2}M} \tilde{\cal D}_{\nu} \phi^{i}
\overline{\chi}^i_L \gamma^\mu \gamma^\nu \psi_\mu
\nonumber \\ &&
-\frac{i}{8M} \overline{\psi}_\mu
\llp \gamma^\nu , \gamma^\rho \rlp \gamma^\mu
\lambda^{(a)} F_{\nu\rho}^{(a)}.
\label{interaction}
\eeq

\begin{figure}[t]
\epsfxsize=14cm

\centerline{\epsfbox{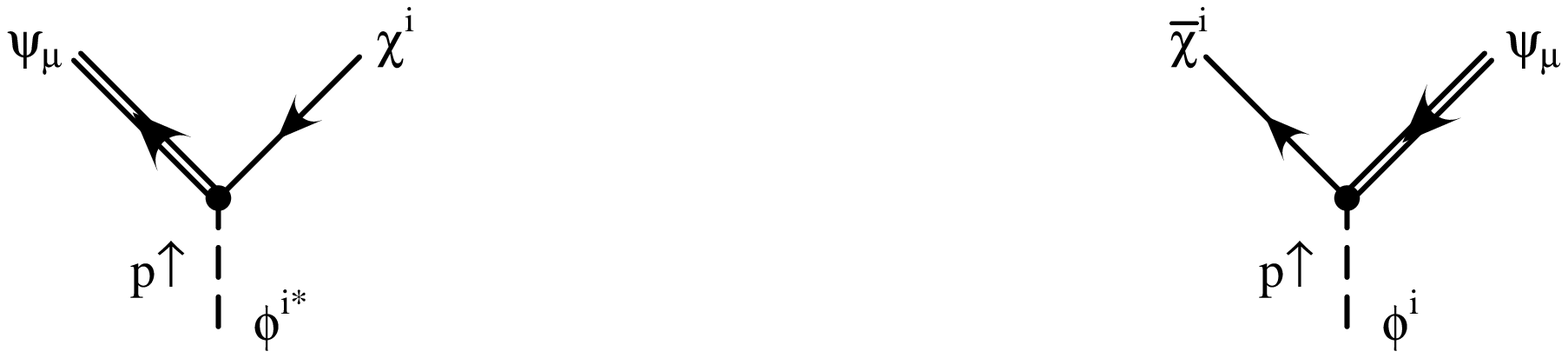}}

{\hspace{15mm}
$\frac{-1}{\sqrt{2}{M}}\gamma_\nu\gamma_\mu (1+\gamma_5) p^\nu$
\hspace{6.1cm}
$\frac{-1}{\sqrt{2}{M}}\gamma_\mu\gamma_\nu (1-\gamma_5) p^\nu$}

\vspace{5mm}

\centerline{\epsfbox{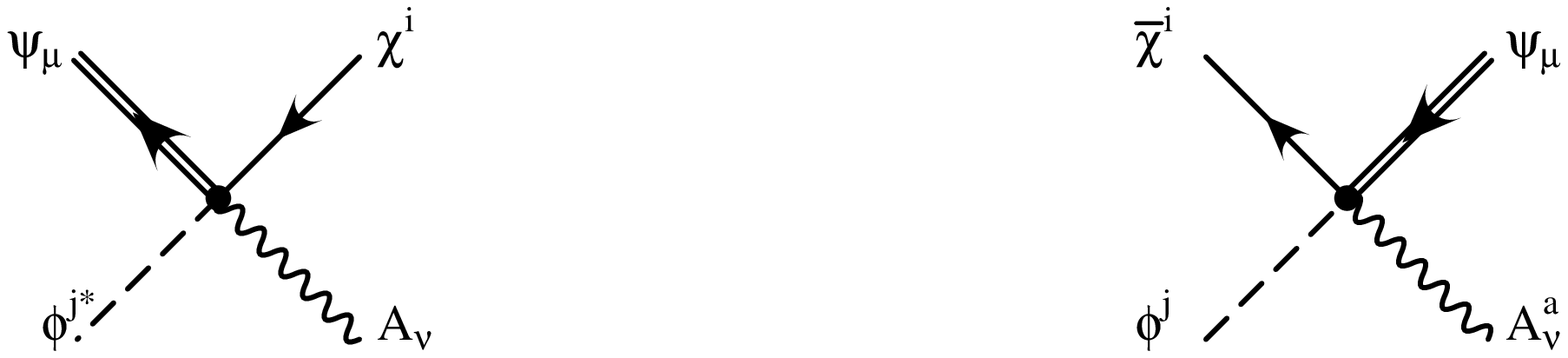}}

{\hspace{12mm}
$\frac{-1}{2\sqrt{2}{M}} g T^a_{ji} \gamma_\nu\gamma_\mu (1+\gamma_5)$
\hspace{5.6cm}
$\frac{1}{2\sqrt{2}{M}} g T^a_{ij} \gamma_\mu\gamma_\nu (1-\gamma_5)$}

\vspace{5mm}

\centerline{\epsfbox{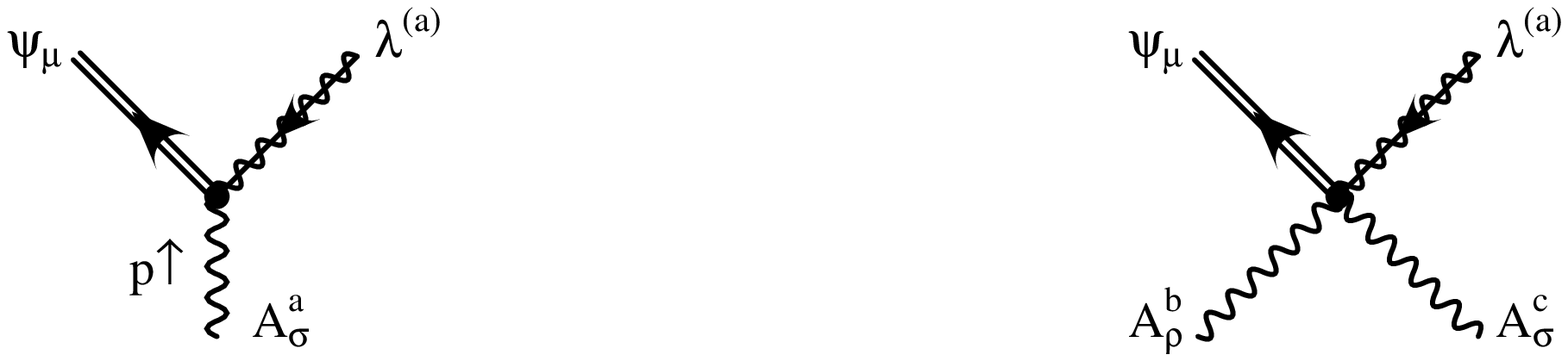}}

{\hspace{18mm}
$\frac{-i}{4M} p_\rho [\gamma^\rho , \gamma_\sigma] \gamma_\mu$
\hspace{6.7cm}
$\frac{-1}{4M} g f^{abc} [\gamma_\rho , \gamma_\sigma] \gamma_\mu$
\hspace{1mm}}

\vspace{10mm}

\caption{Feynman rules for the interactions of gravitino.}
\label{fig:F-rule1}
\end{figure}

{}From this, we construct Feynman rules for the interaction of the
gravitino field $\psi_\mu$ with matter fields $\phi$, $\chi$, $A_\mu$
and $\lambda$. In Fig.~\ref{fig:F-rule1}, we show the Feynman
rules derived from the lagrangian (\ref{interaction}). In the following
analysis, we use these rules and ignore other higher dimensional
interactions.
%
%
%
%

\begin{figure}[t]
\epsfxsize=14cm

\centerline{\epsfbox{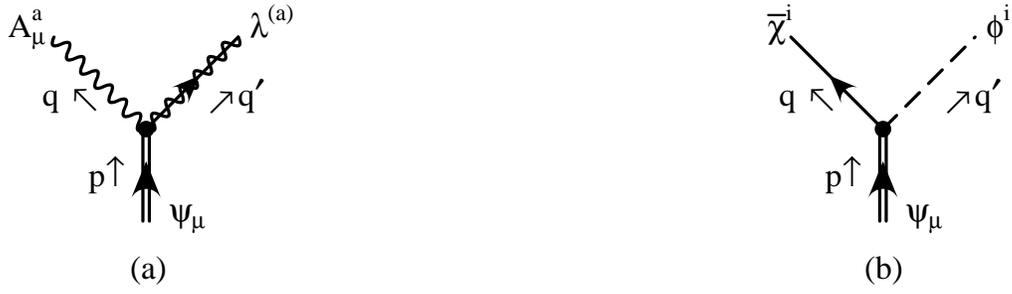}}

\vspace{5mm}

\caption{Feynman diagrams for the processes (a)
$\psi_\mu \rightarrow \lambda + A_\mu$, and (b) $\psi_\mu
\rightarrow \phi^i + \overline{\chi}^i$.}
\label{fig:F-decay}
\end{figure}

As a simple example of applications of these Feynman rules, we calculate
the decay rates of the gravitino. The dominant decay modes of the
gravitino are $\psi_\mu \rightarrow \lambda + A_\mu$ and $\psi_\mu
\rightarrow \phi^i + \overline{\chi}^i$ (and $\psi_\mu \rightarrow
\phi^{i*} + \chi^i$) if these processes are kinematically allowed.
Feynman diagrams for these processes are shown in
Fig.~\ref{fig:F-decay}.
%
%
%
%

For the process $\psi_\mu \rightarrow \lambda + A_\mu$, the invariant
amplitude can be obtained as
\beq
{\cal M}(\psi_\mu \rightarrow \lambda + A_\nu) =
\frac{i}{M} q_\rho \epsilon_\nu
\overline{\lambda}
\lsp g^{\mu\rho} \gamma^\nu - g^{\mu\nu} \gamma^\rho \rsp
\tilde{\psi}_\mu.
\label{amp_gradecay}
\eeq
By using the mode sum for $\tilde{\psi}_\mu$ given in eq.(\ref{P_munu}),
eq.(\ref{amp_gradecay}) becomes
\beq
\overline{\labs {\cal M}(\psi_\mu \rightarrow \lambda + A_\nu) \rabs^2}
&\equiv&
\frac{1}{4} \sum_{\lambda s m}
\labs {\cal M}(\psi_\mu \rightarrow \lambda + A_\nu) \rabs^2
\nonumber \\
&=&
\frac{\mgra^4}{2M^2} \lmp 1 - \lsp \frac{m_\lambda}{\mgra} \rsp^2 \rmp
\lmp 1 + \frac{1}{3} \lsp \frac{m_\lambda}{\mgra} \rsp^2 \rmp,
\label{m-bar-decay}
\eeq
where $m_\lambda$ is the gauge fermion mass, and we have assumed that
the gauge field $A_\mu$ is massless. Notice that in deriving
eq.(\ref{m-bar-decay}), we have taken an average of the helicity of
initial gravitino. By using eq.(\ref{m-bar-decay}), the decay rate is
given by
\beq
\Gamma(\psi_\mu \rightarrow \lambda + A_\nu) &=&
\frac{|{\bf p}_f|}{32\pi^2} \int d\Omega~\overline{\labs{\cal M}\rabs^2}
\nonumber \\
&=&
\frac{1}{32\pi} \frac{\mgra^3}{M^2}
\lmp 1 - \lsp \frac{m_\lambda}{\mgra} \rsp^2 \rmp^3
\lmp 1 + \frac{1}{3} \lsp \frac{m_\lambda}{\mgra} \rsp^2 \rmp,
\label{gamma_g=>al}
\eeq
where ${\bf p}_f$ is the three momentum of the particle in final state.
Notice that if the gauge group $G$ is non-abelian, the above decay rate
should be multiplied by the rank of $G$ in order to calculate a total
decay rate. Especially in the case $\mgra\gg m_\lambda$, the gravitino
lifetime $\tau_{3/2}$ is approximately given by
\beq
\tau_{3/2} (\psi_\mu \rightarrow \lambda + A_\mu) \simeq
4.0\times 10^{8} \times
\lsp\frac{\mgra}{\rm 100GeV}\rsp^{-3} {\rm sec}.
\eeq

For the process $\psi_\mu \rightarrow \phi^i + \overline{\chi}^i$,
the invariant amplitude is given by
\beq
{\cal M}(\psi_\mu \rightarrow \phi^i + \overline{\chi}^i) =
\frac{1}{2\sqrt{2}M} \overline{\chi} \gamma^\mu \qslash^{'} (1+\gamma_5)
\tilde{\psi}_\mu,
\eeq
and by using a similar method to in the previous case, the decay rate is
obtained as
\beq
\Gamma(\psi_\mu \rightarrow \phi^i + \overline{\chi}^i) =
\frac{1}{384\pi} \frac{\mgra^3}{M^2}
\lmp 1 - \lsp \frac{m_\phi}{\mgra} \rsp^2 \rmp^4,
\label{gamma_g=>pc}
\eeq
where $m_\phi$ is the mass of $\phi$ field, and we have assumed that
$\chi$ is a massless fermion. Notice that the decay rate for the process
$\psi_\mu\rightarrow\phi^{i*} + \chi^i$ is equal to the one given in
eq.(\ref{gamma_g=>pc}).

Another application is to calculate the gravitino production cross
sections.  As one can easily see, the most important processes are those
of one gravitino production since the vertices having a gravitino are
suppressed by $M^{-1}$ (or its higher power). Amplitudes for the
gravitino production processes are obtained by combining the Feynman
rules given in Fig.~\ref{fig:F-rule1} with those derived from the
ordinary (global) SUSY lagrangian. We have calculated the total cross
section for the dominant processes of helicity $\pm\frac{3}{2}$
gravitino production, and the results are shown in
Table~\ref{table:cs3/2}.

%
%
\begin{table}[t]
\begin{center}
\begin{tabular}{cc|l} \hline \hline
\multicolumn{2}{l|}{Process} &
{$\sigma = ( g^{2} / 64 \pi M^{2} ) \times$}
\\ \hline
{${\rm (A)}$} & {$\gb{a} + \gb{b} \rightarrow \psi + \gf{c}$} &
{$ (8/3) \ff{c}$}
\\
{${\rm (B)}$} & {$\gb{a} + \gf{b} \rightarrow \psi + \gb{c}$} &
{$ 4 \ff{c}
   \lmp -(3/2) + 2\ln (2/\eps) + \eps - (1/8)\eps^{2} \rmp$}
\\
{${\rm (C)}$} & {$\gb{a} + \cb{i} \rightarrow \psi + \cf{j}$} &
{$ 4 \tt{j}$}
\\
{${\rm (D)}$} & {$\gb{a} + \cf{i} \rightarrow \psi + \cb{j}$} &
{$ 2 \tt{j}$}
\\
{${\rm (E)}$} & {$\cf{i} + \cb{j}^{*} \rightarrow \psi + \gb{a}$} &
{$ 4 \tt{a}$}
\\
{${\rm (F)}$} & {$\gf{a} + \gf{b} \rightarrow \psi + \gf{c}$} &
{$ \ff{c}
   \lmp -(62/3) + 16\ln [(2-\eps)/\eps]
        +22\eps - 2\eps^{2} + (2/3)\eps^{3} \rmp$}
\\
{${\rm (G)}$} & {$\gf{a} + \cf{i} \rightarrow \psi + \cf{j}$} &
{$ 4 \tt{j}
   \lmp -2 + 2\ln (2/\eps) + \eps \rmp$}
\\
{${\rm (H)}$} & {$\gf{a} + \cb{i} \rightarrow \psi + \cb{j}$} &
{$ \tt{j}
   \lmp -6 + 8\ln (2/\eps) + 4\eps -(1/2)\eps^{2}\rmp$}
\\
{${\rm (I)}$} & {$\cf{i} + \overline{\chi}_j \rightarrow \psi + \gf{a}$}
& {$ (8/3) \tt{a}$}
\\
{${\rm (J)}$} & {$\cb{i} + \cb{j}^{*} \rightarrow \psi + \gf{a}$} &
{$ (16/3) \tt{a}$}
\\ \hline \hline
\end{tabular}
\caption{Total cross sections for the helicity $\pm\frac{3}{2}$
gravitino production processes. Spins of the initial states are averaged
and those of the final states are summed. $f^{abc}$ and $T^{a}_{ij}$
represent the structure constants and the generators of the gauge
groups, respectively.  Notice that for the processes (B), (F), (G) and
(H), we cut off the infrared singularities due to the $t$-, $u$-channel
exchanges of gauge bosons, taking a small but a positive parameter
$\delta = (1\pm\cos\theta)_{min}$ where $\theta$ is the scattering angle
in the center-of-mass frame.}
\label{table:cs3/2}
\end{center}
\end{table}
%
%

\section{Effective lagrangian for light gravitino}

\hspace*{\parindent}
In spontaneously broken SUSY models, the massless gravitino field
$\psi_\mu$ acquires mass by absorbing goldstino modes. Before becoming
massive, the gravitino only possesses helicity $\pm\frac{3}{2}$ modes,
and the goldstino provides helicity $\pm\frac{1}{2}$ modes of the
massive gravitino field. This fact suggests that the helicity
$\pm\frac{1}{2}$ mode of the gravitino field behaves like a goldstino.
In fact, if the gravitino mass $\mgra$ is much smaller than the mass
differences between bosons and fermions in the chiral and gauge
multiplets, the above argument is valid and we can obtain an effective
lagrangian for the relativistic gravitino field of helicity
$\pm\frac{1}{2}$ components. In this section, we will derive the
effective lagrangian for the light gravitino field~\cite{PLB175-471} and
apply it for calculating some processes.

For the case $\sqrt{s}\gg \mgra$, the wave function of the gravitino of
helicity $\pm\frac{1}{2}$ components is approximately proportional to
$p_\mu/\mgra$ where $p_\mu$ is a momentum of the gravitino. In this
case, the helicity $\pm\frac{1}{2}$ component of the gravitino field can
be written as
\beq
\psi_\mu \sim
i \sqrt{\frac{2}{3}} \frac{1}{\mgra} \partial_\mu \psi,
\label{h=1/2ofpsi}
\eeq
where $\psi$ represents the spin $\frac{1}{2}$ fermionic field which can
be interpreted as the goldstino. Substituting eq.(\ref{h=1/2ofpsi}) into
the gravitino interaction lagrangian (\ref{interaction}), we obtain the
effective interaction lagrangian for the goldstino components $\psi$.

Using the replacement (\ref{h=1/2ofpsi}), the gravitino interaction
lagrangian (\ref{interaction}) becomes
\beq
{\cal L}_{\rm eff} &=&
\llp \frac{i}{\sqrt{3}M\mgra}
\lmp \lsp \overline{\psi}
\chi_R^i \rsp \partial_\mu\partial^\mu \phi^{i*} -
\lsp \overline{\psi} \partial_\mu\partial^\mu
\chi_R^i \rsp \phi^{i*} \rmp + h.c. \rlp
\nonumber \\ &&
+ \frac{1}{4\sqrt{6}M\mgra}
\overline{\psi} \llp \gamma^\mu , \gamma^\nu \rlp
\gamma^\rho \partial_\rho \lambda^{(a)} F^{(a)}_{\mu\nu}
\nonumber \\ &&
+({\rm total~derivative}),
\label{L_eff_step1}
\eeq
where we have used the relation $\dslash\psi_\mu =0$ since we are
considering processes with a light gravitino in the external line.

%
%
\begin{figure}[t]
\epsfxsize=14cm

\centerline{\epsfbox{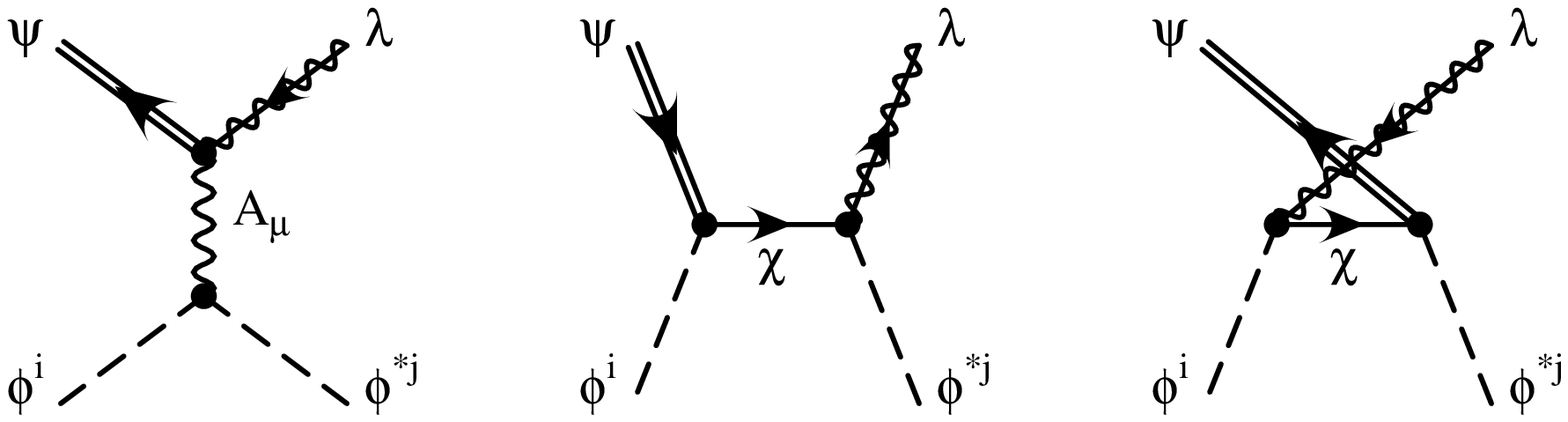}}

{\hspace{28mm}
${\cal M}_s$ \hspace{4.0cm} ${\cal M}_t$ \hspace{4.0cm} ${\cal M}_u$}

\caption{Feynman diagrams for the process
$\phi^i+\phi^{*j}\rightarrow\psi_\mu +\lambda^{(a)}$.}
\label{fig:scat_exam}
\end{figure}
%
%
The terms in the right-hand side of eq.(\ref{L_eff_step1}) are
dimension six operators and one would expect that the processes
involving helicity $\pm\frac{1}{2}$ gravitinos have bad high energy
behaviors. In supergravity, however, this is not the case since the
leading high energy behavior cancels out in the total amplitude. For
example, for the helicity $\pm\frac{1}{2}$ gravitino production process
$\phi^i+\phi^{*j}\rightarrow\psi_\mu^{(1/2)}+\lambda^{(a)}$, there are
$s$-, $t$- and $u$-channel diagrams (see Fig.~\ref{fig:scat_exam}),
whose leading terms are given by
\beq
{\cal M}_s &\simeq&
i\sqrt{\frac{2}{3}} \frac{g}{\mgra M} (T^a)_{ji}
\overline{\psi} \pslash' \lambda^C,
\\
{\cal M}_t &\simeq&
-i \frac{1}{2} \sqrt{\frac{2}{3}} \frac{g}{\mgra M} (T^a)_{ji}
\overline{\psi} \pslash' (1-\gamma_5) \lambda^C,
\\
{\cal M}_u &\simeq&
-i \frac{1}{2} \sqrt{\frac{2}{3}} \frac{g}{\mgra M} (T^a)_{ji}
\overline{\psi} \pslash' (1+\gamma_5) \lambda^C,
\eeq
where $p'$ is the momentum of $\phi^*$.  As one can see, they cancel out
in the total amplitude. This can be understood in the following
way. The helicity $\pm\frac{1}{2}$ components of the gravitino field is
the unphysical in the symmetric phase, and hence total amplitudes with
helicity $\pm\frac{1}{2}$ gravitino at external line should vanish
unless SUSY is broken. That is, the total amplitude for the helicity
$\pm\frac{1}{2}$ gravitino production should be proportional to some
SUSY breaking parameters. Thus, the leading high energy behavior, which
is independent of SUSY breaking parameters, cancels out in the total
amplitude.

%
%
\begin{figure}[t]
\epsfxsize=14cm

\centerline{\epsfbox{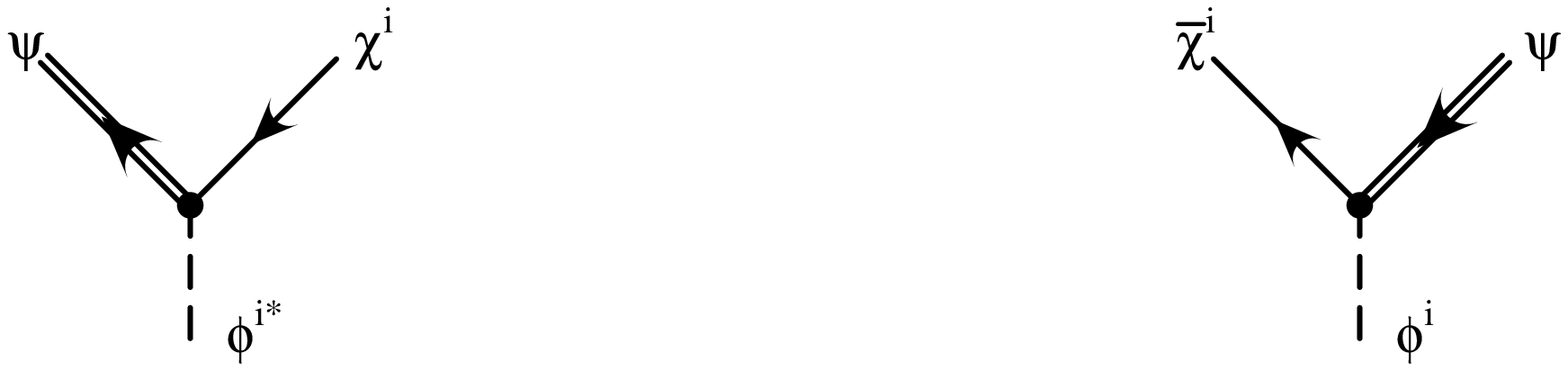}}

{\hspace{16mm}
$\frac{m_\chi^2-m_\phi^2}{2\sqrt{3}M\mgra}(1+\gamma_5)$
\hspace{6.3cm}
$\frac{m_\chi^2-m_\phi^2}{2\sqrt{3}M\mgra}(1-\gamma_5)$}

\vspace{5mm}

\centerline{\epsfbox{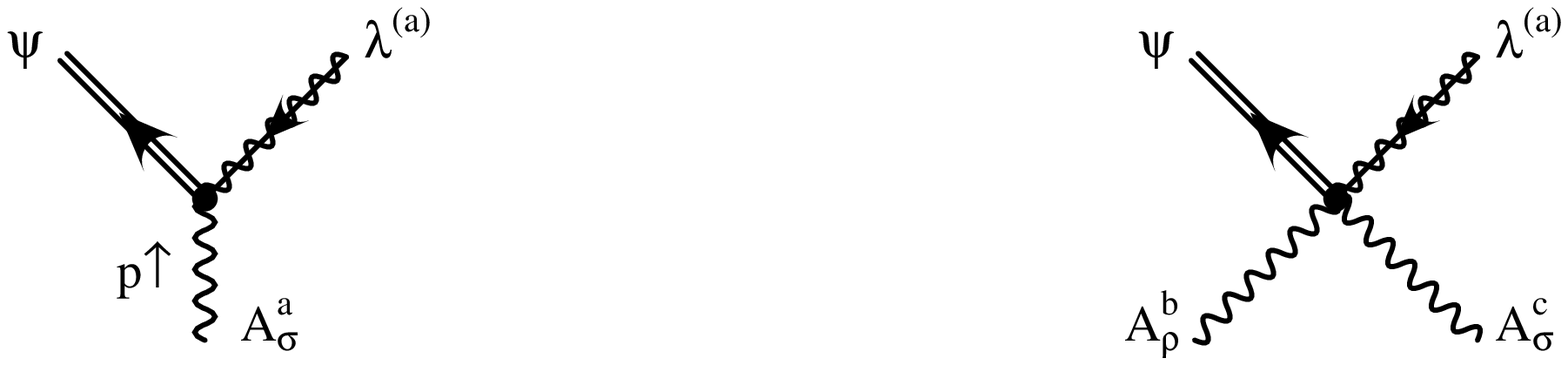}}

{\hspace{15mm}
$\frac{-im_\lambda}{2\sqrt{6}M\mgra}[\gamma_\rho,\gamma_\sigma]p^\rho$
\hspace{6.1cm}
$\frac{-m_\lambda}{2\sqrt{6}M\mgra} gf^{abc}[\gamma_\rho,\gamma_\sigma]$}

\vspace{10mm}

\caption{Feynman rules for the interactions of a light gravitino.}
\label{fig:F-rule2}
\end{figure}
%
%
{}From this fact, we can obtain an effective lagrangian for the helicity
$\pm\frac{1}{2}$ light gravitino field by replacing all the derivatives
operated on $\phi$, $\chi$, and $\lambda$ in eq.(\ref{L_eff_step1}) by
the masses $m_\phi$, $m_\chi$, and $m_\lambda$ of the corresponding
fields. When these derivatives are operated on external lines, this
replacement is easily justified. On the other hand, if they are operated
on propagators, they become internal momenta and seem to make the high
energy behavior worse. As mentioned above, however, this bad high energy
behavior cancels out, and hence the leading high energy behaviors can be
subtracted from the propagators with derivatives in the total
amplitude;%
\beq
\frac{p^2}{p^2 -m_\phi^2} &\rightarrow&
\frac{p^2}{p^2 -m_\phi^2} - 1
= \frac{m_\phi^2}{p^2 -m_\phi^2},
\\
P_L \frac{p^2 \lsp \pslash + m_\chi \rsp}{p^2 -m_\chi^2} P_R
&\rightarrow&
P_L \frac{p^2 \lsp \pslash + m_\chi \rsp}{p^2 -m_\chi^2} P_R
- P_L \pslash P_R
= P_L
\frac{m_\chi^2 \lsp \pslash + m_\chi \rsp}{p^2 -m_\chi^2} P_R ,
\\
\frac{\pslash \lsp \pslash + m_\lambda \rsp}
{p^2 -m_\lambda^2} &\rightarrow&
\frac{\pslash \lsp \pslash + m_\lambda \rsp}
{p^2 -m_\lambda^2}
- 1
= \frac{m_\lambda \lsp \pslash + m_\lambda \rsp}
{p^2 -m_\lambda^2}.
\eeq
Thus, the derivatives in the interaction lagrangian (\ref{L_eff_step1})
can be replaced by the appropriate masses related to the SUSY breaking,
and the lagrangian (\ref{L_eff_step1}) becomes
\beq
{\cal L}_{\rm eff} =
\frac{i(m_{\phi}^{2} - m_{\chi}^{2})}{\sqrt{3} \mgra M}
\lsp \overline{\psi} \chi_R \rsp \phi^{*}
+ \frac{-i m_\lambda}{8 \sqrt{6} \mgra M}
\overline{\psi}
\llp \gamma^{\mu},\gamma^{\nu} \rlp \lambda^{(a)} F^{(a)}_{\mu \nu}
+ h.c.
\label{L_eff}
\eeq
As one can see, in the SUSY limit (\ie $m_{\chi}^{2} - m_{\phi}^{2}
\rightarrow 0$ and $m_\lambda \rightarrow 0$), the above lagrangian vanishes
and the helicity $\pm\frac{1}{2}$ modes of the gravitino field decouple
from the theory.

%
%
%
\begin{table}[t]
\begin{center}
\begin{tabular}{cc|l} \hline \hline
\multicolumn{2}{l|}{Process} &
{$\sigma = (g^{2} m_{G}^{2} / 24 \pi M^{2} \mgra^{2}) \times$}
\\ \hline
{${\rm (A)}$} & {$\gb{a} + \gb{b} \rightarrow \psi + \gf{c}$} &
{$(1/3) \ff{c}$}
\\
{${\rm (B)}$} & {$\gb{a} + \gf{b} \rightarrow \psi + \gb{c}$} &
{$(1/16) \ff{c}
   \lmp -12 + 16\ln (2/\eps) + 8\eps - \eps^{2} \rmp$}
\\
{${\rm (C)}$} & {$\gb{a} + \cb{i} \rightarrow \psi + \cf{j}$} &
{$(1/2) \tt{j}$}
\\
{${\rm (D)}$} & {$\gb{a} + \cf{i} \rightarrow \psi + \cb{j}$} &
{$(1/4) \tt{j}$}
\\
{${\rm (E)}$} & {$\cf{i} + \cb{j}^{*} \rightarrow \psi + \gb{a}$} &
{$(1/2) \tt{a}$}
\\
{${\rm (F)}$} & {$\gf{a} + \gf{b} \rightarrow \psi + \gf{c}$} &
{$(1/12) \ff{c}
   \lmp -22 + 24\ln \{(2-\eps)/\eps\}
        +24\eps - 3\eps^{2} + \eps^{3} \rmp$}
\\
{${\rm (G)}$} & {$\gf{a} + \cf{i} \rightarrow \psi + \cf{j}$} &
{$(1/2) \tt{j}
   \lmp -2 + 2\ln (2/\eps) + \eps \rmp$}
\\
{${\rm (H)}$} & {$\gf{a} + \cb{i} \rightarrow \psi + \cb{j}$} &
{$(1/32) \tt{j}
   \lmp -28 + 32\ln (2/\eps) + 16\eps -\eps^{2}\rmp$}
\\
{${\rm (I)}$} & {$\cf{i} + \overline{\chi}_j \rightarrow \psi + \gf{a}$} &
{$(1/3) \tt{a}$}
\\
{${\rm (J)}$} & {$\cb{i} + \cb{j}^{*} \rightarrow \psi + \gf{a}$} &
{$(1/6) \tt{a}$}
\\ \hline \hline
\end{tabular}
\caption{Total cross sections for the helicity $\pm\frac{1}{2}$ gravitino
production. Spins of the initial states are averaged and those of the
final states are summed. $f^{abc}$ and $T^{a}_{ij}$ represent the
structure constants and the generators of the gauge groups,
respectively.  Notice that for the processes (B), (F), (G) and (H), we
cut off the infrared singularities due to the $t$-, $u$-channel exchange
of gauge bosons, taking small but positive $\delta =
(1\pm\cos\theta)_{min}$ where $\theta$ is the scattering angle in the
center-of-mass frame.}
\label{table:cs1/2}
\end{center}
\end{table}
%
%

In the case of a light gravitino, it is more convenient to use the
effective lagrangian (\ref{L_eff}) rather than the full lagrangian
(\ref{interaction}). Feynman rules derived from the lagrangian
(\ref{L_eff}) are shown in Fig.~\ref{fig:F-rule2}.

By using the effective lagrangian (\ref{L_eff}), we have calculated decay
rates and cross sections for some processes including a light gravitino.
The decay rate of the process $\lambda\rightarrow\psi + A_\mu$ is given by
\beq
\Gamma(\lambda \rightarrow \psi + A_{\mu}) =
\frac{1}{48 \pi}
\frac{m_{\lambda}^{5}}{\mgra^{2} M^{2}}
\lmp 1 - \lsp \frac{\mgra}{m_{\lambda}} \rsp ^{2} \rmp^{3} ,
\label{gamma_l=>ga}
\eeq
and that of the process $\phi \rightarrow \psi + \chi$,
\beq
\Gamma(\phi \rightarrow \psi + \chi) =
\frac{1}{48 \pi}
\frac{m_{\phi}^{5}}{\mgra^{2} M^{2}}
\lmp 1 - \lsp \frac{\mgra}{m_{\phi}} \rsp ^{2} \rmp^{2}.
\label{gamma_p=>gc}
\eeq
Furthermore, total cross sections of some light gravitino production
processes are calculated and the results are shown in
Table~\ref{table:cs1/2}. Notice that the interaction of the helicity
$\pm\frac{1}{2}$ gravitino becomes stronger as the gravitino mass
$\mgra$ becomes lighter, as one can see in eq.(\ref{gamma_l=>ga}),
eq.(\ref{gamma_p=>gc}) and the total cross sections given in
Table~\ref{table:cs1/2}.

The results obtained above can be used when the gravitino mass $\mgra$
is much smaller than the SUSY mass splitting in observable sector. Such
a situation seems to conflict with the super-trace formula obtained in
the previous section. But the super-trace formula is a tree level
relation, and the mass splitting in the observable sector may become
larger than the gravitino mass due to various radiative corrections. In
fact, in some models based on no-scale type K\"ahler
potential~\cite{PLB147-99}, the gravitino mass may become much smaller
than the electroweak scale. In any case, if the gravitino mass is
smaller than the mass splitting in other multiplets, we can use the
results obtained in this section.

%
%

\chapter{Phenomenology of the gravitino : overview}
\label{chap:overview}

\hspace*{\parindent}
Having Feynman rules with the gravitino field in the previous chapter,
we are now at the point to discuss phenomenology of the gravitino.  As
we will see below, the gravitino is almost nothing to do with collider
experiments because its interactions are extremely weak. But if we think
of cosmology, effects of the gravitino may become very significant. Our
main purpose is to investigate the effects of the gravitino on the
inflationary universe quantitatively. Before doing this, we will survey
the phenomenological implications of the massive gravitino.

\section{Collider experiments with the gravitino}

\hspace*{\parindent}
In this section, we will give brief comments on collider experiments
with the massive gravitino. If the gravitino mass is comparable to the
masses of SUSY particles (\ie squarks, sleptons and gauginos), gravitino
production cross sections are extremely small since the interaction
terms for the gravitino are suppressed by powers of $M^{-1}$. For
example, the helicity $\pm\frac{3}{2}$ gravitino production cross
section in $e^+e^-$ collider experiments is estimated to be
\beq
\sigma (e^+ + e^- \rightarrow \psi_\mu + \lambda) =
\frac{1}{96\pi M^2} \lsp \frac{1}{2}g_2^2 + \frac{5}{4}g_1^2 \rsp
\sim 1 \times 10^{-67} {\rm cm}^2,
\eeq
where we have assumed that $\sqrt{s}$ is much larger than the gravitino
and gaugino masses. Compared with the luminosity of LEP $\sim
10^{31}{\rm cm}^{-2}{\rm sec}^{-1}$, or even with a design luminosity of
JLC $\sim 10^{34}{\rm cm}^{-2}{\rm sec}^{-1}$, we have no hope to have a
signal of gravitino production events.

In the case of light gravitino, however, helicity $\pm\frac{1}{2}$ modes
of the gravitino interact strongly, and hence one may expect that some
signals of the gravitino may be detectable if the gravitino mass is
sufficiently small. Bound on the light gravitino mass from collider
experiments depends on the mass spectrum of superparticles. The most
stringent bound on the gravitino mass is derived for the case where
there exists a neutralino lighter than $Z^0$-boson. In this case,
$Z^0$-decay may contain a single photon balanced by missing transverse
momentum in the opposite hemisphere~\cite{PLB258-231}. Such a single
photon is expected to arise from the process $Z^0\rightarrow\psi_\mu
+\chi$, $\chi\rightarrow\psi_\mu +\gamma$, where $\chi$ is the
neutralino. Both of the decay rates for these processes are proportional
to $\mgra^{-2}$, and hence such a single photon event is sizable if the
gravitino mass is extremely small. This process is analyzed in
ref.\cite{PLB258-231}, and the gravitino mass smaller than $\sim
10^{-13}\GEV$ is excluded by the LEP experiments. Notice that this bound
($\mgra\gsim 10^{-13}\GEV$) is comparable to that obtained from
cosmological considerations. If the neutralino mass is larger than the
$Z^0$-boson mass $m_Z$, we cannot use the above arguments, and less
stringent bound may be derived.

\section{Cosmology}

\hspace*{\parindent}
Contrary to collider experiments, the mass of gravitino is severely
constrained if we assume the standard big-bang cosmology.\footnote
{In the ``standard cosmology'', we adopt the following two points. The
first point is that the light nuclei were synthesized through the
(almost) standard big-bang nucleosynthesis, and the second is that the
density parameter $\Omega$ of the present universe is smaller than
$O(1)$. (For the big-bang nucleosynthesis, see Appendix~\ref{ap:bbn}.)}
If the gravitino is unstable, it may decay after the big-bang
nucleosynthesis (BBN) and produces an unacceptable amount of entropy,
which conflicts with the predictions of BBN. In order to keep the
success of BBN, the gravitino mass should be larger than $\sim 10\TEV$
as Weinberg first pointed out~\cite{PRL48-1303}.  In a model with the
gravitino mass larger than $\sim 10\TEV$, the gravitino decay processes
produce a large amount of entropy and dilutes the baryon number density
of the universe.  This requires that the baryon-to-photon ratio before
the decay of the gravitino should be extremely large so that the present
value of the baryon-to-photon ratio is given by $O(10^{-9}$ --
$10^{-10})$.  Furthermore, in a model with $R$-parity invariance, the
gravitino decay also produces an unacceptable amount of LSP which
conflicts with the observations of the present mass density of the
universe~\cite{NPB277-556}. (These problems are sometimes called
``gravitino problem''.)

Meanwhile, in the case of stable gravitino, its mass should be smaller
than $\sim 1\KEV$ not to overclose the universe~\cite{PRL48-223}.
Therefore, the gravitino mass between $\sim 1\KEV$ and $\sim 10\TEV$
conflicts with the standard big-bang cosmology.

However, if the universe went through inflation~\cite{PRD23-347}, we may
avoid the above constraints~\cite{PLB118-59} since the initial abundance
of the gravitino is diluted by the exponential expansion of the
universe. But even if the initial gravitinos are diluted, the above
problems still potentially exist since gravitinos are reproduced by
scattering processes off the thermal radiation after the universe has
been reheated.  The number density of the secondary gravitino is
proportional to the reheating temperature and hence upperbound on the
reheating temperature should be imposed not to overproduce gravitinos.
Therefore, even assuming the inflation, a detailed analysis must be made
to obtain the upperbound on the reheating temperature.

For the case of the unstable gravitino, the cosmological considerations
on regeneration and decay of the gravitino has been done in many
articles~\cite{PLB127-30,PLB138-265,PLB145-181,PLB158-463,NPB259-175,PLB189-23,NPB373-399,TU457}.
These previous works show that the most stringent upperbound on the
reheating temperature comes from the photo-dissociation of light nuclei
(D, T, $^{3}$He, $^{4}$He).  Once gravitinos are produced in the early
universe, most of them decay after BBN since the lifetime of the
gravitino of mass $O(100\GEV-10\TEV)$ is $O((10^{8}-10^{2}){\rm sec})$.
If the gravitinos decay radiatively, emitted high energy photons induce
cascade processes and affect the result of BBN.  Not to change the
abundance of light nuclei, we must have a restriction on the number
density of gravitinos, and this constraint is translated into the
upperbound on the reheating temperature.

In order to analyze the photo-dissociation processes, we must calculate
the following two quantities precisely; the number density of the
gravitinos produced after the reheating, and the high energy photon
spectrum induced by radiative decay of the gravitinos. But the previous
estimations of these values are incomplete. As for the number density of
gravitino, most of the previous works follow the result of
ref.\cite{PLB145-181}, where the number density is underestimated by
factor $\sim$4.  Furthermore, in many articles the spectrum of high
energy photon, which determines the photo-dissociation rates of light
elements, is calculated by using a simple fitting formula. In this
thesis, we calculate these two quantities precisely and find more
precise upperbound on the reheating temperature.

If the gravitino is stable, the situation changes. If a very light
gravitino of mass $m_{3/2}\sim O(1~\KEV)$ was once thermalized, the
critical density of the universe is easily obtained, and hence it
constitutes a dark matter.  For a heavier gravitino, one has to consider
inflation.  We will find that the requirement that the gravitinos
produced after the inflation should not overclose the universe (the
closure limit) places a severe upperbound on the reheating
temperature~\cite{PLB303-289}.  Another stringent constraint comes from
the photo-dissociations of light nuclei by high energy photons produced
through radiative decays of the next-to-the-lightest superparticle into
the gravitinos, which excludes the certain range of the gravitino mass.

In the following chapters, we will investigate these effects
quantitatively.

\section{The Polonyi problem}

\hspace*{\parindent}
As we have seen in Chapter~\ref{chap:sugra}, there exists a scalar boson
with mass of the order of the gravitino mass $\mgra$ in a wide class of
models based on supergravity. (Below, we call this scalar field
``Polonyi field'' $\phi_P$ and denote its mass $m_{\phi_P}$.) If one
assumes that the gravitino mass is of the order of the electroweak scale
from the naturalness point of view, it has been pointed out that the
Polonyi field $\phi_P$ causes serious cosmological
difficulties~\cite{PLB131-59} (so-called ``Polonyi problem'') as
gravitino does. Though this is not directly related to our present
purpose, we briefly discuss this problem because of its significance.

The Polonyi problem stems from the fact that the Polonyi field takes its
amplitude of order $M$ at the end of inflation, \ie there is a Bose
condensation. Such a condensation cannot be eliminated by inflation
since the equation of motion for the Polonyi field is coupled to that of
the inflaton field. Thus, in general the potential minimum of the
Polonyi field at zero temperature is not a stationary point during
inflation, and hence the Polonyi field does not sit at the origin.
Furthermore, it should be noted that scalar fields deviate from the
origin due to the quantum fluctuation during the inflation.

As the temperature drops and the expansion rate $H$ of the universe
becomes comparable to $m_{\phi_P}$, the Polonyi field starts its
oscillation, and subsequently dominates the energy density of the
universe until it decays. The decay rate of the Polonyi field is
estimated as $O(m_{\phi_P}^3/M_{pl}^2)\sim O(\mgra^3/M_{pl}^2)$, and
hence it may cause cosmological difficulties as in the gravitino case.
That is, the Polonyi field decay releases tremendous amount of entropy
and dilutes primordial baryon asymmetry much below what is observed
today (which is sometimes called the entropy crisis). Furthermore, the
decay of the Polonyi field destroys the successful scenario of the BBN
if it occurs after the BBN starts. In fact, the Polonyi problem is more
serious than the gravitino problem, since it cannot be solved by
inflation. As a result, a wide class of models with the SUSY breaking in
hidden sector are excluded by cosmological arguments.

It has been pointed out~\cite{TU467} that the first problem (\ie the
entropy crisis) can be cured if the Affleck-Dine
mechanism~\cite{NPB249-361} for baryogenesis works in the early
universe. Furthermore, the second one can be solved by raising the mass
of the Polonyi field up to $O(10\TEV)$ so that the reheating temperature
by the $\phi_P$ decay is larger than $O(1\MEV)$. Then, the BBN starts
after the decay of $\phi_P$ has completed.

In order to raise Polonyi mass $m_{\phi_P}$ without raising the SUSY
breaking scalar masses in the observable sector, several ideas have been
proposed. One of the attractive proposal is to assume a dynamical SUSY
breaking scenario. If the SUSY is broken by some dynamics at the energy
scale $M_I$ much below the gravitational one, the cut-off scale of the
model may become $O(M_I)$.  In this case, K\"ahler potential may contain
(non-renormalizable) interaction terms which are suppressed by powers of
$M_I^{-1}$ instead of $M^{-1}$. Then, the Polonyi field may have a mass
much larger than that of the gravitino, and hence decays before
the BBN starts.

Alternative approach is to consider no-scale type supergravity
models~\cite{PLB143-410,PRD50-2356}. In some class of models with
no-scale type K\"ahler potential, the gravitino mass is not directly
related to the SUSY breaking masses of the squarks, sleptons and Higgs
bosons. In this case, the gravitino mass (and the mass of the Polonyi
field) can become much larger than the electroweak scale without raising
the SUSY breaking scalar masses in the observable sector.

Furthermore, in some class of models with fine tuning, mixing between
the Polonyi field and the scalar field concerning the fine tuning
becomes substantially large~\cite{PTP91-1277}. In this case, the Polonyi
field decays through the mixing, and the decay rate can be enhanced
without increasing the Polonyi mass $m_{\phi_P}$. If the mixing is
sufficiently large, the Polonyi field decays before the BBN starts. An
example of this type of models is the minimal SUSY SU(5)
model~\cite{NPB193-150,ZPC11-153} with a small self-coupling of the {\bf
24}-Higgs. In this model, the Polonyi field decays into the flavor Higgs
doublets through the mixing to {\bf 24}-Higgs.

As we have discussed, there exist some approach to the Polonyi problem.
However, they still have uncertainties, and it is unclear whether
these scenarios really works well. Therefore, more efforts are needed in
order to check the validity of each scenarios. Finally, it should be
noted that the cosmological evolution of the Polonyi field may affect
the arguments on the gravitino problem. In this thesis, however, we
assume that the Polonyi field plays no significant role in cosmology due
to some (unknown) mechanism, and hence we ignore its effects in the
following chapters.

%
%

\chapter{Cosmology with unstable gravitino}
\label{chap:heavy}

\hspace*{\parindent}
As mentioned in the previous chapter, the gravitino may affect
cosmology. Especially the constraints from the BBN and the present mass
density of the universe strongly suggest the inflation if there exists
the gravitino. In this chapter, we will consider the effects of the
gravitino on the inflationary universe mainly assuming that the
gravitino decays only into photon and photino. We will also give the
analysis for the case where the gravitino only decays into neutrino and
sneutrino.\footnote
{This chapter is based on the work in a collaboration with
M.~Kawasaki~\cite{TU457,TU463}.}

\section{Gravitino production in the early universe}

\hspace*{\parindent}
We will first calculate the number density of the gravitino after the
inflation. Once the universe has reheated, gravitinos are reproduced by
scattering processes of the thermal radiations and they decay with the
decay rate of the order of $\mgra^{3}/M_{pl}^{2}$. Since the
interactions of gravitino are very weak, the gravitinos cannot be
thermalized if the reheating temperature $T_{R}$ is less than
$O(M_{pl})$. In this case, the Boltzmann equation for the gravitino
number density $n_{3/2}$ can be written as
\beq
\frac{d n_{3/2}}{d t} + 3H n_{3/2} =
\vev{\Sigma_{tot} v_{rel}} n_{rad}^{2}
- \frac{\mgra}{\langle E_{3/2}\rangle} \frac{n_{3/2}}{\tau_{3/2}},
\label{bol-ngra}
\eeq
where $H$ is the expansion rate of the universe,
$\langle\cdot\cdot\cdot\rangle$ represents thermal average, $n_{rad}$
the number density of the scalar boson in the thermal bath,
\beq
n_{rad} = \frac{\zeta (3)}{\pi^{2}}T^{3},
\eeq
$v_{rel}$ the relative velocity of the scattering radiations
($\langle v_{rel}\rangle=1$ in our case), and the factor $\mgra/\langle
E_{3/2}\rangle$ the averaged Lorenz factor.  For the radiation
dominated universe, the expansion rate $H$ of the universe is given by
\begin{eqnarray}
H \equiv \frac{\dot{R}}{R} =
\sqrt{\frac{N_{*} \pi^2}{90 M^{2}}}~T^{2} ,
\label{hubble}
\end{eqnarray}
where $R$ is the scale factor and $N_{*}$ the effective total number of
degrees of freedom for effectively massless particles. For the particle
content of the MSSM, $N_{*}(T_{R})\sim228.75$ if $T_{R}$ is much larger
than the masses of the superpartners, and $N_{*}(T \ll 1\MEV) \sim
3.36$. The total cross section $\Sigma_{tot}$ in the thermal bath is
defined as
\beq
\Sigma_{tot} = \frac{1}{2}~\sum_{x,y,z} \eta_{x} \eta_{y}
               ~\sigma_{( x + y \rightarrow \psi_{\mu} + z )}~,
\label{stot}
\eeq
where $\sigma_{( x + y \rightarrow \psi_{\mu} + z )}$ is the cross
section for the process $x + y \rightarrow \psi_{\mu} + z$ (see
Table~\ref{table:cs3/2}), and $\eta_{x}=1$ for incoming bosons,
$\eta_{x}=\frac{3}{4}$ for incoming fermions. For the MSSM particle
content, $\Sigma_{tot}$ is given by
\beq
\Sigma_{tot} = \frac{1}{M^{2}}
               \lmp 2.50 g_{1}^{2}(T) +
                    4.99 g_{2}^{2}(T) +
                    11.78 g_{3}^{2}(T) \rmp ,
\eeq
where $g_{1}$, $g_{2}$ and $g_{3}$ are the gauge coupling constants of
the gauge group ${\rm U(1)}_{Y}$, ${\rm SU(2)}_{L}$ and ${\rm
SU(3)}_{C}$, respectively. Notice that in high energy scattering
processes, the effect of the renormalization of the gauge coupling
constants should be taken into account. Using the one loop
$\beta$-function of the MSSM, the solution to the renormalization group
equation of gauge coupling constants is given by
\beq
g_{i}(T) \simeq
\lmp
g_{i}^{-2}(m_{Z}) - \frac{b_{i}}{8\pi^{2}} \ln \lsp \frac{T}{m_{Z}} \rsp
\rmp^{-1/2},
\eeq
with $b_{1}=11$, $b_{2}=1$, $b_{3}=-3$. In this thesis, we use this
formula.

At the time right after the end of the reheating, the first term
dominates the right-hand side of eq.(\ref{bol-ngra}) since gravitinos
have been diluted by the de~Sitter expansion of the universe during the
inflation.  As a first step to solve eq.(\ref{bol-ngra}), we assume a
``naive'' adiabatic expansion of the universe;
\beq
RT={\rm const}.
\label{rt=const}
\eeq
Then using the yield variable $Y_{3/2} \equiv n_{3/2} / n_{rad}$ and
ignoring the decay contributions, eq.(\ref{bol-ngra}) becomes
\begin{eqnarray}
\frac{d Y_{3/2}}{d T} =
- \frac{\vev{\Sigma_{tot} v_{rel}} n_{rad}}{HT}.
\label{bol-ygra}
\end{eqnarray}
Notice that the right-hand side of this equation is (almost)
independent of $T$, and hence we can easily integrate
eq.(\ref{bol-ygra}).

However, eq.(\ref{bol-ygra}) does not give a correct value of $Y_{3/2}$
since the conserved quantity is not $RT$ but the entropy per comoving
volume;
\beq
R^3 S = {\rm const},
\eeq
with $S$ being the entropy density. Therefore, if the number of the
gravitino per comoving volume is conserved, the yield variables
$Y_{3/2}$ for different temperature $T_1$ and $T_2$ are related as
\beq
Y_{3/2}(T_1) &=&
\frac{S(T_1)/n_{rad}(T_1)}{S(T_2)/n_{rad}(T_2)} Y_{3/2}(T_2)
\nonumber \\
&=& \frac{N_S(T_1)}{N_S(T_2)} Y_{3/2}(T_2),
\eeq
where $N_S\equiv S/S_0$ with $S_0\equiv (2\pi/45)T^3$. (The prefactor
$N_S(T_1)/N_S(T_2)$ is sometimes called a dilution factor.)

Taking this effect into account, the yield of the gravitino at $T$ is
given by\footnote
{In the recent article, Fischler~\cite{PLB332-277} have proposed a new
mechanism to produce gravitinos in the thermal bath. If one adopts his
mechanism, the number density of the gravitino becomes much larger than
the results obtained in this thesis, and hence the constraints obtained
below must become more stringent.  However, it is not clear to us if his
new mechanism is relevant.}
\beq
Y_{3/2} (T) =
\frac{N_{S} (T)}{N_{S}(T_{R})} \times
\frac{n_{rad}(T_{R}) \vev{\Sigma_{tot} v_{rel}}}{H(T_{R})}.
\label{ygra}
\eeq
For the MSSM particle content, $N_{S}(T_{R}) \sim 228.75$ and
$N_{S}(T\ll 1\MEV) \sim 3.91$. Eq.(\ref{ygra}) shows that $Y_{3/2}$ is
proportional to $T_{R}$, \ie as the reheating temperature increases
the yield of the gravitino becomes larger. From eq.(\ref{ygra}), we can
derive a simple fitting formula for $Y_{3/2}$;
\begin{eqnarray}
Y_{3/2} (T\ll1\MEV) \simeq 2.14 \times 10^{-11} ~ \lsp
\frac{T_{R}}{10^{10}\GEV} \rsp
\lmp 1 - 0.0232 \ln \lsp \frac{T_{R}}{10^{10}\GEV} \rsp \rmp,
\label{fitting}
\end{eqnarray}
where the logarithmic correction term comes from the renormalization of
the gauge coupling constants. The difference between the exact formula
(\ref{ygra}) and the above approximated one is within $\sim$ 5\% for
$10^{6}$ GeV $\lsim$ $T_{R}$ $\lsim$ $10^{14}$ GeV (and within $\sim$
25\% for $10^{2}$ GeV $\lsim$ $T_{R}$ $\lsim$ $10^{19}$ GeV).  Notice
that the gravitino abundance derived here is about four times larger
than the one obtained in ref.\cite{PLB145-181}.

As the temperature of the universe drops and the Hubble time $H^{-1}$
approaches $\tau_{3/2}$, the decay term becomes the dominant part of the
right-hand side of eq.(\ref{bol-ngra}).  Ignoring the scattering term,
eq.(\ref{bol-ngra}) can be rewritten as
\begin{eqnarray}
\frac{d Y_{3/2}}{d t} =
- \frac{Y_{3/2}}{\tau_{3/2}},
\label{bol-decay}
\end{eqnarray}
where we have taken $\mgra/\langle E_{3/2}\rangle=1$ since gravitinos
are almost at rest. Using eq.(\ref{ygra}) as a boundary condition, we
can solve eq.(\ref{bol-decay}) and the solution is given by
\begin{eqnarray}
Y_{3/2} (t) =  \frac{n_{3/2}(t)}{n_{rad}(t)} =
\frac{N_{S} (T)}{N_{S}(T_{R})} \times
\frac{n_{rad}(T_{R}) \vev{\Sigma_{tot} v_{rel}}}{H(T_{R})}
e^{-t/\tau_{3/2}},
\label{ygra2}
\end{eqnarray}
where the relation between $t$ and $T$ can be obtained by solving
eq.(\ref{hubble}) with eq.(\ref{rt=const});
\begin{eqnarray}
t =
\frac{1}{2} \sqrt{ \frac{90 M^{2}}{N_{*}\pi^{2}} } T^{-2}.
\label{tt}
\end{eqnarray}

\section{Radiative decay of the gravitino}

\hspace*{\parindent}
Radiative decays of the gravitino may affect BBN.  First, we analyze
this effect assuming that the gravitino $\psi_{\mu}$ mainly decays into
a photon $\gamma$ and a photino $\tilde{\gamma}$.

In order to investigate the photo-dissociation processes, we must know the
spectra of the high energy photon and electron induced by the gravitino
decay. In this section, we will derive these spectra by solving the
Boltzmann equations numerically.

Once high energy photons are emitted in the gravitino decay, they induce
cascade processes. In order to analyze these processes, we take the
following radiative processes into account.
\begin{itemize}
\item
The high energy photon with energy $\epho$ produces an $e^{+}$ $e^{-}$
pair by scattering off the background photon if the energy of the
background photon is larger than $m_{e}^{2}/\epho$. We call this process
the double photon pair creation. For sufficiently high energy photons,
this is the dominant process since the cross section or the number
density of the target is much larger than that of other processes.  A
numerical calculation shows that this process determines the shape of
the spectrum of the high energy photon for $\epho\gsim m_{e}^{2}/22T$.
\item
Below the effective threshold of the double photon pair creation, high
energy photons lose their energy by the photon-photon scattering.  But
in the limit of $\epho \rightarrow 0$, the total cross section for the
photon-photon scattering is proportional to $\epho^{3}$ and this process
loses its significance.
\item
Finally, photons lose their energy by scattering off charged particles
in the thermal bath. The dominant processes are pair creation in the
nuclei and the Compton scattering off the thermal electron.
\item
Emitted high energy electrons and positrons lose their energy by the
inverse Compton scattering off the background photon.
\item
The source of these cascade processes are the high energy photons
emitted in the decay of gravitinos. Notice that we only consider the
decay channel $\psi_{\mu}\rightarrow\gamma +\tilde{\gamma}$ and hence
the energy of the incoming photon $\epsilon_{\gamma 0}$ is monochromatic.
\end{itemize}

The Boltzmann equations for the photon and the electron distribution
function $f_{\gamma}$ and $f_{e}$ are given by
\begin{eqnarray}
\frac{\partial f_{\gamma}(\epho)}{\partial t}
&=&
\left. \frac{\partial f_{\gamma}(\epho)}{\partial t} \right |_{\rm DP}
+ \left. \frac{\partial f_{\gamma}(\epho)}{\partial t} \right |_{\rm PP}
+ \left. \frac{\partial f_{\gamma}(\epho)}{\partial t} \right |_{\rm PC}
+ \left. \frac{\partial f_{\gamma}(\epho)}{\partial t} \right |_{\rm CS}
\nonumber \\
&&
+ \left. \frac{\partial f_{\gamma}(\epho)}{\partial t} \right |_{\rm IC}
+ \left. \frac{\partial f_{\gamma}(\epho)}{\partial t} \right |_{\rm DE},
\label{bol-fp} \\
\frac{\partial f_{e}(\eele)}{\partial t}
&=&
\left. \frac{\partial f_{e}(\eele)}{\partial t} \right |_{\rm DP}
+ \left. \frac{\partial f_{e}(\eele)}{\partial t} \right |_{\rm PC}
+ \left. \frac{\partial f_{e}(\eele)}{\partial t} \right |_{\rm CS}
+ \left. \frac{\partial f_{e}(\eele)}{\partial t} \right |_{\rm IC},
\label{bol-fe}
\end{eqnarray}
where DP (PP, PC, CS, IC, and DE) represents double photon pair creation
(photon-photon scattering, pair creation in nuclei, Compton scattering,
inverse Compton scattering, and the contribution from the gravitino
decay). Full details are shown in Appendix~\ref{ap:spectrum}.

%
%
\begin{figure}[p]
\epsfxsize=14cm

\centerline{\epsfbox{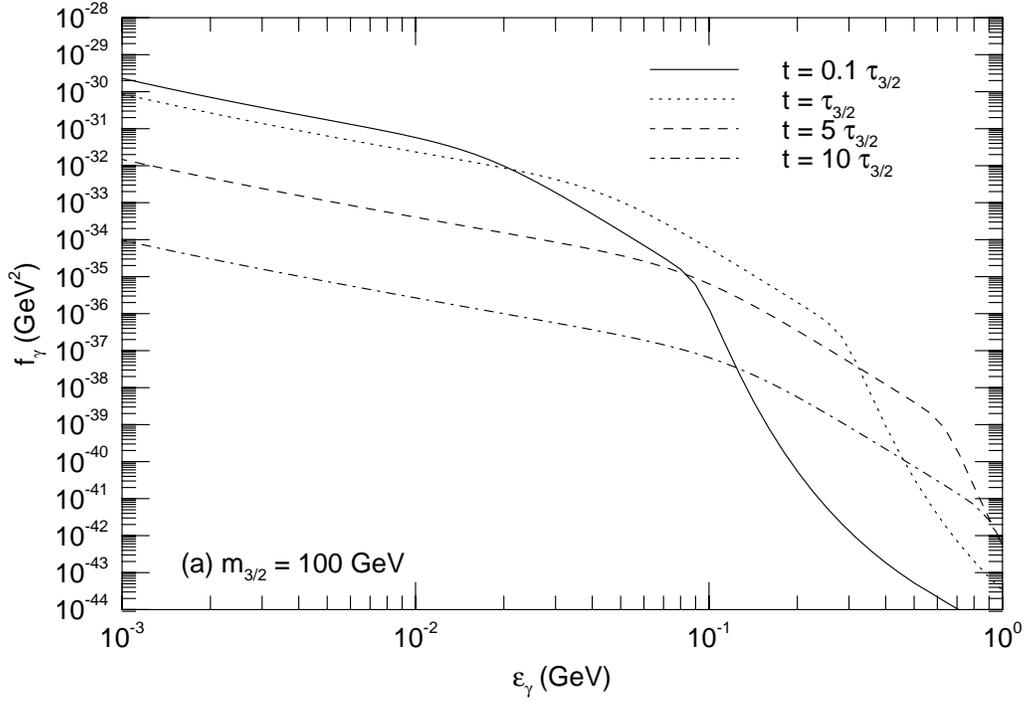}}

\vspace{1.0cm}

\centerline{\epsfbox{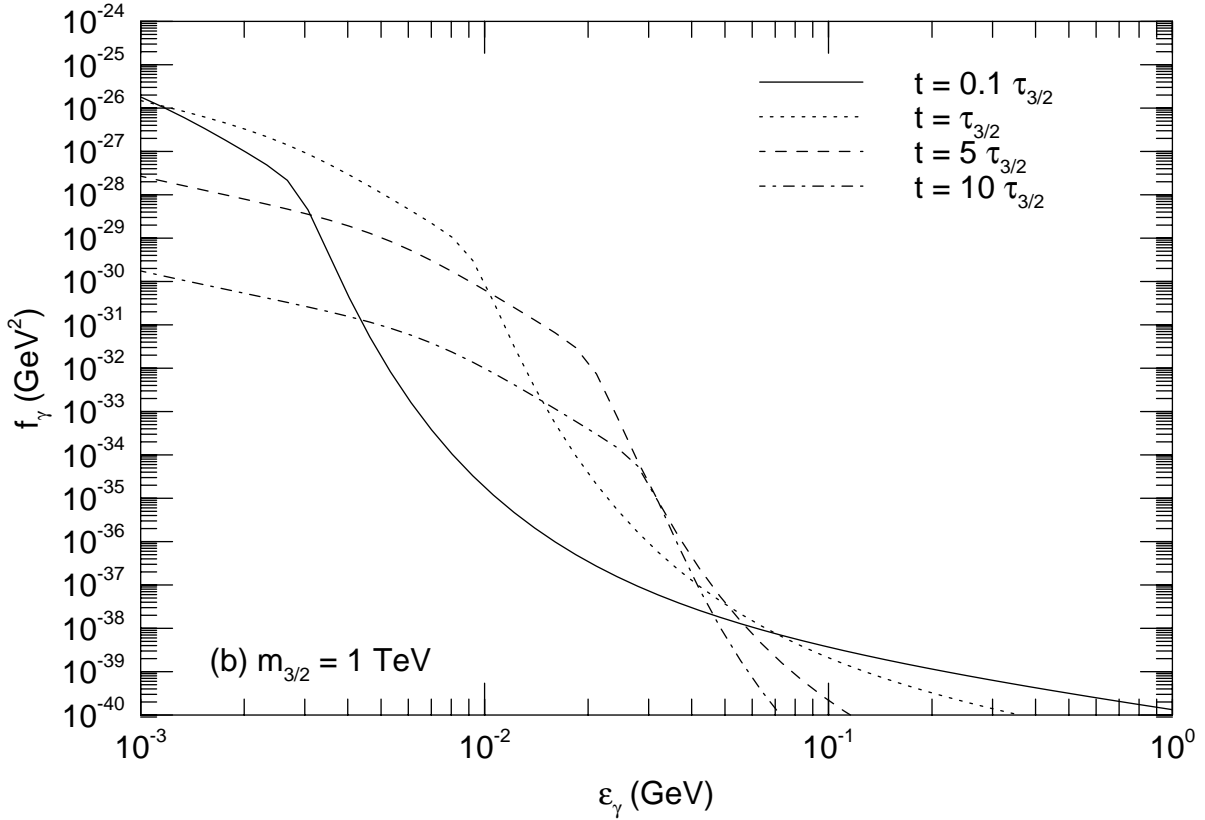}}

\caption{Time evolution of the photon spectrum with the case of
$T_R=10^{10}\GEV$, and (a) $\mgra =100\GEV$, and (b) $\mgra =1\TEV$. The
solid lines (the dotted lines, the dashed lines, and the dotted-dashed
lines) correspond to the photon spectra at the time $t=0.1\tau_{3/2}$
($t=\tau_{3/2}$, $t=5\tau_{3/2}$, and $t=10\tau_{3/2}$).}
\label{fig:f_time}
\end{figure}
%
%

By solving the Boltzmann equations (\ref{bol-fp}) and (\ref{bol-fe})
numerically, we obtain the time evolution of the photon and the
electron spectra. The typical time evolutions of the photon spectrum are
shown in Figs.\ref{fig:f_time}.

\section{BBN with high energy photon injection}

\hspace*{\parindent}
BBN is one of great successes of the standard big bang cosmology. It is
believed that light elements with atomic number less than 7 are produced
when the cosmic temperature is between 1MeV and 10keV. Theoretical
predictions for abundances of light elements are excellently in good
agreement with the observations if the baryon-to-photon ratio $\eta_B$
is about $3\times 10^{-10}$~\cite{APJ376-51}. (For a review, see
Appendix~\ref{ap:bbn}.)

However, the presence of the gravitino might destroy this success of the
BBN.  Gravitino may have three effects on BBN. First the energy density
of the gravitino at $T\sim 1$MeV speeds up the cosmic expansion and
leads to increase the $n/p$ ratio and hence $^4$He abundance also
increases.  Second, the radiative decay of gravitino reduces the
baryon-to-photon ratio and results in too baryon-poor universe. Third,
the high energy photons emitted in the decay of gravitino destroy the
light elements.  Among three effects, photo-dissociation by the high
energy photons is the most dangerous for the gravitino with mass less
than $\sim$ 1TeV. In the following we investigate the photo-dissociation
of light elements.

%
%
\begin{table}
\begin{center}
    \begin{tabular}{l|rr} \hline\hline
        Reaction & Threshold (MeV) & Reference \\ \hline
        D $+ \gamma \rightarrow n + p$  & 2.225 & \cite{Evans} \\
        T $+ \gamma \rightarrow n +$ D & 6.257
        & \cite{ZP208-129,PRL44-129}\\
        T $+ \gamma \rightarrow  p + n + n$ & 8.482 & \cite{PRL44-129}\\
        $^3$He $+ \gamma \rightarrow p + $D & 5.494 & \cite{PL11-137}\\
        $^3$He $+ \gamma \rightarrow p + $D & 7.718 & \cite{PL11-137}\\
        $^4$He $+ \gamma \rightarrow p + $ T & 19.815
        &\cite{SJNP19-589} \\
        $^4$He $+ \gamma \rightarrow n + $$^3$He & 20.578 &
	\cite{CJP53-802,PLB47-433} \\
        $^4$He $+ \gamma \rightarrow p + n + $ D & 26.072 &
        \cite{SJNP19-589}  \\ \hline\hline
    \end{tabular}
    \caption{Photo-dissociation reactions.}
    \label{table:dist-reaction}
\end{center}
\end{table}
%
%
The high energy photons emitted in the decay of gravitinos lose their
energy during multiple electro-magnetic processes described in the
previous section. Surviving soft photons can destroy the light elements
(D, T, $^3$He, $^4$He) if their energy are greater than the threshold of
the photo-dissociation reactions.  In this thesis, we consider the
photo-dissociation reactions listed in Table~\ref{table:dist-reaction}.
For the process D($\gamma$,$n$)$p$, we use the cross section in analytic
form which is given in ref.\cite{Evans}, and the cross sections for
other reactions are taken from the experimental data. (For references,
see Table~\ref{table:dist-reaction}). We neglect the reactions
$^4$He($\gamma$, D)D and $^4$He($\gamma$, 2p 2n) since their cross
sections are small compared with the other reactions. Furthermore, we do
not include the photo-dissociation processes for $^7$Li and $^7$Be
because the data of the cross section for $^7$Be is not available, and
hence we cannot predict the abundance of $^7$Li a part of which comes
from $^7$Be.

The time evolution of the light elements are described by
\begin{eqnarray}
    \frac{d n_{\rm D}}{d t} + 3Hn_{\rm D} & = &
    - n_{\rm D}\sum_i \int_{E_i}d\epsilon_{\gamma}
    \sigma^i_{ {\rm D}\rightarrow a}(\epsilon_{\gamma})
    f_{\gamma}
    (\epsilon_{\gamma})
    \nonumber \\ &&
    + \sum_i \int_{E_i}d\epsilon_{\gamma}
    \sigma^i_{a\rightarrow {\rm D}}(\epsilon_{\gamma})n_{a}
    f_{\gamma}
    (\epsilon_{\gamma}),
    \label{evol-h2}\\
    \frac{d n_{\rm T}}{d t} +3Hn_{\rm T} & = &
    - n_{\rm T}\sum_i
    \int_{E_i}d\epsilon_{\gamma}
    \sigma^i_{ {\rm T}\rightarrow a}(\epsilon_{\gamma})
    f_{\gamma}
    (\epsilon_{\gamma})
    \nonumber \\ &&
    + \sum_i \int_{E_i}d\epsilon_{\gamma}
    \sigma^i_{a\rightarrow {\rm T}}(\epsilon_{\gamma})n_{a}
    f_{\gamma}
    (\epsilon_{\gamma}),
    \label{evol-h3}\\
    \frac{d n_{^3{\rm He}}}{d t} +3Hn_{^3{\rm He}} & = &
    - n_{^3{\rm He}}\sum_i
    \int_{E_i}d\epsilon_{\gamma}
    \sigma^i_{ {^3{\rm He}}\rightarrow a}(\epsilon_{\gamma})
    f_{\gamma}
    (\epsilon_{\gamma})
    \nonumber \\ &&
    + \sum_i \int_{E_i}d\epsilon_{\gamma}
    \sigma^i_{a\rightarrow {^3{\rm He}}}(\epsilon_{\gamma})n_{a}
    f_{\gamma}
    (\epsilon_{\gamma}),
    \label{evol-he3}\\
    \frac{d n_{^4{\rm He}}}{d t} +3Hn_{^4{\rm He}} & = &
    - n_{^4{\rm He}}\sum_i
    \int_{E_i}d\epsilon_{\gamma}
    \sigma^i_{ {^4{\rm He}}\rightarrow a}(\epsilon_{\gamma})
    f_{\gamma}
    (\epsilon_{\gamma})
    \nonumber \\ &&
    + \sum_i \int_{E_i}d\epsilon_{\gamma}
    \sigma^i_{a\rightarrow {^4{\rm He}}}(\epsilon_{\gamma})n_{a}
    f_{\gamma}
    (\epsilon_{\gamma}),
    \label{evol-he4}
\end{eqnarray}
where $\sigma^i_{a\rightarrow b}$ is the cross section of the
photo-dissociation process $i$; $a + \gamma \rightarrow b +
\cdot\cdot\cdot$, and $E_i$ is the threshold energy of reaction $i$.
When the energy of the high energy photon is relatively low, \ie $2\MEV
\lsim \epho\lsim 20\MEV$, D, T and $^3$He are destroyed and their
abundances decrease. On the other hand, if the photons have energy high
enough to destroy $^4$He, it seems that such high energy photons
decrease the abundance of all light elements.  However, since D, T and
$^3$He are produced by the photo-dissociation of $^4$He whose abundance
is much higher than the other elements, their abundances increase or
decrease depending on the number density of high energy photon. When the
number density of high energy photons with their energy greater than
$\sim$ 20MeV (\ie the threshold energy for $^4$He destruction) is
extremely high, all light elements are destroyed. But as the photon
density becomes lower, there is some range of the high energy photon
density in which the overproduction of D, T and $^3$He becomes
significant.  And if the density is sufficiently low, the high energy
photon does not affect the BBN at all.

{}From various observations, the primordial abundances of light elements
(D, $^3$He, $^4$He) are estimated~\cite{APJ376-51} as
\begin{eqnarray}
    &&  0.22 < Y_p \equiv
    \llp \frac{\rho_{^{4}{\rm He}}}{\rho_{B}}\rlp_p < 0.24 ,
    \label{obs-he4}\\
    && \llp \frac{n_{\rm D}}{n_{\rm H}} \rlp_p > 1.8\times 10^{-5},
    \label{obs-h2}\\
    && \llp\frac{n_{\rm D}+n_{^3{\rm He}}}{n_{\rm H}} \rlp_p
    <  1.0 \times 10^{-4},
    \label{obs-h23}
\end{eqnarray}
where ${\rho_{^{4}{\rm He}}}$ and ${\rho_{B}}$ are the mass densities of
$^4$He and baryon. (For details, see Appendix~\ref{ap:bbn}.) The
abundances of light elements modified by the gravitino decay must
satisfy the observational constraints above. In order to make precise
predictions for the abundances of light elements, the evolution
equations (\ref{evol-h2}) -- (\ref{evol-he4}) should be incorporated in
the nuclear network calculation of BBN.  Therefore, we modify
the Kawano computer code~\cite{Kawano} to include the photo-dissociation
processes.

{}From the above arguments it is clear that there are at least three free
parameters, \ie the mass of gravitino $m_{3/2}$, the reheating
temperature $T_R$ and the baryon-to-photon ratio $\eta_B$.  Furthermore,
we also study the case in which the gravitino has other decay channels.
Here, we do not specify other decay channel.  Instead, we introduce
another free parameter $B_{\gamma}$ which is the branching ratio for the
channel $\psi_{\mu} \rightarrow \gamma +\tilde{\gamma}$. Therefore, we
must study the effect of gravitino decay on BBN in four dimensional
parameter space. However, in the following arguments it will be shown
that the baryon-to-photon ratio $\eta_B$ is not an important parameter
in the present calculation because the allowed value for $\eta_{B}$ is
almost the same as that in the standard case (\ie without gravitino).

%
%
\begin{figure}[p]
\epsfxsize=14cm

\centerline{\epsfbox{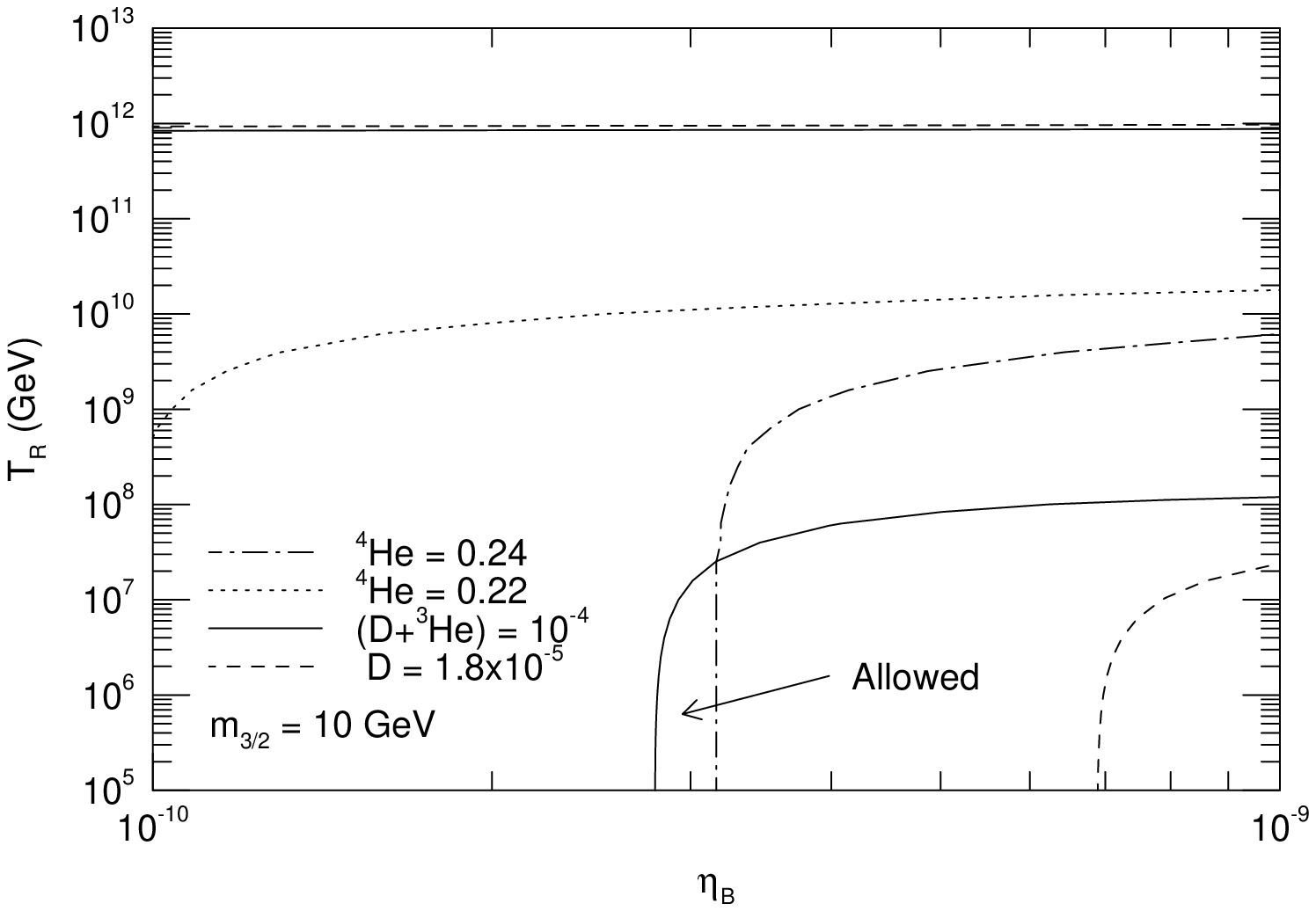}}

\caption{Contours for critical abundance of light elements in
the $\eta_B$ -- $T_R$ plane for $m_{3/2} = 10\GEV$. The solid line
(dashed line, dotted line, dotted-dashed line) represents the
constraints from overproduction of (D+$^3$He) (overdestruction of D,
overdestruction of $^4$He, overproduction of $^4$He).}
\label{fig:eta-T1}

\vspace{0.7cm}

\centerline{\epsfbox{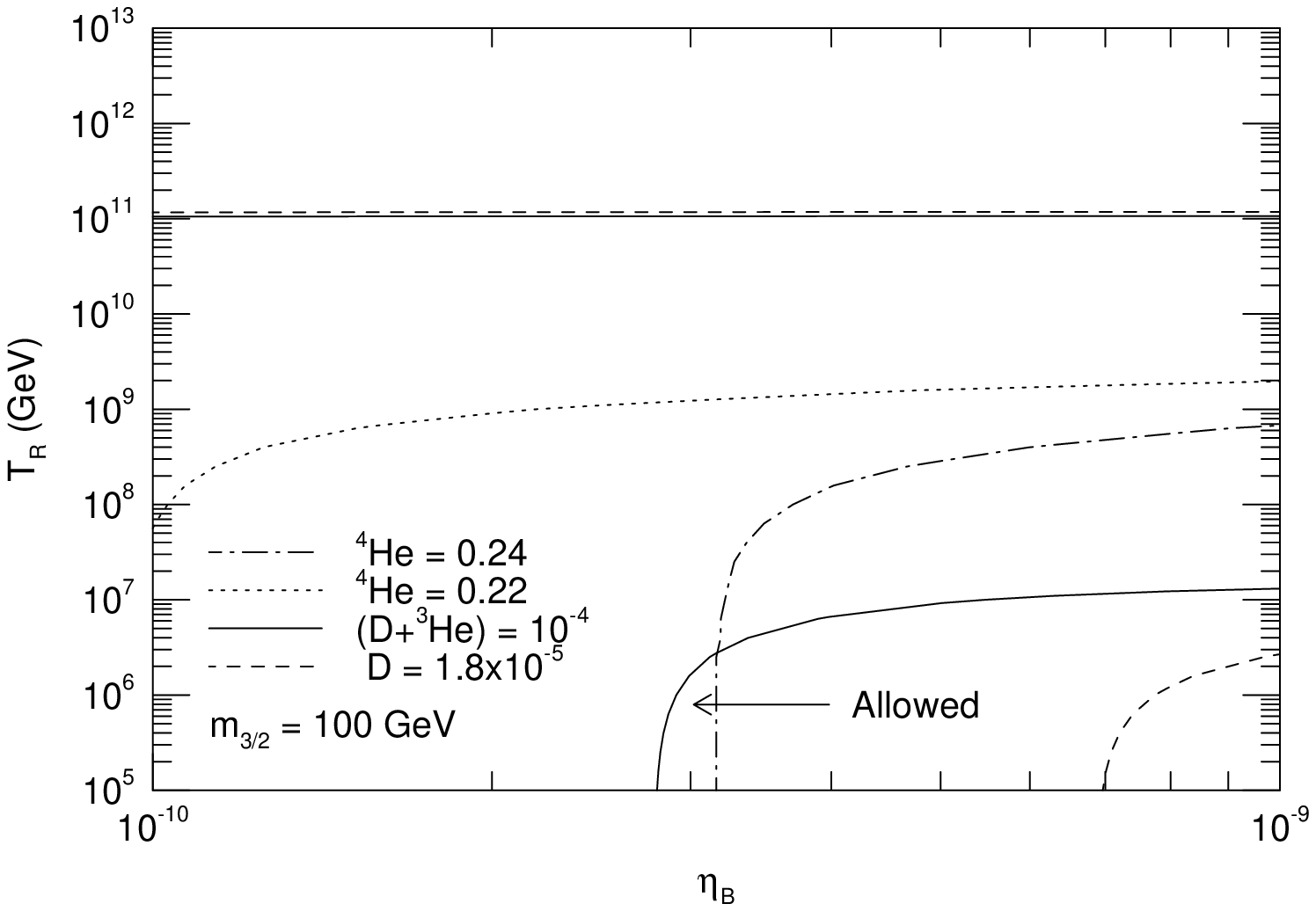}}

\caption{Same as Fig.~\protect\ref{fig:eta-T1} except for
$m_{3/2} = 100\GEV$.}
\label{fig:eta-T2}

\end{figure}

\begin{figure}[p]
\epsfxsize=14cm

\centerline{\epsfbox{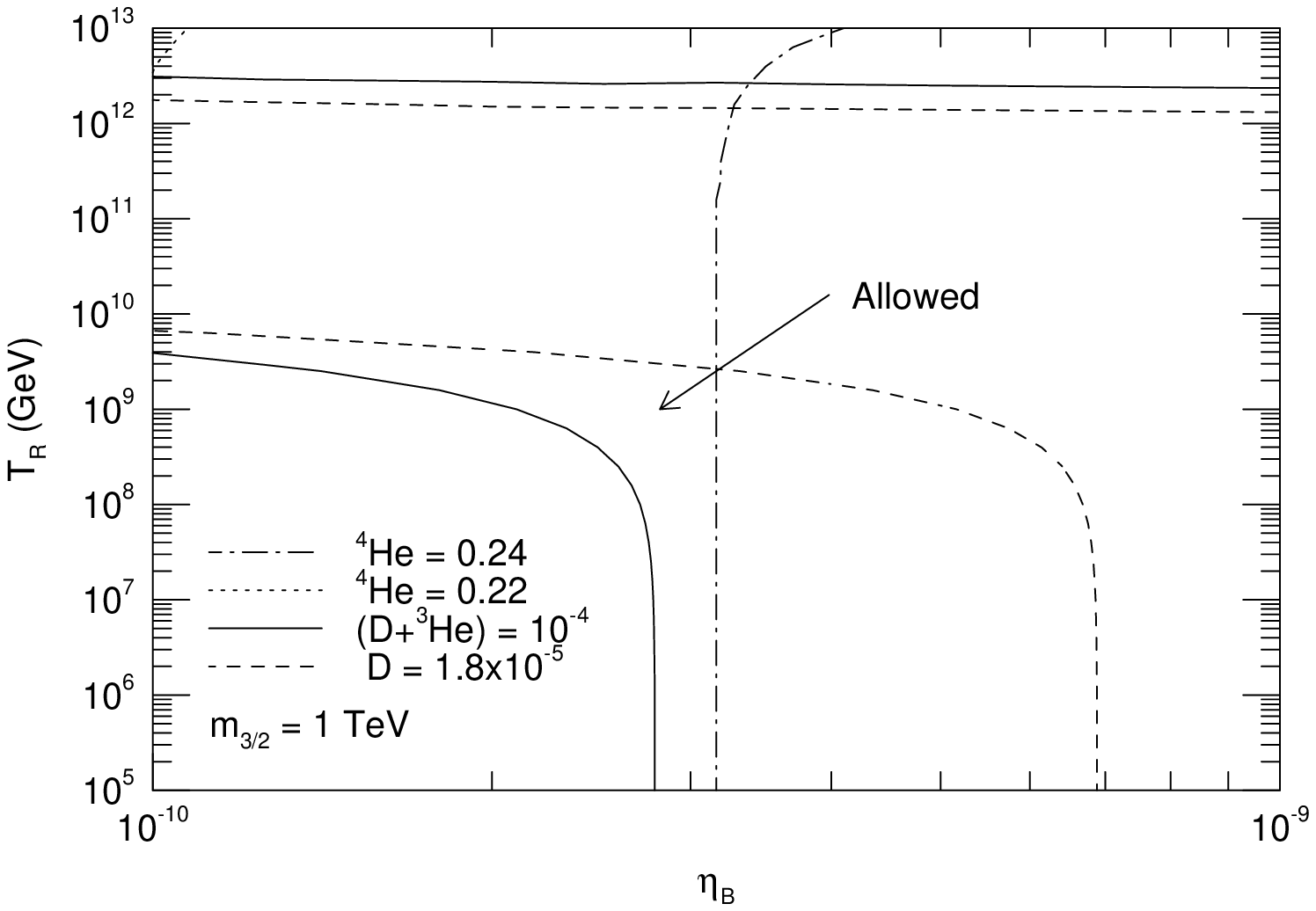}}

\caption{Same as Fig.~\protect\ref{fig:eta-T1} except for
$m_{3/2} = 1\TEV$.}
\label{fig:eta-T3}

\vspace{0.7cm}

\centerline{\epsfbox{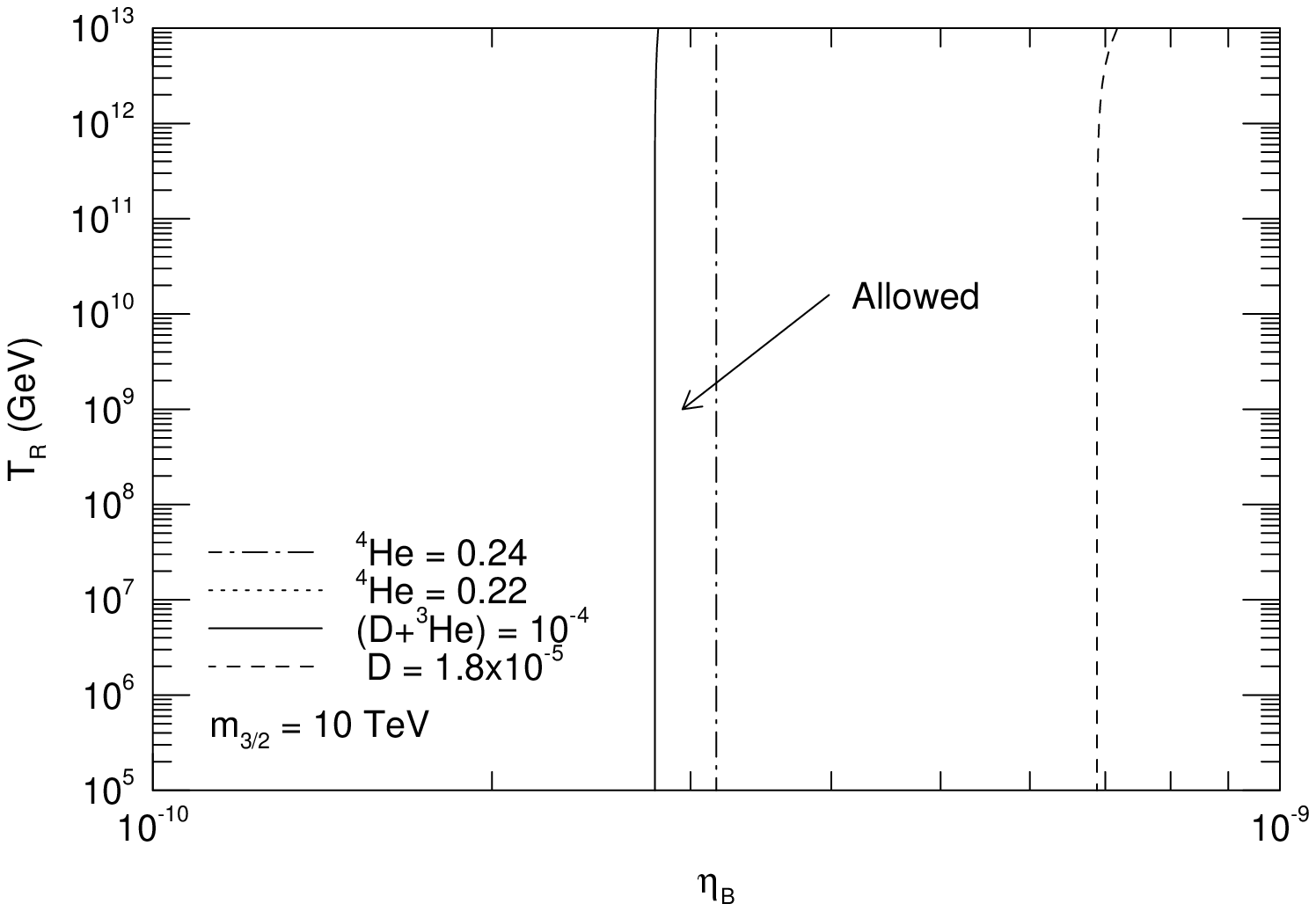}}

\caption{Same as Fig.~\protect\ref{fig:eta-T1} except for
$m_{3/2} = 10\TEV$.}
\label{fig:eta-T4}

\end{figure}
%
%

First we investigate the photo-dissociation effect when all
gravitinos decay into photons and photinos ($B_{\gamma} =1 $). We
take the range of three free parameters as $10\GEV \le m_{3/2} \le
10\TEV$, $10^{5}\GEV \le T_R \le 10^{13}\GEV$ and $10^{-10} \le \eta_B
\le 10^{-9}$. In this calculation, we assume that the photino is
massless.\footnote
{Constraints from the photo-dissociation of the light elements are
almost independent of the photino mass if the photino is sufficiently
lighter than the gravitino.}
The contours for the critical abundances of the light elements D,
(D+$^3$He) and $^4$He in the $\eta_B$ -- $T_R$ plane are shown in
Fig.~\ref{fig:eta-T1} -- Fig.~\ref{fig:eta-T4} for $m_{3/2} = 10\GEV$,
$100\GEV$, $1\TEV$ and $10\TEV$. For lower reheating temperatures ($T_R
\lsim 10^6\GEV$), the number density of the gravitino is very low and
hence the number density of the induced high energy photons is too low
to affect the BBN. Therefore, the resultant abundances of light elements
are the same as those in the standard BBN.  The effect of the
photo-dissociation due to gravitino decay becomes significant as the
reheating temperature increases.

As seen in Fig.~\ref{fig:eta-T1} -- Fig.~\ref{fig:eta-T4}, the allowed
range of the baryon-to-photon ratio is almost same as that without
gravitino for $m_{3/2} \lsim 1\TEV$, \ie a very narrow range around
$\eta_B \sim 3\times 10^{-10}$ is allowed.  However, for the cases of
$m_{3/2}\sim 1\TEV$ and $T_R\sim 10^{9}\GEV$ or $m_{3/2}\sim 1 \TEV$ and
$T_R \sim 10^{12}\GEV$ , lower values of $\eta_B$ are allowed
(Fig.~\ref{fig:eta-T3}). In this case, the critical photon energy ($\sim
m_e^2/22T$) for the double photon pair creation process is lower than the
threshold of photo-dissociation reaction of $^4$He. Therefore, for $T_R
\lsim 10^{12} \GEV$, the abundance of $^4$He is not affected by the
gravitino decay. Then, the emitted photons only destroy $^3$He and D
whose abundances would be larger than the observational constraints for
low baryon density if gravitino did not exist.  Therefore one sees the
narrow allowed band at $T_R\simeq 10^9\GEV$ where only a small number of
$^3$He and D are destroyed to satisfy the constraints (\ref{obs-h2}) and
(\ref{obs-h23}).  For $T_R\gsim 10^{12}\GEV$, since a large number of
high energy photons are produced even above the threshold of double
photon pair creation, a part of $^4$He are destroyed to produce $^3$He
and D, which leads to the very narrow allowed region at $T_R \sim
10^{12}\GEV$.  However, even in this special case the upper limit of
allowed reheating temperature changes very little between $\eta_B =
10^{-10}$ and $\eta_B \sim 3\times 10^{-10}$. This allows us to fix
$\eta_{B}=3.0\times 10^{-10}$ in deriving the upperbound on the
reheating temperature.

%
%
\begin{figure}[t]
\epsfxsize=14cm
\centerline{\epsfbox{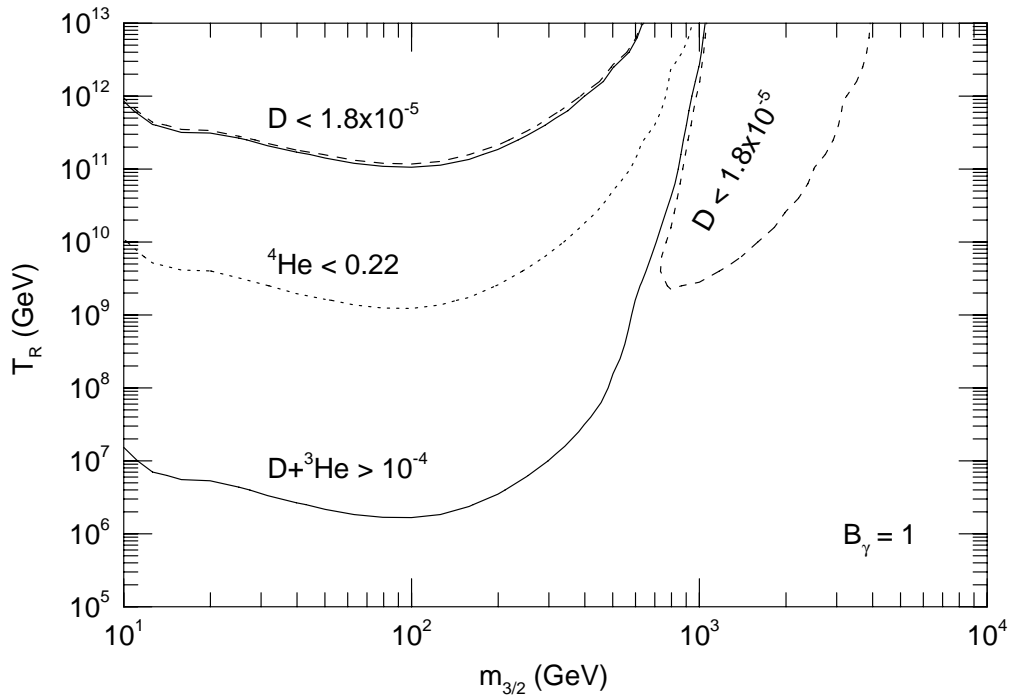}}
\caption{Upperbound on $T_R$ as a function of $\mgra$. Here, we
take $B_{\gamma} =1$.  In the region above the solid curve $^3$He and D
are overproduced, the abundance of $^4$He is less than 0.22 above the
dotted curve and the abundance of D is less than $1.8\times 10^{-5}$
above the dashed curve.}
\label{fig:bound0}
\end{figure}
%
%

The allowed regions that satisfy the observational constraints
(\ref{obs-he4}) -- (\ref{obs-h23}) are also shown in Fig.~\ref{fig:bound0}
in the $m_{3/2}$ -- $T_R$ plane for the case of $\eta_B = 3\times
10^{-10}$ and $B_\gamma =1$. In Fig.~\ref{fig:eta-T1} --
Fig.~\ref{fig:eta-T4} and Fig.~\ref{fig:bound0} one can see four typical
cases depending on $T_R$ and $m_{3/2}$.
\begin{itemize}
\item $m_{3/2} \lsim 1\TEV$, $T_R \lsim 10^{11}\GEV$;\\ In this case
the lifetime of the gravitino is so long that the critical energy for
the double photon process ($\sim m_{e}^{2}/22T$) is higher than the
threshold of the photo-dissociation reactions for $^4$He at the decay
time of gravitino.  Thus $^4$He is destroyed to produce T, $^3$He and D.
(Since T becomes $^3$He by $\beta$-decay, hereafter we denote T and
$^3$He by the word ``$^3$He''.) Since the reheating temperature is not
so high, the number density of gravitino is not high enough to destroy
all the light elements completely. As a result, $^3$He and D are
produced too much and the abundance of $^4$He decreases.  To avoid the
overproduction of ($^3{\rm He}+{\rm D}$), the reheating temperature
should be less than $\sim$ $(10^{6}-10^{9})\GEV$.
\item $m_{3/2} \lsim 1\TEV$, $T_R \gsim 10^{11}\GEV$;\\ The
lifetime is long enough to destroy $^4$He and the gravitino abundance is
so large that all the light elements are destroyed since the reheating
temperature is high enough. This parameter region is strongly excluded
by the observation.
\item $1\TEV \lsim m_{3/2} \lsim 3\TEV$;\\ The lifetime becomes
shorter as the mass of gravitino increases, and the decay occurs
when the double photon pair creation process works well.  If the cosmic
temperature at $t\sim\tau_{3/2}$ is greater than $\sim$
$m_e^2/22E_{^4{\rm He}}$ (where $E_{^4{\rm He}} \sim 20\MEV$ represents
the typical threshold energy of $^4$He destruction processes), $^4$He
abundance is almost unaffected by the high energy photons as can be seen
in Fig.~\ref{fig:eta-T3}. In this parameter region, the overproduction
of (D+$^3$He) cannot occur since $^4$He is not destroyed. In this case,
the destruction of D is the most important to set the limit of the
reheating temperature. This gives the constraint of $T_R \lsim
(10^{9}-10^{12})\GEV$.
\item $m_{3/2}\gsim 3\TEV$;\\ In this case the gravitinos decay when
the temperature of the universe is so high that all high energy photons
are quickly thermalized by the double photon process before they destroy
the light elements.  Therefore, the effect on the BBN is negligible.
Fig.~\ref{fig:eta-T4} is an example of this case. The resultant contours
for abundances of light elements are almost identical as those without
the decaying gravitino.
\end{itemize}
So far we have assumed that all gravitinos decay into photons and
photinos. But if other superpartners are also lighter than the
gravitino, the decay channels of gravitino increases and the branching
ratio for the channel $\psi_{\mu} \rightarrow \gamma + \tilde{\gamma}$
becomes less than 1. In this case, various decay products affect the
evolution of the universe and BBN. In this thesis, instead of studying
all decay channels, we consider only the $\gamma + \tilde{\gamma}$
channel with taking the branching ratio $B_{\gamma}$ as another free
parameter.  With this simplification, the effect of all possible decay
products other than photon is not taken into account.  Therefore, the
resultant constraints on the reheating temperature and the mass of
gravitino should be taken as the conservative constraints since other
decay products may destroy more light elements and make the constraints
more stringent.

Although we have four free parameters in the present case, the result
for $B_{\gamma} =1$ implies that the allowed range of $T_R$ and
$m_{3/2}$ is obtained taking the baryon-to-photon ratio to be $3\times
10^{-10}$. Since our main concern is to set the constraints on $T_R$ and
$m_{3/2}$, we can safely fix $\eta_B$ as $3\times 10^{-10}$.

%
%
\begin{figure}[p]
\epsfxsize=14cm

\centerline{\epsfbox{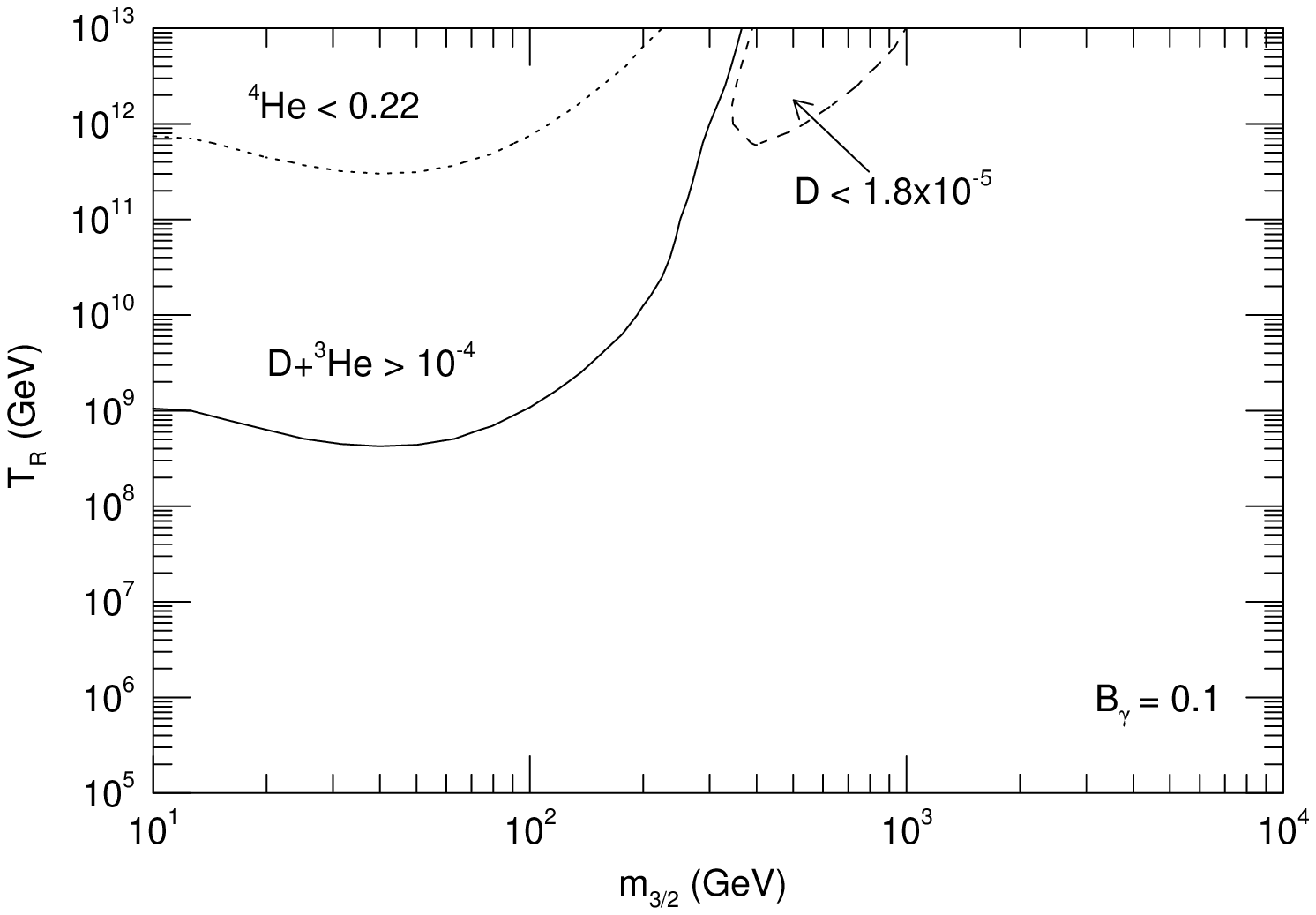}}

\caption{Same as Fig.~\protect\ref{fig:bound0} except for
$B_\gamma = 0.1$.}
\label{fig:bound1}

\vspace{0.7cm}

\centerline{\epsfbox{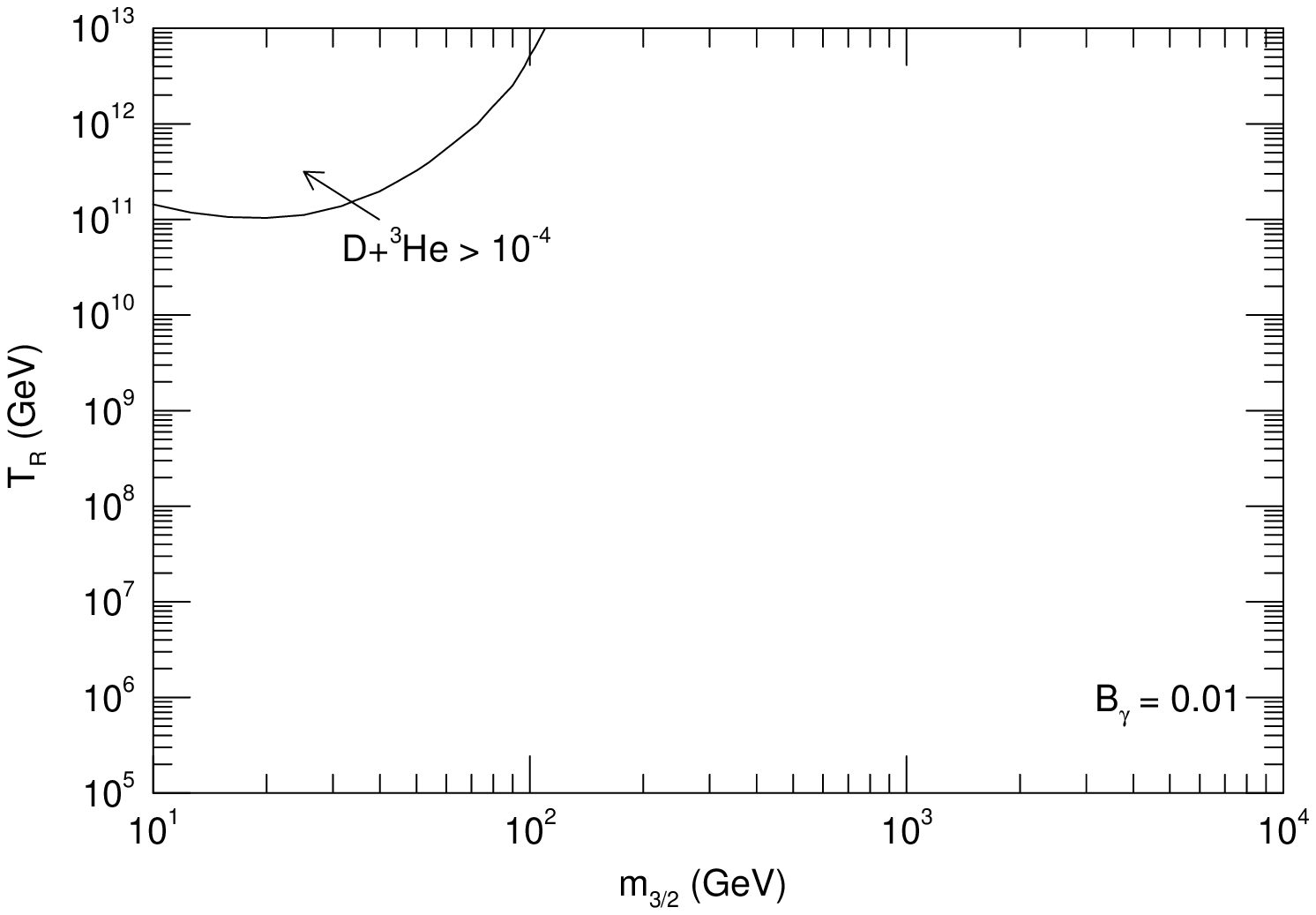}}

\caption{Same as Fig.~\protect\ref{fig:bound0} except for
$B_\gamma = 0.01$.}
\label{fig:bound2}

\end{figure}

%
%

The constraints for $B_{\gamma} = 0.1$ and $B_{\gamma} = 0.01$ are shown
in Fig.~\ref{fig:bound1} and Fig.~\ref{fig:bound2} which should be
compared with Fig.~\ref{fig:bound0} (\ie the case of $B_{\gamma} =1$).
Since the number density of the high energy photons is proportional to
$B_{\gamma}$, the constraint on the reheating temperature becomes less
stringent as $B_{\gamma}$ decreases. In addition, the total lifetime of
gravitino is given by
\begin{equation}
    \tau_{3/2} = \tau(\psi_{\mu} \rightarrow \gamma +
    \tilde{\gamma} ) \times B_{\gamma}.
\end{equation}
Thus the gravitinos decay earlier than for the $B_{\gamma}=1$ case, and
hence the constraints from ($^3$He $+$ D) overproduction becomes less
stringent.  This effect can be seen in Fig.~\ref{fig:bound1}, where the
constraint due to the overproduction of ($^3$He $+$ D) has a cut at
$\mgra \simeq 400\GEV$ compared with $\sim 1\TEV$ for $B_{\gamma} = 1$.

%
%
\begin{figure}[t]
\epsfxsize=14cm

\centerline{\epsfbox{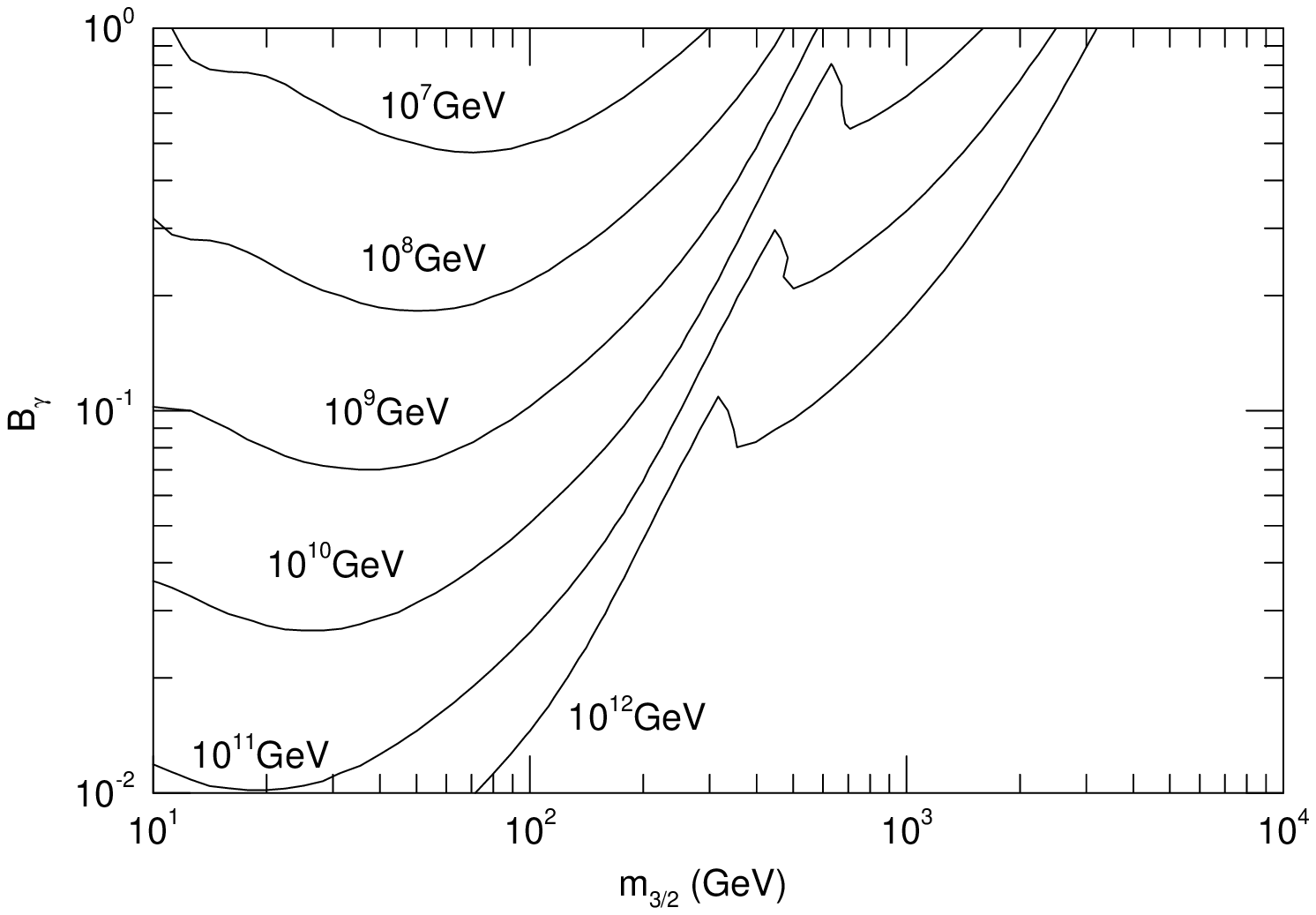}}

\caption{Contours for the upper limits of the reheating
temperature in the $m_{3/2}$ -- $B_{\gamma}$ plane. The numbers in the
figure denote the limit of the reheating temperature.}
\label{fig:mgra-Br}

\end{figure}

%
%

In Fig.~\ref{fig:mgra-Br}, the contours for the upperbound on reheating
temperature are shown in the $m_{3/2}$ -- $B_{\gamma}$ plane.  One can
see that the stringent constraint on $T_R$ is obtained for $m_{3/2}\lsim
100\GEV$ even if the branching ratio is small. Notice that the actual
constraint may become more stringent if we take the effects of other
decay products into account.

\section{BBN with high energy neutrino injection}

\hspace*{\parindent}
As seen in the previous section, the existence of the unstable gravitino
which decays into a photon and a photino set a stringent upperbound on
the reheating temperature $T_R$. However, the constraints might become
much weaker when the gravitino decays only into weakly interacting
particles. In the particle content of the MSSM, the only candidate is
the decay into a neutrino and a sneutrino.

Even if the gravitino decays only into a neutrino and a sneutrino, the
emitted high energy neutrino may scatter off the background neutrino and
produce an electron-positron (or muon-anti-muon) pair, which then
produces many soft photons through electro-magnetic cascade processes
and destruct light elements.  Since the interaction between the high
energy neutrinos and the background neutrinos is weak, it seems that the
destruction of the light elements is not efficient.  In fact, Gratsias,
Scherrer and Spergel~\cite{PLB262-198} showed that the constraint is not
so stringent for the case where the gravitino decays into a neutrino and
a sneutrino.  However the previous analysis seems to be incomplete in a
couple of points.  First, Gratsias et~al.~\cite{PLB262-198} totally
neglected the secondary high energy neutrinos which are produced by the
neutrino-neutrino scattering. The effect of the secondary neutrino may
be important for the heavy gravitino case ($m_{3/2}\gsim 1$TeV).
Second, they only studied the case where the destruction of $^4$He
results in the overproduction of $(^3{\rm He} + {\rm D})$.  However, for
the heavy gravitino which decays in early stage of the BBN, the
destruction of D is more important since the electro-magnetic cascade
process is so efficient that the energy of soft photons becomes less
than the threshold of $^4$He destruction.  Furthermore, as pointed out
in the previous section, the previous estimation of the gravitino
production in the reheating epoch after the inflation is underestimated.
Those effects which are not taken into account in ref.\cite{PLB262-198}
may lead to a more stringent constraint on the reheating temperature.
Therefore, in this section we reexamine the effects of the decay of the
gravitino into a neutrino and a sneutrino ($\psi_{\mu}\rightarrow \nu +
\tilde{\nu}$) with taking all relevant processes into account~\cite{TU463}.

The high energy neutrinos $\nu$ produced in the gravitino decay scatter
off the thermal neutrino $\nu_b$ in the background through the following
processes;
\begin{eqnarray}
    \nu_{i} + \nu_{i,b} &\rightarrow & \nu_i + \nu_i ,\\
    \nu_{i} + \bar{\nu}_{i,b} &\rightarrow &\nu_i + \bar{\nu}_i,\\
    \nu_{i} + \bar{\nu}_{i,b} &\rightarrow &\nu_j + \bar{\nu}_j ,\\
    \nu_{i} + \nu_{j,b} &\rightarrow &\nu_i + \nu_j,\\
    \nu_{i} + \bar{\nu}_{j,b} & \rightarrow &\nu_i + \bar{\nu}_j,\\
    \nu_{i} + \bar{\nu}_{i,b} &\rightarrow & e^{-}  + e^{+},
	\label{nu2e}\\
    \nu_{i} + \bar{\nu}_{i,b} & \rightarrow & \mu^{-}  + \mu^{+},
	\label{nu2mu}
\end{eqnarray}
where index $i$ and $j$ represent $e$, $\mu$ and $\tau$ with $i\neq j$.
The primary and secondary high energy neutrinos scatter off the
background neutrinos and produce charged lepton pairs. Then, the charged
leptons produced in the processes (\ref{nu2e}) and (\ref{nu2mu})
induce the electro-magnetic cascade processes. By the same procedure
as in the previous section, we obtain the high energy photon and the
electron spectra. The detailed analysis of the neutrino spectrum is
given in Appendix~\ref{ap:spectrum}.

The photo-dissociation of the light elements are analyzed in the same
way as in the previous section. In the present calculation, there are at
least three free parameters, \ie the mass of the gravitino $m_{3/2}$,
the reheating temperature $T_R$ and the baryon-to-photon ratio $\eta_B$.
However as shown in the previous section, the baryon-to-photon
ratio $\eta_B$ is not important parameter because the allowed value for
$\eta_{B}$ is almost the same as that in the standard case (\ie without
gravitino). Therefore, we fix $\eta_B = 3\times 10^{-10}$ in the following
analysis.

%
%
\begin{figure}[t]
\epsfxsize=14cm
\centerline{\epsfbox{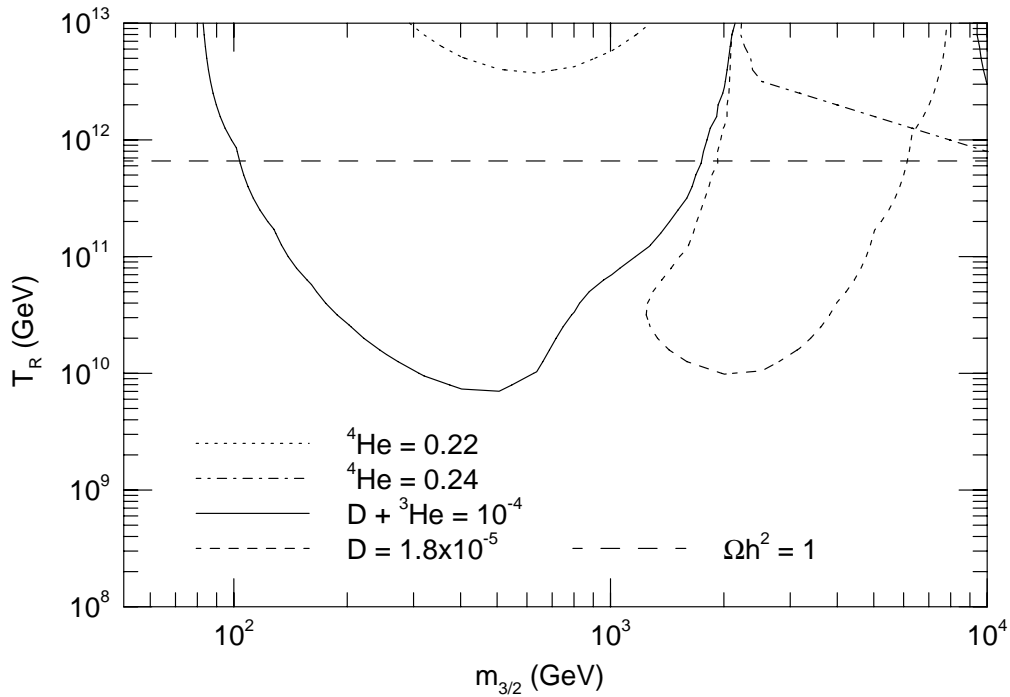}}
\caption{Upperbound on the reheating temperature from BBN and the
present mass density of the sneutrino. The region above the curves is
excluded.}
\label{fig:snu_lsp}
\end{figure}
%
%

The allowed regions that satisfy the observational constraints
(\ref{obs-he4}) -- (\ref{obs-h23}) are shown in the $m_{3/2}$ -- $T_R$
plane in Fig.~\ref{fig:snu_lsp}. In Fig.~\ref{fig:snu_lsp} one can see
that for the gravitino of mass between 100GeV and 1TeV, the
overproduction of ${\rm D}$ and $^3{\rm He}$ gives the most stringent
constraint, while the upperbound on the reheating temperature is
determined from the destruction of D for $\mgra \simeq (1-3)$TeV.
Notice that D destruction was not considered in the previous
work~\cite{PLB262-198}. Furthermore, the constraint from (D+$^3$He)
overproduction is more stringent. The reasons why we obtain the more
stringent constrain are (i) the gravitino abundance is $(4-5)$ times
larger than the one that the previous authors used, (ii) the secondary
neutrinos are taken into account in our calculation, and (iii) the
gravitino lifetime for $\psi_{\mu} \rightarrow \nu + \tilde{\nu}$ is
longer by a factor 2 than the lifetime for $\psi_{\mu} \rightarrow
\gamma + \tilde{\gamma}$. (In ref.\cite{PLB262-198} it is presumed that
$\Gamma_{3/2}(\psi_{\mu}\rightarrow \nu +\tilde{\nu}) =
\Gamma_{3/2}(\psi_{\mu} \rightarrow \gamma + \tilde{\gamma})$.)

\section{Discussion about hadron injection}

\hspace*{\parindent}
In the previous sections, we have assumed that the gravitino only decays
into non-hadronic particles, and derived constraints on the reheating
temperature. If the gravitino produces hadronic decay products, however,
the previous constraints may change. In this section, we will briefly
discuss the effect of such strongly interacting decay products.

The effects of the hadronic decay products on the BBN are investigated
by several authors~\cite{PRD37-3441,APJ330-545,NPB311-699}, and the
standard BBN scenario may be affected by the following processes.
\begin{itemize}
\item
Injecting high energy particles at the time $(1-10^2){\rm sec}$ cause
non-standard $p\leftrightarrow n$ conversion processes. The major effect
is to induce $p\rightarrow n$ reactions because there are more target
protons than target neutrons. The extra neutrons produced in this period
results in overproduced $^4$He abundance.
\item
Hadronic decay of the gravitino with lifetime $\tau_{3/2}\gsim 10^2~{\rm
sec}$ causes dissociation processes of light elements. Especially even
in the case $\tau_{3/2}\lsim 10^4{\rm sec}$ in which the light element
photo-dissociation processes are not effective, $^4$He may be
significantly destroyed to produce D and $^3$He.
\end{itemize}
In the following, we discuss the above effects.

The $p\leftrightarrow n$ conversion induced by hadronic injection was
studied by Reno and Seckel~\cite{PRD37-3441} in details. In the standard
BBN scenario, the neutron fraction in the thermal bath changes from
$\sim 0.16$ (at $T\sim 0.7\MEV$) to $\sim 0.12$ (at $T\sim 0.08\MEV$).
Hadronic injection at this period $(0.7\MEV\lsim T\lsim 0.08\MEV)$
affects this evolution. Reno and Seckel claimed that the yield of the
gravitino $Y_{3/2}$ should be smaller than $O(10^{-11}-10^{-12})$ in
order not to overproduce $^4$He. From this bound, we can estimate that
the reheating temperature $T_R$ should be lower than
$O((10^{10}-10^{11})\GEV)$ for the case $10\TEV\lsim\mgra
\lsim 100\TEV$.

If the gravitino lifetime is longer than $\sim 100{\rm sec}$, gravitinos
decay after the ``deuterium bottleneck'' breaks. In this case, emitted
hadrons induce light element destruction processes, which give more
stringent constraint than that from $^4$He overproduction. These
processes are analyzed in ref.\cite{PRD37-3441} (for the case
$\tau_{3/2}\lsim 10^{4}{\rm sec}$), and in
refs.\cite{APJ330-545,NPB311-699} (for the case $\tau_{3/2}\gsim
10^{4}{\rm sec}$).

For the case $10^2{\rm sec}\lsim\tau_{3/2}\lsim 10^{4}{\rm sec}$,
(D+$^3$He) overproduction induced by $^4$He destruction gives a
constraint. According to ref.\cite{PRD37-3441}, the reheating
temperature $T_R$ should be lower than $O((10^{8}-10^{11})\GEV)$ if the
gravitino mass is given by (1--10)TeV.

If the gravitino lifetime is longer than $\sim 10^4{\rm sec}$,
photo-dissociation of light elements becomes effective. In this case, we
should take the effects of high energy photons as well as injected
hadrons into account. As discussed in
refs.~\cite{APJ330-545,NPB311-699}, the main effects of the hadrinic
injection in this case is the destruction of $^4$He, and the creation of
D, $^3$He, $^6$Li and $^7$Li. According to
refs.~\cite{APJ330-545,NPB311-699}, the hadronic branching ratio must be
very small in order not to overproduce (D+$^3$He), and hence the
constraint is almost the same as in the case $B_\gamma =1$, except for
the case $\tau_{3/2}\sim 10^{5-6}{\rm sec}$. In a small parameter region
$\tau_{3/2}\sim 10^{5-6}{\rm sec}$, one may avoid the severe constraint
even in the case $B_\gamma =1$ by assuming the non-vanishing hadronic
branching ratio, since the effect of photo-dissociation of D is
compensated by the supply of D through the $^4$He destruction processes.
However in this case, $^6$Li and $^7$Li seem to be overproduced.
Furthermore, as discussed in ref.\cite{APJ330-545} some uncertainties
exist in the analysis of refs.\cite{APJ330-545,NPB311-699}. The most
important source of uncertaintie is in the experimental data of hadron
scattering processes, especially those concerning Li. Therefore, it is
unclear to us if the scenario proposed in
refs.\cite{APJ330-545,NPB311-699} works well.

\section{Other constraints}
\label{sec:others}

\hspace*{\parindent}
In the previous sections, we have considered the constraints from the
photo-dissociation of light elements. But as we have seen, if the mass
of the gravitino is larger than a few TeV, the gravitino decay does not
induce light element photo-dissociation and no constraints has been
obtained.  In this case, we must consider other effects of the
gravitino.

If we consider the present mass density of the LSP produced by the
gravitino decay, we can get the upperbound on the reheating temperature.
In SUSY models with $R$-invariance (which is an usual assumption), the
LSP (in the previous cases, photino or sneutrino) is stable. Thus the
LSPs produced by the decay of gravitinos survive until today, and they
contribute to the energy density of the present universe.  Since one
gravitino produces one LSP, we can get the present number density of
the LSP $n_{LSP}$ as
\begin{eqnarray}
n_{LSP} =
Y_{3/2} (T \ll 1 \MEV) \times \frac{\zeta(3)}{\pi^{2}} T_{NOW}^{3},
\label{n_LSP}
\end{eqnarray}
where $T_{NOW}\simeq 2.7{\rm K}$ is the present temperature of the
universe.\footnote
{In calculating eq.(\ref{n_LSP}), we have ignored the effect of the pair
annihilation of the LSP ($\tilde{\gamma}+\tilde{\gamma}\rightarrow f +
\bar{f}$, or $\tilde{\nu}+\tilde{\nu}^* \rightarrow
f + \bar{f}$). For the pair annihilation processes, the cross section is
roughly estimated as $\sigma v\lsim\alpha_W/m_{LSP}^2$ (where $\alpha_W$
is the coupling factor and $v$ is the relative velocity), and the
condition for sufficiently large pair annihilation rate ($n_{LSP}\sigma
v \gsim H$, with $n_{LSP}$ being the number density of the LSP) reduces
to $n_{LSP}/n_{\gamma}\gsim 10^{-7}$ (with $Y_{LSP}$ being the yield
variable for the LSP), which is less stringent than the constraint
(\ref{ub-omega}). However, if the LSP mass is small enough to hit the
pole of the $Z^0$-boson propagator, the constraint (\ref{ub-omega}) may
become weaker, and the constraint from the $^4$He overproduction may
become significant for the large gravitino mass case ($\mgra\gsim
3\TEV$).}
The density parameter of the LSP
\beq
\Omega_{LSP} \equiv \frac{m_{LSP}n_{LSP}}{\rho_{c}},
\eeq
can be easily calculated, where $m_{LSP}$ is the mass of the LSP,
$\rho_{c}\simeq 8.1\times 10^{-47} h^2 \GEV^4$ is the critical density
of the universe and $h$ is the Hubble parameter in units of
100km/sec/Mpc. If we impose $\Omega_{LSP}\leq 1$ in order not to
overclose the universe, the upperbound on the reheating temperature is
given by
\begin{eqnarray}
T_{R} \leq
2.7
\times 10^{11} \lsp \frac{m_{LSP}}{100\GEV} \rsp^{-1} h^{2}~\GEV
,
\label{ub-omega}
\end{eqnarray}
where we have ignored the logarithmic correction term of $\Sigma_{tot}$.

%
%
\begin{figure}[t]
\epsfxsize=14cm
\centerline{\epsfbox{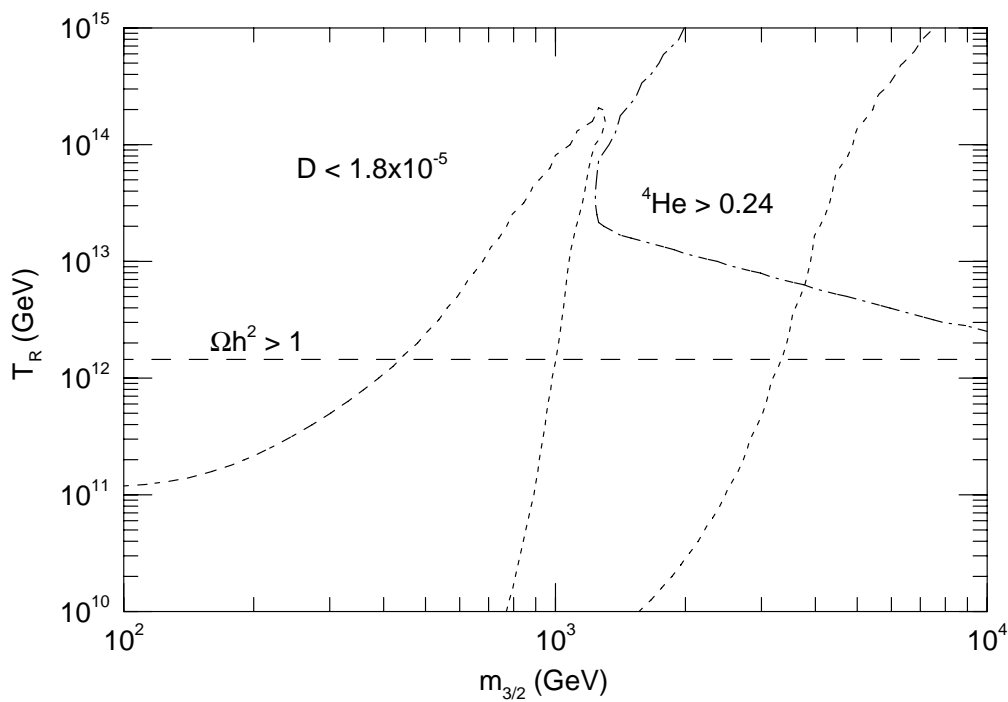}}
\caption{Upperbound of the reheating temperature. Dashed line
represents the constraint from the present mass density of photino.
Dotted-dashed line represents the upperbound requiring $^4$He $< 0.24$.
Constraints from D photo-dissociation is also shown by dotted line.}
\label{fig:expansion}
\end{figure}
%
%

To set the upperbound on the reheating temperature, we need to know the
mass of the LSP. First let us consider the case where the LSP is photino.
In this case, if one assumes the gaugino-mass unification condition, the
lower limit of the mass of photino is $18.4\GEV$~\cite{PRD44-927}.  Then
we can get the following upperbound on the reheating temperature;
\begin{equation}
T_R \leq 1.5 \times 10^{12} h^{2}\GEV
{}~~~({\rm if~the~LSP~is~photino}).
\end{equation}
This constraint is shown in Fig.~\ref{fig:expansion}. Notice that this
bound is independent of the gravitino mass and branching ratio. If the
LSP is sneutrino, the present limit on sneutrino mass 41.8GeV~\cite{PDG}
sets the upperbound on the reheating temperature;
\begin{eqnarray}
T_{R} \leq 6.6\times 10^{11}  h^{2}~\GEV
{}~~~({\rm if~the~LSP~is~sneutrino}),
\end{eqnarray}
which is also shown in Fig.~\ref{fig:snu_lsp}.

Another important constraint comes from the effect on the cosmic
expansion rate at the BBN. As mentioned before, if the density of
gravitino in the nucleosynthesis epoch becomes higher, the expansion of
the universe increases, which leads to more abundance of $^4$He. We
study this effect by using the modified Kawano code and show the result in
Fig.~\ref{fig:expansion}. In the calculation, we take $\eta_B =
2.8\times 10^{-10}$ and $\tau_{n}=887\sec$ (where $\tau_n=(889\pm
2.1)$sec is the neutron lifetime~\cite{PDG}) so that the the predicted
$^4$He abundance is minimized without conflicting the observational
constraints on other light elements. The resultant upperbound on the
reheating temperature is approximately given by
\beq
T_R \lsim 2\times 10^{13}\GEV \left(\frac{\mgra}{1\TEV}\right)^{-1},
\label{4he_synth}
\eeq
for $\mgra\geq 1\TEV$.\footnote
{For the case $\mgra\leq 1\TEV$, primordial $^4$He is destroyed by the
high energy photons induced by the decay of the gravitino, and hence the
constraint (\ref{4he_synth}) becomes insignificant. However, if the
gravitino decays into a neutrino and a sneutrino, this is not the case.}
This bound is also shown in Fig.~\ref{fig:expansion} by the solid line.
Notice that this bound is derived irrespective of the species of the
LSP.

%
%

\chapter{Cosmology with stable gravitino}
\label{chap:light}

\hspace*{\parindent}
In the previous chapter, we have studied cosmology with unstable
gravitino and derived constraints on the reheating temperature after
inflation. However, there is another possibility that the gravitino is
stable and hence it is the LSP. In this case, the constraints obtained
in the previous chapter are not appropriate. In this chapter, we study
cosmological constraints when the gravitino is the
LSP.\footnote
{This chapter is based on the work in a collaboration with H.~Murayama
and M.~Yamaguchi~\cite{PLB303-289}.}

\section{Constraints from the mass density of the universe}

\hspace*{\parindent}
In the case of the stable gravitino, the Boltzmann equation for the
gravitino number density $n_{3/2}$ is given by
\beq
\frac{d n_{3/2}}{d t} + 3H n_{3/2} =
\vev{\Sigma_{tot}^{(1/2)} v_{rel}} n_{rad}^{2} +
\sum_{\tilde{i}} n_{\tilde{i}}
\frac{m_{\tilde{i}}}{\vev{E_{\tilde{i}}}} \Gamma_{\tilde{i}},
\label{bol_light}
\eeq
where $n_{\tilde{i}}$ is the number density of the superparticle
$\tilde{i}$, $\Gamma_{\tilde{i}}$ the decay rate of $\tilde{i}$ into its
superpartner and a gravitino, and $m_{\tilde{i}}/\langle E_{\tilde{i}}
\rangle$ the Lorentz factor. Notice that the first term in the
right-hand side of eq.(\ref{bol_light}) represents the contribution from
the scattering processes off thermal radiations, while the second one is
that from the decay of superparticles into a gravitino and some ordinary
particles.\footnote
{In the case of heavy unstable gravitino analyzed in the previous
chapter, we have ignored the contribution from the decay of the
superparticles (or its inverse processes). In fact, these contributions
are significant only for the case of extremely light gravitino
($\mgra\lsim 10^{-4}\GEV$, as we will see later), and hence our
approximation is justified.}

Here we comment on $\Sigma_{tot}^{(1/2)}$. If the gravitino mass is
small compared with the typical mass splitting in the matter sector,
interactions of the helicity $\pm\frac{1}{2}$ modes of the gravitino
become stronger than that of the helicity $\pm\frac{3}{2}$ modes. From
this fact, we conclude that the helicity $\pm\frac{1}{2}$ gravitino
production processes are the most significant in this case. As we can
see in eq.(\ref{L_eff}), the interactions of the gravitino to gauge
multiplets are described by dimension-five operators and those to chiral
multiplets by dimension-four ones. Therefore, the former dominates the
gravitino production at high temperatures. For the MSSM particle content,
$\Sigma_{tot}^{(1/2)}$ is given by
\beq
\Sigma_{tot}^{(1/2)} &=&
\frac{1}{2}~\sum_{x,y,z}
\eta_{x} \eta_{y} \sigma(x + y \rightarrow \psi + z)
\nonumber \\ &=&
\frac{1}{24 \pi \mgra^{2} M^{2}}
\lsp 2.44 g_{1}^{2} m_{G1}^{2} +
9.16 g_{2}^{2} m_{G2}^{2} +
26.00 g_{3}^{2} m_{G3}^{2} \rsp,
\eeq
where $\sigma(x + y \rightarrow \psi + z)$ is the helicity
$\pm\frac{1}{2}$ gravitino production cross section (see
Table~\ref{table:cs1/2}), $m_{G3}$ -- $m_{G1}$ are the gauge fermion
masses, and $g_3$ -- $g_1$ are the gauge coupling constants.

Using the yield variable $Y_X\equiv n_X/n_{rad}$ and the condition of
``naive'' adiabatic expansion (\ref{rt=const}), eq.(\ref{bol_light}) is
rewritten as
\beq
\frac{d Y_{3/2}}{d T} =
- \frac{n_{rad} \vev{\Sigma_{tot} v_{rel}}}{H T}
- \sum_{\tilde{i}}
\frac{m_{\tilde{i}}/\vev{E_{\tilde{i}}}}{H T}
\Gamma_{\tilde{i}} Y_{\tilde{i}}.
\label{bolty_light}
\eeq
Integrating this equation from $T_R$ to $T$ ($T_R\gg T$) and taking
the effects of the dilution factor $(N_{S}(T)/N_{S}(T_{R}))$ into
account, we get
\beq
Y_{3/2}(T) = \frac{N_{S} (T)}{N_{S}(T_{R})} \times
\lmp \overline{Y}_{scatt}(T) + \overline{Y}_{decay}(T) \rmp.
\label{ygra_light}
\eeq
with
\beq
\overline{Y}_{scatt} &=&
\frac{n_{rad}(T_{R}) \vev{\Sigma_{tot}^{(1/2)} v_{rel}}}{H(T_{R})} ,
\label{yscatt} \\
\overline{Y}_{decay} &=&
\int_{T}^{T_{R}} \frac{dT}{T}
\sum_{\tilde{i}}
\frac{m_{\tilde{i}}\Gamma_{\tilde{i}}}{\vev{E_{\tilde{i}}}H}
Y_{i}.
\label{ydecay}
\eeq
{}From eq.(\ref{yscatt}), we can see that $\overline{Y}_{scatt}$ is
proportional to the reheating temperature $T_R$. On the other hand,
$\overline{Y}_{decay}$ is almost independent of $T_{R}$ as far as the
reheating temperature is higher than the masses of the superparticles.
Notice that effects of the gravitino annihilation processes cannot be
ignored in eq.(\ref{bol_light}) if the $\overline{Y}_{scatt}$ in
eq.(\ref{yscatt}) or the $\overline{Y}_{decay}$ in eq.(\ref{ydecay})
becomes $O(1)$. We take these effects into account by the following
simple method; in the case where the sum of the right hand sides of
eq.(\ref{yscatt}) and eq.(\ref{ydecay}), which is the yield of the
thermal helicity $\pm\frac{1}{2}$ gravitino, is larger than
$\frac{3}{2}$, we consider that the gravitinos are thermalized and take
$\overline{Y}_{scatt}+\overline{Y}_{decay}=\frac{3}{2}$.

Using eq.(\ref{yscatt}) and eq.(\ref{ydecay}), we estimate the present
gravitino number density and determine the upperbound on the reheating
temperature by using the closure limit. In our numerical calculation, we
take all the squark and slepton masses to be 1~TeV and the GUT relation
on the gaugino masses;
\beq
\frac{3}{5}\frac{m_{G1}}{g_{1}^{2}}
= \frac{m_{G2}}{g_{2}^{2}}
= \frac{m_{G3}}{g_{3}^{2}} .
\label{gut}
\eeq
Energy density of the gravitino $\rho_{3/2}$ is given by
\beq
\rho_{3/2} (T)= \mgra Y_{3/2}(T) n_{rad}(T) ,
\label{egra}
\eeq
and we determine the upperbound on $T_{R}$ by the condition
\beq
\rho_{3/2} (T_{NOW}) \leq \rho_c ,
\label{cond1}
\eeq
where $T_{NOW}\simeq 2.7{\rm K}$ and $\rho_c\simeq 8.1\times 10^{-47} h^{2}~
\GEV^{4}$ $(0.4\lsim h \lsim 1$) is the critical density of the
present universe.

%
%
\begin{figure}[t]
\epsfxsize=14cm
\centerline{\epsfbox{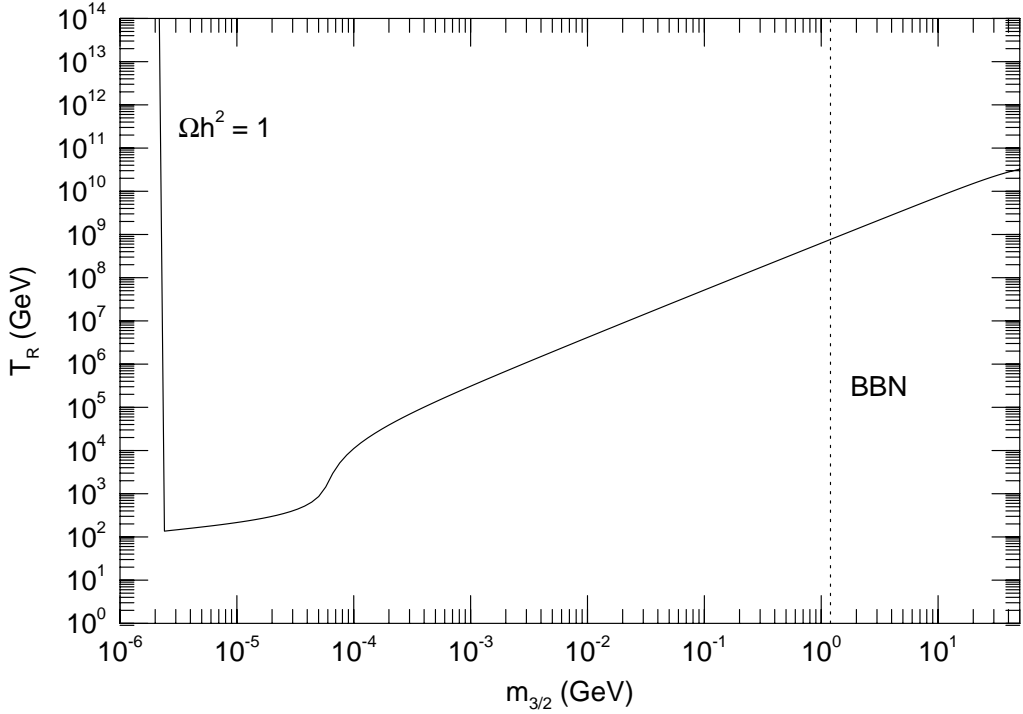}}

\caption{Cosmological constraints on the gravitino mass $\mgra$ and
the reheating temperature $T_R$ in the framework of the MSSM when the
gravitino is the LSP.  We take all the squark and slepton masses to be
1~TeV, $m_{G1}=m_{NSP}=50\GEV$ and the GUT relation on the gauge fermion
masses is assumed. The solid line denotes the upperbound on the
reheating temperature from the closure limit. The dotted line is the
upperbound on the gravitino mass from light element photo-dissociation.
We assume the NSP relic density as given in
eq.(\protect\ref{mass*yield}), and require the NSP lifetime to be
shorter than $2.6\times 10^{6}$ sec.}
\label{fig:omega_l}

\end{figure}
%
%
The result is shown in Fig.~\ref{fig:omega_l}. Here we take
$m_{G1}=m_{NSP}=50~\GEV$ with $m_{NSP}$ being the mass of the
next-to-the-lightest superparticle (NSP), and to get a conservative
bound we take $h=1$. For $\mgra\gsim10^{-4}~\GEV$, the upperbound
on $T_{R}$ is approximately proportional to $\mgra$. When
$2\times10^{-6}~\GEV$$\lsim$$\mgra$$\lsim10^{-4}~\GEV$, the upperbound
on $T_{R}$ is around $O(100~\GEV)$. And for the very light gravitino
case ($\mgra\lsim2\times10^{-6}~\GEV$), there is no constraint on
$T_{R}$.  These features can be understood in the following way. When
the gravitino mass is larger than about $10^{-4}~\GEV$, interactions are
so weak that decaying processes cannot produce sufficient gravitinos to
overclose the universe. Therefore, it is the scattering process that is
important to estimate the number density of the gravitino. In this case,
\beq
Y_{3/2}(T_{NOW}) = \frac{N_{S}(T_{NOW})}{N_{S}(T_{R})}
                   \frac{\sqrt{90} \zeta(3) M}{\pi^{3} \sqrt{N_{*}}}
                   ~T_{R} \vev{\Sigma_{tot} v_{rel}} ,
\label{y-heavy}
\eeq
from eq.(\ref{yscatt}). Combining eq.(\ref{y-heavy}) with
eq.(\ref{cond1}), we get the upperbound on the reheating temperature,
which is approximately proportional to the gravitino mass. On the
other hand, if
$2\times10^{-6}~\GEV$$\lsim$$\mgra$$\lsim$$10^{-4}~\GEV$, the decay
processes become significant. In this case, $\rho_{3/2}$ is larger
than $\rho_c$ unless the reheating temperature is smaller than the
squark and slepton masses.  Therefore, it is necessary to lower the
reheating temperature below the squark and slepton mass scale in order
not to overclose the universe.  And when $\mgra \lsim 2\times10^{-6}~
\GEV$, the gravitino mass is so small that $\rho_{3/2}$ cannot exceed
$\rho_c$ even if the gravitino is thermalized.

\section{Constraint from BBN}

\hspace*{\parindent}
Next, let us consider the constraint from the light element
photo-dissociation. If a decay of a heavy particle produces high energy
photons after the primordial nucleosynthesis, we must require that these
photons do not change the abundance of the light elements. Here we
consider the decay of the NSP. Since the gravitino is the LSP, the NSP
can decay only into a gravitino and some ordinary particles through the
supergravity interaction. Therefore, the NSP have much longer lifetime
than other superparticles and may affect the predictions of the BBN.

If the NSP were stable, it would survive until today. Its relic density
in this case has been calculated \cite{PLB283-80,PLB298-120,PRD47-376}.
For the neutralinos, in a wide range of parameters, the relic
density is larger than $10^{-3}$ to the critical one. This relic density
can be translated into $m_{NSP}Y_{NSP} \geq 5.0
\times10^{-11}~\GEV$,\footnote
{It is plausible that this bound is also valid when a slepton or a
chargino is the lightest.}
where $m_{NSP}$ and $Y_{NSP}$ are the mass and yield of the NSP. In
the following analysis, we conservatively take
\begin{eqnarray}
m_{NSP} Y_{NSP} = 5.0 \times 10^{-11}~\GEV ,
\label{mass*yield}
\end{eqnarray}
and assume that the NSP decay produces high energy photons.
Photo-dissociation of light elements by the NSP decay can be analyzed in
the same way as in the case of unstable gravitino. Detailed analysis
shows that the energy density of the NSP as large as the one given in
eq.(\ref{mass*yield}) will overproduce $(\rm{^{3}He + D})$ unless the
lifetime of the NSP is shorter than about $2.6\times10^{6}$sec. (See
Fig.~\ref{fig:my10tev} -- Fig.~\ref{fig:my10mev} in the next chapter.)
Therefore, we impose
\begin{eqnarray}
\llp \frac{1}{48 \pi}
     \frac{m_{NSP}^{5}}{\mgra^{2} M^{2}}
     \lmp 1 - \lsp \frac{\mgra}{m_{NSP}} \rsp ^{2} \rmp^{3}
\rlp^{-1}
\leq 2.6 \times 10^{6} {\rm sec} .
\label{cond2}
\end{eqnarray}
Here, we assume that the NSP is the ${\rm U(1)}_{Y}$ gauge fermion
(bino) and used eq.(\ref{gamma_l=>ga}) for the decay rate of the
NSP.\footnote
{If the bino is the NSP, it decays into a gravitino and a photon, and
into a gravitino and a $Z^0$-boson. But when the bino is lighter than the
$Z^0$-boson, the latter decay channel is forbidden kinematically and the
decay rate of the bino is $\sin^{2}\theta_{W} \simeq 0.23$ times
smaller than the value of eq.(\ref{gamma_l=>ga}). For the case $m_{NSP}
= ~50\GEV$, we have considered this effect.}
The right hand side of eq.(\ref{cond2}) strongly depends on the NSP mass
and especially when the NSP mass is small, a severe upperbound on the
gravitino mass is obtained. The bound we obtained is $\mgra \leq
1.2~\GEV$ (6.7~GeV, 244.4~GeV, 711.0~GeV) for $m_{NSP} = 50~\GEV$
(100~GeV, 500~GeV, 1~TeV). This bound for the case of $m_{NSP} =
50~\GEV$ is also shown in Fig.~\ref{fig:omega_l}. If the reheating
temperature is sufficiently low compared to the NSP mass, the NSP is
not produced significantly and the above constraint can be avoided.
Furthermore, the constraint is not significant when the NSP decay
does not produce sufficient photons. This is the case if the NSP is a
sneutrino.

\section{Remarks}

\hspace*{\parindent}
Before closing this chapter, we would like to consider phenomenological
implications of the gravitino LSP. As we discussed previously, the
lightest among the superpartners of the standard model particles is no
longer stable. Therefore, this particle (NSP) can be charged or even
colored. In a supergravity model with a no-scale like K\"{a}hler
potential \cite{NPB247-373}, squarks and sleptons are massless at the
tree-level. Although they acquire their masses from gaugino masses
through radiative corrections, a right-handed charged slepton tends to
be the lightest. Indeed with the GUT relation of the gaugino masses, the
mass of the right-handed slepton is very close to the ${\rm U(1)}_Y$
gaugino mass. Detailed analysis showed that a neutral superparticle is
the lightest only when its mass is lighter than $\sim 150$ GeV
\cite{PRD45-328,NPB398-3}. The gravitino LSP can cure this
difficulty of the fascinating class of supergravity model which might be
derived from superstring compactification~\cite{PLB155-151}.  It can
also solve the conflict~\cite{PRD46-3966,PLB291-255,PRD47-2468} between
the constraints from the mass density of the neutralino LSPs and those
from the proton decay in the minimal SUSY SU(5) GUT, because the former
gets meaningless when there is a superparticle lighter than the
neutralino.  In accelerator experiments, the most drastic change occurs
when the NSP is charged.  Then, we have a chance to find a spectacular
track of the charged NSP in a detector.

Finally, we we comment on the case where the gravitino is extremely light
($\mgra\ll 1~\KEV$) and interacts strongly. In this case, the gravitino
decouples from the thermal bath at temperature well below the
electroweak scale.  The standard nucleosynthesis scenario constrains the
number of the species of the neutrino-like light particles to be smaller
than 3.3~\cite{APJ376-51}.\footnote
{In a recent paper, Kernan and Krauss~\cite{PRL72-3309} has claimed that
the number of the species of neutrinos $N_\nu$ should be smaller than
3.04. In this thesis, however, we use $N_\nu\leq 3.3$ in order to derive
a conservative constraint.}
Therefore, in order to get a sufficient dilution, the gravitino must
decouple before $T \simeq 200~\MEV$.  After the QCD phase transition,
$l+\bar{l}\leftrightarrow\psi +\psi$ scatterings become the most
important in the thermalization processes of the gravitino, where $l$
denotes $e$, $\mu$, $\nu_{e}$, $\nu_{\mu}$ or $\nu_{\tau}$, and
$\bar{l}$ its anti-particle.  Comparing the gravitino production rate of
these processes with the expansion rate of the universe, we get
\begin{eqnarray}
\mgra \gsim 10^{-13}~\GEV \times
            \lsp \frac{m_{\tilde{l}}}{100~\GEV} \rsp ,
\label{bound3}
\end{eqnarray}
where $m_{\tilde{l}}$ is the slepton mass.\footnote
{One can also consider the constraints from the collider
experiments~\cite{PLB175-471,PLB258-231} or from the cooling of the red
supergiant star~\cite{PLB114-23}. These arguments give similar bound on
the gravitino mass.}
Remember that the interactions of the light gravitino are proportional to
$\mgra^{-1}$ as we have seen in Chapter~\ref{chap:feynman}, and hence
the lowerbound on the gravitino mass is obtained here.

%
%

\chapter{Conclusions and discussion}
\label{chap:summary}

\section{Summary of conclusions}

\hspace*{\parindent}
We summarize the effects of the massive gravitino on the inflationary
universe.  If the gravitino is unstable and decays only into a photon
and a photino, stringent constraints on the reheating temperature are
obtained.  In the case where the gravitino mass $\mgra$ is less than
$\sim 1\TEV$, $({\rm D} + ^3{\rm He})$ overproduction due to the
photo-dissociation of $^4$He set the upperbound on the reheating
temperature $T_R$, $T_R\lsim 10^{6-9} \GEV$.  Constraints in the case
$1\TEV\lsim\mgra\lsim 3\TEV$, the photo-dissociation of D gives the
upperbound on the reheating temperature, $T_R\lsim 10^{9-12} \GEV$. If
the gravitino mass is heavier than $\sim 3\TEV$, the reheating
temperature is constrained to be lower than $\sim 10^{12}\GEV$ not to
overclose the universe. These constraints are shown in
Fig.~\ref{fig:expansion};
\beq
T_R & \lsim & 10^{6-7}\GEV ~~~ (m_{3/2} \lsim 100\GEV),
\\
T_R & \lsim & 10^{7-9}\GEV ~~~ (100\GEV \lsim m_{3/2} \lsim 1\TEV),
\\
T_R & \lsim & 10^{9-12}\GEV ~~~ (1\TEV \lsim m_{3/2} \lsim 3\TEV),
\\
T_R & \lsim & 10^{12}\GEV ~~~ (3\TEV \lsim m_{3/2} \lsim 10\TEV).
\eeq

If most of the decay products of the gravitino are charged particles,
above constraints are expected to be valid since the spectrum of the
high energy photon is fixed mainly by the total amount of energy
injection. An exception is the case where the gravitino only decays into
a neutrino and a sneutrino. In this case, the constraints are less
stringent than in the case of high energy photon injection, and the
constraints are given by
\beq
T_R & \lsim & 10^{12}\GEV ~~~ (m_{3/2} \lsim 100\GEV),
\\
T_R & \lsim & 10^{10-12}\GEV ~~~ (100\GEV \lsim m_{3/2} \lsim 5\TEV),
\\
T_R & \lsim & 10^{12}\GEV ~~~ (5\TEV \lsim m_{3/2} \lsim 10\TEV),
\eeq
which are also shown in Fig.~\ref{fig:snu_lsp}.

If the gravitino is light and stable, interactions of helicity
$\pm\frac{1}{2}$ gravitino become strong and a more stringent bound on
the reheating temperature is derived from the closure limit;
\beq
T_R \lsim 10^{12}\GEV \times
\lsp \frac{\mgra}{100\GEV} \rsp.
\eeq
Notice that this constraint is not appropriate for the case $\mgra\lsim
1\KEV$.  Furthermore, the effects of the NSP decay on BBN set the
upperbound on the gravitino mass. The detailed value of the upperbound
depends on the mass of the NSP. For example, the gravitino mass larger
than $\sim 1\GEV$ is excluded for the case of $m_{NSP} = 50\GEV$.

\section{Discussion}

\hspace*{\parindent}
We have derived constraints on the reheating temperature after inflation
by assuming the existence of the massive gravitino field. As we have seen,
cosmological arguments give stringent upperbound on the reheating
temperature.

Some comments on our results are in order. When the gravitino is
unstable and its mass is smaller than $\sim 1\TEV$, the upperbound on
the reheating temperature is given by $(10^{6}-10^{9})\GEV$. This
constraint seems to be very stringent since such a low reheating
temperature requires very small decay rate of the inflaton field. For
example, in a chaotic inflation~\cite{PLB129-177} with a inflaton whose
interactions are suppressed by $M^{-1}$, the decay rate of the inflaton
is expected to be $\Gamma_{\rm inf} \sim m_{\rm inf}^3/M_{pl}^2$ with
$m_{\rm inf}$ being the inflaton mass, and hence the reheating
temperature is estimated as~\cite{Kolb&Turner}
\beq
T_R \sim 0.1\sqrt{\Gamma_{\rm inf}M_{pl}} \sim 10^{9}\GEV,
\label{tr}
\eeq
requiring that the inflaton field should produce the density
perturbations observed by COBE ($m_{\rm inf} \sim
10^{13}\GEV$)~\cite{PRL69-3602}. Notice that if the interaction of teh
inflaton becomes stronger, the decay rate becomes larger and hence a
higher reheating temperature is obtained. Thus, in the case that the
gravitino is lighter than $\sim 1\TEV$ (which is favored from the
naturalness point of view), we have to adopt an inflaton with extremely
small decay rate.

Let us compare our results with those in other literatures. The
gravitino number density and the photon spectrum are the essential
quantity for the analysis. Our number density of gravitino produced in
the reheating epochs of the inflationary universe is about four times
larger than one given in ref.\cite{PLB145-181}. Since the authors of
ref.\cite{PLB145-181} neglected the interactions between the
gravitino and chiral multiplets (which is the second term in
eq.(\ref{interaction})), they might underestimate the total cross
section for the production of gravitino.  All previous works concerning
the gravitino problem were based on the gravitino number density
given in ref.\cite{PLB145-181}.  Therefore our constraints are more
stringent than others.

Furthermore, our photon spectrum is different from that in
ref.\cite{NPB373-399} as shown in Fig.\ref{fig:ellis}. The spectrum
adopted by ref.\cite{NPB373-399} has more power to destroy light
elements above threshold for the photon-photon scattering and less power
below the threshold.  In refs.\cite{APJ335-786,APJ349-415}, the Compton
scattering process was not taken into account in calculating the photon
spectrum which the authors of ref.\cite{NPB373-399} used to derive
a fitting formula for the high energy photon spectrum.  Therefore, it is
expected that the difference comes mainly from the neglect of Compton
scattering off the thermal electron, which is the most dominant process
for the relatively low energy photons.

%
%
\begin{figure}[t]
\epsfxsize=14cm
\centerline{\epsfbox{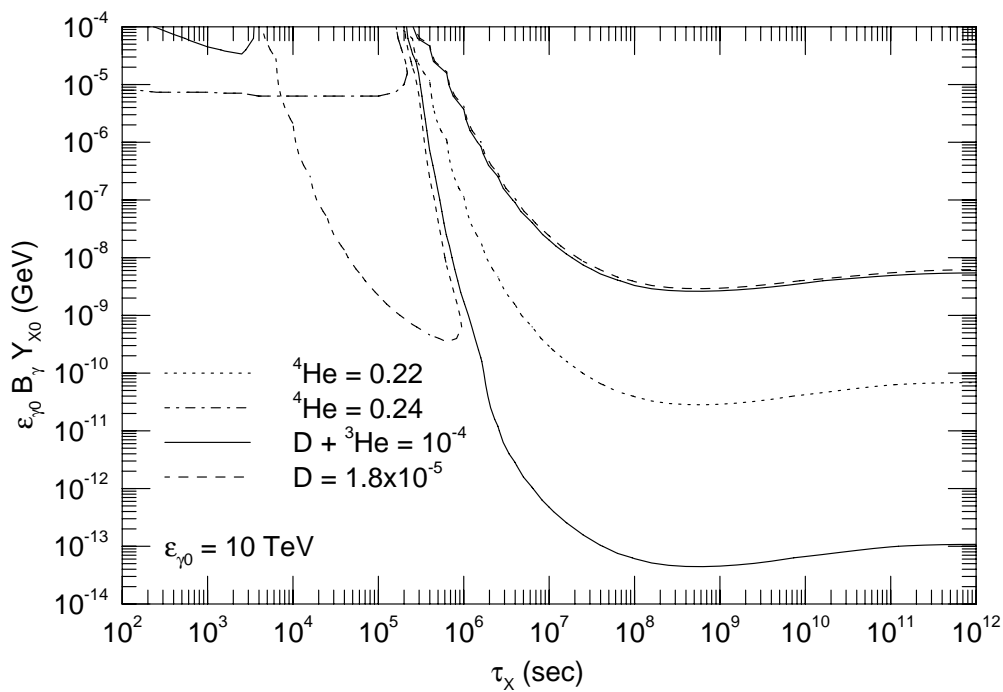}}
\caption{Contours for critical abundance of light elements in
the $\tau_X$ -- $\epsilon_{\gamma 0}B_\gamma Y_{X0}$ plane for
$\epsilon_0= 10\TEV$. The solid line (dashed line, dotted line,
dotted-dashed line) is a bound from overproduction of (D+$^3$He)
(overdestruction of D, overdestruction of $^4$He, and overproduction of
$^4$He). In deriving the mass density of $X$, we take $B_\gamma =1$ and
$m_X =2\epsilon_{\gamma 0}$.}
\label{fig:my10tev}
\end{figure}

\begin{figure}[p]
\epsfxsize=14cm
\centerline{\epsfbox{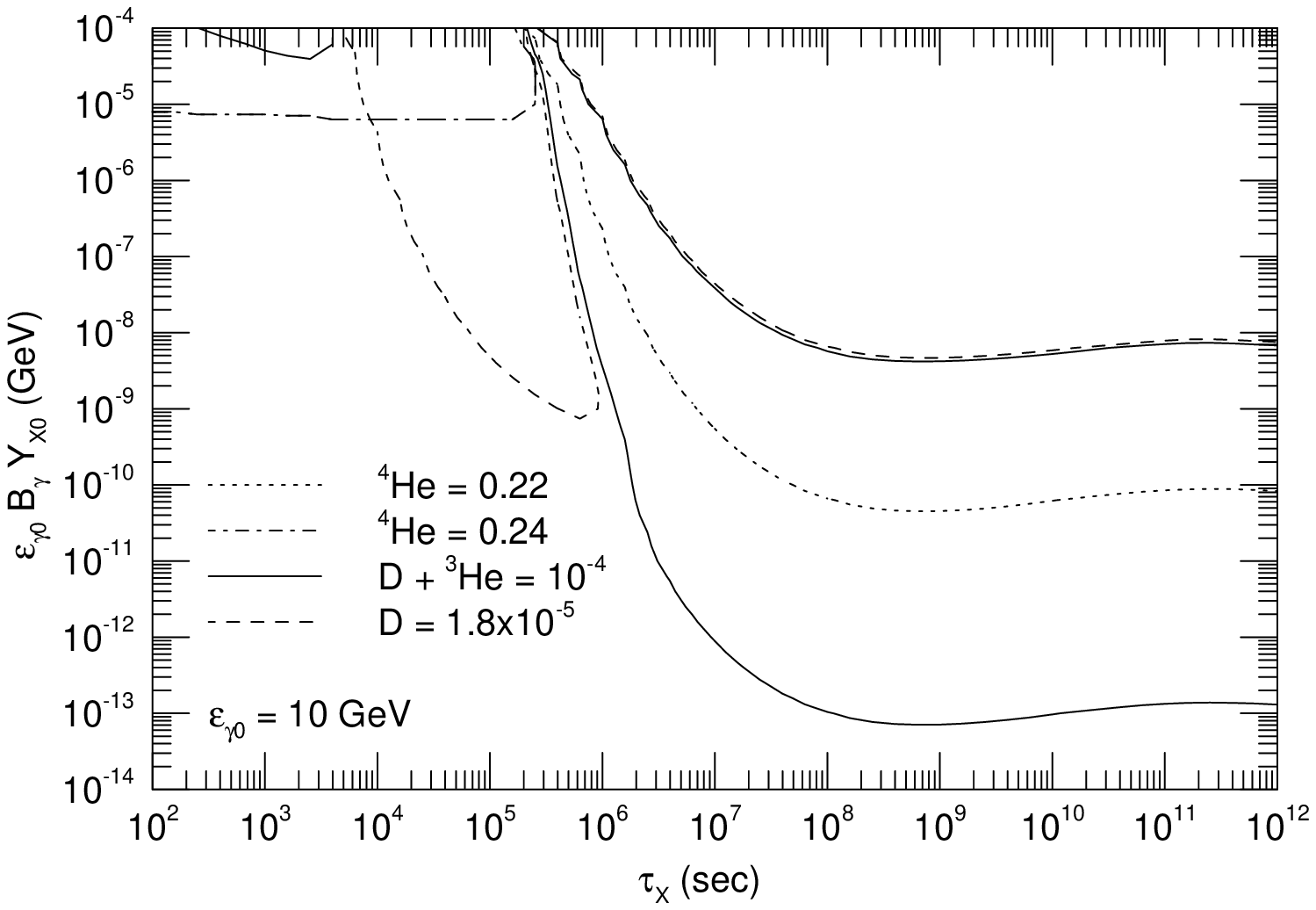}}
\caption{Same as Fig.~\protect\ref{fig:my10tev} except for
$\epsilon_{\gamma 0}= 10\GEV$.}
\label{fig:my10gev}
\vspace{0.7cm}
\centerline{\epsfbox{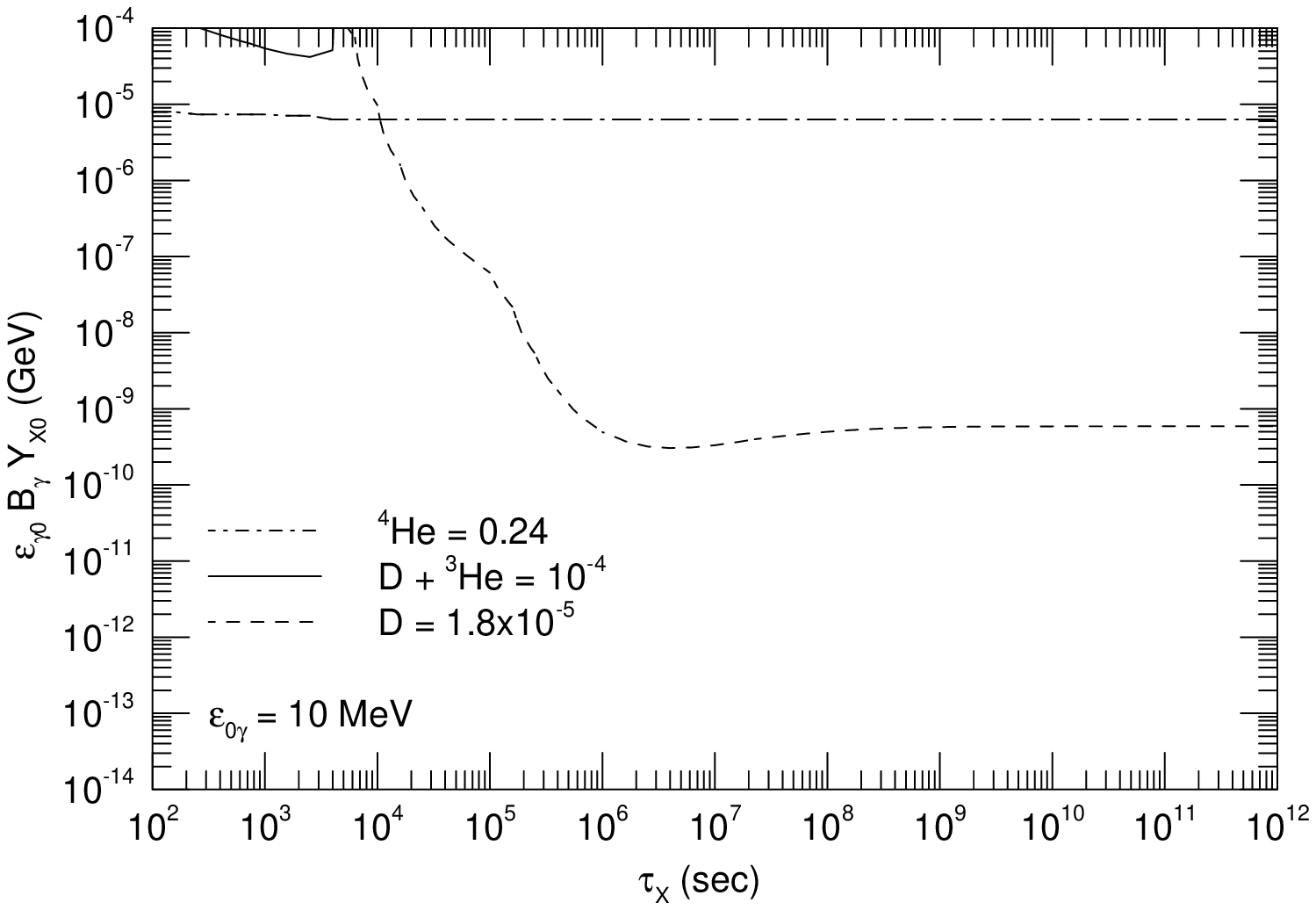}}
\caption{Same as Fig.~\protect\ref{fig:my10tev} except for
$\epsilon_{\gamma 0}= 10\MEV$.}
\label{fig:my10mev}
\end{figure}
%

Finally, we discuss applications of the cosmological arguments to
other cases. If some exotic particle decays radiatively with a lifetime
longer than $\sim 1{\rm sec}$, the standard BBN scenario may be
affected. The constraints we have used in the case of massive gravitino
can be applied to other cases of exotic particle $X$. Using the similar
method as in the gravitino case, we derive constraints from the
BBN on radiatively decaying weakly interacting particles. The results
are given in Fig.\ref{fig:my10tev} -- Fig.\ref{fig:my10mev}, where
$\tau_X$ is the lifetime of $X$, $Y_{X0}$ the yield of $X$ before $X$
decays, $\epsilon_{\gamma 0}$ the energy of the primary photon, and
$B_{\gamma}$ the branching ratio of radiative decay channel.  The
evolution of the yield variable $Y_X\equiv n_X/n_{rad}$ of $X$ is given
by $Y_X(t) = Y_{X0}e^{-t/\tau_X}$. Notice that we assume that the
emitted photon is monochromatic, and that the number of photons emitted
from one decay process is one. As we can see, the constraint on the
combination $\epsilon_{\gamma 0}B_{\gamma}Y_{X0}$ is almost the same as
in the cases of $\epsilon_{\gamma 0}=10\GEV$ and $\epsilon_{\gamma
0}=10\TEV$ since the spectrum of the energetic photon depends almost
only on the total amount of energy injection, while for the case of
$\epsilon_{\gamma 0}=10\MEV$ the constraint from $({\rm D} + ^3{\rm
He})$ is negligible since the energy of primary photon is below the
threshold of $^4$He destruction.

Constraints from the present mass density of the universe may give
another information on exotic particles. Especially in the SUSY models,
the present mass density of the relic LSP is calculated in detail, and
constraints on SUSY parameters are derived. Furthermore, particles with
lifetime longer than $\sim 10^{10}{\rm sec}$ may distort the cosmic
microwave background.

These arguments have been applied to the LSP in SUSY models, the Polonyi
field which is responsible for SUSY breaking~\cite{PLB131-59}, the
massive neutrino~\cite{PRL39-165,NPB302-697}, the
axion~\cite{PLB120-127,PLB120-133,PLB120-137} and so on as well as the
gravitino. Thus, cosmological considerations provide us with a great
insight into exotic particles which we cannot investigate by any
collider experiments because of the weakness of their interactions.

%
%

\newpage
\section*{Acknowledgement}

\hspace*{\parindent}
The author wishes to express his sincere gratitude to T.~Yanagida for
various suggestions, stimulating discussions and careful reading of
this manuscript, and to M.~Kawasaki, H.~Murayama and M.~Yamaguchi, who
are collaborators in various parts of this thesis, for fruitful
discussions. He is also grateful to M.~Yoshimura for helpful comments.
All of them helped the author greatly completing this thesis through
valuable discussions and by making useful comments. Without their hearty
encouragement, this work would not be completed.

He would like to thank Y.~Shimizu for useful comments and careful
reading of this manuscript. He is also grateful to T.~Ataku, and also to
Y.~Shimizu for taking part in the maintenance of the computers he used
for numerical calculations. Finally, he expresses his gratitude for all
the members of the high energy theory group of Tohoku University for
their hospitality.

\appendix

%
%

\chapter{Notations}
\label{ap:notation}

\section{Conventions}

\hspace*{\parindent}
We use the metric (in flat space-time)
\beq
g_{\mu\nu} = {\rm diag}(1,-1,-1,-1),
\eeq
and for totally anti-symmetric tensor (in flat space-time), we use the
convention $\epsilon_{0123}=-1$.

\section{Two component notation}

\hspace*{\parindent}
For two component spinor, we essentially follow the notation used in
ref.\cite{Wess&Bagger}, except for the convention of the metric.

$\sigma$-matrices are defined as
\beq
\sigma_0 = \bar{\sigma}_0 =
\lsp \begin{array}{cc}
1 & 0 \\ 0 & 1
\end{array} \rsp, &~~~&
\sigma_1 = - \bar{\sigma}_1 =
\lsp \begin{array}{cc}
0 & 1 \\ 1 & 0
\end{array} \rsp,
\nonumber \\
\sigma_2 = - \bar{\sigma}_2 =
\lsp \begin{array}{cc}
0 & -i \\ i & 0
\end{array} \rsp, &~~~&
\sigma_3 = - \bar{\sigma}_3 =
\lsp \begin{array}{cc}
1 & 0 \\ 0 & -1
\end{array} \rsp.
\eeq
For the anti-symmetric symbols $\epsilon_{\alpha\beta}$ and
$\epsilon^{\alpha\beta}$, we use the following conventions;
\beq
&& \epsilon^{12} = -\epsilon^{21} = \epsilon_{21} = -\epsilon_{12} =1,
\\
&& \epsilon^{11} = \epsilon^{22} = \epsilon_{11} = \epsilon_{22} =0.
\eeq
Notice that $\epsilon_{\alpha\beta}$ and $\epsilon^{\alpha\beta}$ are
related as $\epsilon_{\alpha\beta}\epsilon^{\beta\gamma} =
\delta_\alpha^\gamma$. By using $\epsilon^{\alpha\beta}$, $\sigma_\mu$
and $\bar{\sigma}_\mu$ are related as
\beq
\bar{\sigma}_\mu^{\dot{\alpha}\alpha} =
\epsilon^{\dot{\alpha}\dot{\beta}} \epsilon^{\alpha\beta}
\sigma_{\mu\beta\dot{\beta}}.
\eeq

{}From these $\sigma$-matrices, the generators of the Lorentz group is the
spinor representation are obtained
\beq
\sigma_{\mu\nu} =
\frac{1}{4}
\lsp \sigma_\mu \bar{\sigma}_\nu - \sigma_\nu \bar{\sigma}_\mu \rsp,~~~
\bar{\sigma}_{\mu\nu} =
\frac{1}{4}
\lsp \bar{\sigma}_\mu \sigma_\nu - \bar{\sigma}_\nu \sigma_\mu \rsp.
\eeq

\section{Four component notation}

\hspace*{\parindent}
$\gamma$-matrices obey the following commutation relation;
\beq
\llp \gamma_\mu , \gamma_\nu \rlp = 2 g_{\mu\nu}.
\eeq
{}From $\gamma^0$ -- $\gamma^3$, $\gamma_5$ matrix is defined as
\beq
\gamma_5 = i \gamma^0 \gamma^1 \gamma^2 \gamma^3.
\eeq

$\gamma$-matrices are transposed by a charge conjugation matrix $C$;
\beq
C^{-1} \gamma_\mu C = - \gamma_\mu^T.
\eeq
Charge conjugation matrix $C$ satisfies the following identities;
\beq
C^\dagger C = 1,
\\
C^T = -C.
\eeq
By using $C$, charge conjugation of four component spinor $\psi$ is
defined as
\beq
\psi^C \equiv C \overline{\psi}^T.
\eeq

\subsection*{Chiral representation}

\hspace*{\parindent}
Four component spinor (in chiral representation) is constructed from two
component spinor $\xi_\alpha$ and $\eta_{\dot{\alpha}}$ as
\beq
\psi \sim
\lsp \begin{array}{c}
\xi_\alpha \\ \overline{\eta}^{\dot{\alpha}}
\end{array} \rsp,~~~
\overline{\psi} \sim \lsp \eta^\alpha, \overline{\xi}_{\dot{\alpha}} \rsp.
\eeq
In this representation, $\gamma$-matrices are given by
\beq
\gamma_\mu =
\lsp \begin{array}{cc}
0 & \bar{\sigma}_\mu \\ \sigma_\mu & 0
\end{array} \rsp,~~~
\gamma_5=
\lsp \begin{array}{cc}
1 & 0 \\ 0 & -1
\end{array} \rsp,
\eeq
and charge conjugation matrix $C$ is given by
\beq
C = i \gamma^2 \gamma^0 = C^*.
\eeq

\subsection*{Dirac representation}

\hspace*{\parindent}
$\gamma$-matrices in Dirac representation are related to those in chiral
representation through unitary transformation;
\beq
\gamma_\mu^{\rm (chiral)} = U \gamma_\mu^{\rm (Dirac)} U^\dagger,
\eeq
with
\beq
U = \frac{1}{\sqrt{2}}
\lsp \begin{array}{cc}
1 & 1 \\ -1 & 1
\end{array} \rsp.
\eeq
Explicit form of the $\gamma$-matrices in Dirac representation are given
by
\beq
\gamma^0 =
\lsp \begin{array}{cc}
1 & 0 \\ 0 & -1
\end{array} \rsp, ~~~
\gamma^i =
\lsp \begin{array}{cc}
0 & \sigma_i \\ -\sigma_i & 0
\end{array} \rsp, ~~~
\gamma_5 =
\lsp \begin{array}{cc}
0 & 1 \\ 1 & 0
\end{array} \rsp,
\eeq
and charge conjugation matrix $C$ is obtained as
\beq
C = i \gamma^2 \gamma^0 = C^*.
\eeq

In Dirac representation, solution to the Dirac equation (in momentum
space);
\beq
\lsp \pslash - m \rsp u({\bf p},s) = 0,
\eeq
with $p^\mu = (p_0 ,~|{\bf p}|\sin\theta\cos\phi,~|{\bf
p}|\sin\theta\sin\phi,~|{\bf p}|\cos\theta)$ takes the
following form;
\beq
u({\bf p},s) =
\lsp \begin{array}{c}
\sqrt{p_0+m} \chi^{(s)} \\
\sqrt{p_0-m} n_i \sigma_i \chi^{(s)}
\end{array} \rsp,
\eeq
where $p_\mu p^\mu = m^2$, ${\bf n} \equiv {\bf p}/|{\bf p}|$, and
\beq
\chi^{(s=+1)} \equiv
\lsp \begin{array}{c}
e^{-i\phi /2} \cos\theta/2 \\
e^{i\phi /2} \sin\theta/2
\end{array} \rsp, ~~~
\chi^{(s=-1)} \equiv
\lsp \begin{array}{c}
-e^{-i\phi /2} \sin\theta/2 \\
e^{i\phi /2} \cos\theta/2
\end{array} \rsp.
\eeq
The above spinor $u({\bf p},s)$ is a eigenstate of helicity ${\bf n}{\bf
\Sigma}$;
\beq
({\bf n}{\bf \Sigma}) u({\bf p},s) = s~u({\bf p},s),
\eeq
where ${\bf \Sigma}$ is given by
\beq
{\bf \Sigma}^i \equiv
\lsp \begin{array}{cc}
\sigma_i & 0 \\
0 & \sigma_i
\end{array} \rsp.
\eeq
%

%
%

\chapter{Photon spectrum}
\label{ap:spectrum}

\hspace*{\parindent}
In order to investigate photo-dissociation processes, we have to calculate
a photon spectrum induced by decay of gravitino. In this appendix, we
will write down Boltzmann equations which determine the high energy
photon spectrum with the case of high energy photon injection, and that
of high energy neutrino injection. We will also solve them numerically
and show the shape of the spectrum.

\section{Boltzmann equations}

\hspace*{\parindent}
In calculating photon spectrum (with high energy photon injection), we
take the following processes into account;
\begin{itemize}
\item double photon pair creation,
\item photon-photon scattering,
\item pair creation in nuclei,
\item Compton scattering off thermal electron,
\item inverse Compton scattering off background photon,
\item radiative decay of the gravitinos.
\end{itemize}
The Boltzmann equations for these cascade processes are formally given
by
\begin{eqnarray}
\frac{\partial f_{\gamma}(\epho)}{\partial t}
&=&
\left. \frac{\partial f_{\gamma}(\epho)}{\partial t} \right |_{\rm DP}
+ \left. \frac{\partial f_{\gamma}(\epho)}{\partial t} \right |_{\rm PP}
+ \left. \frac{\partial f_{\gamma}(\epho)}{\partial t} \right |_{\rm PC}
+ \left. \frac{\partial f_{\gamma}(\epho)}{\partial t} \right |_{\rm CS}
\nonumber \\
&&
+ \left. \frac{\partial f_{\gamma}(\epho)}{\partial t} \right |_{\rm IC}
+ \left. \frac{\partial f_{\gamma}(\epho)}{\partial t} \right |_{\rm DE}
+ \left. \frac{\partial f_{\gamma}(\epho)}{\partial t} \right |_{\rm EXP},
\label{fdot_pho} \\
\frac{\partial f_{e}(\eele)}{\partial t}
&=&
\left. \frac{\partial f_{e}(\eele)}{\partial t} \right |_{\rm DP}
+ \left. \frac{\partial f_{e}(\eele)}{\partial t} \right |_{\rm PC}
+ \left. \frac{\partial f_{e}(\eele)}{\partial t} \right |_{\rm CS}
+ \left. \frac{\partial f_{e}(\eele)}{\partial t} \right |_{\rm IC}
\nonumber \\
&&
+ \left. \frac{\partial f_{e}(\eele)}{\partial t} \right |_{\rm DE}
+ \left. \frac{\partial f_{e}(\eele)}{\partial t} \right |_{\rm EXP},
\label{fdot_ele}
\end{eqnarray}
where terms with the index DP (PP, PC, CS, IC, DE, and EXP) represents
the contribution from the double photon pair creation process
(photon-photon scattering, pair creation in nuclei, Compton scattering,
inverse Compton scattering, contribution from the gravitino decay, and
the effects of the expansion of the universe). Notice that in the main
part of this thesis, the expansion terms and the decay term in the
Boltzmann equation for the electron distribution function are ignored.
Below, we will see contributions from each processes in detail.

\subsection*{Double photon pair creation
[$~\gamma + \gamma \rightarrow e^{+} + e^{-}$~]}

\hspace*{\parindent}
For high energy photon whose energy is larger than $\sim$
$m_{e}^{2}/22T$, double photon pair creation is the most dominant
process.

The total cross section $\sigma_{\rm DP}$ for the double photon pair
creation process is given by
\begin{eqnarray}
\sigma_{\rm DP} \lsp \beta \rsp =
\frac{1}{2} \pi r_{e}^{2} \lsp 1 - \beta^{2} \rsp
\lmp \lsp 3 - \beta^4 \rsp \ln \frac{1+\beta}{1-\beta}
     - 2 \beta \lsp 2 - \beta^{2} \rsp \rmp ,
\label{s-DP}
\end{eqnarray}
where $r_{e}$ is the classical radius of electron which is given by
\beq
r_{e}=\frac{\alpha}{m_{e}},
\eeq
with $\alpha\simeq 1/137$ being the fine structure constant, and $\beta$
is the electron (or positron) velocity in the center of mass frame.
Using this formula, one can write down $(\partial f_{\gamma} /
\partial t)|_{\rm DP}$ as
\begin{eqnarray}
\left.
\frac{\partial f_{\gamma}(\epho)}{\partial t}
\right |_{\rm DP} =
- \frac{1}{8} \frac{1}{\epho^2} f_\gamma ( \epho )
\int_{m_{e}/\epho}^\infty d\ebg \frac{1}{\ebg^2}
\bar{f}_\gamma ( \ebg )
\left. \int_{4m_e^2}^{4\epho\ebg} ds ~s\sigma_{DP} (\beta)
\rabs_{\beta = \sqrt{1-(4m_e^2/s)}}.
\label{DP-p}
\end{eqnarray}

The spectrum of final state electron and positron is obtained in
ref.\cite{AAN}, and $(\partial f_{e} / \partial t)|_{\rm DP}$ is
given by
\begin{eqnarray}
\left.
\frac{\partial f_{e}(\eele)}{\partial t}
\right |_{\rm DP} =
\frac{1}{4} \pi r_{e}^{2} m_{e}^{4}
\int_{\eele}^{\infty}
d\epho \frac{f_{\gamma}(\epho)}{\epho^{3}}
\int_{0}^{\infty}
d\ebg
\frac{\bar{f}_{\gamma}(\ebg)}{\ebg^{2}}
G(\eele, \epho, \ebg),
\label{DP[-1z-e}
\end{eqnarray}
where function $G(\eele,\epho,\ebg)$ is given by
\begin{eqnarray}
G(\eele, \epho, \ebg)
&=&
\frac{4 \lsp \eele + \eele^{'} \rsp^{2}}{\eele \eele^{'}}
\ln \frac{4 \ebg \eele \eele^{'}}
          {m_{e}^{2} \lsp \eele + \eele^{'} \rsp}
- \lmp 1 - \frac{m_{e}^{2}}{\ebg \lsp \eele + \eele^{'} \rsp} \rmp
  \frac{\lsp \eele + \eele^{'} \rsp^{4}}{\eele^{2} \eele^{'~2}}
\nonumber \\ &&
+ \frac{2 \lmp 2 \ebg \lsp \eele +  \eele^{'} \rsp - m_{e}^{2} \rmp
        \lsp \eele +  \eele^{'} \rsp^{2}}
       {m_{e}^{2} \eele \eele^{'}}
- 8 \frac{\ebg \epho}{m_{e}^{2}},
\label{fn-G}
\end{eqnarray}
with $\eele^{'} = \epho + \ebg - \eele$, and $\bar{f}_{\gamma}$
represents the distribution function of the background photon at
temperature $T$,
\begin{eqnarray}
\bar{f}_{\gamma} (\ebg) =
\frac{\ebg^{2}}{\pi^{2}} \times
\frac{1}{\exp(\ebg / T) - 1 }.
\label{fbg}
\end{eqnarray}

\subsection*{Photon-photon scattering
[~$\gamma + \gamma \rightarrow \gamma + \gamma$~]}

\hspace*{\parindent}
If the photon energy is below the effective threshold of the double
photon pair creation, photon-photon scattering process becomes
significant. This process is analyzed in ref.\cite{APJ349-415} and for
$\epho^{'}\lsim O(m_{e}^{2}/T)$, $(\partial f_{\gamma} / \partial
t)|_{\rm PP}$ is given by
\begin{eqnarray}
\left.
\frac{\partial f_{\gamma}(\epho^{'})}{\partial t}
\right |_{\rm PP} &=&
\frac{35584}{10125 \pi} \alpha^2 r_{e}^{2} m_{e}^{-6}
\int_{\epho^{'}}^{\infty}
d\epho f_{\gamma}(\epho) \epho^{2}
\lmp
1 - \frac{\epho^{'}}{\epho} + \lsp \frac{\epho^{'}}{\epho} \rsp^{2}
\rmp^{2}
\int_{0}^{\infty}
d\ebg \ebg^{3} \bar{f}_{\gamma}(\ebg)
\nonumber \\ &&
- \frac{1946}{50625 \pi} f_{\gamma}(\epho^{'})
\alpha^2 r_{e}^{2} m_{e}^{-6} \epho^{'~3}
\int_{0}^{\infty}
d\ebg \ebg^{3} \bar{f}_{\gamma}(\ebg).
\label{PP-p}
\end{eqnarray}
For a larger value of $\epho^{'}$, we cannot use this formula. But for
high energy photons, photon-photon scattering is not significant because
double photon pair creation determines the shape of the photon spectrum.
Therefore, instead of using the exact formula, we take
$m_{e}^{2}/T$ as a cutoff scale of $(\partial f_{\gamma} / \partial
t)|_{\rm PP}$, \ie for $\epho^{'}\leq m_{e}^{2}/T$ we use
eq.(\ref{PP-p}) and for $\epho^{'}>m_{e}^{2}/T$ we take
\begin{eqnarray}
\left.
\frac{\partial f_{\gamma}(\epho^{'}>m_{e}^{2}/T)}{\partial t}
\right |_{\rm PP} = 0.
\end{eqnarray}
Notice that we have checked the cutoff dependence of spectra is
negligible.

\subsection*{Pair creation in nuclei
[~$\gamma + N \rightarrow e^{+} + e^{-} + N$~]}

\hspace*{\parindent}
Scattering off the electric field around nucleon, high energy photon
can produce electron positron pair if the photon energy is larger than
$2m_{e}$. Denoting total cross section of this process $\sigma_{\rm PC}$,
$(\partial f_{\gamma} / \partial t)|_{\rm NP}$ is given by
\begin{eqnarray}
\left.
\frac{\partial f_{\gamma}(\epho)}{\partial t}
\right |_{\rm NP} =
- \bar{n}_{N} \sigma_{\rm PC}(\epho) f_{\gamma}(\epho),
\label{PC-p}
\end{eqnarray}
where $\bar{n}_{N}$ is the nucleon number density. For $\sigma_{\rm
PC}$, we use the approximate formula derived by Maximon~\cite{Maximon}.
For $\epho$ near the threshold ($\epho < 4m_e$), the approximate formula
is given by
\beq
\sigma_{\rm PC}(\epho)|_{k<4} &=& \frac{2\pi}{3} Z^2 \alpha r_e^2
\lsp \frac{k-2}{k} \rsp^3
\nonumber \\ && \times
\lsp 1 + \frac{1}{2} \rho + \frac{23}{40} \rho^2
+ \frac{11}{60}\rho^3 + \frac{29}{960} \rho^4 +O(\rho^5) \rsp,
\eeq
where
\beq
k \equiv \frac{\epho}{m_e},~~~
\rho \equiv \frac{2k-4}{k+2+2\sqrt{2k}},
\eeq
and $Z$ is the charge of nuclei. For large $\epho$ ($\epho\geq 4m_e$),
$\sigma_{\rm PC}$ are expanded in the parameter $k^{-1}$, which is given
by
\beq
\sigma_{\rm PC}(\epho)|_{k\geq 4} &=& Z^2 \alpha r_e^2 \Bigg[
\frac{28}{9} \ln 2k - \frac{218}{27}
\nonumber \\ &&
+ \lsp \frac{2}{k} \rsp^2 \lmp
\frac{2}{3} \lsp \ln 2k \rsp^3
- \lsp \ln 2k \rsp^2 + \lsp 6 -  \frac{\pi^2}{3} \rsp \ln 2k
+ 2\zeta (3) + \frac{\pi^2}{6} - \frac{7}{2} \rmp
\nonumber \\ &&
- \lsp \frac{2}{k} \rsp^4
\lmp \frac{3}{16} \ln 2k + \frac{1}{8} \rmp
\nonumber \\ &&
- \lsp \frac{2}{k} \rsp^6
\lmp \frac{29}{2304} \ln 2k - \frac{77}{13824} \rmp
+O(k^{-8})\Bigg] .
\label{maximon}
\eeq

Differential cross section for this process $d \sigma_{\rm PC} / dE_+$ is
given in ref.\cite{BLP};
\beq
\frac{d \sigma_{\rm PC}}{d E_+} &=&
Z^2 \alpha r_e^2 \lsp \frac{p_+p_-}{\epho^3} \rsp \Bigg[
- \frac{4}{3} -2 E_+E_- \frac{p_+^2 + p_-^2}{p_+^2 p_-^2}
\nonumber \\ &&
+ m_e^2 \lmp l_- \frac{E_+}{p_-^3} + l_+ \frac{E_-}{p_+^3}
- l_+ l_- \frac{1}{p_+p_-} \rmp
\nonumber \\ &&
+ L \Bigg\{ - \frac{8E_+E_-}{3p_+p_-}
+ \frac{\epho^2}{p_+^3p_-^3}
\lsp E_+^2E_-^2 + p_+^2p_-^2 - m_e^2 E_+E_- \rsp
\nonumber \\ &&
- \frac{m_e^2 \epho}{2 p_+p_-}
\lsp l_+ \frac{E_+E_- - p_+^2}{p_+^3} + l_- \frac{E_+E_- - p_-^2}{p_-^3} \rsp
\Bigg\} \Bigg],
\eeq
where
\beq
p_{\pm} &\equiv& \sqrt{E_\pm^2 - m_e^2},
\\
L &\equiv& \ln \frac{E_+E_- + p_+p_- + m_e^2}{E_+E_- - p_+p_- + m_e^2},
\\
l_\pm &\equiv& \ln \frac{E_\pm + p_\pm}{E_\pm - p_\pm},
\eeq
with $E_-$ ($E_+$) being the energy of electron (positron) in final
state.  By using this formula, $(\partial f_{e} / \partial t)|_{\rm NP}$
is given by
\begin{eqnarray}
\left.
\frac{\partial f_{e}(\eele)}{\partial t}
\right |_{\rm NP} =
\bar{n}_{N} \int_{\eele+m_{e}}^{\infty} d\epho
\frac{d \sigma_{\rm PC}}{d \eele} f_{\gamma}(\epho).
\label{PC-e}
\end{eqnarray}

\subsection*{Compton scattering
[~$\gamma + e^{-} \rightarrow \gamma + e^{-}$~]}

\hspace*{\parindent}
Compton scattering is one of the processes by which high energy photons
lose their energy. Since the photo-dissociation of light elements occurs
when the temperature drops below $\sim$ 0.1MeV, we can consider the
thermal electrons to be almost at rest. Using the total and differential
cross sections at the electron rest frame $\sigma_{\rm CS}$ and
$d\sigma_{\rm CS}/d\eele$, one can derive
\begin{eqnarray}
\left.
\frac{\partial f_{\gamma}(\epho^{'})}{\partial t}
\right |_{\rm CS} &=&
\bar{n}_{e}
\int_{\epho^{'}}^{\infty} d \epho f_{\gamma}(\epho)
\frac{d\sigma_{\rm CS}(\epho^{'}, \epho)}{d\epho^{'}}
- \bar{n}_{e} \sigma_{\rm CS} f_{\gamma}(\epho^{'}),
\label{CS-p} \\
\left.
\frac{\partial f_{e}(\eele^{'})}{\partial t}
\right |_{\rm CS} &=&
\bar{n}_{e}
\int_{\eele^{'}}^{\infty} d \epho f_{\gamma}(\epho)
\frac{d\sigma_{\rm CS}(\epho+m_{e}-\eele^{'}, \epho)}{d\epho^{'}},
\label{CS-e}
\end{eqnarray}
where $\bar{n}_{e}$ is the number density of background electron.
Explicit forms of $\sigma_{\rm CS}$ and $d\sigma_{\rm CS}/d\eele$ are
given as
\beq
\sigma_{\rm CS} &=& 2 \pi r_e^2 \frac{1}{x}
\Bigg\{ \lsp 1 - \frac{4}{x} - \frac{8}{x^2} \rsp \ln (1+x)
%
%
+ \frac{1}{2} + \frac{8}{x} - \frac{1}{2(1+x)^2} \Bigg\},
\\
\frac{d\sigma_{\rm CS}(\epho^{'},\epho)}{d\epho^{'}} &=&
\pi r_e^2\frac{m_e}{\epho^2} \Bigg\{
\frac{\epho}{\epho^{'}} + \frac{\epho^{'}}{\epho}
%
%
+ \lsp \frac{m_e}{\epho^{'}} - \frac{m_e}{\epho} \rsp^2
- 2 m_e \lsp \frac{1}{\epho^{'}} - \frac{1}{\epho} \rsp
\Bigg\},
\eeq
with
\beq
x \equiv \frac{s - m_e^2}{m_e^2} = \frac{2 \epho}{m_e}.
\eeq

\subsection*{Inverse Compton scattering
[~$e^{\pm} + \gamma \rightarrow e^{\pm} + \gamma$~]}

\hspace*{\parindent}
Formula for the inverse Compton process is given by
Jones~\cite{PR167-1159}, and $(\partial f / \partial t)|_{\rm IC}$ is
given by
\begin{eqnarray}
\left.
\frac{\partial f_{\gamma}(\epho)}{\partial t}
\right |_{\rm IC} &=&
2 \pi r_{e}^{2} m_{e}^{2}
\int_{\epho+m_{e}}^{\infty}
d\eele \frac{2 f_{e}(\eele)}{\eele^{2}}
\int_{0}^{\infty}
d\ebg
\frac{\bar{f}_{\gamma}(\ebg)}{\ebg}
F(\epho, \eele, \ebg),
\label{IC-p} \\
\left.
\frac{\partial f_{e}(\eele^{'})}{\partial t}
\right |_{\rm IC} &=&
2 \pi r_{e}^{2} m_{e}^{2}
\int_{\eele^{'}}^{\infty}
d\eele \frac{f_{e}(\eele)}{\eele^{2}}
\int_{0}^{\infty}
d\ebg
\frac{\bar{f}_{\gamma}(\ebg)}{\ebg}
F(\eele+\ebg-\eele^{'}, \eele,\ebg)
\nonumber \\ &&
- 2 \pi r_{e}^{2} m_{e}^{2}
\frac{f_{e}(\eele^{'})}{\eele^{'~2}}
\int_{\eele^{'}}^{\infty} d\epho
\int_{0}^{\infty}
d\ebg
\frac{\bar{f}_{\gamma}(\ebg)}{\ebg}
F(\epho, \eele^{'}, \ebg),
\label{IC-e}
\end{eqnarray}
where function $F(\epho,\eele,\ebg)$ is given by
\begin{eqnarray}
\left.F(\epho,\eele,\ebg)\rabs_{0\geq q \geq 1} &=&
2 q \ln q + ( 1 + 2 q ) ( 1 - q ) +
\frac{(\Gamma_{\epsilon} q)^{2}}{2 \lsp 1 - \Gamma_{\epsilon} q \rsp}
( 1 - q ),
\\
\left.F(\epho,\eele,\ebg)\rabs_{\rm otherwise} &=& 0,
\label{fn-F}
\end{eqnarray}
with
\begin{eqnarray*}
\Gamma_{\epsilon} = \frac{4 \ebg \eele}{m_{e}^{2}},
{}~~~
q = \frac{\epho}{\Gamma_{\epsilon} (\eele - \epho)}.
\end{eqnarray*}

\subsection*{Radiative decay of the gravitino
[~$\psi_{\mu} \rightarrow \gamma + \tilde{\gamma}$~]}

\hspace*{\parindent}
Source of the non-thermal photon and electron spectra is radiative decay
of gravitino. Since gravitinos are almost at rest when they decay and we
only consider two body decay process, incoming high energy photons have
monochromatic energy $\epsilon_{\gamma 0}$, which is given by
\begin{eqnarray}
\epsilon_{\gamma 0} =
\frac{\mgra^{2} - m_{\tilde{\gamma}}^{2}}{2 \mgra}.
\label{e0}
\end{eqnarray}
In this case, $(\partial f_{\gamma} / \partial t)|_{\rm DE}$ can be written
as
\begin{eqnarray}
\left.
\frac{\partial f_{\gamma}(\epho)}{\partial t}
\right |_{\rm DE} =
B_\gamma \delta \lsp \epho - \epsilon_{\gamma 0} \rsp
\frac{n_{3/2}}{\tau_{3/2}},
\label{DE-p}
\end{eqnarray}
where $B_\gamma$ is the branching ratio for the process $\psi_\mu
\rightarrow \gamma + \tilde{\gamma}$. If the gravitino decays into an
electron and a selectron, $(\partial f_e / \partial t)|_{\rm DE}$ cannot
be ignored. Denoting the branching ratio for this decay process $B_e$,
the formula for $(\partial f_e / \partial t)|_{\rm DE}$ is given by
\begin{eqnarray}
\left.
\frac{\partial f_e(\eele)}{\partial t}
\right |_{\rm DE} =
B_e \delta \lsp \eele - E_{e0} \rsp
\frac{n_{3/2}}{\tau_{3/2}},
\label{DE-e}
\end{eqnarray}
where $E_{e0}$ is the energy of the injecting electron.

\section{Spectrum with high energy photon injection}

\hspace*{\parindent}
Having obtained the explicit forms for the Boltzmann equations, we solve
them in this section. In deriving high energy photon and electron
spectra, we have to use numerical method since the Boltzmann equations
(\ref{fdot_pho}) and (\ref{fdot_ele}) are very much complicated. Before
solving them numerically, we explain our approach to the Boltzmann
equations~\cite{TU474}.

At first, we classify the right-hand side of eq.(\ref{fdot_pho}) and
eq.(\ref{fdot_ele}) into the ``outgoing'' parts and ``incoming'' ones.
Below, we ignore the expansion terms $(\partial f_\gamma/\partial
t)|_{\rm EXP}$ and $(\partial f_e/\partial t)|_{\rm EXP}$ in the
Boltzmann equations (\ref{fdot_pho}) and (\ref{fdot_ele}) since the
expansion rate of the universe is much smaller than the scattering rates
of electro-magnetic processes. Then, eq.(\ref{fdot_pho}) and
eq.(\ref{fdot_ele}) can be written as
\beq
\frac{\partial f_\gamma (\epho)}{\partial t} &=&
- \Gamma_\gamma(\epho;T) f_\gamma (\epho)
+ \dot{f}_{\gamma,{\rm IN}}(\epho),
\\
\frac{\partial f_e (\eele)}{\partial t} &=&
- \Gamma_e(\eele;T) f_e (\eele)
+ \dot{f}_{e,{\rm IN}}(\eele),
\eeq
where ``incoming'' terms $\dot{f}_{\gamma ,{\rm IN}}$ and
$\dot{f}_{e,{\rm IN}}$ can be obtained as the sums of the contributions
from the decay terms and the functionals of the distribution functions
of the photon and the electron;
\beq
\dot{f}_{\gamma,{\rm IN}}(\epho) &=&
\left. \frac{\partial f_\gamma(\epho)}{\partial t} \right|_{\rm DE}
+ \int_{\epho}^{\infty} d\epho' K_{\gamma,\gamma}(\epho,\epho';T)
f_\gamma (\epho')
\nonumber \\ &&
+ \int_{\epho}^{\infty} d\eele' K_{\gamma,e}(\epho,\eele';T) f_e (\eele'),
\label{Gamma-K_g}
\\
\dot{f}_{e,{\rm IN}}(\eele) &=&
\left. \frac{\partial f_e(\eele)}{\partial t} \right|_{\rm DE}
+ \int_{\eele}^{\infty} d\epho' K_{e,\gamma}(\eele,\epho';T)
f_\gamma (\epho')
\nonumber \\ &&
+ \int_{\eele}^{\infty} d\eele' K_{e,e}(\eele,\eele';T) f_e (\eele').
\label{Gamma-K_e}
\eeq
The explicit forms of $\Gamma_A$ and $K_{AB}$ ($A, B = \gamma, e$) can be
obtained from the full details of the Boltzmann equations given in the
previous section.

As mentioned before, the strategy we use here is to calculate a
stationary solution to the Boltzmann equations (\ref{fdot_pho}) and
(\ref{fdot_ele}), which obey
\beq
\frac{\partial f_\gamma (\epho)}{\partial t} =
\frac{\partial f_e (\eele)}{\partial t} = 0.
\eeq
In our approach, the distribution functions $f_\gamma$ and $f_e$ of the
photon and the electron can be formally given as
\beq
f_\gamma (\epho) &=&
\frac{\dot{f}_{\gamma,{\rm IN}}(\epho)}{\Gamma_\gamma(\epho;T)},
\label{solution_fp} \\
f_e (\eele) &=&
\frac{\dot{f}_{e,{\rm IN}}(\eele)}{\Gamma_e(\eele;T)}.
\label{solution_fe}
\eeq

The important thing is that the ``incoming'' terms $\dot{f}_{\gamma
,{\rm IN}}(\epho)$ and $\dot{f}_{e,{\rm IN}}(\eele)$ depend only on
$f_\gamma(\epsilon')$ and $f_e(\epsilon')$ with $\epsilon'>\epho,
\eele$. Therefore, if the distribution functions $f_\gamma(\epho)$
and $f_e(\eele)$ with $\epho, \eele > \epsilon$ (and the source terms)
are known, we can obtain $f_\gamma(\epsilon)$ and $f_e(\epsilon)$ from
eq.(\ref{solution_fp}) and eq.(\ref{solution_fe}). By using this fact,
we solve the Boltzmann equations with monochromatic high energy photon
injection.  As one will see, extensions to the cases with
non-monochromatic sources or the high energy electron source are
trivial.

In our numerical calculations, we use the mesh points $E_i$
($0\leq i\leq N_{\rm mesh}+1$) with $E_0 = \epsilon_{\rm min}$ and
$E_{N_{\rm mesh}}=\epsilon_0$, where $\epsilon_{\rm min}$ is
some minimum energy we are concerning (which we take $\epsilon_{\rm
min}=1\MEV$) and $\epsilon_0$ is the energy of the primary
injecting photons. The energy range between $\epsilon_{\rm min}$ and
$\epsilon_0$ is divided logarithmically, {\it i.e.} the $i$-th
mesh point $E_i$ is given by
\beq
E_i = \epsilon_{\rm min}
\lsp\frac{\epsilon_0}{\epsilon_{\rm min}}\rsp^{i/N_{\rm mesh}}.
\eeq
By using the fact that the incoming photons into the mesh point
$E_{N_{\rm mesh}}$ are supplied only by the photon source,
incoming terms for $i = N_{\rm mesh}$ is given by
\beq
\dot{f}_{\gamma,{\rm IN}} (E_{N_{\rm mesh}}) &\simeq&
\dot{n}_{\gamma} \times \frac{1}{\Delta_{N_{\rm mesh}}},
\label{fin_p}
\\
\dot{f}_{e,{\rm IN}} (E_{N_{\rm mesh}}) &\simeq& 0,
\label{fin_e}
\eeq
where $\dot{n}_{\gamma}$ is the production rate of monochromatic
photons, and $\Delta_i\equiv (E_{i+1} - E_{i-1})/2$. Combining
eq.({\ref{fin_p}) and eq.({\ref{fin_e}) with eq.(\ref{solution_fp}) and
eq.(\ref{solution_fe}), we can obtain the distribution functions at the
$N_{\rm mesh}$-th mesh point $f_\gamma(E_{N_{\rm mesh}})$ and
$f_e(E_{N_{\rm mesh}})$.

Next we determine the distribution functions for lower energy region.
Essentially, ``incoming'' terms for the $i$-th mesh point are derived
from the distribution functions $f_\gamma(E_j)$ and $f_e(E_j)$ with
$j>i$.  Discretizing eq.(\ref{Gamma-K_g}) and eq.(\ref{Gamma-K_e}),
``incoming'' terms are (approximately) given by
\beq
\dot{f}_{\gamma,{\rm IN}} (E_i) &\simeq&
\sum_{j>i} \Delta_j
\lmp K_{\gamma\gamma}(E_i,E_j) f_\gamma(E_j) +
K_{\gamma e}(E_i,E_j) f_e(E_j) \rmp,
\label{fin_p_disc}
\\
\dot{f}_{e,{\rm IN}} (E_i) &\simeq&
\sum_{j>i} \Delta_j
\lmp K_{e\gamma}(E_i,E_j) f_\gamma(E_j) +
K_{e e}(E_i,E_j) f_e(E_j) \rmp,
\label{fin_e_disc}
\eeq
from which we can obtain $f_\gamma(E_i)$ and $f_e(E_i)$ by using
eq.(\ref{solution_fp}) and eq.(\ref{solution_fe}).\footnote
{In fact, eq.(\ref{fin_p}) -- eq.(\ref{fin_e_disc}) receive corrections
of the order of $\Delta$, which is due to the discretization of the
energy range. In our numerical calculations, we take these corrections
into account.}
In a way explained above, we calculate the photon and electron
distribution functions $f_\gamma(E_i)$ and $f_e(E_i)$ at each mash
points from $i = N_{\rm mesh}-1$ to $i = 0$ in order.

%
%
\begin{figure}[p]
\epsfxsize=14cm
\centerline{\epsfbox{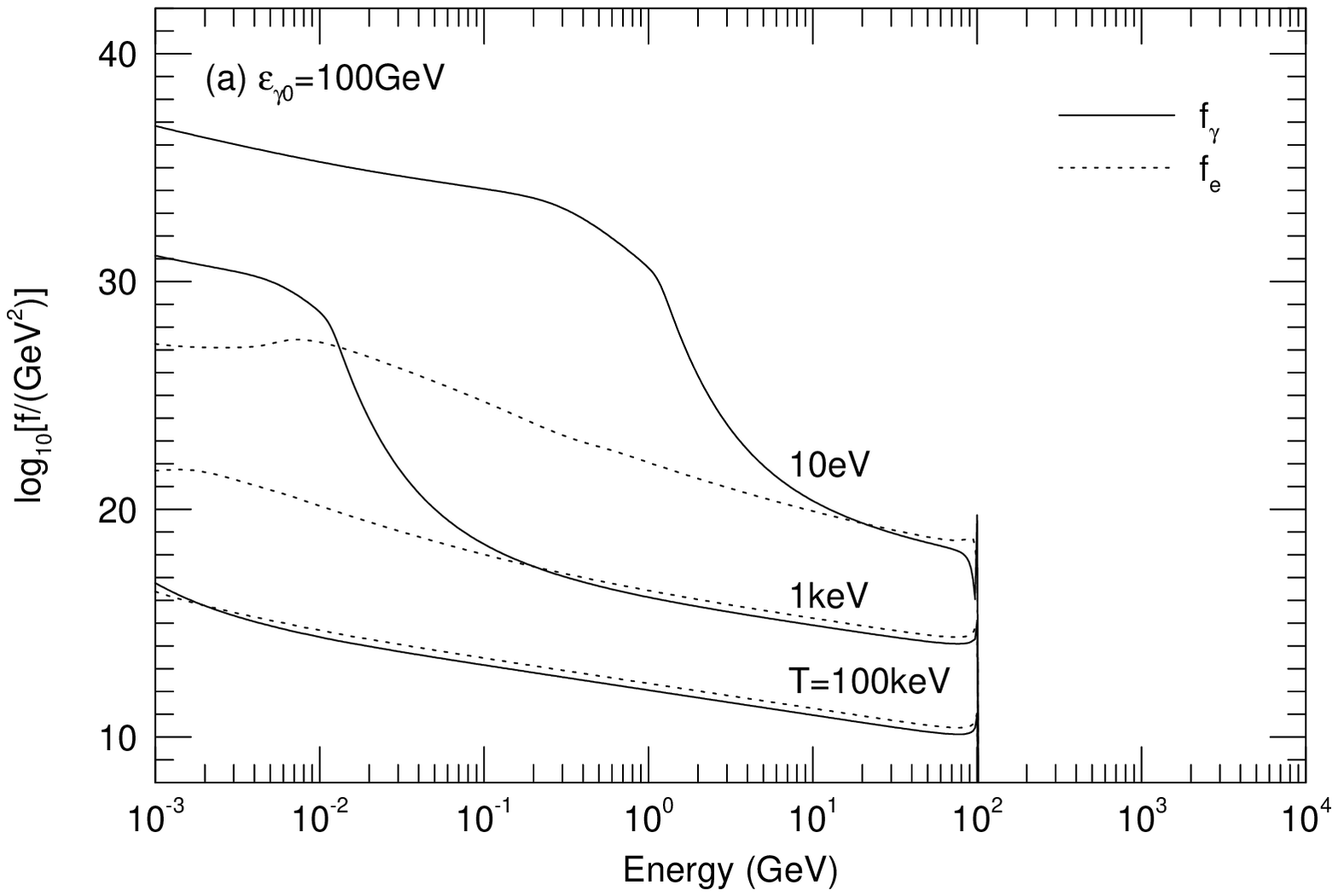}}

\vspace{1.5cm}

\centerline{\epsfbox{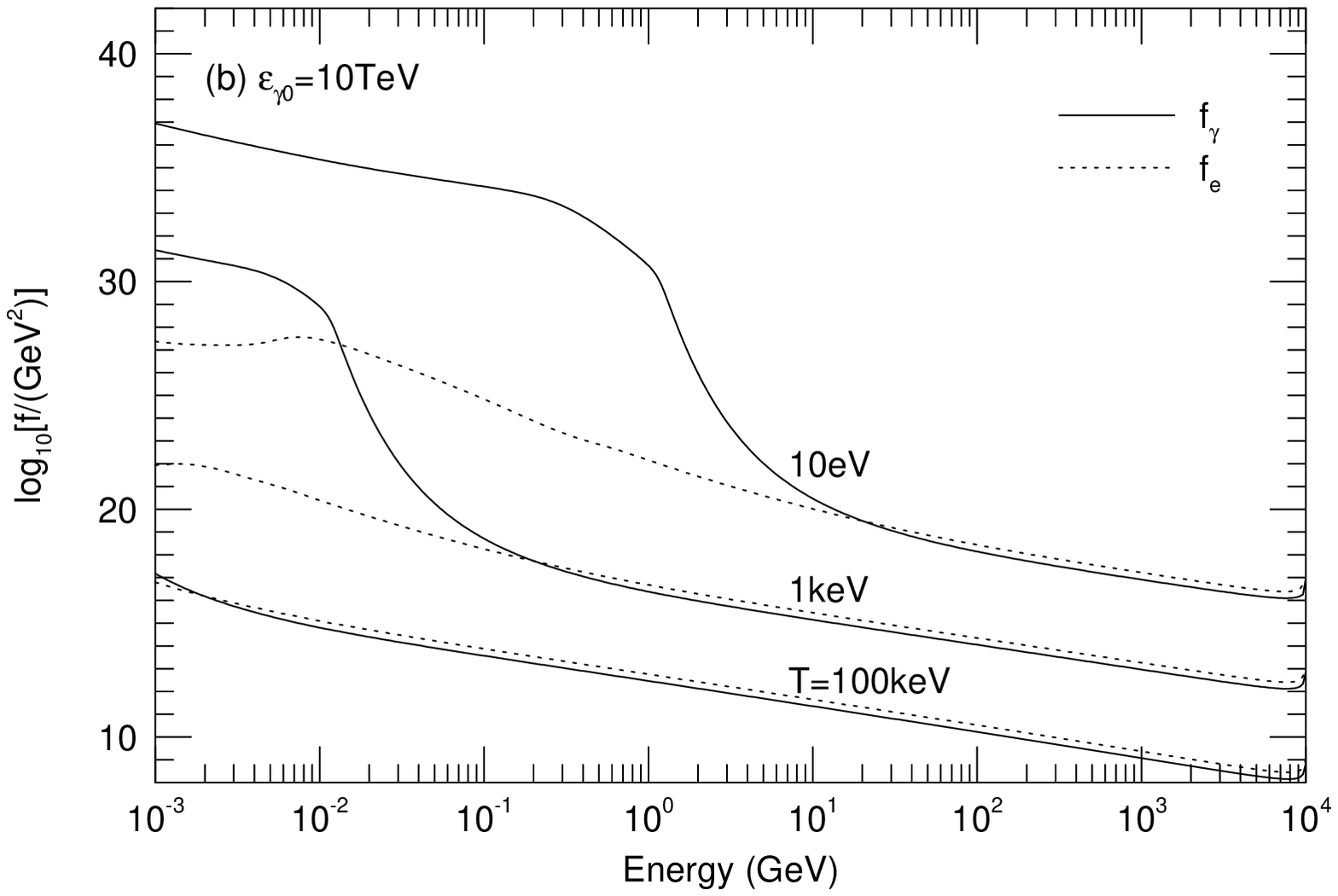}}

\caption{Typical spectra of photon (the solid lines) and electron
(the dotted lines). We take the temperature of the background
photon to be $T=100\KEV, 1\KEV, 10\EV$, and the energy of the incoming
high energy photon $\epsilon_{\gamma 0}$ is (a) 100GeV and (b) 10TeV.
Normalization of the initial photon is given by $\epsilon_{\gamma
0}\times( \partial f_{\gamma}(\epho)/\partial t)|_{\rm
DE}=\delta(\epho - \epsilon_{\gamma 0})~\GEV^{5}$.}
\label{fig:ref_spectra}
\end{figure}
%
%

For each $T$ and $\epsilon_{\gamma 0}$, we calculate the
spectra $f_{\gamma}(\epho)$ and $f_{e}(\eele)$ by
solving eq.(\ref{fdot_pho}) and eq.(\ref{fdot_ele}) numerically. Typical
spectra are shown in Fig.~\ref{fig:ref_spectra} in which we show
the case with $\epsilon_{\gamma 0}=100\GEV$ and 10TeV, $T=100\KEV,
1\KEV, 10\EV$, and the incoming flux of the high energy photon is
normalized to be
\begin{eqnarray}
\epsilon_{\gamma 0} \times
\left. \frac{\partial f_{\gamma}(\epho)}
            {\partial t} \right |_{\rm DE} =
\delta(\epho - \epsilon_{\gamma 0})~\GEV^{5}.
\label{nomalization}
\end{eqnarray}
Notice that eq.(\ref{fdot_pho}) and
eq.(\ref{fdot_ele}) are linear in $f_{\gamma}$ and $f_{e}$. Therefore,
once the solution $f_{\gamma,{\rm ref}}$ is obtained with some reference
value of the decay term $(\partial f_{\gamma} / \partial t)|_{\rm
DE,ref}$, we can reconstruct the photon spectrum for arbitrary value of
$(\partial f_{\gamma} / \partial t)|_{\rm DE}$ with temperature $T$ and
the incident photon energy $\epsilon_{\gamma 0}$ fixed;
\begin{eqnarray}
f_{\gamma} (\epho)=
f_{\gamma,{\rm ref}} (\epho) \times
\frac{(\partial f_{\gamma} / \partial t)|_{\rm DE}}
     {(\partial f_{\gamma} / \partial t)|_{\rm DE,ref}}.
\label{arbitrary-fp}
\end{eqnarray}

The behaviors of the photon spectrum can be understood in the following
way. In the region $\epho\gsim m_{e}^{2}/22T$, the photon number density
is extremely suppressed since the rate of double photon pair creation
process is very large. Just below this threshold value, the shape of the
photon spectrum is determined by the photon-photon scattering process,
and if the photon energy is sufficiently small, the Compton scattering
with the thermal electron is the dominant process for photons.

%
%
\begin{table}[t]
\begin{center}
\begin{tabular}{r|cccc}
\multicolumn{5}{l}
{$\epsilon_{\gamma 0} = 10~\TEV$} \\ \hline \hline
{Temperature} & {$P_{\rm low}$}
& {$N_{\rm low}~\GEV^2$} & {$P_{\rm pp}$} & {$N_{\rm pp}~\GEV^2$}
\\ \hline
{1~eV} &
{$-1.57$} & {$1.6\times 10^8$} & {$-5.10$} & {$6.9\times 10^{-18}$}
\\
{10~eV} &
{$-1.34$} & {$5.4\times 10^8$} & {$-5.20$} & {$6.0\times 10^{-18}$}
\\
{100~eV}&
{$-1.22$} & {$1.7\times 10^9$} & {$-4.84$} & {$1.1\times 10^{-17}$}
\\ \hline\hline
\end{tabular}

\vspace*{3mm}

\begin{tabular}{r|cccc}
\multicolumn{5}{l}
{$\epsilon_{\gamma 0} = 1~\TEV$} \\ \hline\hline
{Temperature} & {$P_{\rm low}$}
& {$N_{\rm low}~\GEV^2$} & {$P_{\rm pp}$} & {$N_{\rm pp}~\GEV^2$}
\\ \hline
{1~eV} &
{$-1.56$} & {$1.4\times 10^8$} & {$-5.07$} & {$6.2\times 10^{-18}$}
\\
{10~eV} &
{$-1.34$} & {$4.9\times 10^8$} & {$-5.17$} & {$5.5\times 10^{-18}$}
\\
{100~eV}&
{$-1.22$} & {$1.4\times 10^9$} & {$-4.79$} & {$1.0\times 10^{-17}$}
\\ \hline\hline
\end{tabular}

\vspace*{3mm}

\begin{tabular}{r|cccc}
\multicolumn{5}{l}
{$\epsilon_{\gamma 0} = 100~\GEV$} \\ \hline\hline
{Temperature} & {$P_{\rm low}$}
& {$N_{\rm low}~\GEV^2$} & {$P_{\rm pp}$} & {$N_{\rm pp}~\GEV^2$}
\\ \hline
{1~eV} &
{$-1.56$} & {$1.4\times 10^8$} & {$-5.01$} & {$5.7\times 10^{-18}$}
\\
{10~eV} &
{$-1.33$} & {$4.7\times 10^8$} & {$-5.15$} & {$5.3\times 10^{-18}$}
\\
{100~eV}&
{$-1.22$} & {$1.3\times 10^9$} & {$-4.74$} & {$1.1\times 10^{-17}$}
\\ \hline\hline
\end{tabular}

\vspace*{3mm}

\begin{tabular}{r|cccc}
\multicolumn{5}{l}
{$\epsilon_{\gamma 0} = 10~\GEV$} \\ \hline\hline
{Temperature} & {$P_{\rm low}$}
& {$N_{\rm low}~\GEV^2$} & {$P_{\rm pp}$} & {$N_{\rm pp}~\GEV^2$}
\\ \hline
{1~eV} &
{---} & {---} & {---} & {---}
\\
{10~eV} &
{$-1.33$} & {$4.5\times 10^8$} & {$-5.12$} & {$5.5\times 10^{-18}$}
\\
{100~eV}&
{$-1.22$} & {$1.3\times 10^9$} & {$-4.77$} & {$9.6\times 10^{-18}$}
\\ \hline\hline
\end{tabular}

\caption
{$P_{\rm low}$, $N_{\rm low}$, $P_{\rm pp}$ and $N_{\rm pp}$ for the
cases of $T= 1~\EV, 10~\EV, 100~\EV$, and $\epsilon_{\gamma 0} =
10~\TEV, 1\TEV, 100\GEV, 10\GEV$. Here, we take $N_{\rm
mesh}=401$.}
\label{table:P&N}
\end{center}
\end{table}
%
%

As one can see in Figs.~\ref{fig:ref_spectra}, both the photon spectrum
for sufficiently low energy region and that for the photon-photon
scattering region obey power-law spectrum; $f_\gamma(\epho) \propto
\epho^P$. We fit the photon spectrum at the sufficiently low energy
region as
\beq
f_\gamma(\epho) \simeq N_{\rm low}
\lsp \frac{\dot{\rho}_{\rm IN}}{\GEV^5} \rsp
\lsp \frac{T}{\GEV}\rsp^{-3}
\lsp\frac{\epho}{\GEV}\rsp^{P_{\rm low}},
\label{fit_low}
\eeq
and that for the photon-photon scattering region as
\beq
f_\gamma(\epho) \simeq N_{\rm pp}
\lsp \frac{\dot{\rho}_{\rm IN}}{\GEV^5} \rsp
\lsp \frac{T}{\GEV}\rsp^{-6}
\lsp\frac{\epho}{\GEV}\rsp^{P_{\rm pp}},
\label{fit_pp}
\eeq
with $\dot{\rho}_{\rm IN}$ being the total amount of the energy
injection from the gravitino decay;
\beq
\dot{\rho}_{\rm IN} = \int d\epho \epho
\left.\frac{\partial f_\gamma(\epho)}{\partial t} \right|_{\rm DE}.
\eeq
Notice that the amplitude of the spectrum is proportional to the mean
free time of the photon. In low energy region the mean free time is
determined by Compton scattering and depends on the background
temperature as $\sim T^{-3}$. For the photon-photon scattering region,
the amplitude is proportional to $\sim T^{-6}$ since the cross section
depends on $\sim T^3$. For the cases $T=$ 1eV, 10eV, 100eV and
$\epsilon_0 =$ 10TeV, 1TeV, 100GeV, 10GeV, we calculate the
fitting parameters $P_{\rm low}$, $P_{\rm pp}$, $N_{\rm low}$ and
$N_{\rm pp}$, and the results are shown in
Table~\ref{table:P&N}.\footnote
{Notice that the formulae (\ref{fit_low}) and (\ref{fit_pp}) are
relevant for the cases when the initial photon energy $\epsilon_{\gamma
0}$ is much larger than the effective threshold of the double photon
pair creation process.}
These variables slightly depend on the background temperature, while
their dependence on the initial photon energy $\epsilon_{\gamma 0}$ is
insignificant.

In Fig.~\ref{fig:ellis}, we compare our photon spectrum with the
results of the simple fitting formula used in ref.\cite{NPB373-399},
which is given by\footnote
{Although the photon spectrum is not explicitly given in
ref.~\cite{NPB373-399}, we obtain it from their photon production
spectrum divided by the cross section for Compton scattering. Since the
cross section has energy dependence, the resultant spectrum $(\propto
\epsilon_{\gamma}^{-0.9})$ becomes softer than that for photon
production $(\propto \epsilon_{\gamma}^{-1.5})$.}
\beq
f_\gamma (\epho) =  \frac{1}{\bar{n}_e \sigma_{\rm CS}(\epho)} \times
\frac{n_{3/2}}{\tau_{3/2}} \frac{dn_E(\epho)}{d\epho},
\eeq
where
\beq
\left.\frac{dn_E(\epho)}{d\epho}
\rabs_{0\leq\epho\leq\epsilon_{\rm max}/2} &=&
\frac{24\sqrt{2}}{55}
\frac{\epsilon_{\gamma 0}}{\sqrt{\epsilon_{\rm max}}}
\epho^{-3/2},
\\
\left.\frac{dn_E(\epho)}{d\epho}
\rabs_{\epsilon_{\rm max}/2\leq\epho\leq\epsilon_{\rm max}} &=&
\frac{3}{55}
\epsilon_{\gamma 0}\epsilon_{\rm max}^3
\epho^{-5},
\\
\left.\frac{dn_E(\epho)}{d\epho}
\rabs_{\epho\geq\epsilon_{\rm max}} &=& 0,
\eeq
with
\beq
\epsilon_{\rm max} \equiv \frac{m_e^2}{22T}.
\eeq

As one can see, not only the absolute value but also the form of the
spectrum differs between them.  The fitting formula in
ref.\cite{NPB373-399} is derived from the numerical results given in
refs.\cite{APJ335-786,APJ349-415} in which, however, the effect of the
Compton scattering is not taken into account. Our results indicate that
the number of Compton scattering events is comparable to that of the
inverse Compton events for such low energy region, since the number
density of the high energy electron is extremely smaller than that of
high energy photon. Therefore, the deformation of the photon spectrum by
Compton scattering is expected below the threshold of the photon-photon
scattering.

%
%
\begin{figure}[t]
\epsfxsize=14cm
\centerline{\epsfbox{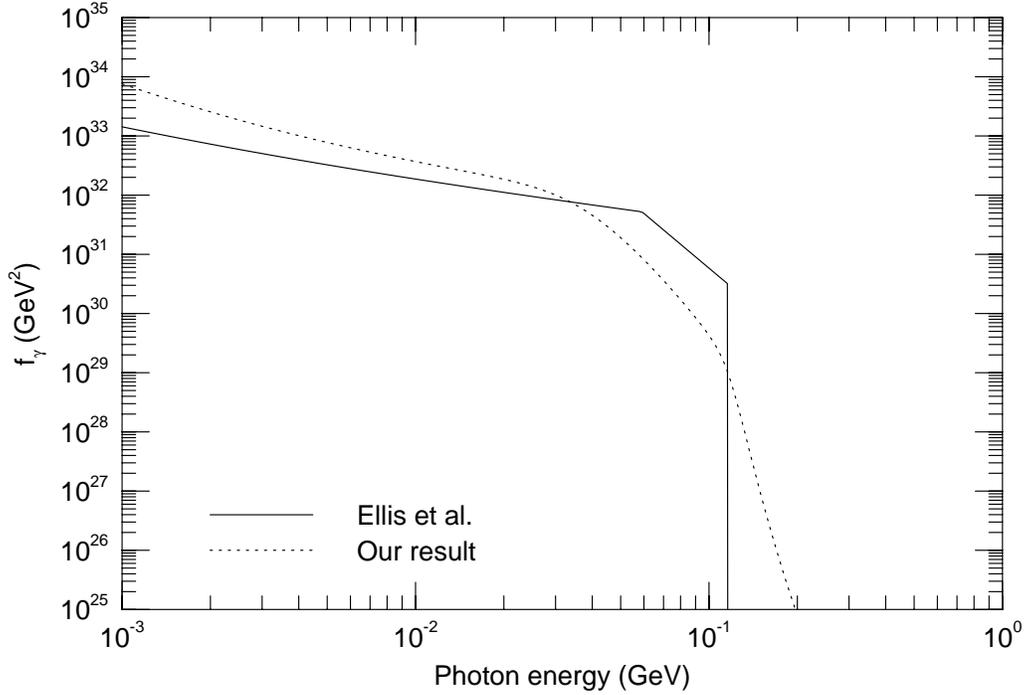}}

\caption{Photon spectrum derived from the fitting formula used in Ellis
et~al.~\protect\cite{NPB373-399} (solid line) is compared with our
result with $\epho =100\GEV$ (dashed line). We take the
temperature of the background photon to be $100$eV and the normalization
of the incoming flux is the same as Fig.~\protect\ref{fig:ref_spectra}.}
\label{fig:ellis}

\end{figure}
%
%

Before closing section, we should comment on the case with high energy
electron sources. By numerical calculations, we have checked that the
spectra in the case with high energy electron injection are almost the
same as those with photon injection if the background temperature and
the total amount of the energy injection are fixed. Numerically, their
differences are at most $O(10\%)$. Therefore, the formulae given in
eq.(\ref{fit_low}) and eq.(\ref{fit_pp}) with Table~\ref{table:P&N} are
well approximated ones if one redifines the $\dot{\rho}_{\rm IN}$ as
\beq
\dot{\rho}_{\rm IN} = \int d\epho \epho
\left.\frac{\partial f_\gamma(\epho)}{\partial t} \right|_{\rm DE} +
\int d\eele \eele
\left.\frac{\partial f_e(\eele)}{\partial t} \right|_{\rm DE}.
\eeq

\section{Spectrum with high energy neutrino injection}

\hspace*{\parindent}
Next we will consider the case of high energy neutrino injection. In
this case, the emitted high energy neutrinos may scatter off the
background neutrinos and produce an $e^+ e^-$ (or $\mu^+ \mu^-$) pairs,
which then produces many soft photons through electro-magnetic
interactions. Since the rates of neutrino-neutrino scattering processes
are not large enough, we cannot ignore the effect of the expansion of
the universe. Therefore we calculate high energy photon spectrum in two
steps; first we determine the time evolution of the distribution function
of high energy neutrinos, and then we calculate the photon (and the
electron) spectrum by regarding the high energy neutrinos as sources of
high energy $e^+ e^-$ (and $\mu^+ \mu^-$) pairs. The procedures for the
second step are essentially equal to those given in the previous
section, and in this section, we will explain the first step.

Let us begin by deriving the Boltzmann equations which determines the
high energy neutrino spectra. Injecting high energy neutrinos scatter
off the thermal neutrinos in the following processes;
\beq
\nu_{i} + \nu_{i,b} &\rightarrow & \nu_i + \nu_i ,
\label{nu_process1} \\
\nu_{i} + \bar{\nu}_{i,b} &\rightarrow &\nu_i + \bar{\nu}_i,
\label{nu_process2} \\
\nu_{i} + \bar{\nu}_{i,b} &\rightarrow &\nu_j + \bar{\nu}_j ,
\label{nu_process3} \\
\nu_{i} + \nu_{j,b} &\rightarrow &\nu_i + \nu_j,
\label{nu_process4} \\
\nu_{i} + \bar{\nu}_{j,b} & \rightarrow &\nu_i + \bar{\nu}_j,
\label{nu_process5} \\
\nu_{i} + \bar{\nu}_{i,b} &\rightarrow & e^{-}  + e^{+},
\label{nu_process6} \\
\nu_{i} + \bar{\nu}_{i,b} & \rightarrow & \mu^{-}  + \mu^{+}.
\label{nu_process7}
\eeq
where index $i$ and $j$ are the generation indices with $i\neq j$, and
$l^\pm$ represents $e^\pm$ and $\mu^\pm$.  All the amplitude squared
$|{\cal M}|^2$ in these reactions take the form given by
\beq
|{\cal M}|^2 = 32G_F^2 \lmp a (pp')^2 + b(pq)^2 + c(pq')^2 + d m^2
(pp') \rmp ,
\label{amplitude}
\eeq
where $G_F \equiv (1/\sqrt{2}v^2) \simeq 1.17\times 10^{-5}\GEV^{-2}$ is
the Fermi constant (with $v\simeq 246\GEV$), $p$ and $p'$ the initial
momenta of high energy neutrino and background neutrino, $q$ and $q'$
the final momenta, $m$ the mass of the fermion in final state, and the
coefficients $a$ -- $d$ depend on individual reaction. Coefficients for
each processes are given in Table~\ref{table:a-d}.
%
%
\begin{table}
\begin{center}
\begin{tabular}{c|lccc} \hline \hline
{Process} &
{~~a~~~~~~~~~} &
{~~~~~~~~~b~~~~~~~~~} &
{~~~~~~~~~c~~~~~~~~~} &
{~~~~~~~~~d~~~~~~~~~} \\ \hline
{$\nu_{i} + \nu_{i,b} \rightarrow \nu_i + \nu_i$} &
{~~2} &  {0} & {0} &  {0} \\
{$\nu_{i} + \bar{\nu}_{i,b} \rightarrow \nu_i + \bar{\nu}_i$} &
{~~0} &  {0} & {9} &  {0} \\
{$\nu_{i} + \bar{\nu}_{i,b} \rightarrow \nu_j + \bar{\nu}_j$} &
{~~0} &  {0} & {1} &  {0} \\
{$\nu_{i} + \nu_{j,b} \rightarrow \nu_i + \nu_j$} &
{~~1} &  {0} & {0} &  {0} \\
{$\nu_{i} + \bar{\nu}_{j,b} \rightarrow \nu_i + \bar{\nu}_j$} &
{~~0} &  {0} & {1} &  {0} \\
{$\nu_{i} + \bar{\nu}_{i,b} \rightarrow l_i^{-}  + l_i^{+}$} &
{~~0} & {$16\sin^4\theta_W$} & {$16\sin^4\theta_W$} & {$16\sin^4\theta_W$} \\
{$\nu_{i} + \bar{\nu}_{i,b} \rightarrow l_j^{-}  + l_j^{+}$} &
{~~0} & {$16\sin^4\theta_W$} &
{$(4\sin^2\theta_W -2)^2$} &
{$(4\sin^2\theta_W -2)^2-4$} \\
\hline \hline
\end{tabular}
\caption{Coefficients $a$ -- $d$ for each processes. Index $i$ and $j$
(with $i\neq j$) represent the generation, $\nu_{i,b}$ is the background
neutrino of $i$-th generation, $l^{\pm}_i$ is the charged lepton of
$i$-th generation (in our case, $e^{\pm}$ or $\mu^{\pm}$), and
$\theta_W$ is the Weinberg angle.}
\label{table:a-d}
\end{center}
\end{table}
%
%

First, we will consider the neutrino scattering processes
(\ref{nu_process1}) -- (\ref{nu_process5}). Let us derive a formula for
the energy distribution of scattered neutrino for the case where a high
energy neutrino with energy $\enu$ runs through a region filled with
neutrino gas with temperature $T_\nu$. For this purpose, it is
convenient to consider the momenta $p$ -- $q'$ in the center-of-mass
frame. In the center-of-mass frame, we parametrize $p$ -- $q'$ by using
parameters $\xi$, $\eta$, $\zeta$ ($0\leq\xi,\eta\leq\pi$, and
$0\leq\zeta\leq 2\pi$) as
\beq
p^\mu_{\rm cm} &=& \omega (1,~0,~\sin\xi,~\cos\xi),
\\
p'^\mu_{\rm cm} &=& \omega (1,~0,~-\sin\xi,~-\cos\xi),
\\
q^\mu_{\rm cm} &=&
\omega (1,~\sin\eta \cos\zeta,~\sin\eta \sin\zeta,~\cos\eta),
\\
q'^\mu_{\rm cm} &=&
\omega (1,~-\sin\eta \cos\zeta,~-\sin\eta \sin\zeta,~-\cos\eta).
\eeq
Then the differential cross section is given by
\beq
\frac{d \sigma}{d\cos\eta d\zeta} =
\frac{1}{2 \pi^2 s} G_F^2
\lmp a (pp')^2 + b(pq)^2 + c(pq')^2 + d m^2(pp') \rmp ,
\label{dsigma4nu}
\eeq
with
\beq
s \equiv ( p + p' )^2 = 4 \omega^2.
\eeq

Boosting this system to $z$-direction with velocity $\beta$, we obtain
\beq
p^\mu &=& \omega
\lsp \gamma ( 1 + \beta\cos\xi ),~0,~\sin\xi,~\gamma ( \cos\xi + \beta )
\rsp,
\\
p'^\mu &=& \omega
\lsp \gamma ( 1 - \beta\cos\xi ),~0,~\sin\xi,~\gamma ( -\cos\xi + \beta )
\rsp,
\eeq
with $\gamma^{-1}\equiv\sqrt{1-\beta^2}$. We identify these momenta as
those in the comoving frame. Then the energy of initial high energy
neutrino $\enu$ and that of thermal neutrino $\enubg$ are given by
\beq
\enu &=& \omega \gamma ( 1 + \beta\cos\xi ),
\\
\enubg &=& \omega \gamma ( 1 - \beta\cos\xi ).
\eeq
Furthermore, the angle $\theta$ between ${\bf p}$ and ${\bf p}'$ (in the
comoving frame) can be obtained from the following relation;
\beq
\cos\theta = 1 - \frac{2 - 2\beta^2}{1 - \beta^2\cos\xi},
\eeq
and the energy of the neutrino in final state $\enu'$ is given by
\beq
\enu' \equiv q^0 = \omega \gamma ( 1 + \beta\cos\eta ).
\eeq
Using these identities and integrating with $\zeta$, differential cross
section (\ref{dsigma4nu}) becomes
\beq
\frac{d\sigma}{d\enu'} &=&
\frac{s}{32\pi} G_F^2 \lsp \beta\gamma\omega \rsp^{-1}
\nonumber \\ &&
\times\Big\{ 8a +
b \lsp 3 - \cos^2\xi - \cos^2\eta - 4\cos\xi\cos\eta +
3\cos^2\xi\cos^2\eta \rsp
\nonumber \\ &&
+ c \lsp 3 - \cos^2\xi - \cos^2\eta + 4\cos\xi\cos\eta +
3\cos^2\xi\cos^2\eta \rsp \Big\}.
\eeq

Now we are ready to calculate the number density of scattered neutrino
with energy $\enu'$ -- $(\enu' + d\enu')$ in unit time. For this
purpose, let us consider the process in which incident high energy
neutrino with energy $\enu$ -- $( \enu + d\enu )$ scatters off the
thermal neutrino with energy $\enubg$ -- $( \enubg + d\enubg )$ and
relative angle $\theta$ -- $( \theta + d\theta )$. For this process,
number density of the target (thermal) neutrino is given by
\beqstar
({\rm target}) =
\frac{1}{2} \bar{f}_\nu ( \enubg ) d\enubg \sin\theta d\theta.
\eeqstar
where $\bar{f}_\nu ( \enubg )$ is the distribution function of the
background neutrino;
\beq
\bar{f}_{\nu} (\enubg) = \frac{\enubg^2}{2\pi^2}
\frac{1}{e^{-\enubg / T_{\nu}}+1},
\eeq
with $T_{\nu}$ being the neutrino temperature. Furthermore, relative
velocity of the two neutrino is $(1-\cos\theta )$ and hence one can
obtain the incident flux as
\beqstar
({\rm flux}) = ( 1-\cos\theta ) f_\nu ( \enu ) d\enu.
\eeqstar
Then one can obtain the contribution to the time derivative of the
neutrino distribution function as
\beq
\left. \frac{\partial f_\nu ( \enu' )}{\partial \enu'} \rabs_{+} &=&
\int ({\rm target}) \times ({\rm flux}) \times \frac{d\sigma}{d\enu'}
\nonumber \\
&=&
\frac{1}{128\pi}G_F^2 \int_{\enu'}^{\infty} d\enu
\int d\enubg \frac{1}{\enu\enubg}
f_\nu ( \enu ) \bar{f}_\nu ( \enubg )
\int_0^{4\enu\enubg} ds~s^2 \lsp E_{\rm in}^2 - s \rsp^{-1/2}
\nonumber \\ &&
\times \Big\{ \lsp 8a + 3b + 3c \rsp - 4 (b-c) \cos\xi \cos\eta
\nonumber \\ &&
+ (b+c) \lsp 3 \cos^2\xi\cos^2\eta - \cos^2\xi - \cos^2\eta \rsp \Big\},
\label{df/dt_step1}
\eeq
with
\beq
E_{\rm in} = \enu + \enubg.
\eeq
Notice that the angles $\xi$ and $\eta$ can be represented by the
variables in the comoving frame through the following relations;
\beq
\cos\xi &=&
\lsp \enu - \enubg \rsp \lsp E_{\rm in}^2 -s \rsp^{-1/2},
\\
\cos\eta &=&
\lsp 2\enu' - E_{\rm in} \rsp \lsp E_{\rm in}^2 -s \rsp^{-1/2}.
\eeq
Define
\beq
H_p \lsp \enu, \enubg \rsp \equiv
\int_0^{4\enu\enubg} ds~s^2 \lsp E_{\rm in}^2 - s \rsp^{-p/2},
\eeq
then eq.(\ref{df/dt_step1}) can be rewritten as
\beq
\left. \frac{\partial f_\nu ( \enu' )}{\partial \enu'} \rabs_{+} &=&
\frac{1}{128\pi}G_F^2 \int d\enu d\enubg \frac{1}{\enu\enubg}
f_\nu ( \enu ) \bar{f}_\nu ( \enubg )
\nonumber \\ &&
\times \Big\{ (8a+3b+3c) H_1 \lsp \enu, \enubg \rsp
\nonumber \\ &&
- (b+c) \lsp \enu - \enubg \rsp^2 H_3 \lsp \enu, \enubg \rsp
\nonumber \\ &&
- (b+c) \lsp 2\enu' - E_{\rm in} \rsp^2 H_3 \lsp \enu, \enubg \rsp
\nonumber \\ &&
-4(b-c) \lsp \enu - \enubg \rsp \lsp 2\enu' - E_{\rm in} \rsp
H_3 \lsp \enu, \enubg \rsp
\nonumber \\ &&
+3(b+c) \lsp \enu - \enubg \rsp^2 \lsp 2\enu' - E_{\rm in} \rsp^2
H_5 \lsp \enu, \enubg \rsp \Big\}.
\eeq
After some complicated calculations, the above equation becomes very
simple form;
\beq
\left. \frac{\partial f_\nu ( \enu' )}{\partial \enu'} \rabs_{+} &=&
\frac{4}{3\pi}G_F^2 \int_{\enu'}^{\infty} d\enu \int d\enubg
\frac{\enubg}{\enu^2} f_\nu ( \enu ) \bar{f}_\nu ( \enubg )
\nonumber \\ &&
\times \Bigg\{ a \enu^2 + b \lsp \enu - \enu' \rsp^2 + c \enu'^2
\nonumber \\ &&
+ \frac{1}{2} a \enu\enubg + \frac{3}{2} b \lsp \enu - \enu' \rsp \enubg
- \frac{1}{2} c \enu'\enubg
\nonumber \\ &&
+ \frac{1}{10} (a+6b+c) \enubg^2 \Bigg\}.
\eeq
In the following analysis, we use the approximation
$\enu,\enu'\gg\enubg$. With this approximation, the above equation
becomes
\beq
\left. \frac{\partial f_\nu ( \enu' )}{\partial \enu'} \rabs_{+} &=&
\frac{4}{3\pi}G_F^2 \int_{\enu'}^{\infty} d\enu \frac{1}{\enu^2}
\lmp a \enu^2 + b \lsp \enu - \enu' \rsp + c \enu'^2 \rmp
f_\nu ( \enu )
\nonumber \\ &&
\times \int_0^\infty d\enubg
\enubg  \bar{f}_\nu ( \enubg ).
\label{fdot_nu+}
\eeq

Next we will calculate how many neutrinos scatter off the background
neutrino in unit time. As is the similar method used above, contribution
to the time derivative of the neutrino distribution function from this
effect can be obtained as
\beq
\left. \frac{\partial f_\nu ( \enu )}{\partial \enu} \rabs_{-} &=&
- \frac{1}{8} \frac{1}{\enu^2} f_\nu ( \enu^2 )
\int_0^\infty d\enubg \frac{1}{\enubg^2}
\bar{f}_\nu ( \enubg ) \int_0^{4\enu\enubg} ds ~s\sigma (s)
\nonumber \\ &=&
- \frac{4}{3\pi}G_F^2
\lsp a + \frac{1}{3}b + \frac{1}{3}c \rsp \enu f_\nu ( \enu )
\int_0^\infty d\enubg \enubg  \bar{f}_\nu ( \enubg ),
\label{fdot_nu-}
\eeq
where $\sigma (s)$ is the total cross section obtained from the
amplitude (\ref{amplitude}). Notice that the condition for the neutrino
number conservation is realized;
\beq
\int_0^\infty d\enu
\left. \frac{\partial f_\nu ( \enu )}{\partial \enu} \rabs_{+}
=
- \int_0^\infty d\enu
\left. \frac{\partial f_\nu ( \enu )}{\partial \enu} \rabs_{-},
\eeq
unless the effects of inerastic channels ($\nu + \bar{\nu} \rightarrow
e^++e^-, \mu^+ + \mu^-$) are taken into account.

Effects of the charged lepton pair creation process can be taken into
account in the same way, and the contribution to the time derivative of
the neutrino distribution function is given by
\beq
\left. \frac{\partial f_\nu ( \enu )}{\partial \enu'}
\rabs_{\nu + \bar{\nu} \rightarrow l^{-}+l^{+}} &=&
- \frac{1}{8} \frac{1}{\enu} f_\nu ( \enu )
\int_0^\infty d\enubg \frac{1}{\enubg^2}
\bar{f}_\nu ( \enubg ) \int_{4m^2}^{4\enu\enubg} ds ~s\sigma (s)
\nonumber \\ &=&
-\frac{G_F^2}{16\pi} \frac{1}{\enu^2} f_{\nu_i}(\enu)
\int_0^\infty d \enubg \bar{f}_\nu (\enubg)
\nonumber  \\ &&
\times \lmp \lsp a + \frac{1}{3} b + \frac{1}{3} c \rsp I_2 +
\lsp 2d - \frac{1}{3} b -  \frac{1}{3} c \rsp m^2  I_1 \rmp,
\label{nu=>charged}
\eeq
with
\beq
I_2 &=& \frac{4}{3} \lmp 4 - \frac{4m^2}{\enu \enubg} \rmp^{1/2}
\enu \enubg \lsp 8 \enu^2 \enubg^2 - 2 m^2 \enu \enubg - 3m^4 \rsp
\nonumber \\ &&
- 4m^6 \ln \llp
\frac{2 \lmp 4 - \lsp 4m^2/\enu \enubg \rsp \rmp^{1/2} \enu \enubg
+ 4 \enu \enubg - 2m^2}{2m^2} \rlp,
\\
I_1 &=& 2 \lmp 4 - \frac{4m^2}{\enu \enubg} \rmp^{1/2}
\enu \enubg \lsp 2 \enu \enubg - m^2 \rsp
\nonumber \\ &&
- 2m^4 \ln \llp
\frac{2 \lmp 4 - \lsp 4m^2/\enu \enubg \rsp \rmp^{1/2} \enu \enubg
+ 4 \enu \enubg - 2m^2}{2m^2} \rlp.
\eeq
Coefficients for the charged lepton production processes are given in
Table~\ref{table:a-d}.

By using the above formulae, one can obtain the Boltzmann
equation describing the evolution of the spectra for the high energy
neutrinos;
\begin{eqnarray}
    \frac{\partial f_{\nu_i}(\enu^\prime)}{\partial t} &=&
    \frac{4G_F^2}{3\pi}
    \int_{\enu^\prime}^{\infty} \frac{d \enu}{\enu^2}
    f_{\nu_i}(\enu)
    \sum_j \lmp a_{in,ij}\enu^2
    + b_{in,ij}(\enu - \enu^\prime)^2
    + c_{in,ij} \enu^{\prime 2} \rmp
    \nonumber \\
    & & \times \int_0^{\infty} d \enubg \enubg
    \bar{f}_{\nu}(\enubg)
    \nonumber \\
    & & -\frac{4G_F^2}{3\pi} \enu^\prime f_{\nu_i}(\enu^\prime)
    \lsp a_{out} + \frac{1}{3}b_{out} + \frac{1}{3}c_{out} \rsp
    \int_0^{\infty}d\bar{\epsilon}_{\nu} \bar{\epsilon}_{\nu}
    \bar{f}_{\nu}(\bar{\epsilon}_{\nu})
    \nonumber\\
    & & + \left.
    \frac{\partial f_{\nu_i}(\enu^\prime)}
    {\partial t}\right|_{\nu_i + \bar{\nu}_i
    \rightarrow e^{-}+e^{+}}
    + \left.\frac{\partial f_{\nu_i}(\enu^\prime)}
    {\partial t}\right|_{\nu_i + \bar{\nu}_i \rightarrow
    \mu^{-}+\mu^{+}}
    \nonumber\\
    & & +
    \frac{1}{6\tau_{3/2}}n_{3/2}
    \delta(\enu^\prime -m_{3/2}/2)
    \nonumber \\
    & & + \enu^\prime H \frac{\partial f_{\nu_i}(\enu^\prime)}
    {\partial \enu^\prime}
    - 2 H f_{\nu_i}(\enu^\prime)\enu',
\end{eqnarray}
where $H$ is the expansion rate of the universe, and the coefficients
$a$ -- $c$ are given by
\begin{eqnarray}
&&
a_{out} = 4,~~~b_{out} = 0,~~~c_{out} = 13,
\\ &&
a_{in,ii}= 6,~~~b_{in,ii}= 9,~~~c_{in,ii}= 11,
\\ &&
a_{in,ij}= 1,~~~b_{in,ij}= 1,~~~c_{in,ij}= 2,~~~(i\neq j).
\end{eqnarray}
The formula for charged lepton pair creation are given in
eq.(\ref{nu=>charged}).

%
%
\begin{figure}[p]
\epsfxsize=14cm

\centerline{\epsfbox{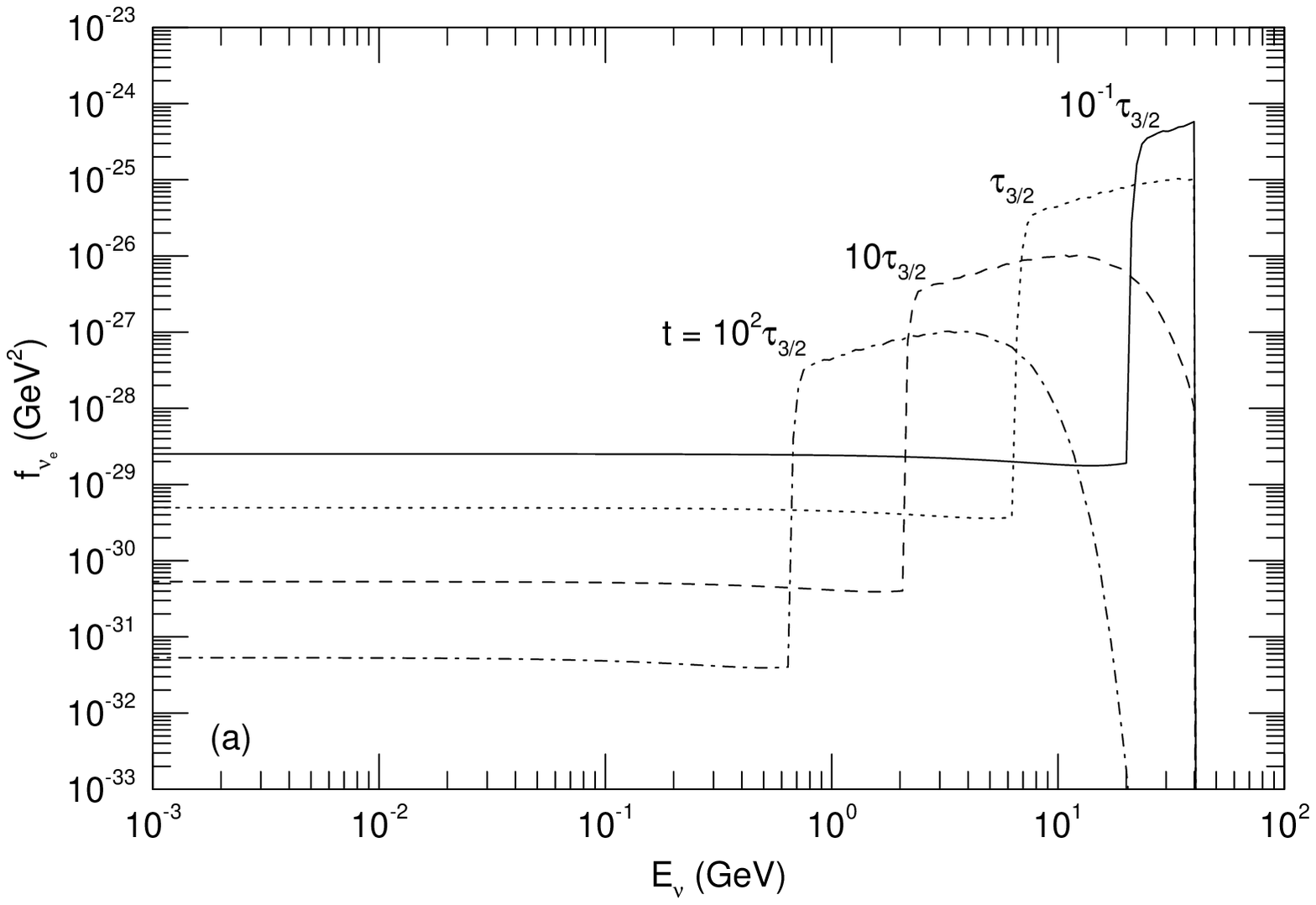}}

\vspace{1.5cm}

\centerline{\epsfbox{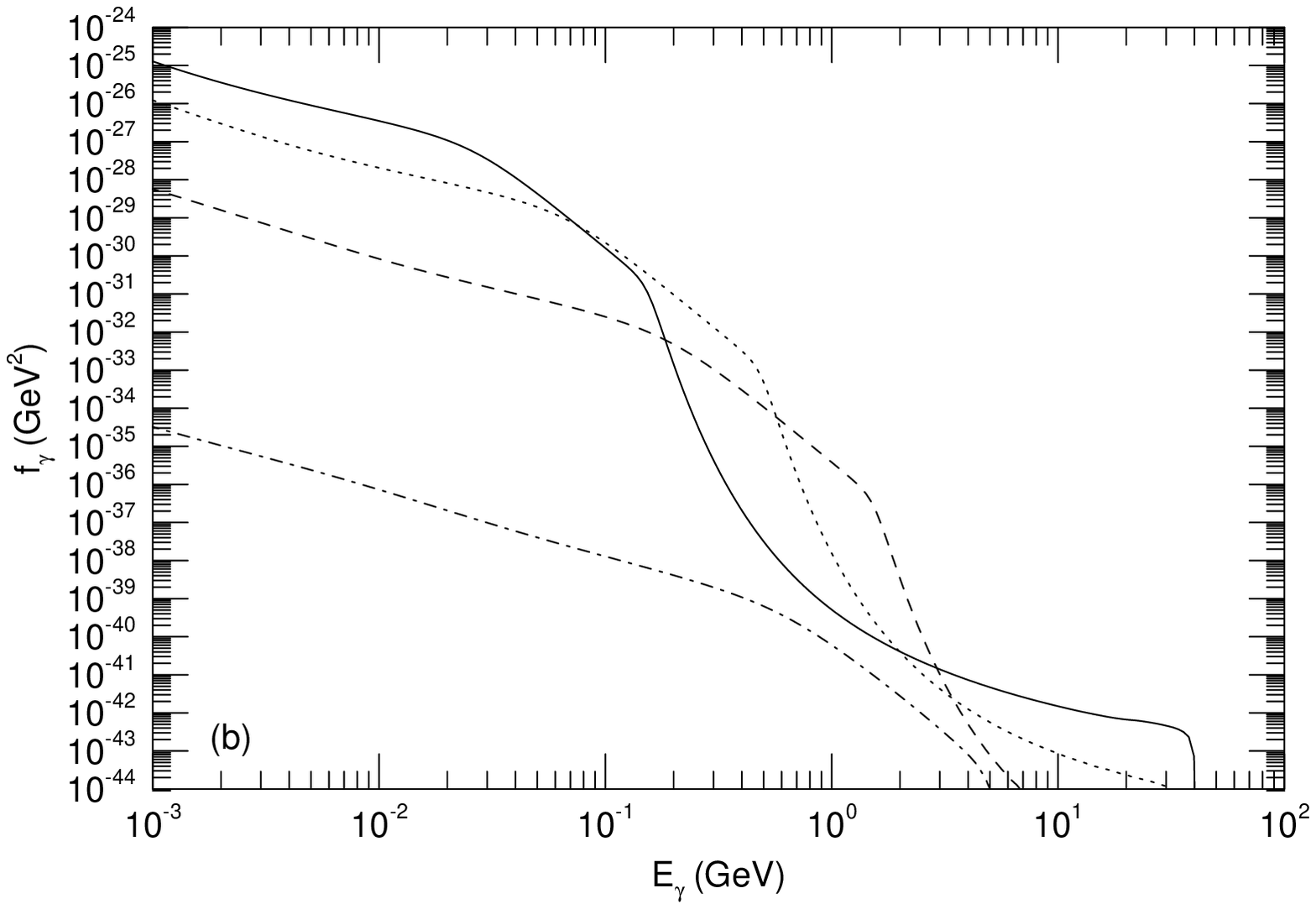}}

\caption{Time evolution of the distribution function of (a) the electron
neutrino, and (b) the photon for the case $\mgra = 100\GEV$. The solid curve
(dotted curve, dashed curve, and dotted-dashed curve) represents the
spectrum at the time $t = 10^{-1}\tau_{3/2}$ ($\tau_{3/2}$,
$10\tau_{3/2}$, and $t = 10^{2}\tau_{3/2}$). Yield of the gravitino is
normalized as $Y_{3/2} = e^{-t/\tau_{3/2}}$.}
\label{fig:fnu2}
\end{figure}

\begin{figure}[p]
\epsfxsize=14cm
\centerline{\epsfbox{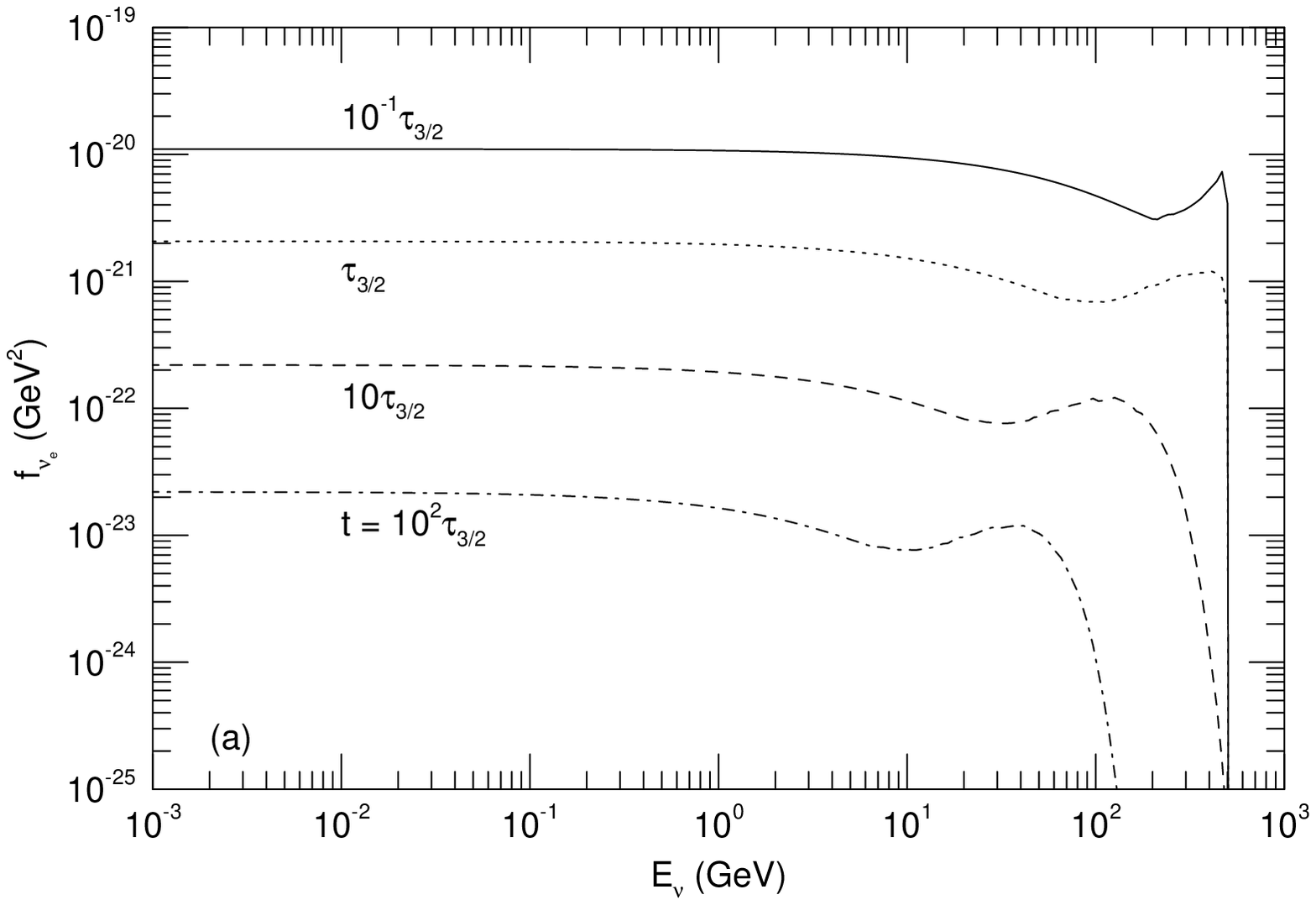}}

\vspace{1.5cm}
\centerline{\epsfbox{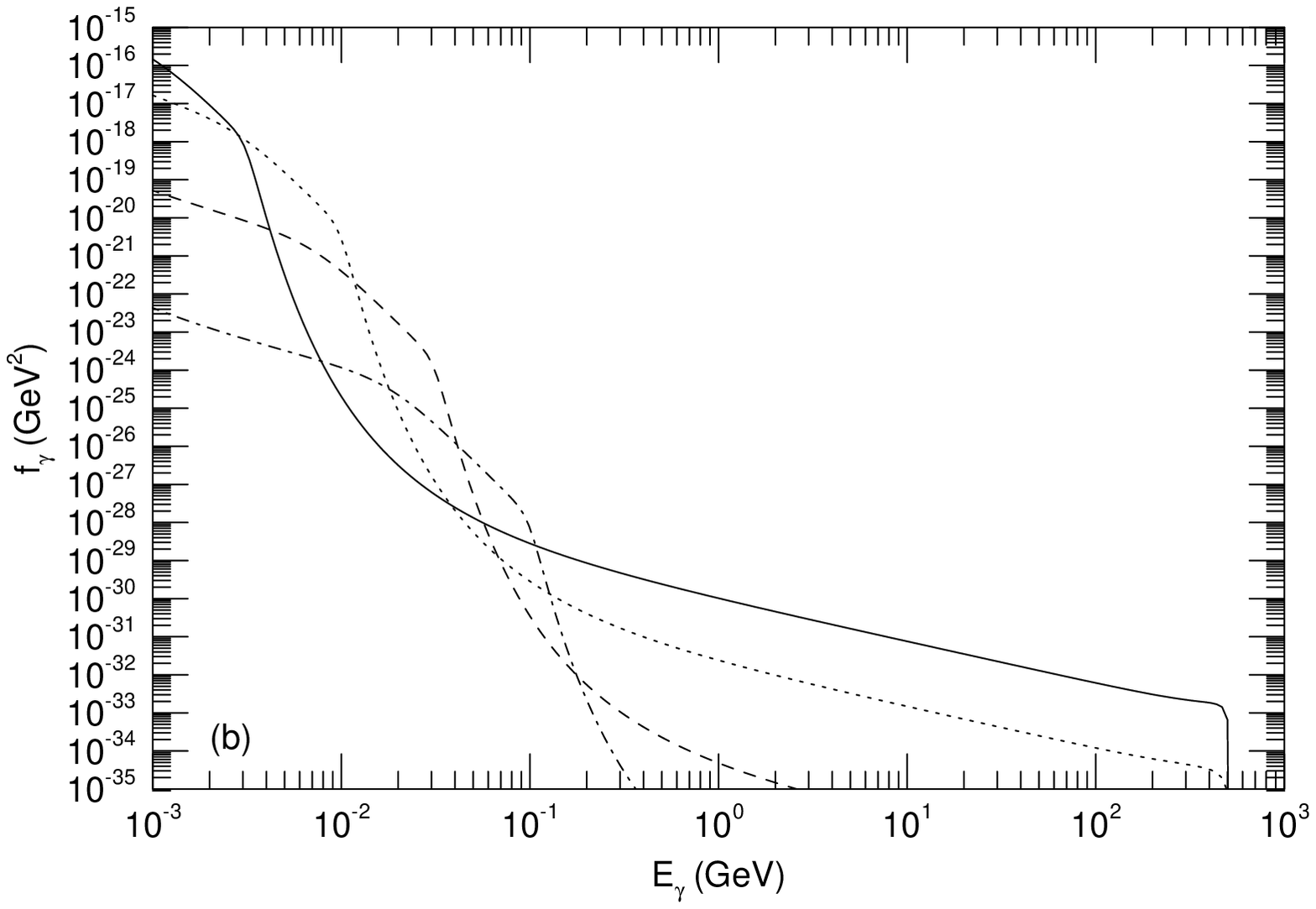}}
\caption{Same as Fig.~\protect\ref{fig:fnu2} except for $\mgra = 1\TEV$.}
\label{fig:fnu3}
\end{figure}

\begin{figure}[p]
\epsfxsize=14cm

\centerline{\epsfbox{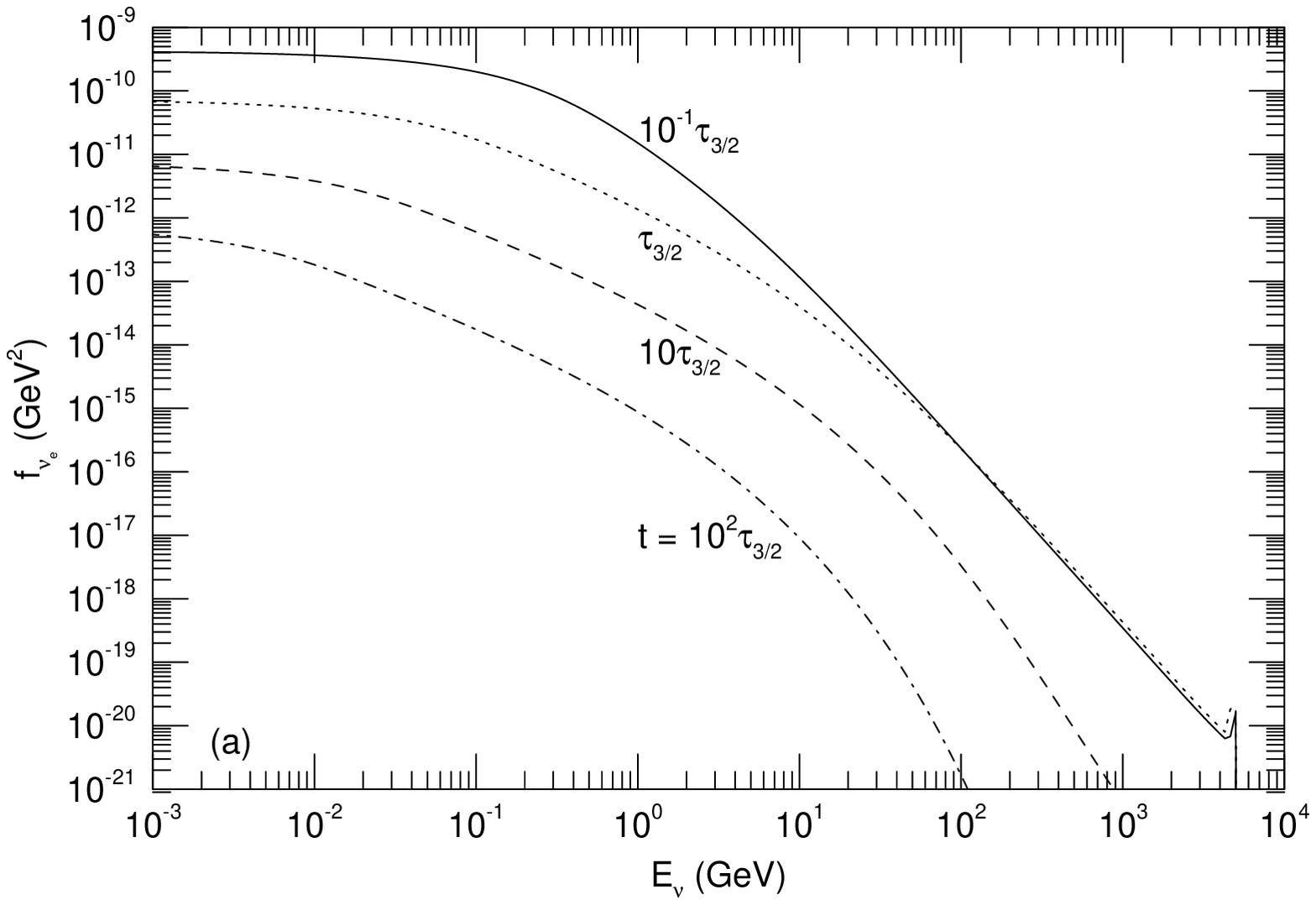}}

\vspace{1.5cm}

\centerline{\epsfbox{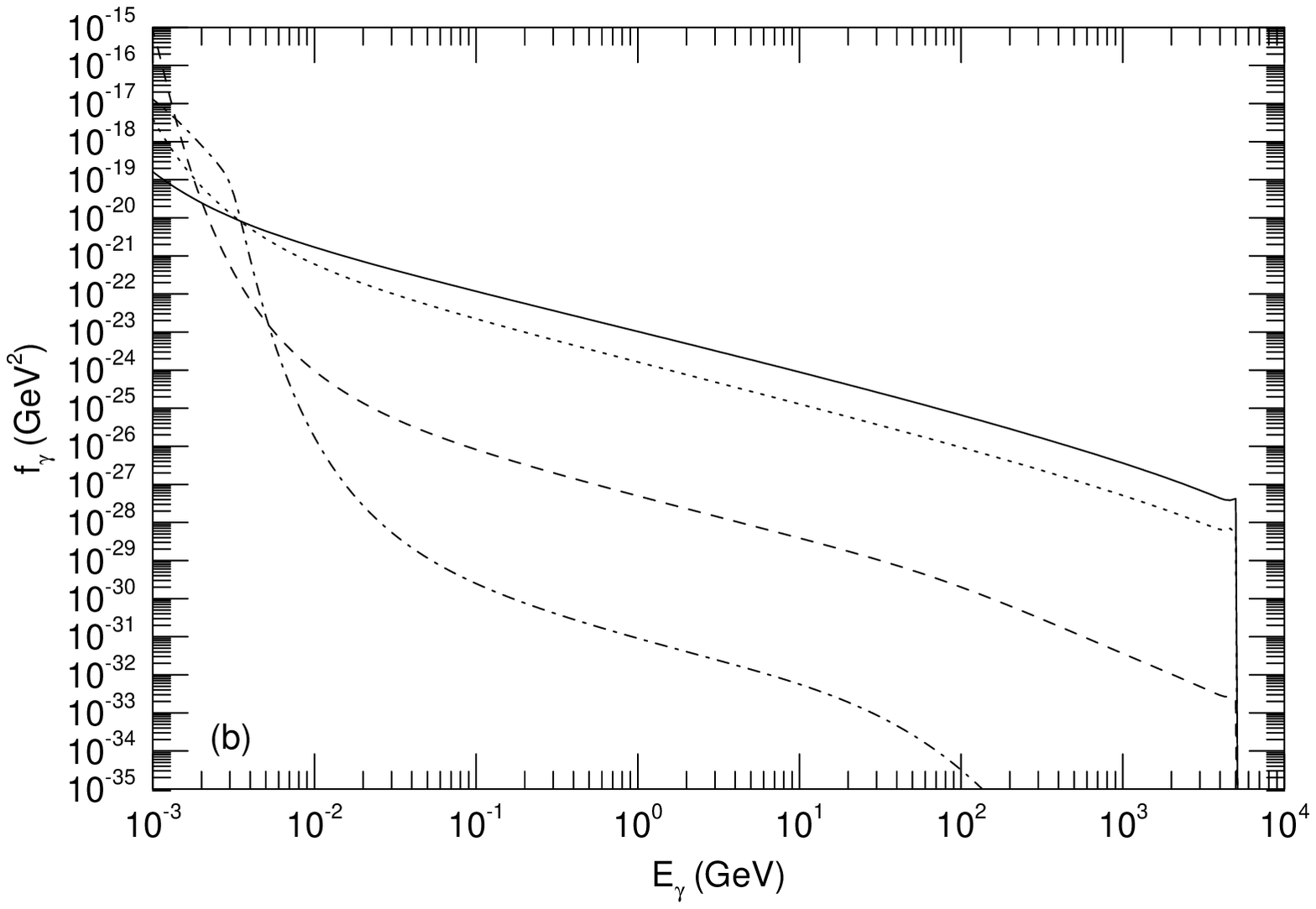}}

\caption{Same as Fig.~\protect\ref{fig:fnu2} except for
$\mgra = 10\TEV$.}
\label{fig:fnu4}
\end{figure}

\begin{figure}[t]
\epsfxsize=14cm
\centerline{\epsfbox{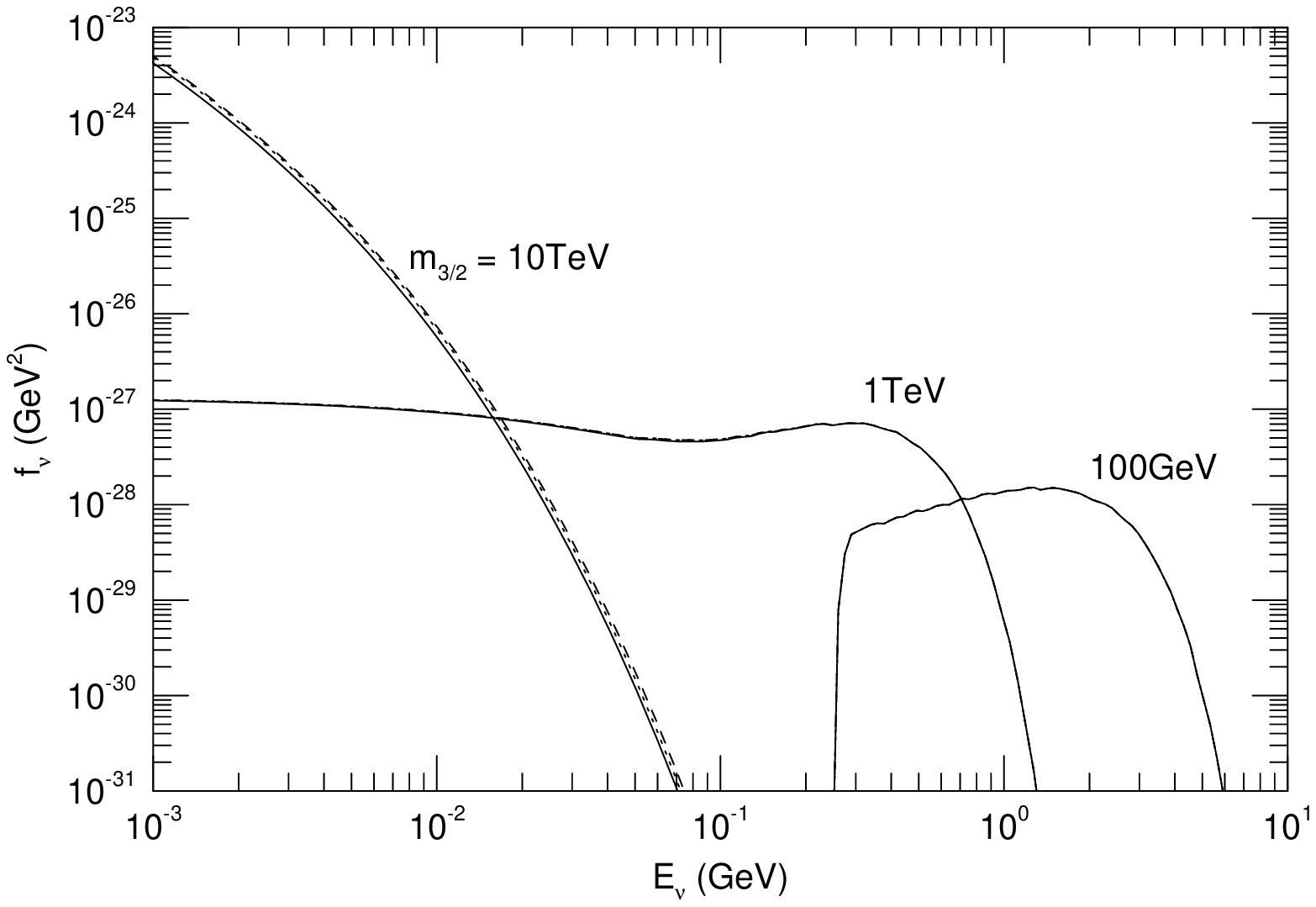}}

\caption{Spectra of high energy electron-neutrino (solid curve),
muon-neutrino (dotted curve) and tau-neutrino (dashed curve) at $T=1$eV
for $m_{3/2} = 100\GEV$, $1\TEV$ and $10\TEV$. The gravitino abundance
is normalized as $Y_{3/2} = e^{-t/\tau_{3/2}}$.}
\label{fig:f1ev}
\end{figure}

%
%

We solve the Boltzmann equations for the distribution function of
$\nu_e$, $\nu_\mu$ and $\nu_\tau$ numerically. In Figs~\ref{fig:fnu2} --
\ref{fig:fnu4}, we shown the time evolution of the distribution function of
the electron neutrino for the cases of $\mgra =$ 100GeV, 1TeV, and
10TeV. We have also checked that the difference among three types of
neutrinos are very small since the differences only comes from the
charged lepton pair creation processes which are sub-dominant compared
with the neutrino-neutrino scatterings. In Fig.~\ref{fig:f1ev}, the
spectra of three types of neutrinos at $T=1\EV$ are shown for $\mgra =$
100GeV, 1TeV, and 10TeV. (The time evolutions of the photon spectra are
also shown in Fig.~\ref{fig:fnu2} -- Fig.~\ref{fig:fnu4}.)

By regarding these high energy neutrinos as sources of high energy
charged leptons (\ie $e^+e^-$ and $\mu^+\mu^-$), we calculate the
high energy photon spectrum induced by the gravitino decay into a
neutrino and a sneutrino. The Boltzmann equations for the photon and
the electron distribution function $f_{\gamma}$ and $f_{e}$ are given by
\begin{eqnarray}
\frac{\partial f_{\gamma}(\epho)}{\partial t}
&=&
\left. \frac{\partial f_{\gamma}(\epho)}{\partial t} \right |_{\rm DP}
+ \left. \frac{\partial f_{\gamma}(\epho)}{\partial t} \right |_{\rm PP}
+ \left. \frac{\partial f_{\gamma}(\epho)}{\partial t} \right |_{\rm PC}
+ \left. \frac{\partial f_{\gamma}(\epho)}{\partial t} \right |_{\rm CS}
\nonumber \\ &&
+ \left. \frac{\partial f_{\gamma}(\epho)}{\partial t} \right |_{\rm IC}
+ \left. \frac{\partial f_{\gamma}(\epho)}{\partial t} \right |_{\rm EXP},
\label{bol-fp_ap} \\
\frac{\partial f_{e}(\eele)}{\partial t}
&=&
\left. \frac{\partial f_{e}(\eele)}{\partial t} \right |_{\rm DP}
+ \left. \frac{\partial f_{e}(\eele)}{\partial t} \right |_{\rm PC}
+ \left. \frac{\partial f_{e}(\eele)}{\partial t} \right |_{\rm CS}
+ \left. \frac{\partial f_{e}(\eele)}{\partial t} \right |_{\rm IC}
\nonumber\\ &&
+ \left. \frac{\partial f_{e}(\eele)}{\partial t} \right |_{\rm NEU}
+ \left. \frac{\partial f_{e}(\eele)}{\partial t} \right |_{\rm EXP},
\label{bol-fe_ap}
\end{eqnarray}
where NEU represents the contribution from the $\nu$ - $\nu$ scatterings.
In our numerical calculations, we neglect the effects of the expansion
of the universe as in the previous case. In solving
eq.(\ref{bol-fp_ap}) and eq.(\ref{bol-fe_ap}), we assume
\beq
\left. \frac{f_{e}(E)}{\partial t} \rabs_{\rm NEU}=
- \sum_i \lsp
\left. \frac{f_{\nu_i}(E)}{\partial t}
\rabs_{\nu_i + \bar{\nu}_i \rightarrow e^{-}+e^{+}} +
\left. \frac{f_{\nu_i}(E)}{\partial t}
\rabs_{\nu_i + \bar{\nu}_i \rightarrow \mu^{-}+\mu^{+}} \rsp,
\eeq
which respects the energy conservation although the actual energy
distribution may be different. This assumption is adequate for our
purpose since the photon (electron) spectrum is determined almost only
by the total amount of the energy injection~\cite{TU457}.  Furthermore we
treat muons (and anti-muons) as electrons with the same energy and
neglect the contribution from tau leptons whose creation rate is much
smaller than the other charged leptons.  Full details for other terms
are shown in the previous section.

%
%

\chapter{Big-bang nucleosynthesis}
\label{ap:bbn}

\hspace*{\parindent}
Along with the existence of the cosmic microwave background, big-bang
nucleosynthesis (BBN) is one of the most important predictions of the
big-bang cosmology. As we will see later, observations of the abundance
of light nuclei with atomic number less than 7 are in good agreements
with theoretical predictions. In this appendix, we will briefly review
the theoretical framework of BBN, and compare its results to the
observations.

\section{Theoretical framework}

\hspace*{\parindent}
BBN occurs when the temperature of the universe drops below $\sim$ 1MeV.
In this section, we will briefly review the BBN scenario by following
the thermal history of the universe of temperature about 1MeV. (More
details, see ref.\cite{Kolb&Turner}.)

\subsection*{{\it T} $\gg$ 1MeV}

\hspace*{\parindent}
When the temperature of the universe is much larger than 1MeV, nuclear
statistical equilibrium (NSE) among the light nuclei is established.
Especially, proton ($p$) and neutron ($n$) are converted to each other
in the following weak interaction processes;
\begin{eqnarray*}
n &\leftrightarrow& p + e^{-} + \overline{\nu}_{e},
\\
n + \nu_{e} &\leftrightarrow& p + e^{-},
\\
n + e^{+} &\leftrightarrow& p + \overline{\nu}_{e}.
\end{eqnarray*}
Since the rate of $p\leftrightarrow n$ conversion is sufficiently large,
chemical potential of $e$, $\nu_{e}$, $p$ and $n$ are related as
\begin{eqnarray}
\mu_{n} + \mu_{\nu} = \mu_{p} + \mu_{e},
\label{chemical-pot}
\end{eqnarray}
where $\mu_{X}$ represents the chemical potential of particle $X$.

Here, we comment on the magnitude of the chemical potentials of leptons.
Since the universe is charge neutral, charge density of electron is
equal to the number density of proton if $T$ is much smaller than the
muon mass. If the temperature is high enough, electron can be regarded
as a relativistic particle. In this case, charge density is
$O(\mu_{e}T^{2})$ and we obtain
\begin{eqnarray}
\left. \frac{\mu_{e}}{T} \rabs_{T\gsim 1\MEV} \sim
\frac{n_{e}}{n_{\gamma}} =
\frac{n_{p}}{n_{\gamma}} \sim
O(10^{-10}).
\end{eqnarray}
In order to estimate the neutrino chemical potential, we must know the
lepton number density of the universe. In the early universe when the
temperature is higher than the electroweak scale, baryon number and
lepton number are converted to each other through the sphaleron
transition. If this is true, lepton number of the universe is as the
same order as the baryon number.  From this fact, we adopt
$\mu_{\nu}/T\sim O(10^{-10})$.

In kinetic equilibrium, number density of nuclei with atomic number $A$
and charge $Z$ is given by
\begin{eqnarray}
n_{A} =
g_{A} \lsp \frac{m_{A}T}{2\pi} \rsp^{3/2}
\exp \lsp \frac{\mu_{A}-m_{A}}{T} \rsp,
\label{na}
\end{eqnarray}
where $g_{A}$ represents the internal degrees of freedom. Especially
number density of $p$ and $n$ can be written as
\begin{eqnarray}
n_{p} &=&
2 \lsp \frac{m_{p}T}{2\pi} \rsp^{3/2}
\exp \lsp \frac{\mu_{p}-m_{p}}{T} \rsp,
\label{np} \\
n_{n} &=&
2 \lsp \frac{m_{n}T}{2\pi} \rsp^{3/2}
\exp \lsp \frac{\mu_{n}-m_{n}}{T} \rsp.
\label{nn}
\end{eqnarray}
If the nucleon with the atomic number $A$ and charge $Z$ can be made out
of $Z$ protons and $(A-Z)$ neutrons rapidly enough, $\mu_{A}$ is
related to $\mu_{p}$ and $\mu_{n}$ in the following way;
\begin{eqnarray}
\mu_{A} = Z \mu_{p} + ( A - Z ) \mu_{n}.
\label{mua}
\end{eqnarray}
Using eq.(\ref{na}) -- eq.(\ref{mua}),
$n_{A}$ can be expressed as
\begin{eqnarray}
n_{A} =
g_{A} A^{3/2} 2^{-A} \lsp \frac{2\pi}{m_{N}T} \rsp^{3(A-1)/2}
n_{p}^{Z} n_{n}^{A-Z} \exp \lsp B_{A}/T \rsp,
\label{na2}
\end{eqnarray}
where $B_{A}$ is binding energy which is defined as
\begin{eqnarray}
B_{A} = Z m_{p} + ( A - Z ) m_{n} - m_{A},
\label{ba}
\end{eqnarray}
and for simplicity all the ``nucleon masses'' ($m_{p}$, $m_{n}$ and
$m_{A}/A$) in the prefactor are replaced by the common mass $m_{N}$
since their differences are not important. Numerical values of $B_{A}$
and $g_{A}$ for some nuclei are shown in Table~\ref{ba&ga}.
%
%
\begin{table}
\begin{center}
\begin{tabular}{c|cc} \hline \hline
{$^{A}Z$} & {$B_{A}(\MEV)$} & {$g_{A}$} \\ \hline
{$D$} & {2.22} &  {3} \\
{$^{3}{\rm H}$} & {6.92} &  {2} \\
{$^{3}{\rm He}$} & {7.72} &  {2} \\
{$^{4}{\rm He}$} & {28.3} &  {1} \\
{$^{12}{\rm C}$} & {92.2} &  {1} \\ \hline  \hline
\end{tabular}
\caption{The binding energy and internal degrees of freedom of light nuclei.}
\label{ba&ga}
\end{center}
\end{table}
%
%

For a later convenience, we define the mass fraction of species $A$;
\begin{eqnarray}
X_{A} = \frac{A n_{A}}{n_{B}},
\label{xa}
\end{eqnarray}
with $n_{B}$ being the baryon number density;
\begin{eqnarray}
n_{B} = n_{p} + n_{n} + \sum_{A,Z} A n_{A}.
\label{nb}
\end{eqnarray}
Then we can get
\begin{eqnarray}
X_{A} &=&
g_{A} \lmp \zeta (3) ^{A-1} \pi^{(1-A)/2} 2^{(3A-5)/2} \rmp
A^{5/2} \lsp T / m_{N} \rsp^{3(A-1)/2}
\nonumber \\ &&
\times \eta_B^{A-1} X_{p}^{Z} X_{n}^{A-Z} \exp \lsp B_{A} / T \rsp ,
\label{xa2}
\end{eqnarray}
where $\eta_B\equiv n_B/n_\gamma$ is the baryon-to-photon ratio. Notice
that $\eta_B$ is related to the present baryonic density parameter
$\Omega_B$ as
\begin{eqnarray}
\eta_B =
\frac{n_{B}}{n_{\gamma}}
\simeq 2.68 \times 10^{-8} \Omega_{B} h^{2}.
\label{eta-omegaB}
\end{eqnarray}
Since $X_{A}$ is proportional to $\eta_B^{A-1}$, mass fraction of
species with large atomic number is extremely small if they are in
chemical equilibrium. For example, with $T=10\MEV$,
\begin{eqnarray*}
X_{p} &\simeq& 0.5,
\\
X_{n} &\simeq& 0.5,
\\
X_{\rm D} &\simeq& 10^{-11},
\\
X_{^{3}{\rm He}} &\simeq& 10^{-23},
\\
X_{^{4}{\rm He}} &\simeq& 10^{-34}.
\end{eqnarray*}

\subsection*{0.3MeV $\lsim$ {\it T} $\lsim$ 1MeV}

\hspace*{\parindent}
Once the temperature of the universe drops below ${\sim}1\MEV$,
situation changes. When the temperature becomes $T_{F}\simeq 0.8\MEV$,
$p{\leftrightarrow}n$ conversion rate becomes smaller than the expansion
rate of the universe. At this freeze out temperature $T_{F}$, ratio of
$n_{p}$ and $n_{n}$ is given by
\begin{eqnarray}
\left. \frac{n_{n}}{n_{p}} \rabs_{T=T_{F}} =
\exp \lsp \frac{m_{p} - m_{n}}{T_{F}} \rsp \simeq
\frac{1}{6}.
\label{nn/np}
\end{eqnarray}
Below $T_{F}$, number density of neutron decreases mainly through the
free neutron decay process. But the decay rate of neutron is smaller
than the expansion rate and hence, mass fraction of neutron is almost
unchanged. On the other hand, nucleon production rate is still
sufficiently large, number density of light nuclei (like D, $^{3}{\rm
He}$, $^{4}{\rm He}$, $\cdot\cdot\cdot$) takes the NSE value.  At this
stage mass fractions of light nuclei are as follows;
\begin{eqnarray*}
X_{p} &\simeq& 1/7,
\\
X_{n} &\simeq& 6/7,
\\
X_{\rm D} &\simeq& 10^{-12},
\\
X_{^{3}{\rm He}} &\simeq& 10^{-23},
\\
X_{^{4}{\rm He}} &\simeq& 10^{-28}.
\end{eqnarray*}

\subsection*{{\it T} $\lsim$ 0.3MeV}

\hspace*{\parindent}
When the temperature becomes ${\sim}0.3\MEV$, NSE value of $X_{^{4}{\rm
He}}$ approaches to 1. This fact suggests that $^{4}{\rm He}$ can be
synthesized effectively at $T\lsim 0.3\MEV$. But for $T\gsim 0.1\MEV$,
D, which is a source of $^{4}{\rm He}$, is easily destroyed through the
scattering processes off the background photon.

Once the temperature drops below ${\sim}0.1\MEV$, D's are effectively
produced and they are rapidly translated into $^{4}$He through the
following processes;
\begin{eqnarray*}
&&
{\rm D} ({\rm D},n) ^{3}{\rm He} ({\rm D},p) ^{4}{\rm He},
\\ &&
{\rm D} ({\rm D},p) ^{3}{\rm H} ({\rm D},n) ^{4}{\rm He},
\\ &&
{\rm D} ({\rm D},\gamma) ^{4}{\rm He}.
\end{eqnarray*}
Through these processes, almost all of neutrons in the universe are
trapped in $^4$He and primordial abundance of D and $^3$He becomes much
less than that of $^4$He. Order of the primordial mass fraction of
$^4$He, $Y_{p}$, can be estimated to be
\begin{eqnarray}
Y_{p} =
\left. \frac{4n_{^4{\rm He}}}{n_{B}} \rabs_{p} \sim
\left. \frac{2n_{n}}{n_{B}} \rabs_{T\sim 0.1\MEV} \sim
0.25,
\label{yp}
\end{eqnarray}
where we have used $X_{n}|_{T\sim 0.1\MEV}\sim$0.125. (Notice that
$Y_{P}$ given in eq.(\ref{yp}) is overestimated and the result of
numerical calculation shows that $Y_{p}\sim$0.24.)

Finally, we will comment on the nucleosynthesis beyond $^4$He. Since
there is no tightly bounded isotope with $A$=5 and 8, nucleus which is
heavier than $^4$He is scarcely synthesized. One important exception is
$^7$Li which can be synthesized in the following two processes;
\begin{eqnarray*}
&&
^{4}{\rm He} (^3 {\rm H},\gamma) ^{7}{\rm Li},
\\ &&
^{4}{\rm He} (^3 {\rm He},\gamma) ^{7}{\rm Be}
(e^{-},\nu_{e}) ^{7}{\rm Li}.
\end{eqnarray*}

\section{Numerical results}

\hspace*{\parindent}
In order to calculate the primordial abundance of light elements, we
use numerical method. The following results are obtained by using
the FORTRAN code written by Kawano~\cite{Kawano}.

Before looking at the numerical results, let us consider about the basic
parameters of BBN. Essentially, primordial abundances of light elements
depend only on one cosmological parameter, $\eta_B$, however predicted
abundances of light nuclei are also affected by uncertainties come from
the uncertainties of the fundamental theory (especially, by those of
weak interaction).

Neutron lifetime $\tau_n$ is the significant parameter of BBN. In order
to calculate the primordial abundance of $^4$He, we must accurately
determine neutron freeze out temperature $T_{F}$, which depends on weak
interaction rates. These rates are calculated from only one matrix
element which is also related to neutron lifetime $\tau_{n}$, and hence
experimental data of neutron lifetime is very important for calculating
the primordial abundances. We will see that the predicted abundance of
$^4$He decreases as neutron lifetime $\tau_{n}$ increases. This can be
understood in the following way. A larger value of $\tau_{n}$ implies
smaller weak interaction rates, and this leads to a lower freeze out
temperature $T_F$. Therefore, neutron lifetime increases, ratio of $n_n$
to $n_p$ at $T_F$ decreases and hence a smaller value of $^4$He
abundance is expected.

%
%
\begin{figure}[p]
\epsfxsize=15cm
\centerline{\epsfbox{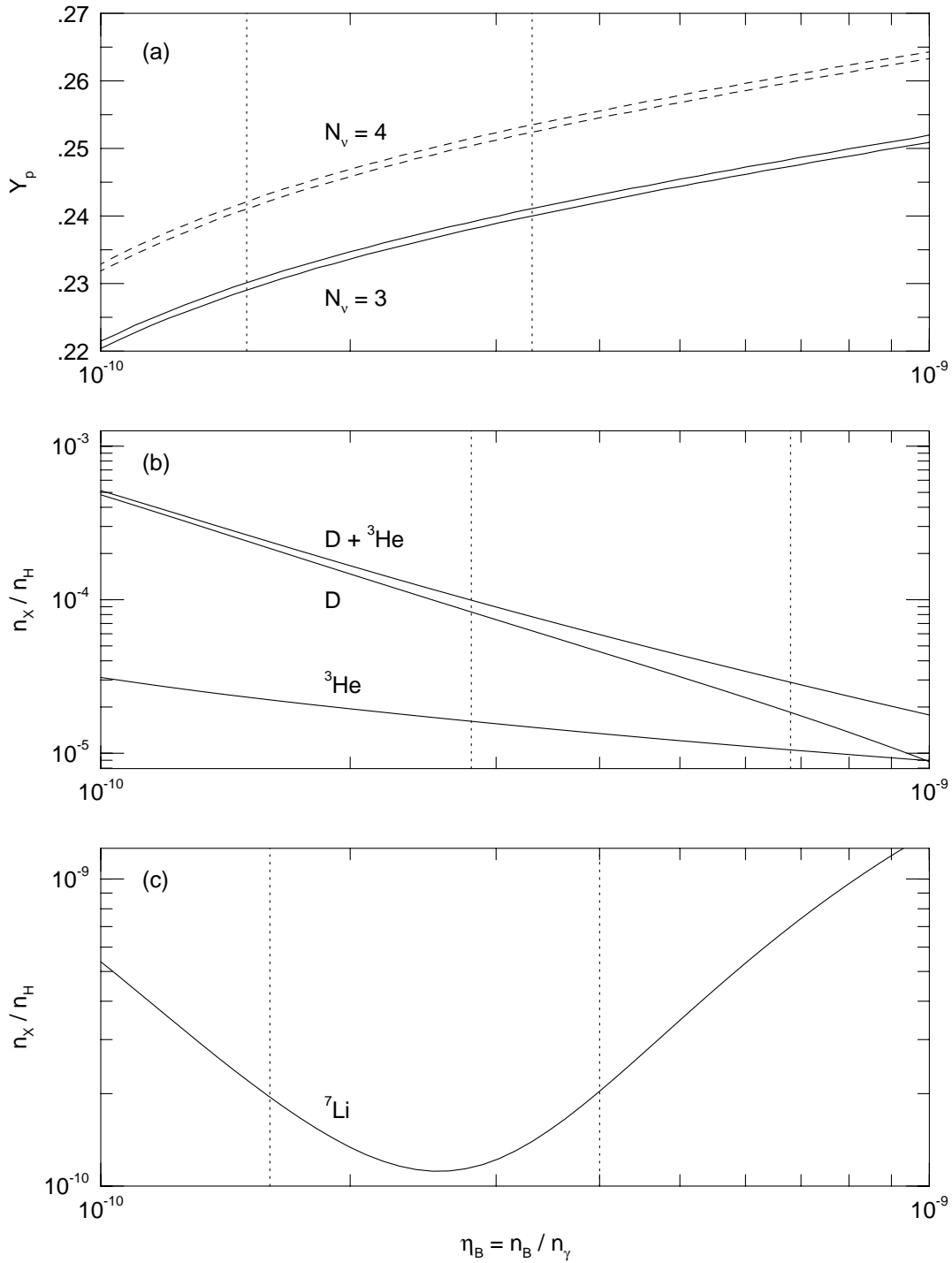}}
\caption{Predicted abundances of (a) $^4$He, (b) D and $^3$He, and
(c) $^7$Li. For a neutron lifetime $\tau_n$, we use $\tau_n =
891.2{\rm sec}$ (upper line) and $\tau_n = 887.0{\rm sec}$ (lower line).
Solid lines corresponds to the results with three neutrino species, and
for $^4$He, predicted abundance with four neutrino species are also
shown in dashed lines. Dotted lines represent bounds on $\eta_B$
obtained from observations.}
\label{fig:bbn}
\end{figure}
%
%
In Fig. \ref{fig:bbn}, we show the results of the numerical
calculations of BBN. For a neutron lifetime $\tau_n$, we use
$\tau_n = 891.2{\rm sec}$ and $\tau_n = 887.0{\rm sec}$. As mentioned
before, for a larger value of $\tau_n$, a smaller value of the
primordial $^4$He abundance is obtained. Primordial abundances also
depend on the baryon-to-photon ratio $\eta_B$. As one can see, predicted
abundance of $^4$He is a increasing function of $\eta_B$ while those of D
and $^3$He decreases as $\eta_B$ increases, and $^7$Li abundance takes its
minimum value at $\eta_B \sim 3\times 10^{-10}$.

In Fig.~\ref{fig:bbn}, we also plot the predicted abundance of
$^4$He in a model with four neutrino species. As one can see, if we add
extra neutrino species, predicted abundance of $^4$He increases. This
can be understood in the following way. If the extra neutrino exists,
the expansion rate of the universe at $T\sim 1\MEV$ gets higher due to
the energy density of extra neutrino, and hence the neutron freeze out
temperature $T_F$ gets higher. In general, if there exists some exotic
particle whose energy density at $T\sim 1\MEV$ is not negligible, it
speeds up the cosmic expansion rate and raise the neutron freeze out
temperature. In this case more $^4$He is synthesized since lager amount
of relic neutron is expected. this argument constrains the properties of
exotic particles. For example, existence if additional light neutrino
species conflicts with observational data of $^4$He abundance.

\section{Observations}

\hspace*{\parindent}
In order to see the validity of the theoretical predictions of big-bang
nucleosynthesis, we must know the actual values of the {\it primordial}
abundances of light elements. But it is not a simple task because the
abundances of light elements evolve with time. Especially some nuclei
are produced or destroyed in stars, which changes the abundances of
light elements. Therefore we must reconstruct primordial abundances from
the observational data. In this section, we will derive the constraints
on the primordial abundances of D, $^3$He, $^4$He and $^7$Li, and
compare them with the theoretical predictions. The arguments and
observational data used in this section mainly follows
ref.\cite{APJ376-51}.

\subsection*{D and $^3$He}

\hspace*{\parindent}
We start with discussing the primordial abundance of D. Since the
binding energy of D is very small ($\sim$2.23MeV), D is easily destroyed
by $(\gamma,p)$ reaction at $T\gsim 6.0\times 10^{5}\K$, and it is hard
to produce D after BBN. Therefore most of D's in the present (or
pre-solar) universe are expected to be produced by the big-bang
nucleosynthesis. In the following we regard the observed pre-solar D
abundance as a lowerbound on the primordial abundance of D.

Let us reconstruct pre-solar D abundance from the observational data.
Pre-solar value of (D+$^3$He) abundance can be found in gas-rich
meteorites;
\begin{eqnarray}
3.84 \times 10^{-4} \leq
\llp \frac{\rm D + {^{3}He}}{\rm ^{4}He} \rlp_{pre\odot} \leq
4.22 \times 10^{-4}.
\label{y23/y4}
\end{eqnarray}
On the other hand, carbonaceous chondrites, which is one of the most
primitive solar system materials, provide us a pre-solar $^3$He
abundance;
\begin{eqnarray}
1.48 \times 10^{-4} \leq
\llp \frac{\rm ^{3}He}{\rm ^{4}He} \rlp_{pre\odot} \leq
1.56 \times 10^{-4}.
\label{y3/y4}
\end{eqnarray}
In order to normalize the abundance of the light elements by hydrogen
abundance, we use the standard solar model prediction on $^4$He
abundance;
\begin{eqnarray}
0.09 \leq
\llp \frac{\rm ^{4}He}{\rm H} \rlp_{pre\odot} \leq
0.11,
\label{y4}
\end{eqnarray}
and we obtain
\begin{eqnarray}
&&
1.8 \times 10^{-5} \leq
\llp \frac{\rm D}{\rm H} \rlp_{pre\odot} \leq
3.3 \times 10^{-5},
\label{y2} \\ &&
1.3 \times 10^{-5} \leq
\llp \frac{\rm ^{3}He}{\rm H} \rlp_{pre\odot} \leq
1.8 \times 10^{-5},
\label{y3} \\ &&
3.3 \times 10^{-5} \leq
\llp \frac{\rm D+ {^{3}He}}{\rm H} \rlp_{pre\odot} \leq
4.9 \times 10^{-5}.
\label{y23}
\end{eqnarray}
As mentioned before, we expect the primordial abundance of D is larger
than the pre-solar value since the abundance of D decreases with the
galactic evolution. Therefore, pre-solar value of D abundance requires
that
\begin{eqnarray}
\llp \frac{\rm D}{\rm H} \rlp_{p} \geq
\llp \frac{\rm D}{\rm H} \rlp_{pre\odot} \geq
1.8 \times 10^{-5}.
\label{lb-y2}
\end{eqnarray}

Next we will consider the upperbound. Since some of D's are destroyed,
it is hard to estimate the primordial D abundance without uncertainty.
But D is burned to $^3$He, abundance of the sum (D+$^3$He) is less
uncertain. Therefore we sill see the upperbound on this quantity below.

Once D is trapped in stars, it either survives or burns to $^3$He.
Furthermore some $^3$He in stars are destroyed. Therefore using the
survival fraction of $^3$He in stars, $g_{\rm ^{3}He}$, primordial
abundance of (D+$^3$He) can be written as
\begin{eqnarray}
\llp \frac{\rm D+ {^{3}He}}{\rm H} \rlp_{p} &=&
\llp \frac{\rm D}{\rm H} \rlp_{pre\odot}
+ g_{\rm ^{3}He}^{-1} \llp \frac{\rm ^{3}He}{\rm H} \rlp_{pre\odot}
\nonumber \\
&=&
\llp \frac{\rm D+ {^{3}He}}{\rm H} \rlp_{pre\odot}
+ \lsp g_{\rm ^{3}He}^{-1} - 1 \rsp \llp \frac{\rm ^{3}He}{\rm H}
\rlp_{pre\odot}
\label{relation-y23}
\end{eqnarray}
Combining eq.(\ref{relation-y23}) with eq.(\ref{y3}) and eq.(\ref{y23}),
and taking $g_{\rm ^{3}He}\geq 0.25$, we can get
\begin{eqnarray}
\llp \frac{\rm D+ {^{3}He}}{\rm H} \rlp_{p} \leq 1 \times 10^{-4}.
\label{ub-y23}
\end{eqnarray}

\subsection*{$^4$He}

\hspace*{\parindent}
In our universe, $^4$He is the most abundant element next to hydrogen,
but some of them have non-primordial origin, \ie some $^4$He's originate
to stars.  In order to see the uncontaminated primordial abundance of
$^4$He, we shall use the observational data taken from the old
environments which are not affected by galactic evolution.
Here, the primordial $^4$He abundance has been estimated from the
observation in metal-poor extragalactic \HII region.
%
%
\begin{figure}[p]
\epsfxsize=14cm
\centerline{\epsfbox{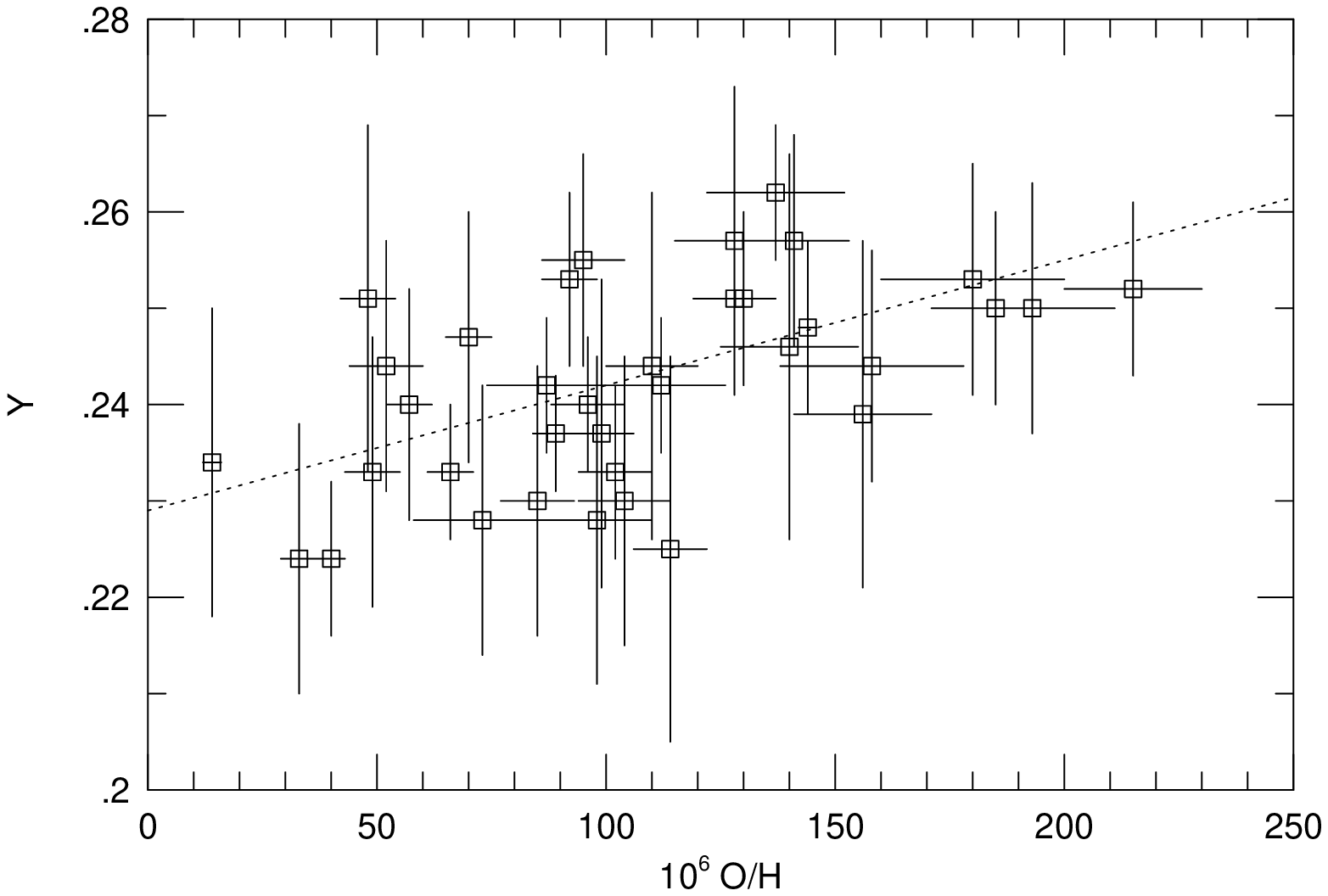}}
\caption{$^4$He mass fraction $Y$ vs. observed oxygen abundance.}
\label{fig:he-o}
\vspace{1.5cm}
\centerline{\epsfbox{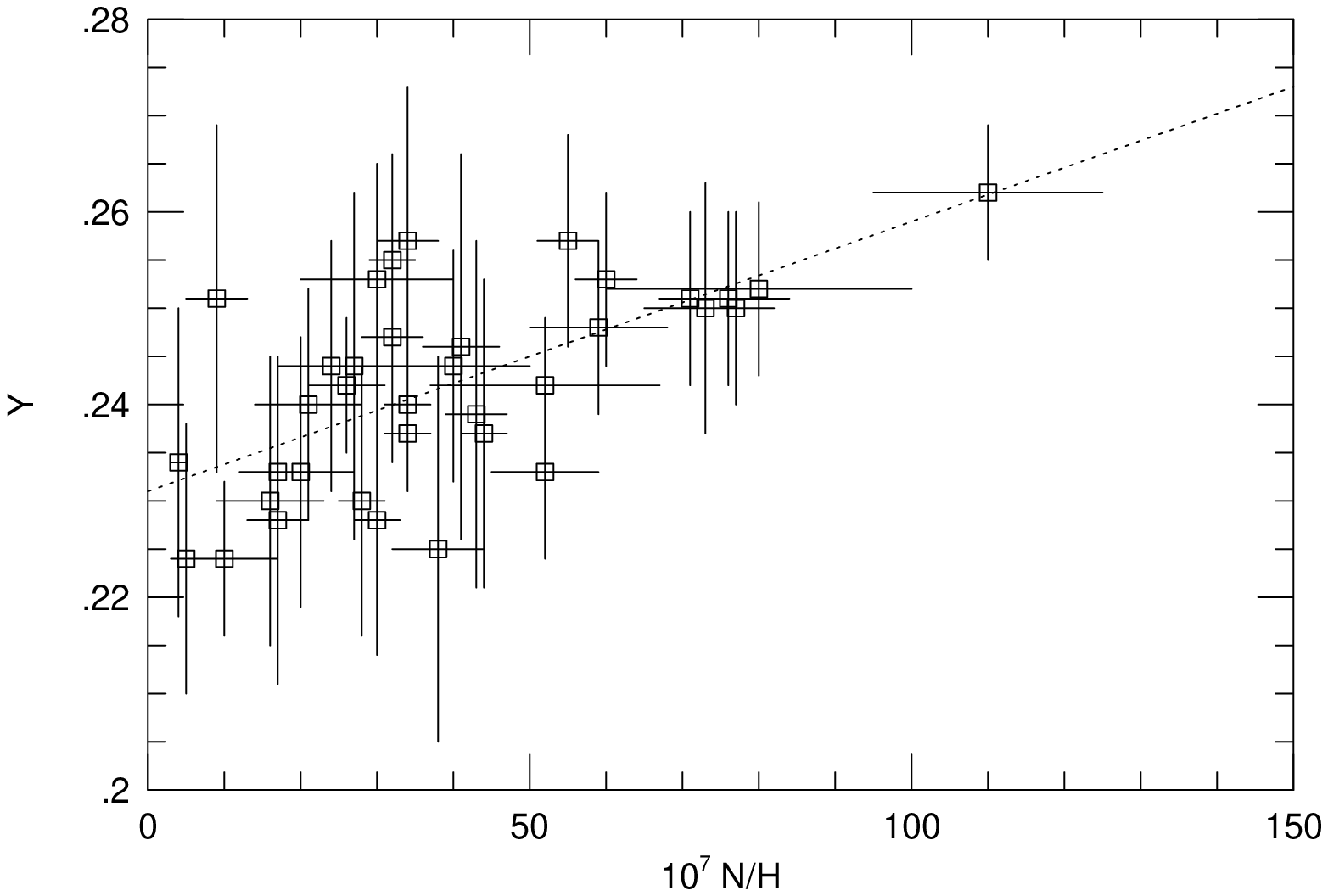}}
\caption{$^4$He mass fraction $Y$ vs. observed nitrogen abundance.}
\label{fig:he-n}
\end{figure}
\begin{figure}[t]
\epsfxsize=14cm
\centerline{\epsfbox{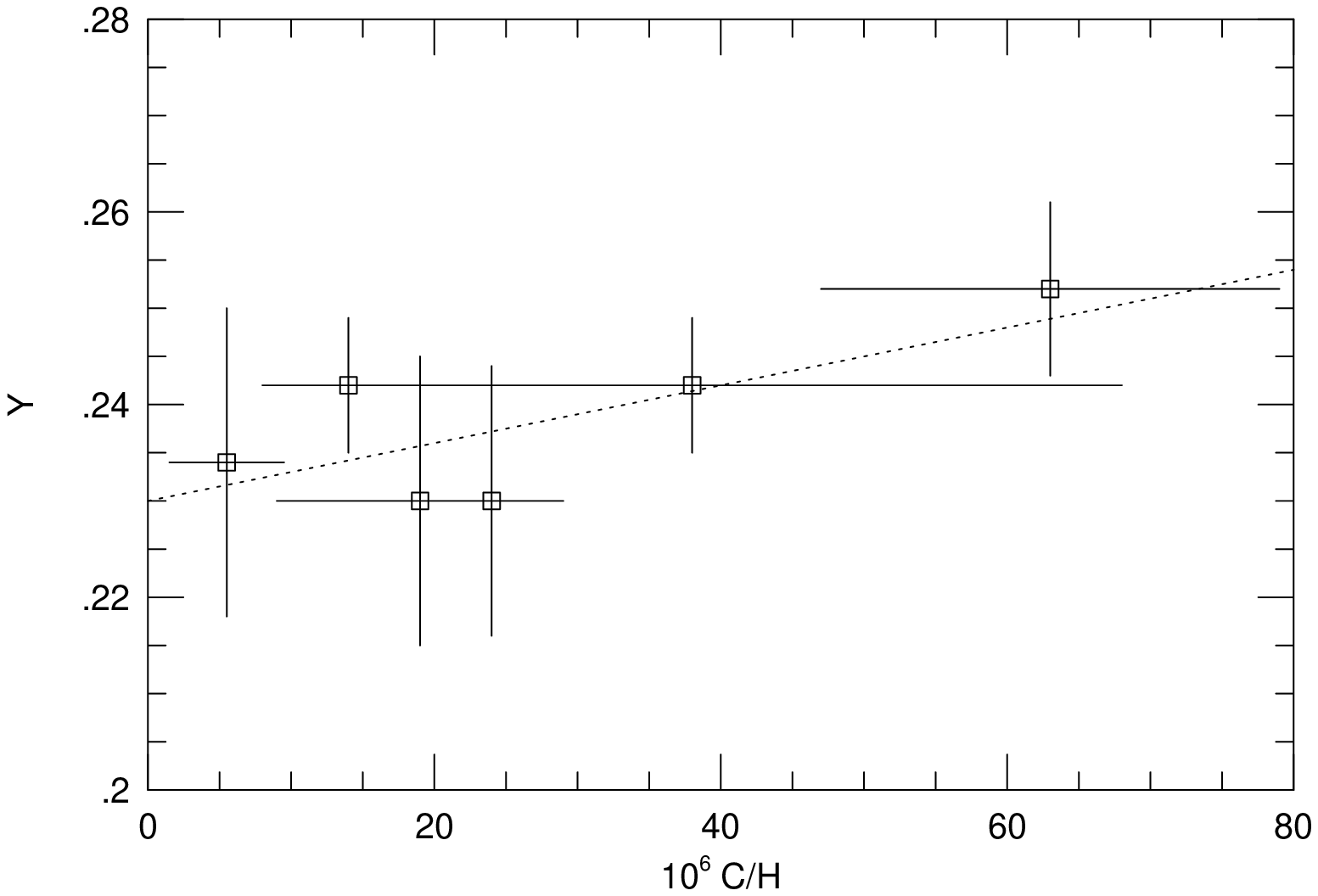}}
\caption{$^4$He mass fraction $Y$ vs. observed carbon abundance.}
\label{fig:he-c}
\end{figure}
%
%

In Fig. \ref{fig:he-o} -- Fig. \ref{fig:he-c}, observational data of
$^4$He abundance, $Y$, obtained from metal-poor extragalactic \HII
region are seen as a function of metalicity (O/H, N/H,
C/H)~\cite{APJ376-51}. Fitting these data linearly, one can get the
following fitting formulae;
\begin{eqnarray}
Y &=& 0.229 \pm 0.004 + (1.3 \pm 0.3) \times 10^{2} ({\rm O/H}),
\label{YO} \\
Y &=& 0.231 \pm 0.003 + (2.8 \pm 0.7) \times 10^{3} ({\rm N/H}),
\label{YN} \\
Y &=& 0.230 \pm 0.007 + (3.1 \pm 2.2) \times 10^{2} ({\rm C/H}).
\label{YC}
\end{eqnarray}

{}From these data, primordial abundance of $^4$He should be read off.
Since the primordial $^4$He in these metal-poor region is still
contaminated by the one originating to stars, we take a limit that
metalicity goes to zero. Then, the primordial abundance is
estimated to be
\begin{eqnarray}
{\rm O} ~&:&~ Y_{p} = 0.229 \pm 0.004,
\label{YOp} \\
{\rm N} ~&:&~ Y_{p} = 0.231 \pm 0.003,
\label{YNp} \\
{\rm C} ~&:&~ Y_{p} =  0.230 \pm 0.007.
\label{YCp}
\end{eqnarray}
Taking the observational uncertainty into account, we adopt the
primordial $^4$He abundance as $Y_{p}=0.23\pm 0.01$.

\subsection*{$^7$Li}

\hspace*{\parindent}
Next we will see the primordial abundance of $^7$Li. Since $^7$Li can be
produced after BBN by cosmic ray spallation or some stellar processes
(like novae outbursts), and it is easily destroyed at $T\geq 2\times
10^{6}$K, we can not regard the present abundance of $^7$Li as
primordial one. This means that we have to observe the old environment
so as to see the primordial $^7$Li abundance.  For this purpose,
observational data of $^7$Li abundance in metal-poor (and $T_{\rm
surf}\gsim 5500$K, as will be explained below) population~II stars are
usually used since they are expected to reflect the uncontaminated
$^7$Li abundance.

In Fig. \ref{fig:li7}, we plot the $^7$Li abundance $[^7{\rm
Li}]\equiv 12+\log_{10}(^7{\rm Li/H})$ in the most metal-poor
population~II stars as a function of surface temperature $T_{\rm surf}$.
As one can see, ``$^7$Li-plateau'' appears at $T_{\rm surf}\gsim$5500K.
Using the data with $T_{\rm surf}\geq 5500$K, the ``plateau'' value of
[$^7$Li] is estimated to be
\begin{eqnarray}
\llp ^{7}{\rm Li} \rlp_{\rm plateau} = 2.08 \pm 0.07.
\label{plateau}
\end{eqnarray}
We identify this value as the primordial abundance of $^7$Li. Notice
that the data taken from stars with of $T_{\rm surf}\lsim 5500$K are not
appropriate for our purpose, because cool stars have thick surface
convective zones which carry $^7$Li to deep region hot enough to burn
it.  Thus, we expect that $^7$Li abundances at the low surface
temperature stars are not primordial one.  Using eq.(\ref{plateau}),
upperbound on the primordial $^7$Li abundance is given by
\begin{eqnarray}
\llp ^{7}{\rm Li} \rlp_{p} \leq 2.15.
\label{ub-y7}
\end{eqnarray}
%
%
\begin{figure}[t]
\epsfxsize=14cm
\centerline{\epsfbox{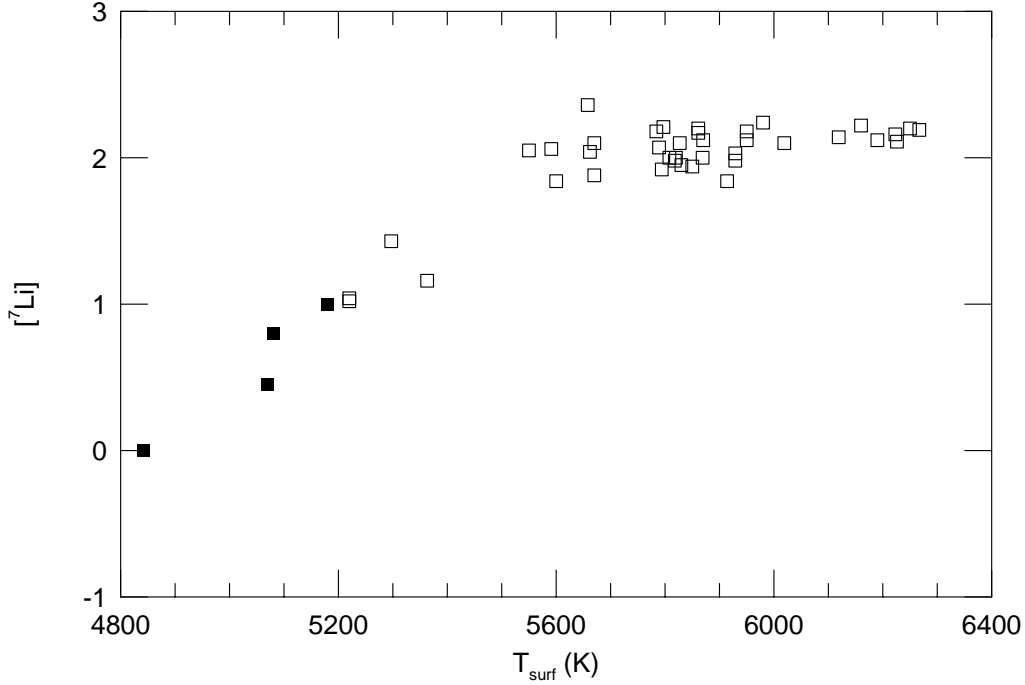}}
\caption{$^7$Li abundance  $[^7{\rm Li}]\equiv 12+\log_{10}(^7{\rm Li/H})$
in the metal-poor population~II stars as a function of surface
temperature $T_{\rm surf}$. The filled marks represent upper limits to
the $^7$Li abundance.}
\label{fig:li7}
\end{figure}
%
%

\subsection*{Comparison with the theoretical predictions}

\hspace*{\parindent}
Now we are at the position to compare the observational data of light
nuclei with the theoretical predictions.  At first, we will constrain
baryon-to-photon ratio $\eta_B$ from the data of D, (D+$^3$He) $^4$He, and
$^7$Li.

Let us begin with D and $^3$He. Lowerbound on the primordial abundance
of D is given in eq.(\ref{lb-y2}) and using that, $\eta_B$ is constrained
to be
\begin{eqnarray}
{\rm D} ~:~ \eta_B \leq 6.8 \times 10^{-10}.
\label{eta2}
\end{eqnarray}
On the other hand, upperbound on the primordial (D+$^3$He) abundance,
eq.(\ref{ub-y23}), requires
\begin{eqnarray}
{\rm D + {^{3}He}}~:~ \eta_B \geq 2.8 \times 10^{-10}.
\label{eta23}
\end{eqnarray}

Theoretical prediction on the primordial abundance of $^4$He is
considerably affected by an uncertainty of the neutron life time ($\tau_n
= 889.1\pm 2.1{\rm sec}$). In order to get a conservative constraint on
the baryon-to-photon ratio $\eta_B$, we use the value $\tau_n
= 891.2{\rm sec}$ in deriving the lowerbound on $\eta_B$, and $\tau_n
= 887.0{\rm sec}$ for upperbound. As a result, we obtain constraints on
$\eta_B$ from $Y_p=0.23\pm 0.01$ as
\beq
{\rm ^4He} ~:~ 1.5 \times 10^{-10} \leq \eta_B \leq 3.3 \times 10^{-10}.
\label{eta4}
\eeq

Upperbound on the primordial $^7$Li abundance is given in
eq.(\ref{ub-y7}); $[^7{\rm Li}]_p\leq 2.15$. If we naively use this value,
$\eta_B$ is constrained to be
$1.9\times{10^{-10}}\leq\eta_B\leq3.3\times{10^{-10}}$. But from the
uncertainties in the nuclear reaction rate, the predicted $^7$Li
abundance is expected to be uncertain by $\sim$40\%. Assuming 40\%
residual uncertainty in the primordial abundance, bound on $\eta_B$ is
found to be
\begin{eqnarray}
{\rm ^{7}Li} ~:~ 1.6 \times 10^{-10} \leq \eta_B \leq 4.0 \times 10^{-10}.
\label{eta7}
\end{eqnarray}

Combining eq.(\ref{eta2}) -- eq.(\ref{eta7}), allowed range of
baryon-to-photon ratio $\eta_B$ is given by
\begin{eqnarray}
{\rm D,~^{3}He,~^{4}He, ~^{7}Li} ~:~
2.8 \times 10^{-10} \leq \eta_B \leq 3.3 \times 10^{-10},
\label{eta2347}
\end{eqnarray}
\ie $\eta_B\sim 3\times 10^{-10}$ is predicted from BBN.

Finally we will comment on the baryonic density of the present universe.
By using eq.(\ref{eta-omegaB}), constraints on $\eta_B$ (\ref{eta2347})
becomes
\beq
1.0 \times 10^{-2} \leq \Omega_{B} h^{2}
\leq 1.2 \times 10^{-2}.
\eeq
Therefore, baryonic dark matter ($\Omega_B\sim 1$) conflicts with the
theoretical predictions of BBN provided $0.5 \lsim h \lsim 1$.

%
%

\newpage

%
%
\newcommand{\Journal}[4]{{\sl #1} {\bf #2} {(#3)} {#4}}
\newcommand{\APJ}{Ap. J.}
\newcommand{\CJP}{Can. J. Phys.}
\newcommand{\NC}{Nuovo Cimento}
\newcommand{\NP}{Nucl. Phys.}
\newcommand{\PL}{Phys. Lett.}
\newcommand{\PR}{Phys. Rev.}
\newcommand{\PRep}{Phys. Rep.}
\newcommand{\PRL}{Phys. Rev. Lett.}
\newcommand{\PTP}{Prog. Theor. Phys.}
\newcommand{\SJNP}{Sov. J. Nucl. Phys.}
\newcommand{\ZP}{Z. Phys.}

\end{document}